\documentclass[twocolumn]{aastex62} 
\usepackage{graphicx}
\usepackage{txfonts}
\usepackage{threeparttable}
\usepackage{pifont}
\newcommand{\cmark}{\ding{51}}%
\newcommand{\xmark}{\ding{55}}%

\received{xxx}
\revised{xxx}
\accepted{}
\submitjournal{ApJ}
\shorttitle{A hard look at local, optically-selected, obscured Seyfert galaxies}
\shortauthors{Kammoun et al.}

\def\xmm{{\it XMM-Newton}}
\def\nustar{{\it NuSTAR}}
\def\chandra{{\it Chandra}}
\def\swift{{\it Swift}}
\def\suzaku{{\it Suzaku}}

\def\cm3{\hbox{cm$^{-3}$}}

\newcommand{\Cahk}{\ifmmode \left[{\rm Ca H+K}\,\textsc{ii}\right\,\lambda3935,3968 \else Ca H+K$\,\lambda3935,3968$\fi}
\newcommand{\Mgb}{\ifmmode \left{\rm Mg}\,\textsc{i}\right\,\lambda5175 \else Mg\,{\sc i}\,$\lambda5175$\fi}
\newcommand{  \caii     }{\ifmmode {\rm Ca}\,\textsc{ii}   \else Ca\,\textsc{ii}\fi}

\begin{document} 

\title{A hard look at local, optically-selected, obscured Seyfert galaxies \footnote{Based on observations made with ESO Telescopes at the La Silla Paranal Observatory under program ID 099.A-0403.}}
\correspondingauthor{Elias Kammoun}
\email{ekammoun@umich.edu}

\author[0000-0002-0273-218X]{E. S. Kammoun}
\affiliation{Department of Astronomy, University of Michigan, 1085 South University Avenue, Ann Arbor, MI 48109-1107, USA}

\author{J. M. Miller}
\affiliation{Department of Astronomy, University of Michigan, 1085 South University Avenue, Ann Arbor, MI 48109-1107, USA}

\author{M. Koss}
\affiliation{Eureka Scientific, 2452 Delmer Street Suite 100, Oakland, CA 94602-3017, USA}

\author{K. Oh}
\affiliation{Korea Astronomy \& Space Science institute, 776, Daedeokdae-ro, Yuseong-gu, Daejeon 34055, Republic of Korea}

\author{A. Zoghbi}
\affiliation{Department of Astronomy, University of Michigan, 1085 South University Avenue, Ann Arbor, MI 48109-1107, USA}

\author{R. F. Mushotzky}
\affiliation{Department of Astronomy and Joint Space-Science Institute, University of Maryland, College Park, MD 20742, USA}

\author{D. Barret}
\affiliation{IRAP, Universit\'{e} de Toulouse, CNRS, UPS, CNES, 9, Avenue du Colonel Roche, BP 44346, 31028 Toulouse Cedex 4, France}

\author{E. Behar}
\affiliation{Department of Physics, Technion, 32000, Haifa, Israel}

\author{W. N. Brandt}
\affiliation{Department of Astronomy and Astrophysics, 525 Davey Lab, The Pennsylvania State University, University Park, PA 16802, USA}
\affiliation{Institute for Gravitation and the Cosmos, The Pennsylvania State University, University Park, PA 16802, USA}
\affiliation{Department of Physics, 104 Davey Lab, The Pennsylvania State University, University Park, PA 16802, USA}

\author{L. W. Brenneman}
\affiliation{Harvard-Smithsonian Center for Astrophysics, 60 Garden St., Cambridge, MA 02138, USA}

\author{J. S. Kaastra}
\affiliation{SRON Netherlands Institute for Space Research, Sorbonnelaan 2, 3584 CA Utrecht, the Netherlands}
\affiliation{Leiden Observatory, Leiden University, PO Box 9513, 2300 RA Leiden, the Netherlands}

\author{A. M. Lohfink}
\affiliation{Department of of Physics, Montana State University, P.O. Box 173840, Bozeman, MT 59717-3840, USA}

\author{D. Proga}
\affiliation{Department of Physics \& Astronomy, University of Nevada Las Vegas, Las Vegas, NV 89154, USA}

\author{D. Stern}
\affiliation{Jet Propulsion Laboratory, California Institute of Technology, 4800 Oak Grove Drive, MS 169-221, Pasadena, CA 91109, USA}

\begin{abstract}
We study the X-ray spectra of a sample of 19 obscured, optically-selected Seyfert galaxies (Sy 1.8, 1.9 and 2) in the local universe ($d \leq 175$~Mpc), drawn from the CfA Seyfert sample. Our analysis is driven by the high sensitivity of \nustar\ in the hard X-rays, coupled with soft X-ray spectra using \xmm, \chandra, \suzaku, and \swift/XRT. We also analyze the optical spectra of these sources in order to obtain accurate mass estimates and Eddington fractions. We employ four different models to analyze the X-ray spectra of these sources, which all result in consistent results. We find that 79-90\% of the sources are heavily obscured with line-of-sight column density $N_{\rm H} > 10^{23}~\rm cm^{-2}$. We also find a Compton-thick ($N_{\rm H} > 10^{24}~\rm cm^{-2}$) fraction of $37-53$\%. These results are consistent with previous estimates based on multi-wavelength analyses. We find that the fraction of reprocessed to intrinsic emission is positively correlated with $N_{\rm H}$ and negatively correlated with the intrinsic, unabsorbed, X-ray luminosity (in agreement with the  Iwasawa-Taniguchi effect). Our results support the hypothesis that radiation pressure regulates the distribution of the circumnuclear material.

\end{abstract}


\section{Introduction}
\label{sec:intro}
It is generally accepted that active galactic nuclei (AGN) are powered by accretion onto supermassive black holes (SMBHs) with masses $M_{\rm BH} \sim 10^{5-9}~M_\odot$ through a geometrically thin, optically thick disk \citep[e.g.,][]{Shak73}. According to the unification scheme, originally proposed by \cite{Antonucci85} \citep[see also][]{Netzer15}, all AGN are relatively similar in terms of physics. However, some key parameters such as orientation \citep[e.g.,][]{Marin16}, mass accretion rate \citep[e.g.,][]{Fanidakis11}, and feedback \citep[see][for a review]{Fabian12} may differ, resulting in the different families of AGN classes \citep[][]{Antonucci93, Padovani17}. Within this general scheme, type-2 objects are the AGN in which the absorber (broad line region and dusty torus), located at distances $\sim 0.1-10$~pc \citep[e.g.,][]{Jaffe04,RamosAlmeida17}, intercepts the line of sight (LOS).  The actual morphology and composition of this material is still uncertain. Several results suggest a clumpy distribution of optically thick clouds rather than a homogeneous structure \citep[e.g.,][]{Honig07, Risaliti07, Balokovic14, Marinucci16}. Recently, \cite{Giustini19} have explored the role of the inclination angle but more importantly the role of the luminosity in driving a disk wind that in turn can affect the AGN appearance (see their figure 5 for a comprehensive summary).

A significant fraction of AGN are obscured by the torus or sometimes by the host galaxy \citep[e.g.,][]{Buchner17,Circosta19}. The most rapid BH growth by accretion likely occurs in Compton-thick (CT; with an equivalent neutral hydrogen column density $N_{\rm H} \geq 1.5\times 10^{24}~\rm cm^{-2}$) quasars at moderate/high redshift \citep[$1 \leq z \leq 6$, e.g.,][]{Draper10,Treister10, Vito18}. Mutli-wavelength studies \citep[e.g.,][]{Risaliti99, Goulding11} suggest that a large fraction ($20-30\%$) of AGN are CT. The values of the CT fraction, quoted from X-ray background synthesis models \citep[e.g.,][]{Gilli07, Treister09, Akylas12, Brightman12, Ueda14, Buchner15}, range between $\sim 9-35\%$ in different models and as a function of redshift. More recently, the cosmic X-ray background modeling by \cite{Ananna19} predicts that $\sim 50 \pm 9 \%$ ($\sim 56 \pm 7 \%$) of AGN within $z = 0.1$ (1.0) are CT. These fractions are higher than the ones observed in hard X-ray surveys \citep[being below $\sim 20\%$; e.g.,][]{Bassani06, Burlon11, Vasudevan13, Ricci17, Masini18}. A major difficulty in identifying CT sources is the attenuation of the direct emission produced by the central engine (in the soft X-rays, utlraviolet, and optical) by the obscuring material. The hard X-rays ($ \geq 15~\rm keV$) and the mid-infrared ($5-50~\rm \mu m$) are the only  spectral bands where this material is optically thin up to high column densities. 

The X-ray emission in non-jetted AGN is widely accepted to be due to Compton up-scattering of ultraviolet (UV)/soft X-ray disk photons off hot electrons \citep[$\sim 10^9~\rm K$; e.g.,][]{Shap76, Haa93}, usually referred to as the ‘X-ray corona’. The resulting energy spectrum can be well described by a power law with a high-energy cutoff. In obscured AGN, this component is heavily attenuated in the soft X-rays (depending on the column density of the obscuring material). However, this attenuated emission is reprocessed by the obscuring material. This reprocessing is imprinted in the energy spectrum in the form of narrow fluorescent emission lines (most prominently the neutral Fe lines at $\sim 6.4~\rm keV$), and the Compton hump peaking at $\sim 20-30~\rm keV$ \citep[e.g.,][]{Ghisellini94}. In order to accurately determine the column density of the reprocessor and the properties of the intrinsic emission, a self-consistent modeling of the physical processes, accounting for the aforementioned features, is required. Several efforts have been made  to model these effects, considering various geometries \citep[e..g, spherical, toroidal, patchy;][]{MYT09, Brightman11, Liu14, Balokovic19, Buchner19}.

Thanks to its unprecedented sensitivity covering the $3-79$~keV band, \nustar\ is playing a key role in identifying the missing fraction of CT sources and determining their properties \citep[e.g.,][and references therein]{Koss16, Annuar17,Marchesi18, Marchesi19, Lamassa19}. In this work, we present a \nustar\ survey observing 21 optically-selected,  obscured Seyfert galaxies. This paper is organized in the following way: in Section~\ref{sec:sample} we present our sample. In Section~\ref{sec:optical} we present the analysis of the optical spectra of the sources. In Section~\ref{sec:datareduction}, we present the X-ray data reduction. In Section~\ref{sec:spectralfit}, we present the various models used to fit the X-ray spectra. The results of the X-ray spectral modeling are presented in Section~\ref{sec:results}. Finally, we present our conclusions in Section~\ref{sec:conclusion}.


\section{The sample}
\label{sec:sample}

The CfA Redshift Survey \citep[CfARS;][]{Huchra83} obtained the optical spectra of a complete sample of 2399 galaxies down to a limiting magnitude of $m_{Zw} \leq 14.5$ in fields limited to $\delta \geq 0\degr$ and $b_{II} \geq 40\degr$, and $\delta \geq -2.5\degr$ and $b_{II} \leq -30\degr$. Its classifications are robust owing to exhaustive high-resolution optical spectroscopy and narrow line classifications \citep{Osterbrock93}. A subset of the CfARS, the CfA Seyfert Sample \citep[CfASyS;][]{Huchra92} originally identified 25 Seyfert 1 (i.e., unobscured) and 23 Seyfert 2 galaxies. However, it was later shown that two of the Seyfert 2 sources (NGC~3227 and Mrk~993) were misclassified \citep[see][and references therein, respectively]{Salamanca94,Corral05, Trippe10}. The remaining 21 obscured sources in the CfASyS were observed through the \nustar\ {\it Obscured Seyferts} Legacy Survey (PI: J. M. Miller), as the first volume-limited ($d \leq 175~\rm Mpc$) sample observed in the hard X-rays with a high-focusing and high-sensitivity instrument. The 21 sources, their coordinates, and redshifts are listed in Table~\ref{table:vdisp}.


\section{Optical spectroscopy}
\label{sec:optical}
From our sample, 9 sources were observed with optical spectroscopy as part of the BAT AGN Spectroscopic Survey \citep{Koss17} DR2 (Koss et al., in prep, where details on the observations and data reduction can be found). In addition to this, another 9 sources were observed as part of the Sloan Digital Sky Survey (SDSS) and two sources (Mrk~573 and Mrk~334) were observed with the Palomar Double Spectrograph (DBSP) on the Hale 200-inch telescope. The latter observations took place on UT 2018 January 14, with both galaxies observed for 300~s at the parallactic angle.  The Palomar observations were taken with the D55 dichroic and the 600/4000 and 316/7500 gratings using a 1.5$\arcsec$ slit, covering the wavelength range of $3200-10200$~\AA.  We measured the sky lines to have a ${\rm FWHM=4.0}$~\AA\ at 5000~\AA\ and ${\rm FWHM=6.0}$~\AA\ at 8500~\AA.

\begin{figure}
\includegraphics[width = 0.95\linewidth]{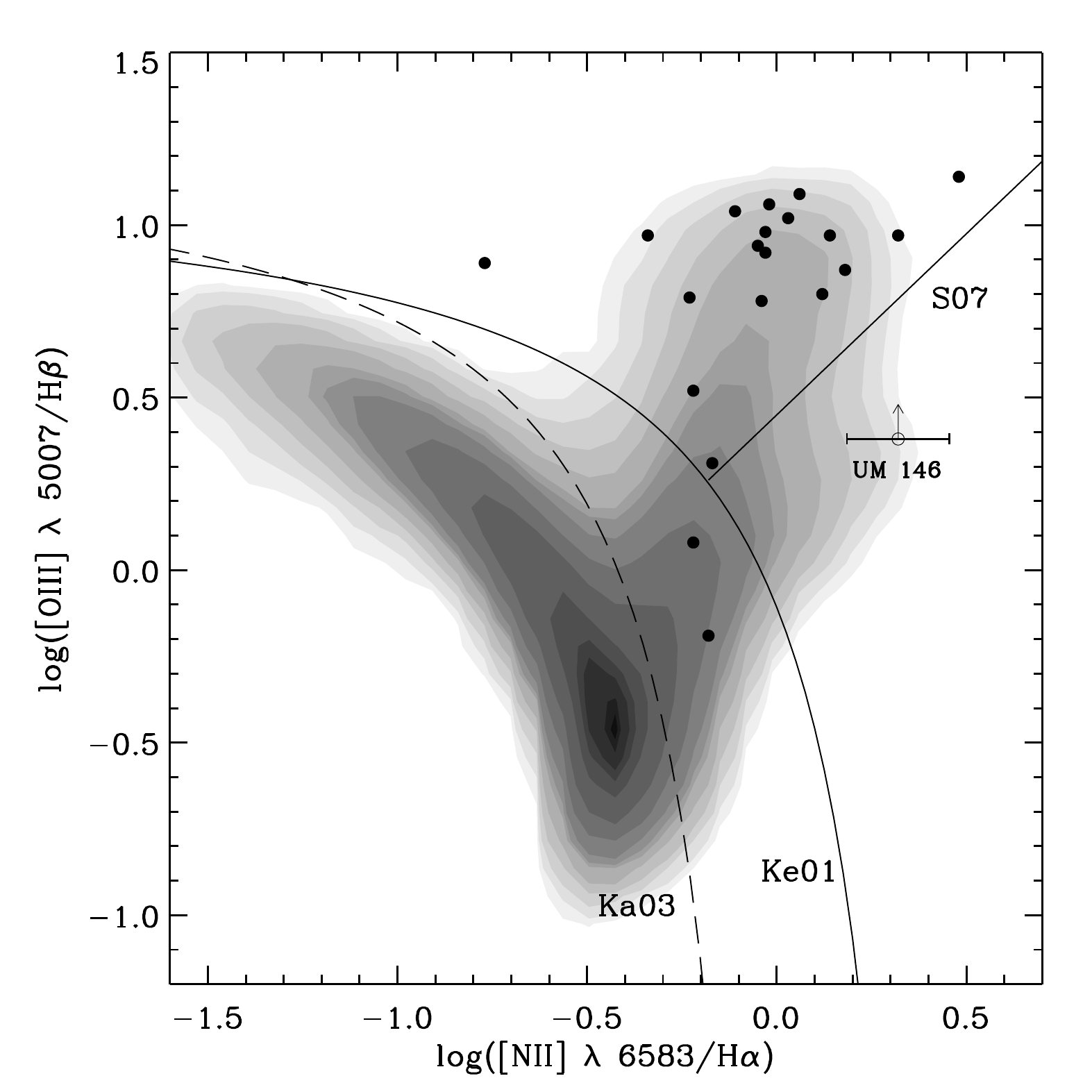}
\caption{Optical excitation diagnostic (BPT) diagram to separate AGN from star-forming galaxies. The solid black curve is the theoretical boundary for the region occupied by starburst derived by the maximum starburst model from \cite{Kewley01}. The dashed line is the empirical SF line from \cite{Kauffmann03}. The solid straight line is the empirical Seyfert-LINER separation from \citep{Schawinski07}.  The Black dots mark our targets. The grey filled contours indicate the distribution of the SDSS narrow emission-line galaxies \citep[$z<0.2$, $N \sim 180000$;][]{Oh11}. Error bars are smaller than the symbol size.} \label{fig:BPT}
\end{figure}

The continuum and the absorption features were fit using the penalized PiXel Fitting software \citep[pPXF;][]{Cappellari04} to measure a central velocity dispersion for the galaxies. A stellar template library from VLT/Xshooter \citep{Chen14} was used to fit the spectra with optimal stellar templates following the general procedure in \cite{Koss17}. These templates have been observed at higher spectral resolution ($R = 10,000$) than the AGN observations and are convolved in pPXF to the spectral resolution of each observation before fitting. When fitting the stellar templates, all prominent emission lines were masked.
  
\begin{deluxetable*}{lllllllll}
\tablecaption{The results obtained from the study of the velocity dispersion. (1): Source name, (2): Right ascension, (3): Declination, (4): redshift, (5): Stellar dispersion, (6): Instrumental dispersion, (7): Mass estimate, (8):Velocity dispersion fit range, (9): AGN classification. The superscripts  in column 1 indicate the instruments used to estimate the velocity dispersions; Palomar ($P$), SDSS ($S$), Xshooter ($X$).}\label{table:vdisp}

\tablehead{\colhead{Source} &  \colhead{RA} & \colhead{Dec} & \colhead{$ z$} & \colhead{$\sigma_{\ast}~(\rm km~s^{-1})$} & \colhead{$\sigma_{\rm inst}~(\rm km~s^{-1})$} & \colhead{$\log M_{\rm BH}/M_\odot$} &   \colhead{Range (\AA)} & \colhead{Type} \\
\colhead{(1)} &  \colhead{(2)} & \colhead{(3)} & \colhead{(4)} & \colhead{(5)} & \colhead{(6)} & \colhead{(7)} &   \colhead{(8)} & \colhead{(9)} }
\startdata
Mrk 573$^P$	&	25.9907	&	2.3499	&	0.0174	&$	128.36	\pm	10.06	$&	116.64	&	7.65	&	3880-5550	&	Sy2	\\
NGC 1144$^X$	&	43.8011	&	-0.1838	&	0.0288	&$	230.01	\pm	6.06	$&	64.17	&	8.76	&	3800-5550	&	Sy2	\\
NGC 3362$^S$	&	161.2155	&	6.5969	&	0.0278	&$	124.25	\pm	4.39	$&	65.22	&	7.58	&	3880-5550	&	Sy2	\\
UGC 6100$^S$	&	165.3917	&	45.6538	&	0.0294	&$	143.81	\pm	4.35	$&	69.78	&	7.86	&	3880-5550	&	Sy2	\\
NGC 3982$^S$	&	179.1174	&	55.1254	&	0.0038	&$	86.19	\pm	2.29	$&	63.75	&	6.89	&	3880-5550	&	Sy2	\\
NGC 4388$^{S}$	&	186.4451	&	12.6621	&	0.0084	&$	113.83	\pm	8.04	$&	64.53	&	6.92$^a$	&	3880-5550	&	Sy2	\\
UGC 8621$^S$	&	204.4166	&	39.1546	&	0.0201	&$	86.00	\pm	5.38	$&	65.67	&	6.88	&	3880-5550	&	Sy1.8	\\
NGC 5252$^{S}$	&	204.5661	&	4.5425	&	0.0231	&$	227.92	\pm	4.84	$&	65.49	&	9.00$^b$	&	3880-5550	&	Sy2	\\
NGC 5347$^S$	&	208.3244	&	33.4908	&	0.0080	&$	92.42	\pm	2.73	$&	65.82	&	7.02	&	3880-5550	&	Sy2	\\
NGC 5695$^X$	&	219.3423	&	36.5678	&	0.0142	&$	145.71	\pm	2.97	$&	72.03	&	7.89	&	3880-5550	&	Sy2	\\
NGC 5929$^S$	&	231.5257	&	41.6707	&	0.0083	&$	128.03	\pm	2.50	$&	62.07	&	7.64	&	3880-5550	&	Sy2	\\
NGC 7674$^X$	&	351.9863	&	8.7790	&	0.0288	&$	133.87	\pm	5.96	$&	28.35	&	7.73	&	3800-5550	&	Sy2	\\
NGC 7682$^X$	&	352.2662	&	3.5333	&	0.0170	&$	114.00	\pm	6.11	$&	65.15	&	7.42	&	3800-5550	&	Sy2	\\
NGC 4395	&	186.4538	&	33.5468	&	0.0011	&$				$&		&	5.56$^c$	&		&	Sy1.8	\\
Mrk 334$^P$	&	0.7901	&	21.9603	&	0.0219	&$				$&		&		&		&	Sy1.8	\\
NGC 5283$^P$	&	205.2739	&	67.6722	&	0.0106	&$	117.44	\pm	3.14	$&	66.28	&	7.48	&	3880-5550	&	Sy2	\\
UM 146 	&	28.8417	&	6.6117	&	0.0174	&$				$&		&	$< 6.7^d$&		&		\\
NGC 5256$^S$	&	204.5742	&	48.2781	&	0.0282	&$	206.14	\pm	11.03	$&	64.89	&	8.55	&	3880-5550	&	Composite	\\
NGC 5674$^P$	&	218.4678	&	5.4583	&	0.0248	&$	116.83	\pm	13.15	$&	82.59	&	7.47	&	Ca Triplet	&	Sy2	\\
NGC 1068$^{P}$	&	40.6699	&	-0.0133	&	0.0037	&$	154.81	\pm	6.26	$&	85.94	&	6.93$^d$	&	Ca Triplet	&	Sy1.9	\\
Mrk 461$^S$	&	206.8241	&	34.1489	&	0.0163	&$	124.78	\pm	4.91	$&	68.78	&	7.59	&	3880-5550	&	Composite	\\
\enddata
\tablecomments{The values of $\sigma_{\ast}$ represent the stellar velocity dispersion after subtracting the instrumental resolution (in quadrature). Black hole mass from the literature: $^a$\cite{Kuo11} (megamaser), $^b$\cite{vandenbosch15} (stellar velocity dispersion), $^c$ \cite{Peterson05} (reverberation mapping), $^d$ \cite{Garcia-Rissmann05} (stellar velocity dispersion). The other masses are from the BASS DR2 (Koss et al., in prep.; see text for more details).}
\end{deluxetable*}

We use the \cite{Kormendy13} relation to estimate the black hole mass from the velocity dispersion (see Table~\ref{table:vdisp}). On the blue end of the Palomar spectrograph we fit the  \Cahk\  absorption features and the \Mgb\ features \citep[e.g.,][]{Greene06} and on the red end we fit the  \caii\ triplet spectral region (8450--8700\,\AA). We note that reverberation mapping \citep[e.g.,][]{Peterson05} and OH megamasers \citep[e.g.,][]{Staveley-Smith92} offer more precise BH mass measurements. We adopted those estimates whenever available (megamaser: NGC~4388, reverberation mapping: NGC~4395). For UM~146, we use a velocity dispersion measurement of $66\pm 6~\rm km~s^{-1}$ from \cite{Garcia-Rissmann05}. However, we found this measurement was very near the instrumental limit ($57\rm ~km~s^{-1}$) where the changes in spectral resolution related to seeing and other observing conditions may significantly affect the measurement.  We therefore assume the velocity dispersion is less than $80\rm~ km~s^{-1}$, corresponding to a black hole mass of less than $\log M_{\rm BH}/M_\odot = 6.7$. We could not obtain velocity dispersion measures for NGC~4395 and Mrk~334. In fact, using the available instrumental resolution, it is hard to measure the stellar velocity dispersion in the dwarf galaxy NGC~4395 \citep{Woo19}. As for Mrk~334, the low AGN obscuration leads to contamination in the form of broad lines and continuum, which makes measuring a velocity dispersion difficult. 

For emission line measurements, we performed spectral line fitting for 21 optical spectra using the Gas AND Absorption Line Fitting ({\tt gandalf}) IDL code \citep{Sarzi06} which has been extensively applied in measuring spectral line strengths of SDSS galaxies \citep{Oh11} and AGN  \citep{Oh15}. Stellar population synthesis model libraries \citep{Bruzual03} and empirical stellar libraries \citep{Sanchez-blazquez06} are used to fit the continuum at rest-frame $\sim 3700$\AA $-7200$\AA. Figure~\ref{fig:BPT} shows the emission-line diagnostic diagram for our sample, plotting the line ratios [\ion{O}{3}] $\rm \lambda5007$/H$\beta$ versus [\ion{N}{2}] $\rm \lambda6583$/H$\alpha$. According to this diagnostic, all sources fall in the Seyfert regime except NGC~5256 and Mrk~461, which are consistent with being in the composite region of the diagram. UM~146 is in the LINER regime, though with a lower-limit on the [\ion{O}{3}] $\rm \lambda5007$/H$\beta$ ratio, which could be in the Seyfert regime. Throughout the rest of this analysis, we limit our study to Seyfert galaxies only, excluding the composite sources (NGC~5256 and Mrk~461). It is also worth noting that most of the sources are found to have broad lines. However, some sources reveal the presence of weak broad H$\alpha$ lines, which may be a signature of outflows.


\section{X-ray data reduction}
\label{sec:datareduction}

In this analysis, the high sensitivity of \nustar\ above 10~keV plays the major role, as it allows us to determine high column densities approaching the CT regime. However, adding soft X-ray spectra is crucial to accurately estimate the spectral slope over the full X-ray band. For that reason we also considered the soft X-ray spectra obtained by \xmm, owing to its high sensitivity and excellent calibration. In cases when \xmm\ observations were not available, we considered either \suzaku\ or \chandra\ observations, if possible. We also note that \swift/XRT snapshots were performed simultaneously with each \nustar\ observation. However, those observations resulted in low signal-to-noise ratio (S/N) spectra, due to the faintness of the sources in our sample. We considered the XRT spectra only in the cases when other soft X-ray spectra are not available, and the quality of the XRT spectra allows us to perform spectral fits. For NGC~1068, we limit our analysis to only one \nustar\ observation. Adding the soft X-ray spectra of this source and accounting for spectral variability has been discussed in further detail in previous studies \citep[e.g.,][]{Bauer15, Zaino20} and is beyond the scope of our analysis. In this section, we present the data reduction of the various instruments used in this analysis. The log of the observations is presented in Table~\ref{table:log}.

\subsection{{\it NuSTAR} observations}
The {\it NuSTAR} \citep{Harrison13} data were reduced following the standard pipeline in the {\it NuSTAR} Data Analysis Software ({\tt NUSTARDAS}\,v1.8.0), and using CALDB v20180814. We cleaned the unfiltered event files with the standard depth correction. We reprocessed the data using the ${\tt saamode = optimized}$ and ${\tt tentacle = yes}$ criteria for a more conservative treatment of the high background levels in the proximity of the South Atlantic Anomaly. We extracted the source and background spectra from circular regions of radii $45-50$\arcsec\ and 100\arcsec, respectively, for both focal plane modules (FPMA and FPMB) using the {\tt HEASOFT} task {\tt Nuproducts}, and requiring a minimum S/N of 4 per energy bin. The spectra extracted from both modules are consistent with each other. The data from FPMA and FPMB are analyzed jointly in this work, but they are not combined together. Recently, \cite{Madsen20} reported on the presence of an occasional thermal blanket tear in the FPMA module leading to an excess at low energy with respect to FPMB. We did not find this effect in any of the sources.


\subsection{\xmm\ observations}

We reduced the \xmm\ data using {\tt SAS v.17.0.0} \citep{Gabriel04} and the latest calibration files. We followed the standard procedure for reducing the data of the EPIC-pn \citep{Stru01} CCD camera\footnote{The inclusion of the EPIC-MOS data would have increased the signal to noise in the soft X-rays (below $\sim 2$~keV. However, the spectra in this range are dominated by the extended diffuse emission (see Section~\ref{sec:spectralfit} for more details). This will not lead to any improvement in measuring the LOS column density or the intrinsic emission. For this reason, and to avoid any uncertainties due to instrument cross-calibration, we decided not to use the MOS data.} The data were processed using EPPROC. Source spectra and light curves were extracted from a circular region with a radius of $25-30$\arcsec. The corresponding background spectra and light curves were extracted from an off-source circular region located on the same chip, with a radius approximately twice that of the source. We filtered out periods with strong background flares. The spectra were then binned requiring a minimum S/N of 4 per energy bin.

\subsection{\textit{ Suzaku} observations}

For NGC~5347 and NGC~5929 we used the XIS \citep{XIS} spectra from {\it Suzaku} \citep{Suzaku07}. The data were reduced following standard procedures using {\tt HEASOFT}. The initial reduction was done with {\tt aepipeline}, using the CALDB calibration release v20160616. Source spectra were extracted using {\tt xselect} from circular regions 3$\arcmin$ and 2.5$\arcmin$ in radius centered on the sources, for NGC~5347 and NGC~5929, respectively. Background spectra were extracted from a source-free region of the same size, away from the calibration source. The response files were generated using {\tt xisresp}. We do not consider the spectrum from XIS1, owing to its poor relative calibration. Spectra from XIS0 and XIS3 were checked for consistency and then combined to form the front-illuminated spectra. The spectra were then binned requiring minimum S/N of 3 per energy bin.

\subsection{\textit{ Chandra} observations}

For NGC~5347 and NGC~5283 we used the archival \chandra\ \citep{CXO00} spectra.  The data were reduced using CIAO version 4.9 and the latest associated calibration files. Source and background spectral files and response files were all created using the CIAO tool {\tt specextract}. We extracted source counts from a circular region, centered on the known source coordinates, with a radius of 5.2\arcsec\ and  3\arcsec\ for NGC~5347 and NGC~5283, respectively. The background spectra were extracted from circular source free regions, with radii equal to the ones of the source region. The resultant data were grouped to require at minimum S/N of 3 per energy bin.


\subsection{\textit{Swift} observations}

For Mrk~334 and NGC 5674 we used the X-ray telescope \citep[XRT;][]{XRT05} spectra from \textit{Neil Gehrels Swift Observatory} \citep[hereafter \swift;][]{Swift04}. We combined all the XRT observations for each source in order to increase the number of counts. The data were reduced following standard procedures using {\tt HEASOFT}. The initial reduction was done with {\tt xrtpipeline}. Source spectra were extracted using {\tt xselect} from circular regions 20$\arcsec$ in radius centered on the source. Background spectra were extracted from an off-source sky region of the same size. We used the default redistribution matrix file (RMF) and ancillary response file (ARF), available in the calibration database. The spectra were then binned requiring minimum S/N of 3 per energy bin.


\section{X-ray spectroscopy}
\label{sec:spectralfit}

Throughout this work, spectral fitting was performed using XSPEC\,v12.10.1o \citep{Arnaud96}. We apply the $\chi^2$ statistic for spectra with more than 20 counts per bin, and the Cash statistic \citep[$C$-stat;][]{Cash79} otherwise. The assumed statistics for each source are presented in the last column of Table~\ref{table:log}. The best-fit parameter values were determined through the minimization of the $\chi^2$ ($C$-stat)\footnote{It has been argued in the literature that $C$-stat, contrary to $\chi^2$-statistic, yields unbiased results \citep[e.g.,][]{Mighell99,Kaastra17}. However, at the limit of 20 counts/bin, the use of $\chi^2$ may produce a bias of the order of 5\% in the best estimate of the flux for those bins. This will have a minor effect on the results of our analysis, which are dominated by statistical errors.}. These values were used for initial values for the prior distributions in Markov chain Monte Carlo (MCMC\footnote{We use the {\tt XSPEC\_EMCEE} implementation of the {\tt PYTHON EMCEE} \citep{EMCEE13} package for X-ray spectral fitting in XSPEC, provided by A. Zoghbi (\url{https://github.com/zoghbi-a/xspec\_emcee}).}). We used the Goodman-Weare algorithm \citep{Goodman10} discarding $300,000$ elements as part of the ‘burn-in’ period. The final chains contain $500,000$ elements. The values presented in this paper represent the mean values across the chain samples. Unless stated otherwise, uncertainties on the parameters are listed at the 1$\sigma$ confidence level, as derived from the chain samples. 

In the following, we present the different models that we used in order to describe the reprocessed emission in the X-ray spectra. We note that the various model set-ups used in this paper have been already presented in \cite{Kammoun19n5347}, where we discuss in detail the case of NGC~5347.

\subsection{Pexmon}
\label{sec:Pexmon}

We initially fit the spectra using the neutral reflection model Pexmon \citep{Nandra07}. This model is based on the Pexrav model \citep{Pexrav95}, but adds the Fe K$\alpha$ fluorescence line based on the Monte Carlo simulations by \cite{Geo91}. In addition,  Fe K$\beta$ and Ni K$\alpha$ fluorescence lines lines are added in Pexmon, in a self-consistent manner, with their fluxes fixed at a fraction of the Fe K$\alpha$ flux. This model accounts also for the Compton shoulder of the Fe K$\alpha$, following the prescription of \cite{Matt02}. The model that we used to fit the spectra can be written (in XSPEC parlance) as follows:

\begin{eqnarray}
\begin{array}{lc}
{\tt model_{Pexmon} } & =  {\tt phabs[1] * ( zphabs[2]*zcutoffpl[3]   }
\end{array}\nonumber\\
\begin{array}{lll}
  {\tt + constant[4]*zcutoffpl[5]  + pexmon[6] } & 
\end{array}\nonumber\\
\begin{array}{lc}
{\tt +Apec_1[7] + Apec_2[8]}  ).&\nonumber 
\end{array}
\label{eq:pexmon_model}
\end{eqnarray}

\noindent In this model, the {\tt phabs[1]} component represents the Galactic absorption, {\tt zcutoffpl[3]} represents the primary emission of the source assumed to be a power-law with a high-energy cutoff (fixed to 500~keV), which is intrinsically absorbed by {\tt zphabs[2]}. A fraction (${\tt constant[4] *  zcutoffpl[5]}$, where $0 \leq {\tt constant[4]} \leq 1$) of the primary emission could be scattered into our LOS by optically thin ionized gas in the polar regions. This fraction could also partly account for unabsorbed emission in the case of a partially covered source. We do not instead assume a partially covering absorber for consistency with other models that are used in this work (see later for more details). In the rest of the work we refer to {\tt constant[4]} as $C_{\rm sc}$. All the parameters of {\tt zcutoffpl[6]} are tied to the ones of {\tt zcutoffpl[3]}. The photon index, cutoff energy and normalization of {\tt pexmon[7]}, which describes the reprocessed emission, are tied to the same parameters of {\tt zcutoffpl[3]}. We leave the reflection fraction $\mathcal{R}$ of {\tt pexmon[7]} free to vary, taking negative values only to account for the reflected emission without the contribution of the power-law component. This component is used to model the reprocessed emission from neutral material without assuming any specific geometry or location. Finally, when needed, we describe the soft emission with ${\tt Apec\_1[8]}$. This component, which mainly arises from extended regions, is most likely due to diffuse thermal emission and/or photoionized emission. In some cases, modeling the soft X-rays require the addition of a second component (${\tt Apec_2[9]}$) with a higher temperature. In the case of multi-epoch observations, we left the normalization of the power-law component free to vary, to account for potential flux variability. If the best-fit values of normalization are consistent between the different epochs, indicating no significant variability, we then tie them and repeat the fit.

NGC 3362 and UGC 8621 were not detected by \nustar. In addition, their \xmm\ spectra are background-dominated above $\sim 5$~keV. For that reason, we removed the {\tt pexmon[7]} component for these two sources. However, to avoid further complexity in presenting our analysis, we consider the models for these sources (i.e., by removing the Pexmon component) in the ``Pexmon" set (as opposed to the MYTorus fits, see next section). The results assuming this model are presented in Table~\ref{table:pexmon}. 

\subsection{MYTorus}
\label{sec:MYT}

In the previous section, we presented a phenomenological model to fit the spectra. However, this model does not assume any geometry. Moreover, it does not account for the scattering within the obscuring material. Thus, we next attempt to model the obscuration and the reprocessed emission using the MYTorus spectral-fitting suite for modeling X-ray spectra from a toroidal reprocessor \citep{MYT09}. This model fixes only the geometry of the reprocessing material without necessarily implying a pc-scale location. We first consider the ``coupled'' configuration of MYTorus (hereafter MYTC). This configuration assumes that intrinsic emission is self-consistently absorbed and reprocessed by toroidal material with a circular cross section and half-opening angle of 60\degr\ and solar abundance. The viewing angle ($\theta$) and the equatorial (global) column density ($N_{\rm H,~eq}$) of the torus are free parameters. We refer the reader to \cite{Yaqoob12} for more details and an extensive discussion about the various configurations of MYTorus and their implications. The model can be written as follows:

\begin{eqnarray}
\begin{array}{lll}
{\tt model_{MYTC} }  &=  {\tt phabs[1] *(  MYTZ[2] * zpowerlaw[3]  } &
\end{array}\nonumber\\
\begin{array}{lll}
  {\tt + constant[4]*zpowerlaw[5]  } & 
\end{array}\nonumber\\
\begin{array}{lll}
{\tt+ constant[6]*(MYTS[7] + MYTL[8]) }
\end{array}\nonumber \\
\begin{array}{lll}
{\tt + Apec_1[9] + Apec_2[10]}). &\nonumber
\end{array}
\label{eq:MYTC}
\end{eqnarray}
\noindent
The ${\tt phabs[1]}$, ${\tt constant[4]*zpowerlaw[5]}$, ${\tt Apec_1[10]}$, and ${\tt Apec_2[11]}$ components are equivalent to the ones in the Pexmon fit. {\tt MYTZ[2]} represents the attenuation of the intrinsic emission. {\tt MYTS[7]} and {\tt MYTL[8]} represent the scattered continuum and the fluorescent emission lines emitted by the torus. The {\tt constant[6]} factor (hereafter, referred to as $A$) corresponds the relative weights of the three MYTorus components and are fixed to unity \citep[as suggested by][]{Yaqoob12}, unless stated otherwise. We remind the reader that $N_{\rm H, LOS}$ can be estimated using using eq. (1) in \cite{MYT09}:
\begin{equation}\label{eq:NHangle}
N_{\rm H, LOS} = N_{\rm H,eq} \left[ 1 - \left( \frac{c}{a} \right)^2 \cos \theta \right]^{1/2},
\end{equation}
\noindent where $c$ is the distance from the center of the torus to the origin of coordinates, and $a$ is the radius of the circular cross section. For a half-opening angle of 60\degr, the ratio $c/a$ is equal to 2. MYTorus does not have a high-energy cutoff. Imposing a cutoff energy to the primary emission would break the self-consistency of the models. Instead, MYTorus assumes various termination energies ($E_{\rm T}$). We used in our analysis the tables with $E_{\rm T} =500$~keV. Using different values did not affect the fits. The results obtained by fitting MYTC are presented in Table~\ref{table:mytc}.

\begin{figure*}
\centering
\includegraphics[width = 0.8\textwidth]{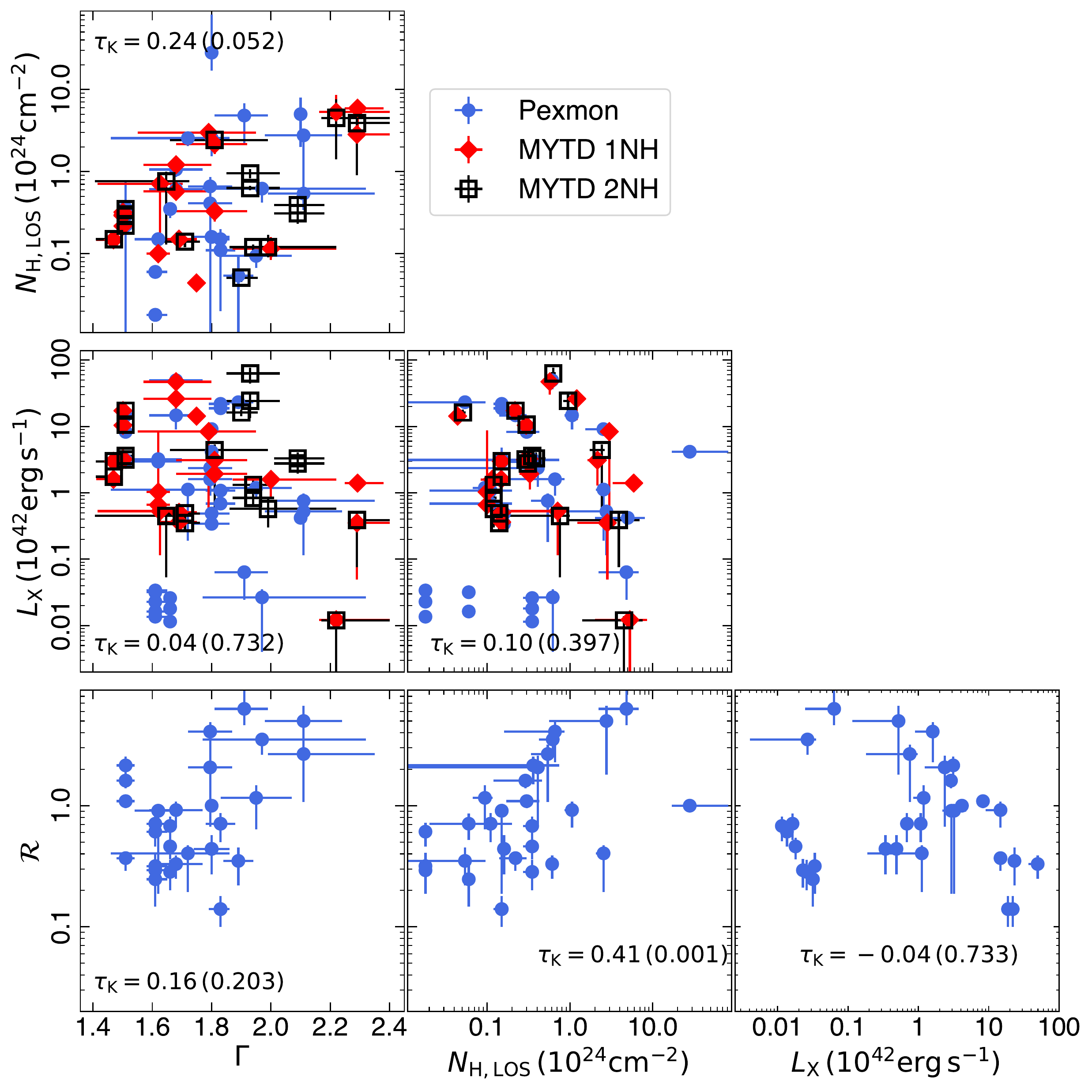}
\caption{Triangle plot showing the best-fit parameters  plotted versus the others, for Pexmon (blue circles), MYTD-1NH (red diamonds), and MYTD-2NH (open black squares). We also show the Kendall correlation coefficient and the corresponding $p$-value between parentheses. We note that objects with multi-epoch observations are reported multiple times.}
\label{fig:triangle}
\end{figure*}

Next, we considered the decoupled configuration of MYTorus (MYTD) which is intended to mimic the Pexmon configuration. In this configuration, the viewing angle of {\tt MYTZ} is fixed to 90\degr, so its $N_{\rm H}$ corresponds to the LOS value. {\tt MYTS} and {\tt MYTL} are decomposed into two components, one from the near side of the torus ($\theta = 90\degr$) and the one from the far side of the torus ($\theta= 0 \degr$). The column densities of these components could be either tied to the one of {\tt MYTZ}, corresponding to a uniform distribution of the material (hereafter MYTD-1NH), or free to vary, corresponding to a patchy structure (hereafter MYTD-2NH). The latter configuration allows us to obtain estimates on both $N_{\rm H,LOS}$ and $N_{\rm H,eq}$. MYTD can be written as follows:

\begin{eqnarray}
\begin{array}{lll}
{\tt model_{MYTD} }  &=  {\tt phabs[1] *(  MYTZ_{90}[2] * zpowerlaw[3]  } &
\end{array}\nonumber\\
\begin{array}{lll}
  {\tt + constant[4]*zpowerlaw[5]  } & 
\end{array}\nonumber\\
\begin{array}{lll}
{\tt + constant[6]*(MYTS_0[7] + MYTL_0[8]) }
\end{array}\nonumber \\
\begin{array}{lll}
{\tt+ constant[9]*(MYTS_{90}[10] + MYTL_{90}[11]) }
\end{array}\nonumber \\
\begin{array}{lll}
{\tt + Apec_1[12] + Apec_2[13]} ). &\nonumber
\end{array}
\label{eq:MYTD}
\end{eqnarray}
\noindent
The relative weights for the {\tt MYTS} and {\tt MYTL} components with $\theta = 0\degr, 90\degr$ ({\tt constant[6,9]}) are tied together, unless otherwise stated. Hereafter, we refer to those constants as $A_0$ and $A_{90}$, respectively. The results obtained by fitting MYTD-1NH and MYTD-2NH are presented in Tables~\ref{table:mytd1nh} and \ref{table:mytd2nh}, respectively.

We did not apply the MYTorus models for the following sources: NGC~5252, NGC~4395, NGC~3362, UGC~6100, and UGC~8621. For NGC~5252 and NGC~4395, the spectra required complex, ionized absorption, in addition to a neutral absorber. Such a configuration is not trivial to implement in the framework of the MYTorus models. A detailed analysis of NGC~4395 is presented in \cite{Kammoun19n4395} \citep[see also][]{Nardini11}. As for NGC~3362, UGC~6100, and UGC~8621, the low quality of the data (for UGC~6100) makes the application of the MYTorus fits non-trivial. As for NGC~3362 and UGC~8621, both sources are not detected by \nustar\ and their \xmm\ spectra are background dominated above $\sim 5$~keV. For that reason we were not able to apply the MYTorus set of models to these sources. It is worth noting that we limited $\Gamma$ to the range 1.4-2.4 because the models we used (Pexmon and MYTorus) are defined in this common range.


\section{Results}
\label{sec:results}

\begin{figure}
\centering
\includegraphics[width = 0.9\linewidth]{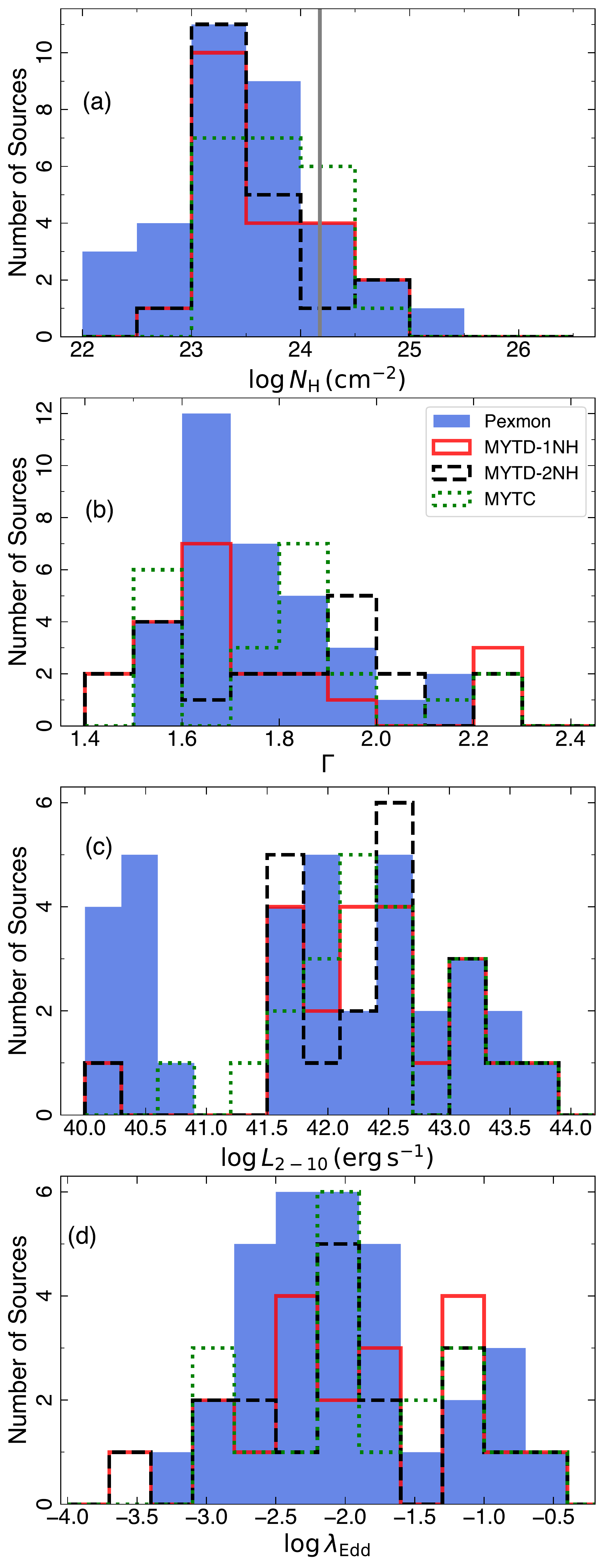}
\caption{The distributions of $N_{\rm H, LOS}$, $\Gamma$, $L_{2-10}$, and $\log \lambda_{\rm Edd}$ (panels a-d) for Pexmon, MYTC, MYTD-1NH, and MYTD-2NH (blue filled, green dotted, red solid, and black dashed histograms, respectively). We note that objects with multi-epoch observations are reported multiple times.}
\label{fig:histogramall}
\end{figure}

All models throughout this paper resulted in statistically acceptable fits. In addition, the results for each source, applying the various models considered in our analysis, agreed in detail (see Section~\ref{sec:consistent} for more details). Thus, in Figures~\ref{fig:spectra1}-\ref{fig:spectra2}, we show only the spectra fitted with Pexmon, and the corresponding residuals. The $\Gamma-N_{\rm H}$ confidence contours for each model are shown in Figures~\ref{fig:mcmc1}-\ref{fig:mcmc4}. For the cases where $\Gamma$ was kept tied, we show only the 1-D posterior distribution $N_{\rm H}$ obtained from the MCMC. We show the results from the MCMC in Tables~\ref{table:pexmon}-\ref{table:mytd2nh}. We note that the best-fit results for the {\tt Apec} components are consistent between all four models. Thus, we show the ones obtained by the Pexmon model only. It is worth mentioning that for all sources the iron abundance in the Pexmon model was fixed to the solar value, except for NGC~1068. For this source, we obtained a best-fit value $A_{\rm Fe} = 1.8 \pm 0.1$ solar. Furthermore, the values of $C_{\rm sc}$ were found to be smaller than 0.1 for all sources, except for NGC~7674. For this source, we found $C_{\rm sc} = 0.1_{-0.04}^{+0.02}$ and $0.6\pm 0.2$ for the \xmm\ and \nustar\ spectra, respectively, fitted with Pexmon.

Throughout the paper, we refer to the intrinsic, unabsorbed, luminosity of the power-law  component in the 2-10~keV range as $L_{2-10}$. Various methods can be used to estimate the bolometric luminosity. One way to do so is by considering luminosity-dependent bolometric corrections \citep[e.g.,][]{Marconi04,Lusso12}. However, in Table~\ref{table:Ledd}, we present the Eddington ratios ($\lambda_{\rm Edd} = L_{\rm bol}/L_{\rm Edd}$) by using the mass estimates from Section~\ref{sec:optical} and adopting a bolometric correction, $\kappa_{\rm bol} = L_{\rm bol}/L_{2-10} = 20$, from \cite{Vasudevan09}. \cite{Koss17} noted that $L_{\rm bol}$ inferred from the X-ray luminosity and the values estimated from the 5100~\AA\ luminosity show a scatter of $\sim 0.45$~dex for type-1 AGN (see their figure 26). Assuming different bolometric correction would result in different bolometric luminosities and Eddington fractions. This would cause some later plots to shift (e.g., Figures~\ref{fig:NH_Edd} and \ref{fig:G_Edd}), but the overall trends identified would remain. The estimated $\lambda_{\rm Edd}$ values are model-dependent, as they rely on the measure of $L_{2-10}$ that may vary between different models. The consistency between the models is further discussed in Section~\ref{sec:consistent}. In addition, $\lambda_{\rm Edd}$ span a large range  $\log \lambda_{\rm Edd} ~ [-4 , -0.5]$ that is different from the one observed in quasars.  Thus any direct comparison between our sample and highly accreting quasars should be addressed carefully as they probe different accretion regimes.


In Figure~\ref{fig:triangle} we show a corner plot for all the relevant parameters for all the observations. The results are shown for Pexmon, MYTD-1NH, and MYTD-2NH, as these models measure the LOS column density.  We plot each measurement for sources showing variability. We show in the same plot the Kendall's coefficient with the null hypothesis probability (i.e., no correlation), between parentheses, for the Pexmon estimates. The results are consistent with the ones obtained for the other two models. Our results suggest moderate positive correlations between the $N_{\rm H,LOS}$ and $\Gamma$, and between $N_{\rm H, LOS}$ and $\mathcal{R}$. We find no correlation between $\Gamma$ and $\mathcal{R}$. This is in agreement with the recent results of \cite{Panagiotou20} who also found no correlation between these two quantities in obscured AGN, while they found a clear correlation in unobscured sources \citep[see also][]{Zdziarski99, Zdziarski03}. In Figure~\ref{fig:histogramall}, we show the histograms for the measured $N_{\rm H, LOS}$, $\Gamma$, $L_{2-10}$, and $\lambda_{\rm Edd}$, for all models. MYTC provides only $N_{\rm H,eq}$. Thus, we used eq.~\ref{eq:NHangle} to derive the $N_{\rm H,LOS}$ assuming the best-fit inclination angle. The results in Figures~\ref{fig:triangle}-\ref{fig:histogramall} are shown for each measurement. Hence, more than one data point will be associated with sources showing variability. Moreover, some sources were not fitted with MYTorus, which result in having less data points for these models compared to Pexmon.


\begin{figure}
\centering
\includegraphics[width = 0.99\linewidth]{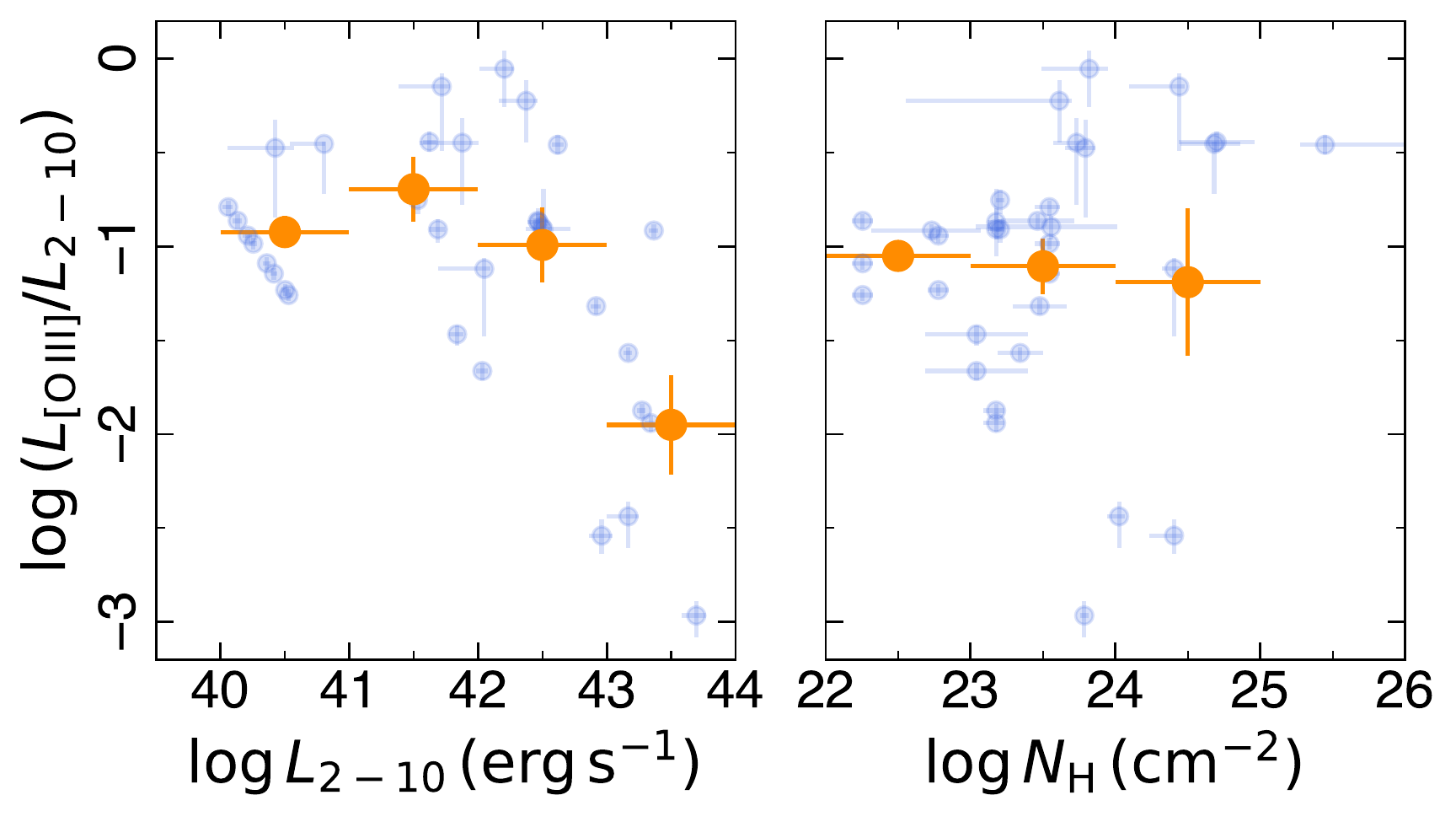}
\caption{Ratio of the extinction corrected [\ion{O}{3}] luminosity over the intrinsic 2-10~keV luminosity versus $L_{2-10}$ and the LOS column density obtained using the Pexmon model (left and right panels, respectively). The results obtained for each source are shown in shaded blue. The orange points correspond to the binned results.}
\label{fig:LO_LX}
\end{figure}

Figure~\ref{fig:LO_LX} shows the scatter plot for the ratio of the extinction-corrected [\ion{O}{3}] luminosity over the intrinsic 2-10~keV luminosity plotted versus $L_{2-10}$ and $N_{\rm H, LOS}$ obtained using the Pexmon model (left and right panels, respectively). The binned data points suggest constant ratios, below $L_{2-10} < 10^{43}~\rm erg~s^{-1}$ with an average value of $\langle \log L_{\rm [O~III]}/L_{2-10} \rangle = -1.09$. A decrease in $L_{\rm [O~III]}/L_{2-10}$ can be seen at high X-ray luminosity, $L_{2-10} > 10^{43}~\rm erg~s^{-1}$. The results obtained by \cite{Berney15} also show a constant luminosity ratio, albeit with a lower average of $\langle \log L_{\rm [O~III]}/L_{2-10} \rangle = -2.01$. Howerver,  \cite{Berney15} studied hard X-ray selected objects from the \swift/BAT AGN survey, with $\log L_{2-10} \sim [42,46]$.

\subsection{Consistency of the models}
\label{sec:consistent}

\begin{figure}
\centering
\includegraphics[width = 0.9\linewidth]{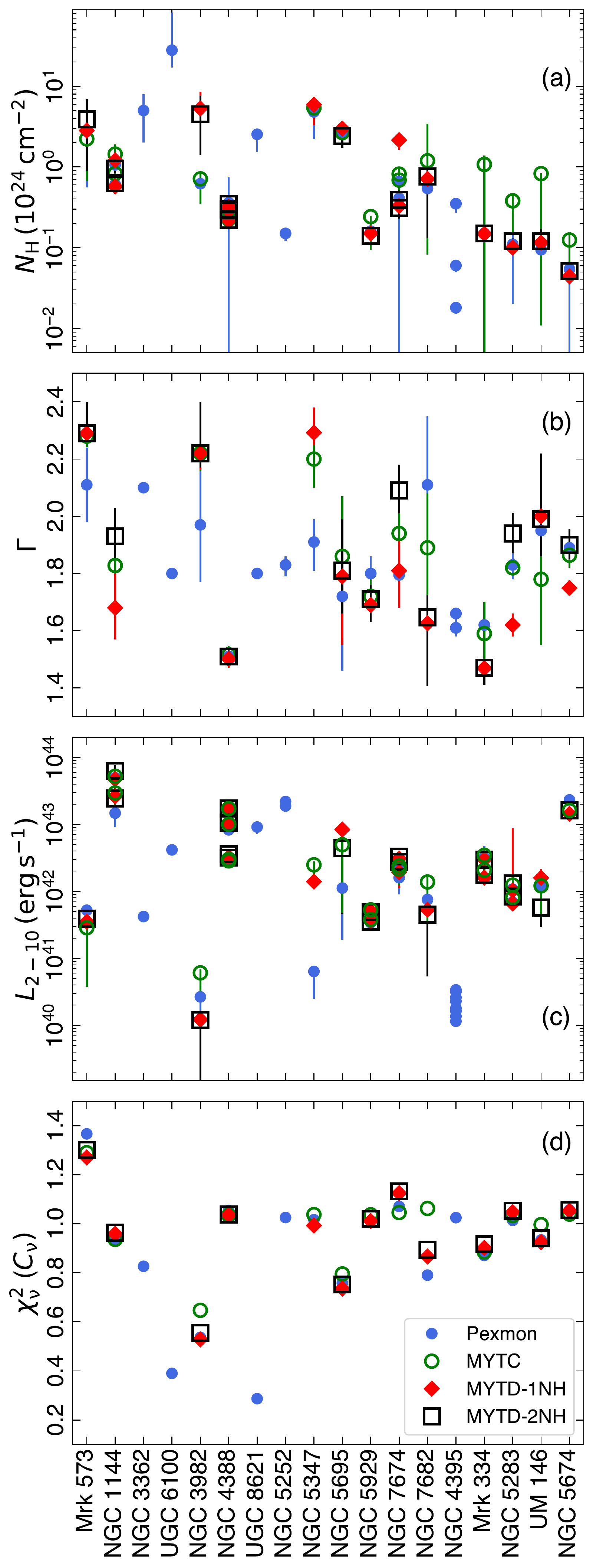}
\caption{The best-fit $N_{\rm H, LOS}$, $\Gamma$, and unabsorbed $L_{2-10}$ (panels a,b and c, respectively) obtained using the Pexmon (blue filled circles) and three MYTorus models (MYTC: green open circles, MYTD-1NH: red diamonds, and MYTD-2NH: black open squares), for all sources. Panel d shows the values  $\chi^2~(C) \rm /dof$ of the best-fit models, using XSPEC, that were used as initial guesses for the MCMC analysis.}
\label{fig:consistency}
\end{figure}

In this section we address the consistency of the results obtained from the different models that are considered in this work. The major difference between these models resides in the way that Pexmon and MYTorus treat the reprocessed emission. Pexmon considers an infinite slab of neutral material that reflects the incident X-ray emission. However, MYTorus considers Compton scattering (in its various regimes) in a torus with a fixed opening angle, providing continuous coverage from the Compton-thin to Compton-thick column densities. In addition, some differences exist between the different configurations of MYTorus, as described in Section~\ref{sec:MYT} \citep[see also][]{MYT09,Yaqoob12}. In Figure~\ref{fig:consistency}, we plot the measured values of $N_{\rm H,LOS}$, $\Gamma$, and $L_{2-10}$ (panels a, b, and c, respectively) obtained from the various models. We note that MYTC measures $N_{\rm H,eq}$, thus we used eq. \ref{eq:NHangle} to obtain the corresponding $N_{\rm H,LOS}$ and assuming the best-fit values of $\theta$, listed in Table~\ref{table:mytc}. In this case, the uncertainty on $N_{\rm H, LOS}$ is estimated using the one on $N_{\rm H,eq}$ only. As for the uncertainty on $L_{2-10}$, we assumed the relative uncertainties obtained from the power law normalization. In panel d of the same figure, we plot the values of $\chi^2/\rm dof$ (or $C/\rm dof$) that are obtained from the best-fit models in XSPEC, and used as an initial guess for the MCMC analysis. This panel demonstrates that all models resulted in statistically good fits.  In the case of Mrk~573, the $\chi^2/\rm dof$ is larger than 1.2. However, the fit is still statistically acceptable with a null hypothesis probability of 0.06. In contrast, UGC~6100, NGC~3982, and UGC~8621 show very low $C$/dof. This is due to the low number of degrees of freedom for these spectra. Only in a few cases, does the MYTorus set of models provide a statistical improvement (in terms of $\chi^2/\rm dof$) with respect to Pexmon.

The column densities are consistent within error bars between the four models in the majority of the cases, except for NGC~3982 and NGC~7674. For NGC~3982, Pexmon predicts a Compton-thin (Cth) column density that is significantly smaller than the CT values obtained by MYTD-1NH and MYTD-2NH. For NGC~7674, MYTD-1NH indicates a large and CT column density, while the other models suggest lower and Cth values. In this case, for MYTD-1NH, the fit is statistically worse than the ones for the other models (see Section~\ref{app:sources} for more details). The photon indices are broadly consistent within uncertainties between the various models, albeit with a large scatter. The intrinsic 2-10~keV luminosity is also consistent between the four models within uncertainties. However, this plot suggests that the luminosity estimates using the various MYTorus configurations tend to be larger than the ones obtained using Pexmon, especially for high values of $N_{\rm H}$. This difference is particularly large in the case of NGC~5347.

\begin{figure}
\centering
\includegraphics[width = 0.9\linewidth]{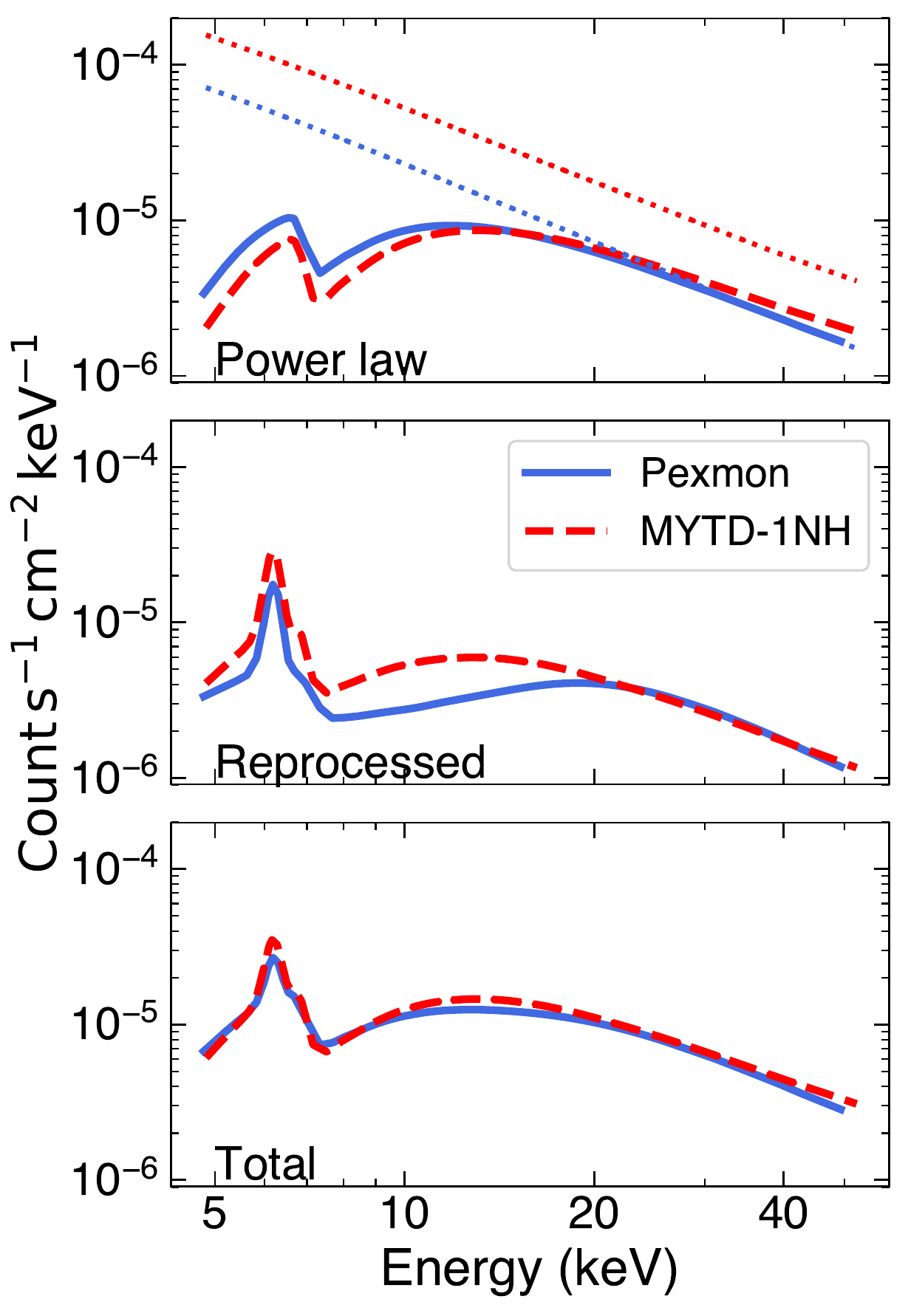}
\caption{The best-fit model components: absorbed power law, reprocessed emission, and total model (top to bottom) obtained by fitting the \nustar\ spectrum of NGC~1144 with Pexmon (solid blue line) and MYTD-1NH (dashed red line). The dotted lines in the top panel correspond to the intrinsic (unabsorbed) power law component for each case. }
\label{fig:model}
\end{figure}

\begin{figure*}
\centering
\includegraphics[width = 0.95\linewidth]{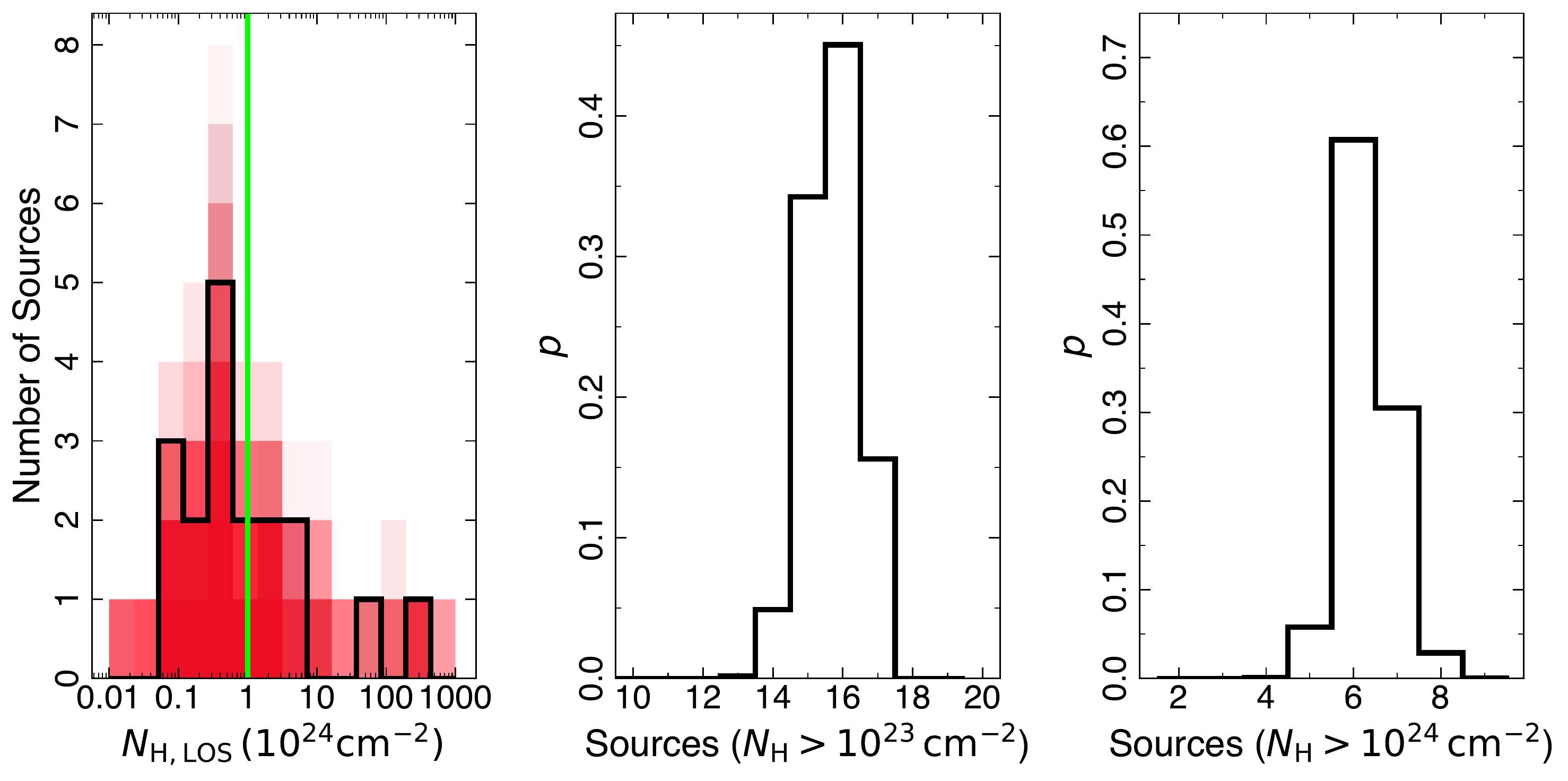}
\caption{Left panel: The $N_{\rm H,LOS}$ distribution obtained from different simulations. The solid black line corresponds to the distributions obtained by averaging the realizations for each source (see Section~\ref{sec:CTfrac} for details). Middle panel: The distribution of the number of heavily obscured sources with $N_{\rm H} > 10^{23}~\rm cm^{-2}$. Right panel: The distribution of the number of CT sources with $N_{\rm H} > 10^{24}~\rm cm^{-2}$.}
\label{fig:simulationCT}
\end{figure*}

In order to assess the reason behind this discrepancy, we plot in Figure~\ref{fig:model} the best-fit Pexmon and MYTD-1NH models obtained by fitting the \nustar\ spectra of NGC~1144. The best-fit results are $N_{\rm H} = 1_{-0.2}^{+0.1} (1.21_{-0.17}^{+0.15})\times 10^{24}~\rm cm^{-2}$ and $L_{2-10} = 1.47 (2.61)\times 10^{43}~\rm erg~s^{-1}$ for Pexmon (MYTD-1NH). In this figure, we show the absorbed and unabsorbed power law component in the top panel, the reprocessed emission in the middle panel, and the total emission in the bottom panel. The reprocessed emission is defined as ${\tt pexmon[6]}$ and ${\tt  A_0*(MYTS_0[7] + MYTL_0[8]) + A_{90}*(MYTS_{90}[10] + MYTL_{90}[11]) }$, for Pexmon and MYTD-1NH, respectively (see Section~\ref{sec:spectralfit}). It is clear from this figure that MYTD-1NH requires larger intrinsic luminosity compared to Pexmon. MYTD-1NH indicates a lower absorbed power-law flux below $\sim 15~\rm keV$ compared to Pexmon. This difference in flux is compensated for when considering the reprocessed emission. As a result, the total emission is consistent between the two models. This effect becomes larger for sources with higher $N_{\rm H}$, in which the intrinsic power-law component cannot be clearly identified.

For NGC~1068, Pexmon resulted in large $N_{\rm H,LOS} = 3.16^{+1.46}_{-1.29} \times 10^{26}~\rm cm^{-2}$. Such high values cannot be achieved by MYTorus which has an upper limit of $10^{25}~\rm cm^{-2}$. For that reason we limit the fits for this source to Pexmon and MYTC only, that resulted in $\chi^2/~\rm dof = 1.07$ and 1.12, respectively. In addition, the fits required a Gaussian emission line at $\sim 6.3~$keV to account for excess seen in the $6-7$~keV range. A similar excess has been reported by \cite{Zaino20} who modeled it with a Gaussian line with a fixed energy at 6~keV and attributed it to instrumental calibration. A detailed analysis of the origin of this excess is beyond the scope of this paper. Due to all the uncertainties and difficulties in modeling the spectra of this source, we excluded all its best-fit parameters from the rest of our analysis, except for $N_{\rm H, LOS}$ that has been well-established in the literature as being CT.

\subsection{Compton-thick fraction}
\label{sec:CTfrac}

Thanks to the high sensitivity of \nustar\ in the hard X-rays we were able to identify three new CT sources: NGC~5347, NGC~5695, and UGC~6100. NGC~5695 was classified by \cite{Lamassa09} as a CT candidate based on its \xmm\ spectrum. In a recent work, \cite{Zhao20} identified Mrk~573 as a CT source, that is also a part of our sample. We identified also two sources (NGC~3362 and UGC~8621) as being CT, based on their \xmm\ spectra. In addition, we analyzed a \nustar\ spectrum of the well-known CT source, NGC~1068. Furthermore, two sources (NGC~3982 and NGC~7674) have been identified as Compton thin when fitted with Pexmon, but CT using MYTD-1NH. Our results for NGC~7674, using Pexmon, are in agreement with the ones found by \cite{Lamassa11} and \cite{Tanimoto20}, but differ from the ones of \cite{Gandhi17} who classified the source as CT. NGC~1144 was found to be at the edge of the CT limit with $N_{\rm H} \simeq 10^{24}~\rm cm^{-2}$. As a result, based on the $N_{\rm H}$ measures obtained from our spectral fits, we identify seven CT sources, and three CT candidates (NGC~3982, NGC~7674, and NGC~1144). In order to find the underlying distribution of $N_{\rm H}$ from our sample, taking into account variability seen in some sources and the uncertainty on our measurements, we performed the following simulation.

Given the general consistency between the different models, we consider the results obtained using Pexmon only, as this model could be applied for all the sources in our sample. Using the $N_{\rm H,LOS}$ distributions obtained from the MCMC analysis, we draw randomly a value for each source. For sources with variable $N_{\rm H, LOS}$, we choose one estimate randomly. We repeated this $10,000$ times and considered the $N_{\rm H}$ distribution for each realization. In the left panel of Figure~\ref{fig:simulationCT} we plot in color the histograms obtained from various realizations. The solid black line correspond to the histograms obtained by averaging the estimated $N_{\rm H,LOS}$ for each source. The middle and right panels of the same figure show the distributions of the number of sources with $N_{\rm H, LOS} > 10^{23}$ and $10^{24}~\rm cm^{-2}$), respectively,  obtained from each realization. As a result, these histograms show that our sample contains $6-7$ CT sources in 90\% of the cases. Furthermore, our results indicate that $15-17$ sources are heavily obscured with $N_{\rm H}>10^{23}~\rm cm^{-2}$ in 94\% of the cases.

\begin{figure}
\centering
\includegraphics[width = 0.95\linewidth]{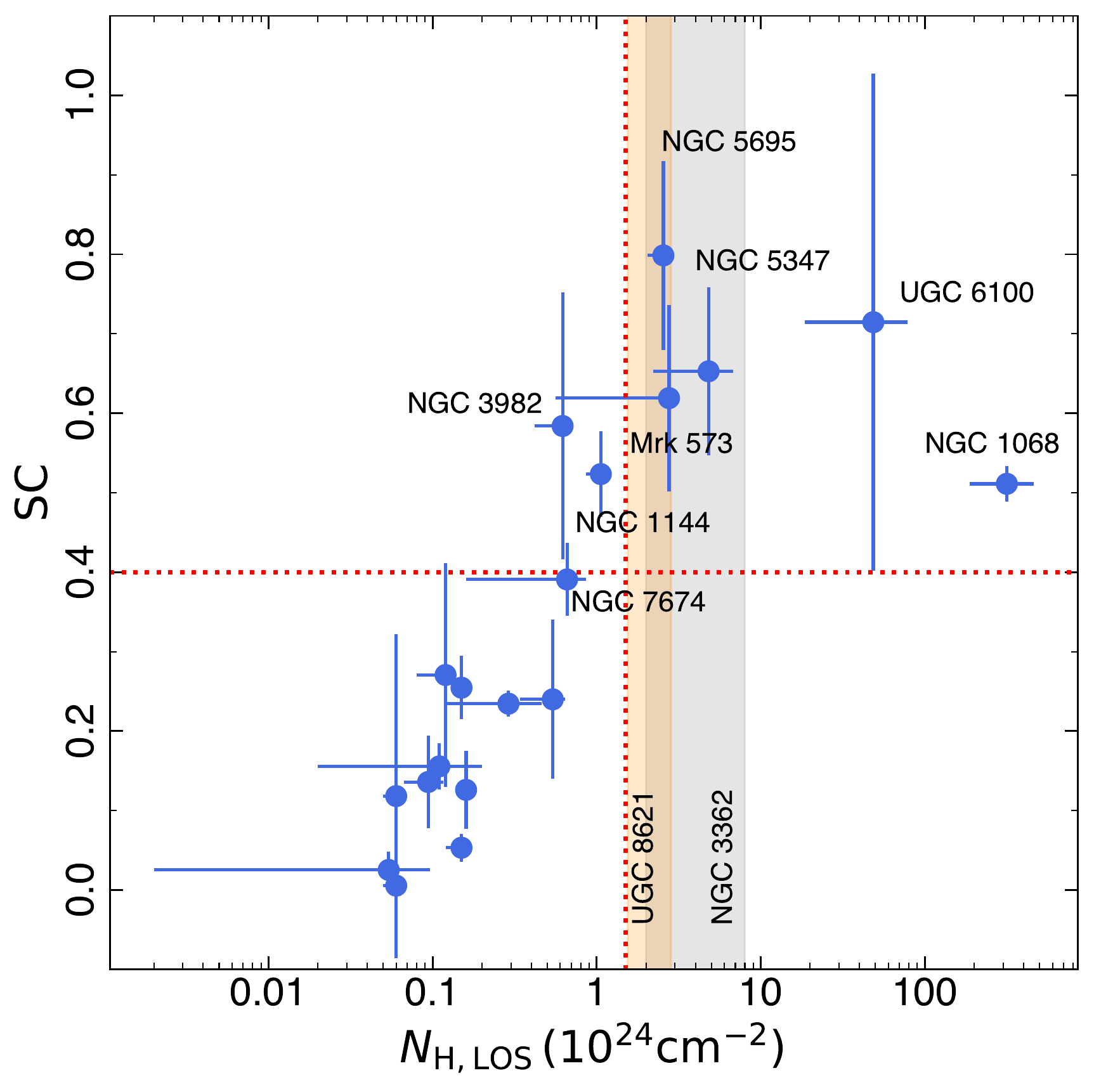}
\caption{Spectral curvature as a function of $N_{\rm H, LOS}$ obtained from Pexmon. The red dotted lines show the CT limit for $SC > 0.4$  and the $N_{\rm H} > 1.5\times 10^{24}~\rm cm^{-2}$. The orange and grey shaded areas correspond to the $N_{\rm H,LOS}$ estimates of UGC~8621 and NGC~3362, respectively.}\label{fig:SC}
\end{figure}

\vfill\null
\startlongtable
\begin{deluxetable}{lllllll}
\tablecaption{The Eddington ratios obtained by fitting the data with Pexmon, MYTD-1NH, and MYTD-2NH. The fourth column corresponds to the spectral curvature estimated from the \nustar\ observations. The last two columns show whether the source is classified as CT (marked with \cmark) or not (marked with \xmark) according to the spectral curvature and $N_{\rm H,LOS}$, respectively. Sources that are considered as CT candidates are marked with a ? mark (see Section~\ref{sec:CTfrac} for details).  \label{table:Ledd}}
\tablehead{\colhead{Source} &  \colhead{} & \colhead{$\log \lambda_{\rm Edd}$} & \colhead{} & \colhead{$SC$}& \colhead{$\rm CT_{SC}$}& \colhead{$\rm CT_{N_{H}}$}\\
\colhead{} &  \colhead{\scriptsize Pexmon} & \colhead{\scriptsize M-1NH} & \colhead{\scriptsize M-2NH} & \colhead{}  & \colhead{} & \colhead{}}
\startdata
Mrk 573	&	-2.73	&	-2.90	&	-2.86	&	$	0.62	\pm	0.12	$	&	\cmark	&	\cmark	\\
NGC 1144	&	-1.86	&	-1.88	&	-1.76	&	$		-		$	&	$-$	&	\xmark	\\
	&	-2.39	&	-2.14	&	-2.17	&	$	0.52	\pm	0.05	$	&	\cmark	&	?	\\
NGC 3362	&	-2.76	&	$-$	&	$-$	&	$		-		$	&	$-$	&	\cmark	\\
UGC 6100	&	-2.04	&	$-$	&	$-$	&	$	0.71	\pm	0.31	$	&	\cmark	&	\cmark	\\
NGC 3982	&	-3.26	&	-3.60	&	-3.61	&	$	0.58	\pm	0.17	$	&	\cmark	&	?	\\
NGC 4388	&	-1.24	&	-1.23	&	-1.19	&	$		-		$	&	$-$	&	\xmark	\\
	&	-0.82	&	-0.72	&	-0.72	&	$		-		$	&	$-$	&	\xmark	\\
	&	-0.57	&	-0.50	&	-0.50	&	$		-		$	&	$-$	&	\xmark	\\
	&	-1.28	&	-1.24	&	-1.24	&	$	0.23	\pm	0.02	$	&	\xmark	&	\xmark	\\
UGC 8621	&	-0.72	&	$-$	&	$-$	&	$		-		$	&	$-$	&	\cmark	\\
NGC 5252	&	-2.53	&	$-$	&	$-$	&	$		-		$	&	$-$	&	\xmark	\\
	&	-2.46	&	$-$	&	$-$	&	$	0.05	\pm	0.02	$	&	$-$	&	\xmark	\\
NGC 5347	&	-3.02	&	-1.67	&	$-$	&	$	0.65	\pm	0.11	$	&	\cmark	&	\cmark	\\
NGC 5695	&	-2.64	&	-1.76	&	-2.04	&	$	0.80	\pm	0.12	$	&	\cmark	&	\cmark	\\
NGC 5929	&	-2.75	&	-2.74	&	-2.75	&	$		-		$	&	$-$	&	\xmark	\\
	&	-2.91	&	-2.88	&	-2.89	&	$	0.13	\pm	0.05	$	&	\xmark	&	\xmark	\\
NGC 7674	&	-2.15	&	-2.24	&	-2.08	&	$		-		$	&	$-$	&	\xmark	\\
	&	-2.32	&	-2.04	&	-2.01	&	$	0.39	\pm	0.05	$	&	?	&	?	\\
NGC 7682	&	-2.34	&	-2.50	&	-2.57	&	$	0.24	\pm	0.10	$	&	\xmark	&	\xmark	\\
NGC 4395	&	-2.12	&	$-$	&	$-$	&	$		-		$	&	$-$	&	\xmark	\\
	&	-1.89	&	$-$	&	$-$	&	$		-		$	&	$-$	&	\xmark	\\
	&	-1.72	&	$-$	&	$-$	&	$		-		$	&	$-$	&	\xmark	\\
	&	-2.04	&	$-$	&	$-$	&	$	0.01	\pm	0.03	$	&	\xmark	&	\xmark	\\
	&	-1.75	&	$-$	&	$-$	&	$	0.12	\pm	0.20	$	&	\xmark	&	\xmark	\\
	&	-2.19	&	$-$	&	$-$	&	$		-		$	&	$-$	&	\xmark	\\
	&	-1.99	&	$-$	&	$-$	&	$		-		$	&	$-$	&	\xmark	\\
	&	-1.83	&	$-$	&	$-$	&	$		-		$	&	$-$	&	\xmark	\\
Mrk 334	&	$-$	&	$-$	&	$-$	&	$	0.25	\pm	0.04	$	&	\xmark	&	\xmark	\\
NGC 5283	&	-2.24	&	-2.26	&	-2.16	&	$	0.16	\pm	0.03	$	&	\xmark	&	\xmark	\\
UM 146	&	-2.44	&	-2.46	&	-2.35	&	$	0.14	\pm	0.06	$	&	\xmark	&	\xmark	\\
NGC 5674	&	-0.90	&	-1.11	&	-1.06	&	$	0.03	\pm	0.02	$	&	\xmark	&	\xmark	\\
NGC 1068	&	$-$	&	$-$	&	$-$	&	$	0.51	\pm	0.02	$	&	\cmark	&	\cmark	\\
\enddata
\end{deluxetable}

In addition, we performed a model-independent test for the number of CT sources using the spectral curvature (SC) metric presented by \cite{Koss16sc}. Following their eq. 2, we estimated SC for all sources with \nustar\ detection as follows:
\begin{equation}\label{eq:SC}
SC = \frac{-0.46 \times E + 0.64 \times F +2.33 \times G}{T},
\end{equation}
\noindent where $E$, $F$, $G$, and $T$ are the count rates in the $8-14$~keV, $14-20$~keV, $20-30$~keV,  and $8-30$~keV ranges, respectively. A source is classified as CT if $SC > 0.4$. The estimated $SC$ are presented in Table~\ref{table:Ledd}. In Figure~\ref{fig:SC}, we plot $SC$ as a function of the measured $N_{\rm H, LOS}$ obtained using Pexmon. According to the $SC$ metric, the five sources (Mrk~573, NGC~5347, UGC~6100, NGC~5695 and NGC~1068) that were detected by \nustar\ and considered as CT, based on the spectral fits, are also found to be CT. Interestingly, the two CT candidates (NGC~3982 and NGC~1144) based on the spectral fits, are found to be CT based on the $SC$ metric. NGC~7674 is found to be at the edge of the CT limit with $SC = 0.39 \pm 0.05$. So we also consider it as a CT candidate based on the $SC$ metric. Two sources (NGC~3362 and UGC~8621) were found to be CT sources based on their spectral fits could not be confirmed by the $SC$ metric because they were not detected with \nustar.

In conclusion, our results suggest that $78.9-89.5\%$ of the sources in our sample are heavily obscured with $N_{\rm H}> 10^{23}~\rm cm^{-2}$. Furthermore, based on both methods we presented above, the number of CT sources in our sample ranges between 6 and 10 sources.  This corresponds to $31.6-52.6\%$ of our sample being CT. Thus for the complete CfASy sample of 46 sources (accounting for all obscured and unobscured Seyferts), our results indicate that $32.6-36.9\%$  ($13-21.7\%$) of the sources are obscured with $N_{\rm H} > 10^{23} ~(>10^{24}~\rm cm^{-2})$. Our results are in full agreement with the ones of \cite{Risaliti99} who found that $77.8 \pm 12.6 \%$ of Seyfert-2 galaxies are heavily obscured with $N_{\rm H} > 10^{23}~\rm cm^{-2}$ and  $47.9 \pm 10 \%$  are CT. \cite{Panessa06} found a CT fraction of $30-50\%$ of the Sy2 sources from the Palomar optical spectroscopic survey of nearby galaxies \citep{Ho95,Ho97}. However, flux-limited hard X-ray surveys have reported a significantly lower observed fraction of CT sources \citep[below $\sim 20\%$, e.g.,][]{Bassani06,Burlon11, Vasudevan13, Masini18}. In particular, \cite{Ricci15} found the CT fraction in the 70-month \swift/BAT catalog to be $7.6_{-2.1}^{+1.1}\%$.  We note that the sources in \cite{Ricci15} span broader redshift and X-ray luminosity ranges than the sources in our sample. In particular, the sources from the  \swift/BAT AGN catalog include more luminous sources compared to the one presented in this work. \cite{Koss16sc}, using a lower redshift range that is similar to ours, found $22\%$ of hard X-ray selected BAT AGN as CT utilizing the spectral curvature metric. Our results confirm the conclusions by \cite{Cappi06} and \cite{Malizia09} who showed that a careful selection of a complete volume limited sample reconciles the fraction of heavily obscured (and CT) sources obtained from multi-wavelength estimates and the ones from X-ray spectral fitting.  It is worth noting that our sample of obscured sources is based on optical selection. However, it might be possible that some elusive objects have been missed by the optical classification due to missing narrow H$\beta$ lines. For example, \cite{Koss17} found that $\sim 30~\%$ of the X-ray selected AGN are missing narrow H$\beta$ lines. This is mainly because X-ray selected AGN have much higher Balmer decrements and more likely to be in edge on galaxies \citep[see e.g.,][]{Koss11}. In this case, it is likely that these sources are highly embedded to be missed within the small volume. This would lead to a higher number of highly obscured sources than the ones we identified. For example, \cite{Smith14} studied four optically elusive AGN and four X-ray bright, optically normal galaxies. Five sources of their sample are at $z < 0.03$, comparable to the redshift range of the sources studied in this work. The authors found that four out of those five sources are highly obscured.

\subsection{Moderately obscured sources}
\label{sec:lowobs}

As mentioned in the previous section, we found that 16 sources are heavily obscured with $N_{\rm H} > 10^{23}~\rm cm^{-2}$. In this section, we focus on the three sources that show moderate obscuration ($10^{22}~\rm cm^{-2} < N_{\rm H} < 10^{23}~\rm cm^{-2}$), namely NGC~4395, UM~146 and NGC~5674.  \cite{Trouille09} have shown that $33\% \pm 4\%$ of their sample of optically-selected obscured sources are X-ray unobscured. In our case, considering the three moderately-obscured sources in our sample, it is well-known that NGC~4395 exhibits strong variability in its obscuring column. Those variations are associated with obscuring clouds in the BLR \citep[see][]{Nardini11, Kammoun19n4395}. Instead, UM~146 and NGC~5674 are two face-on galaxies. Thus, it is less likely that the obscuration is due to gas or dust in their host galaxies. \cite{Trippe10} classified these sources as ``true'' Sy~1.8 and Sy~2, respectively. \cite{Risaliti02} analyzed the \textit{BeppoSAX} spectrum of NGC~5674 (observed in February 2000, i.e., 14 years prior to the \nustar\ observation). They reported an obscuring column density of the order of $6\times 10^{22}~\rm cm^{-2}$ that is consistent with our measurement. Further monitoring observations of these sources will be then necessary to search for any possible variability. This will allow us to understand the nature of the moderate obscuration in these two sources.  

\subsection{On the obscurer structure}
\label{sec:geometry}

\begin{figure*}
\centering
\includegraphics[width = 0.9\linewidth]{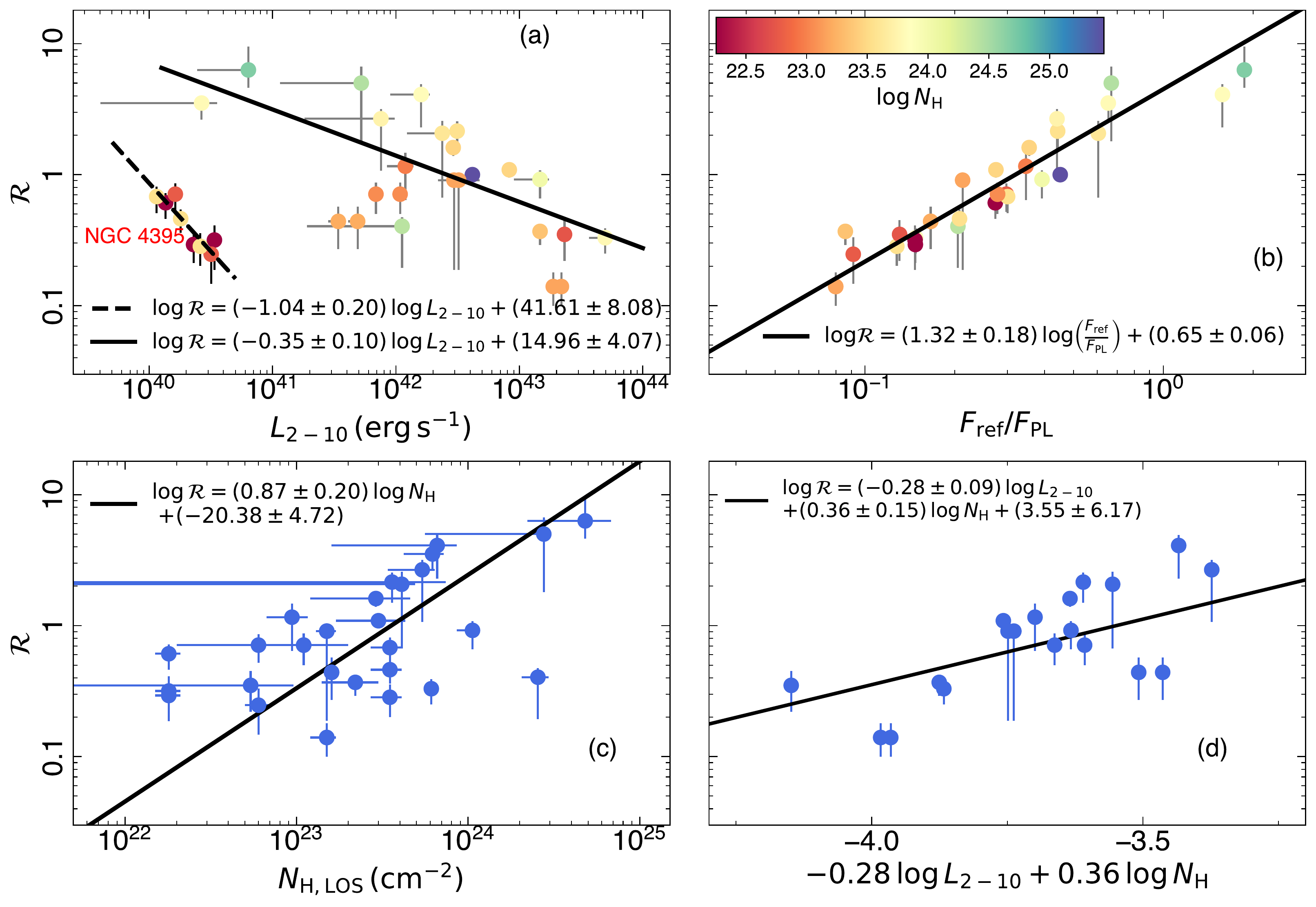}
\caption{Panel a: Reflection fraction as a function of the intrinsic (unabsorbed) X-ray luminosity in the $2-10$~keV range. Two linear fits in the log-log space has been performed: the data points of NGC~4395 (obtained from flux-resolved spectra extracted from four observations) and the rest of the sources (dashed and solid lines, respectively). Panel b: Reflection fraction as a function of the reflected flux over power-law flux ratio, obtained using Pexmon. The data is fitted with a linear model in the log-log plot (solid line) that results in a slope consistent with unity. The color map used in this plot corresponds to the fitted value of $N_{\rm H,LOS}$ obtained using Pexmon. Panel c: Reflection fraction as a function of the LOS column density, fitted with a linear relation in the log-log space. Panel d: The best-fit $\mathcal{R}-L_{2-10}-N_{\rm H}$ plane obtained by excluding NGC~4395.}
\label{fig:R-LX}
\end{figure*}
It is not straightforward to obtain constraints on the structure of the obscuring material (e.g., geometry, filling factor, opening angle, orientation) using our models. This task becomes harder given the fact that all models provide statistically good fits. However, some valuable information can still be extracted from our results. In particular, a more careful look at the $\mathcal{R}$ vs $L_{2-10}$ plot, shown in Figure~\ref{fig:R-LX}a (also in Figure~\ref{fig:triangle}), reveals the presence of a correlation between these two quantities. In fact, one can see that the dwarf galaxy NGC~4395 does not follow the rest of the sources. Considering the full data set results in a Kendall correlation coefficient  $\tau_{\rm K} =  -0.04$ with a null hypothesis probability of 0.73. However, splitting the data set in two results in $\tau_{\rm K} = -0.64/-0.41$ with null hypothesis probability of $0.026/0.007$ for NGC~4395 and the rest of the sources, respectively. We fit the two subsets with a linear fit in the log-log space, of the form,
\begin{equation}\label{eq:R-Lfit}
\log \mathcal{R} = \alpha + \beta \log L_{2-10},
\end{equation}
\noindent using the Orthogonal Distance Regression (ODR\footnote{\url{https://docs.scipy.org/doc/scipy/reference/odr.html}}) method. The best-fits result in consistent slopes $\beta = -1.04 \pm 0.2$ and $-0.35 \pm 0.1$ for NGC~4395 and the rest of the sources, respectively. The reflection fraction in Pexmon is defined as the strength of the reflection component relative to the one expected from a slab spanning a solid angle of $2\pi$ \citep{Nandra07}. In other terms, $\mathcal{R} = 1$ corresponds to a reflector that covers half of the solid angle as seen by the source. In Figure~\ref{fig:R-LX}b, we show $\mathcal{R}$ as a function of the ratio ($F_{\rm ref}/F_{\rm PL}$) of reflection flux over the power-law flux in the $3-30$~keV. Both quantities are positively correlated with a slope of $1.32 \pm 0.18$, consistent with unity within less than $2\sigma$. Fitting $\mathcal{R}$, obtained from Pexmon, vs. the reprocessed-to-power-law ratio, obtained from the MYTorus models, for the sources that could be fitted with both models, results in consistent slopes, albeit with a different normalization. \cite{Geo91} showed that $F_{\rm ref}/F_{\rm PL}$ positively correlates with the equivalent width of the Fe K$\alpha$ line ($\rm EW({\rm Fe~ K\alpha})$; see their figure 16). As a consequence, the anti-correlation seen between the $\mathcal{R}$ and $L_{2-10}$ can be translated in an anti-correlation between  $\rm EW({\rm Fe~ K\alpha})$ and $L_{2-10}$, known as the Iwasawa-Taniguchi effect \citep[or the X-ray Baldwin effect;][]{Iwasawa93}. This effect has been confirmed in Type-2 AGN by \cite{Ricci14}, and later confirmed in a large sample of CT sources by \cite{Boorman16}. The fact that the slope of the $\mathcal{R}-L_{2-10}$ correlation for NGC~4395 is consistent with unity is suggestive of a constant reprocessed flux despite the intrinsic variability \citep[see][]{Kammoun19n4395}. This may indicate that despite the change in the LOS column density in this source, the overall covering fraction of the clouds remains constant. This also suggests that a large fraction of the reprocessed emission arises from distant material.

Furthermore, Figure~\ref{fig:triangle} shows a positive correlation between $\mathcal{R}$ and $N_{\rm H,LOS}$ with $\tau_{\rm K} = 0.4$ and a null-hypothesis probability of 0.002. Similar correlations have been reported in the literature \citep[e.g.,][]{Ricci11, DelMoro17, Panagiotou19}. Fitting $\mathcal{R}$ versus $N_{\rm H,LOS}$ in the log-log space, similar to eq. (\ref{eq:R-Lfit}), for all sources, resulted in a slope of $0.87 \pm 0.2$ (See Figure~\ref{fig:R-LX}c). Thus, we attempt to perform a fit  in three dimensions, using the Scipy package {\tt curve\_fit}\footnote{\url{https://docs.scipy.org/doc/scipy/reference/generated/scipy.optimize.curve_fit.html}}, of the form, 
\begin{equation}\label{eq:R-NH-Lfit}
\log \mathcal{R} =  \hat{\alpha} + \hat{\beta} \log L_{2-10} + \hat{\gamma }\log N_{\rm H,LOS},
\end{equation}
\noindent excluding NGC~4395. We show the best-fit plane in Figure~\ref{fig:R-LX}d, with $\hat{\beta} = -0.28 \pm 0.09$ and $\hat{\gamma} = 0.36 \pm 0.15$. This clearly demonstrates that the reprocessed emission depends simultaneously on both the intrinsic X-ray luminosity and the column density of the reprocessing material.

\begin{figure}
\centering
\includegraphics[width = 0.99\linewidth]{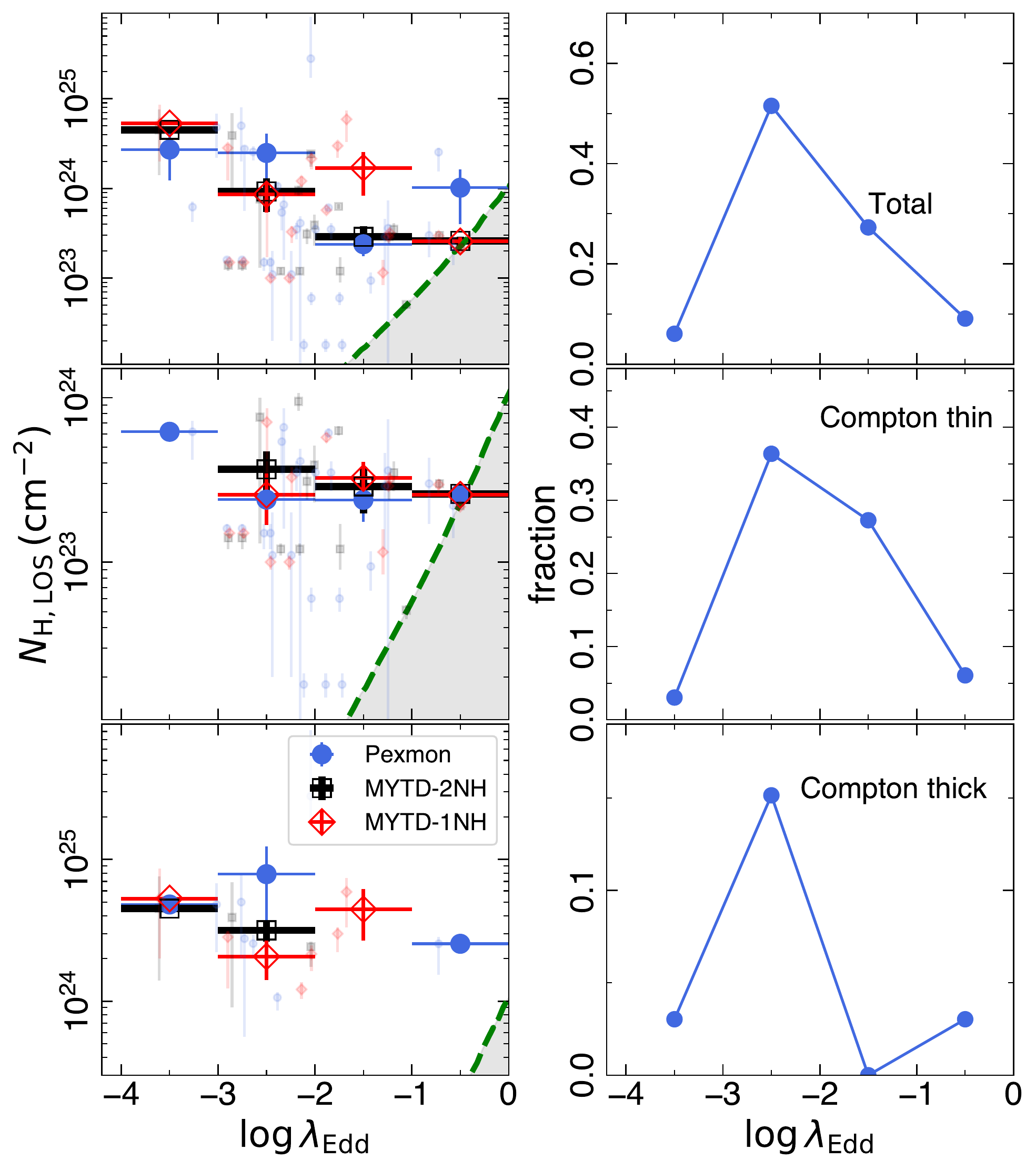}
\caption{Left: $N_{\rm H,LOS}$ as a function of $\log \lambda_{\rm Edd}$ for the full sample, the Cth observations, and the CT observations(top to bottom). The results are shown for Pexmon (blue circles), MYTD-1NH (red diamonds), and MYTD-2NH (black squares). The shaded data points correspond to the individual measurements. The large thick symbols correspond to the binned data. The green dashed lines show the effective Eddington limit below which dusty clouds \citep[with standard ISM grain abundance; adapted from][]{Fabian09} close to the BH see the AGN as being effectively above the Eddington limit (known as the forbidden region, grey area). Long-lived absorbing clouds can only occur above the dashed line. Right: The corresponding fraction of measurements in the various $\log \lambda_{\rm Edd}$ bins (for the Pexmon model only), normalized to  the total number of observations.}
\label{fig:NH_Edd}
\end{figure}

Figure~\ref{fig:NH_Edd} shows the scattered and binned measurements of $N_{\rm H, LOS}$ as a function of $\log \lambda_{\rm Edd}$. We plot the measurements for the full sample, the Cth subset, and the CT subset (top to bottom, respectively). We removed NGC~1068 from this plot due to the large discrepancy in its $L_{2-10}$ estimates using different models. Despite the fact that the size of our sample is small, our results broadly agree with the ones of \cite{Ricci17} based on the 70-month \swift/BAT AGN sample. $N_{\rm H,LOS}$ peaks in the range of $\log \lambda_{\rm Edd} \sim [-4 ;-2.5]$ then decreases at larger Eddington ratios, for both the full sample and the Cth subset. The decrease at high $\lambda_{\rm Edd}$ in our data is shallower than the one found by \cite{Ricci17}. This trend depends on the way we estimated the Eddington fraction, i.e., spectral modeling and the bolometric correction we adopt. The similarity between our results and the ones from the BAT AGN sample are clearer for the fraction of the sources versus the Eddington ratio (right-hand side panels in the same figure). This is suggestive that radiation pressure regulates the distribution of the circumnuclear material \citep[see e.g., figure 5 in][]{Kurosawa09}. As the accretion rate increases, less dense material ($\log N_{\rm H}/\rm cm^{-2} < 24$) is swept away, leaving only the CT material, thus decreasing the torus covering factor \citep[e.g.,][]{Fabian06, Fabian09}. We show in the $N_{\rm H}-\lambda_{\rm Edd}$ plane the effective Eddington limit assuming the standard ISM grain abundance from \cite{Fabian09}. The effective Eddington limit can be lower for dusty gas than for ionized dust-free gas \citep[e.g.,][]{Laor93, Scoville95}. This means that a sub-Eddington AGN may be seen as super-Eddington by the dusty torus with substantial column densities. As a result, long-lived stable clouds can survive radiation pressure only in a regime lower than the effective Eddington limit \citep[upper left part in Figure~\ref{fig:NH_Edd}; see also][]{Fabian08}. All our measurements reside in the long-lived cloud area. 

Within the context of the radiation-driven accretion disk winds presented by \cite{Giustini19}, most of our sources fall in the regime of small BH mass ($M_{\rm BH} < 10^8~ M_\odot$) and moderate accretion rate. This category of sources is expected to have weak/moderate line-driven disk winds. The authors show that failed continuum radiation-driven dusty wind on radial scales of the order of the dust sublimation radius will be extended in the equatorial plane. Thus, at high inclination (which corresponds to the sources we study in this work), this failed wind will obscure the innermost regions of the system. Further investigations will be needed to confirm this hypothesis.


\begin{figure}
\centering
\includegraphics[width = 0.99\linewidth]{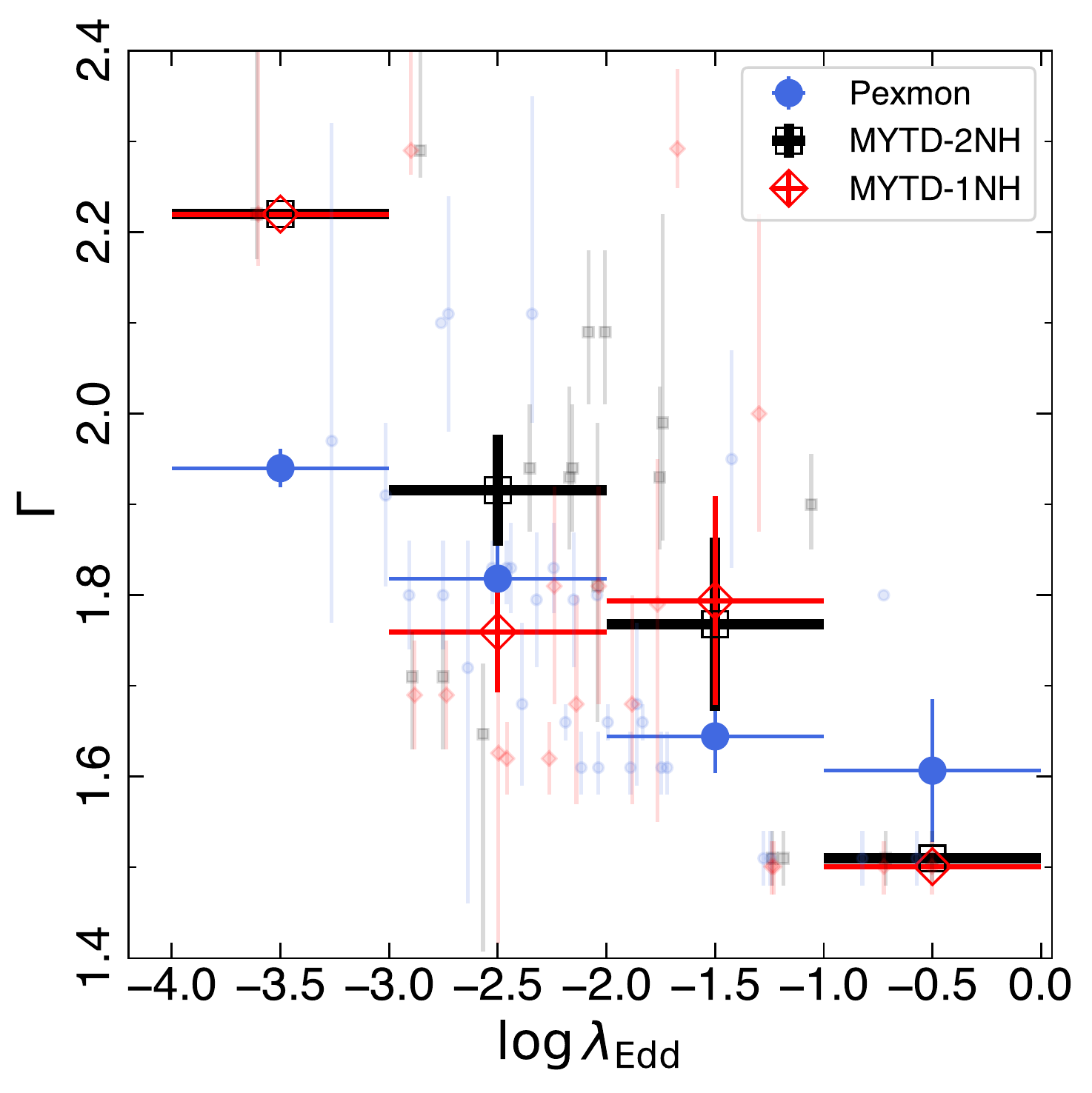}
\caption{$\Gamma$ as a function of $\log \lambda_{\rm Edd}$. The results are shown for the Pexmon (blue circles), MYTD-1NH (red diamonds), and MYTD-2NH (black squares). The shaded data points correspond to the individual measurements. The large thick symbols correspond to the binned data.}
\label{fig:G_Edd}
\end{figure}

In Figure~\ref{fig:G_Edd} we show $\Gamma$ as a function of $\log \lambda_{\rm Edd}$ for Pexmon, MYTD-1NH, and MYTD-2NH. A negative correlation between these two quantities can be seen. Applying the Kendall tau's correlation to the scattered data (for the Pexmon model), we found a correlation coefficient of 0.56 with a $p-$value of $1.7\times 10^{-5}$. Binning the data points leads to a clearer correlation. This is in agreement with the results of \cite{Winter09} and \cite{Younes11} (the latter studied a sample of 13 LINERs). Our results differ from the positive correlation found for high-redshift and more luminous quasars \citep[e.g.,][and references therein]{Shemmer08, Risaliti09qso,Brightman13, Huang20}. This is most likely due to a systematic deviation of lower luminosity AGN from the correlation. In fact, a ``V" shape in the $\Gamma$ vs. $L_{\rm X}$  has been reported in various X-ray binaries by \cite{Liu19} and \cite{Yan19}, indicating a transition at a certain Eddington ratio between the ``harder when brighter" and the ``softer when brighter'' states.

\begin{figure}
\centering
\includegraphics[width = 0.99\linewidth]{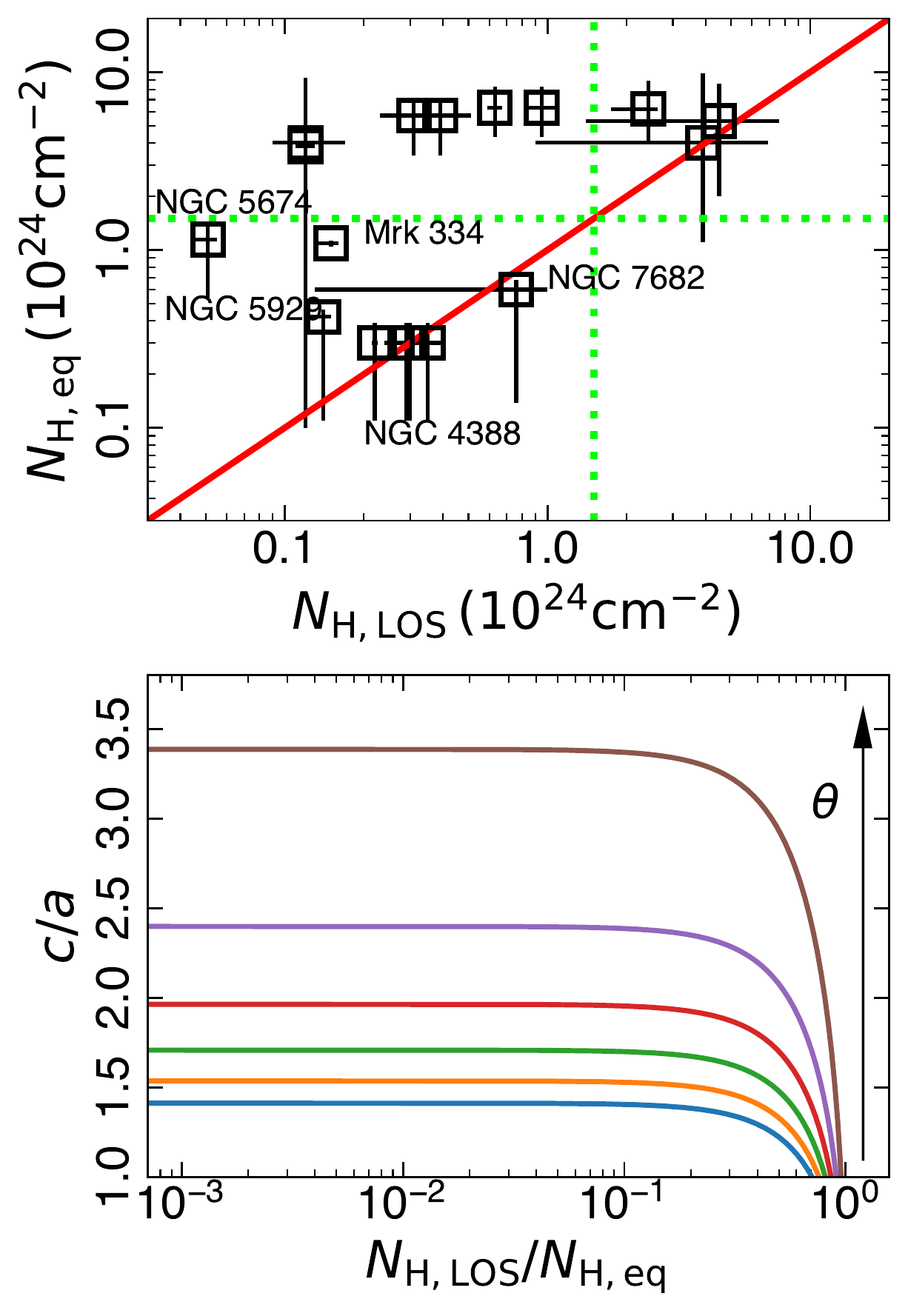}
\caption{Top: The torus (equivalent) column density plotted versus the LOS column density obtained by fitting the spectra with MYTD-2NH. The green dotted lines correspond to the CT limit. The red solid line corresponds to the identity line. Bottom: The ratio $c/a$ (indicating the torus opening angle) versus the column density ratio  $N_{\rm H,LOS}/N_{\rm H,eq}$, for inclinations between 60\degr\ and 85\degr\ (from bottom to top, with an increase by 5\degr).}
\label{fig:NH_LOS}
\end{figure}

Finally, in the upper panel of Figure~\ref{fig:NH_LOS}, we show $N_{\rm H,eq}$, obtained by fitting the data with MYTD-2NH, plotted versus the corresponding $N_{\rm H,LOS}$ values. Five sources  (namely, NGC~4388, NGC~7682, Mrk~334, NGC 5929, and NGC~5674) out of the 13 fitted with this model show a Cth $N_{\rm H,eq}$. For NGC~5674 and Mrk~334, $N_{\rm H,eq}$ is higher than $10^{24}~\rm cm^{-2}$. However, for the other three sources $N_{\rm H,eq}$ is significantly lower though consistent with $N_{\rm H,LOS}$. This may be an indication of a uniform distribution of  absorbing/reprocessing material in these two sources. Interestingly, none of the sources requires a Cth $N_{\rm H,eq}$ while its $N_{\rm H,LOS}$ is CT. In addition, none of the sources require $N_{\rm H,LOS}$ to be significantly higher than $N_{\rm H,eq}$. This result adds to the rigorousness of our analysis. It supports the general scheme in which most of the sources are surrounded by Compton-thick material that reprocesses the intrinsic X-ray light into our LOS \citep[see e.g., figure 1 in][for a recent schematic representation]{Marin16}. The alignment of this material with our LOS will result in observed column densities that are smaller or equal to the global one, being mostly $> 10^{23}~\rm cm^{-2}$ for the sources that are observed at high inclination. In a toroidal geometry, for a given inclination angle, the ratio $N_{\rm H,LOS}/N_{\rm H,eq}$ may be indicator of the opening angle (in other terms the covering fraction; see eq.~\ref{eq:NHangle}). In the lower panel of Figure~\ref{fig:NH_LOS}, we plot the ratio $c/a$ (indicating the half opening angle) as a function of $N_{\rm H,LOS}/N_{\rm H,eq}$ following eq.~\ref{eq:NHangle} for a range of inclinations between 60\degr\ and 85\degr. This plot indicates that large values of $N_{\rm H,LOS}/N_{\rm H,eq}$ (close to 1) indicate small opening angles (i.e., large covering fractions), while small $N_{\rm H,LOS}/N_{\rm H,eq}$ values indicate a wider range of opening angles, depending on the inclination. Alternatively, some recent models that account for different geometries \citep[e.g.,][]{Buchner15,Borus18,Tanimoto19} can be used to derive direct constraints on the torus properties (e..g, equatorial column density, covering fraction, clumpiness). However, given the fact that the rather simple models that we used in this paper give reasonable fits, and given the relatively modest data quality in several cases, the application of additional models, with more free parameters, is not be justified by the data. Thus, this approach will result in many unconstrained parameters. In addition, the consistency between the various parameters that we obtained by applying the four Pexmon and MYTorus models strongly suggests that the application of an additional model will not alter any of the main results and conclusions of our work \citep[see e.g.,][for a comparison between various models]{Kammoun19n5347}.

\section{Conclusions}
\label{sec:conclusion}

We present analysis of optical and X-ray spectra of 21 obscured sources in the CfASyS through the \nustar\ {\it Obscured Seyferts} Legacy Survey. This is the first optically-selected and volume-limited survey of AGN performed in the hard X-rays. Our results are summarized below.

\begin{itemize}
\item Analyzing the optical spectra of those sources allowed us to obtain accurate estimates of their BH masses. Furthermore, we identified two sources in the composite regime of the BPT diagram (Mrk~461 and NGC~5256). Those two sources were excluded from the analysis, restricting the final sample to 19 Seyfert galaxies only.

\item We have fitted the X-ray spectra of the 19 sources using four models. The results from all the models are consistent with each other, except for intrinsic luminosity in the cases of CT sources. In these sources, the intrinsic power law emission is heavily suppressed, and thus could not be well constrained. Different models then result in different values of the intrinsic luminosity due to various physical assumptions of each model.

\item Our results indicate that $82-89\%$ of the sources are heavily obscured with $N_{\rm H} > 10^{23}~\rm cm^{-2}$.  In addition,  $36.8-52.6\%$ of the sources in our sample are CT. Our results are in agreement with the ones found by \cite{Risaliti99} based on multi-wavelength estimates of $N_{\rm H}$ for obscured sources in the local universe, reconciling the long-lasting discrepancy between volume-limited values and those observed from flux-limited hard X-ray surveys.

\item We found a tight anti-correlation between the reflection fraction (representing the ratio of reflected over intrinsic flux) and the intrinsic X-ray luminosity, in agreement with the Iwasawa-Taniguchi effect. Furthermore, our results showed a positive correlation between the reprocessed emission and the LOS column density.

\item Our results support the hypothesis that radiation pressure regulates the distribution of the circumnuclear material.

\end{itemize}

Our results demonstrate the unique power of \nustar\ in unveiling the properties of obscured AGN thanks to its high sensitivity above 10~keV. Deeper observations in the soft and hard X-rays are required, especially for the sources that could not be detected with \nustar\ in order to confirm our conclusions, and to explore possible intrinsic and/or absorption-induced variability. We note that future missions carrying micro-calorimeters such as \textit{XRISM}/Resolve \citep{Xrism} and \textit{Athena}/X-IFU \citep{XIFU} will allow us to investigate in more detail the nature, composition, and the dynamic properties of the obscuring material in AGN. Coupling the high resolution of \textit{XRISM} in the soft X-rays to the high sensitivity in hard X-rays of \nustar\ through simultaneous observations will be of a particular interest. Exploring the role of future missions in studying obscured AGN will be addressed in future work.

\begin{acknowledgements}

We thank the anonymous reviewer for their feedback that improved the quality of our manuscript. KO acknowledges support from the National Research Foundation of Korea (NRF-2020R1C1C1005462). This work made use of data from the \nustar\ mission, a project led by the California Institute of Technology, managed by the Jet Propulsion Laboratory, and funded by the National Aeronautics and Space Administration. This work also made use\xmm, an ESA science mission with instruments and contributions directly funded by ESA Member States and NASA. Results presented in this paper are also based on data obtained with the {\it Suzaku} observatory; and the {\it Chandra} X-ray Observatory. We acknowledge the use of public data from the \swift\ data archive. ESK and JMM thank Tahir Yaqoob for useful discussions regarding the implementation of MYTorus.  We would like to thank Karl Forster for scheduling all of  the \textit{NuSTAR} observations in our program. We would like to thank Xavier Barcons, Fiona Harrison, Tim Kallman, Kirpal Nandra and John Raymond for useful conversations. E.B. is supported by a Center of Excellence of THE ISF (grant No. 2752/19). The figures were generated using matplotlib \citep{Hunter07}, a {\tt PYTHON} library for publication of quality graphics. The MCMC results were presented using the GetDist {\tt PYTHON} package.

\end{acknowledgements}

\facilities{ \chandra, \nustar, \swift, \suzaku, \xmm}

\software{pPXF \citep{Cappellari04}, CIAO \citep[v9.4][]{Ciao}, HEASoft \citep{HEASoft}, NUSTARDAS (v1.8.0, \url{https://heasarc.gsfc.nasa.gov/docs/nustar/analysis/}, SAS  \citep[v17.0.0][]{Gabriel04}, XSPEC \citep{Arnaud96}, XSPEC\_EMCEE ((\url{https://github.com/zoghbi-a/xspec\_emcee}), Matplotlib \citep{Hunter07}, GetDist (\url{https://getdist.readthedocs.io/en/latest/})}

\appendix

\section{Notes on individual sources}
\label{app:sources}

\begin{figure*}
\centering

    	\includegraphics[width = 0.3\textwidth]{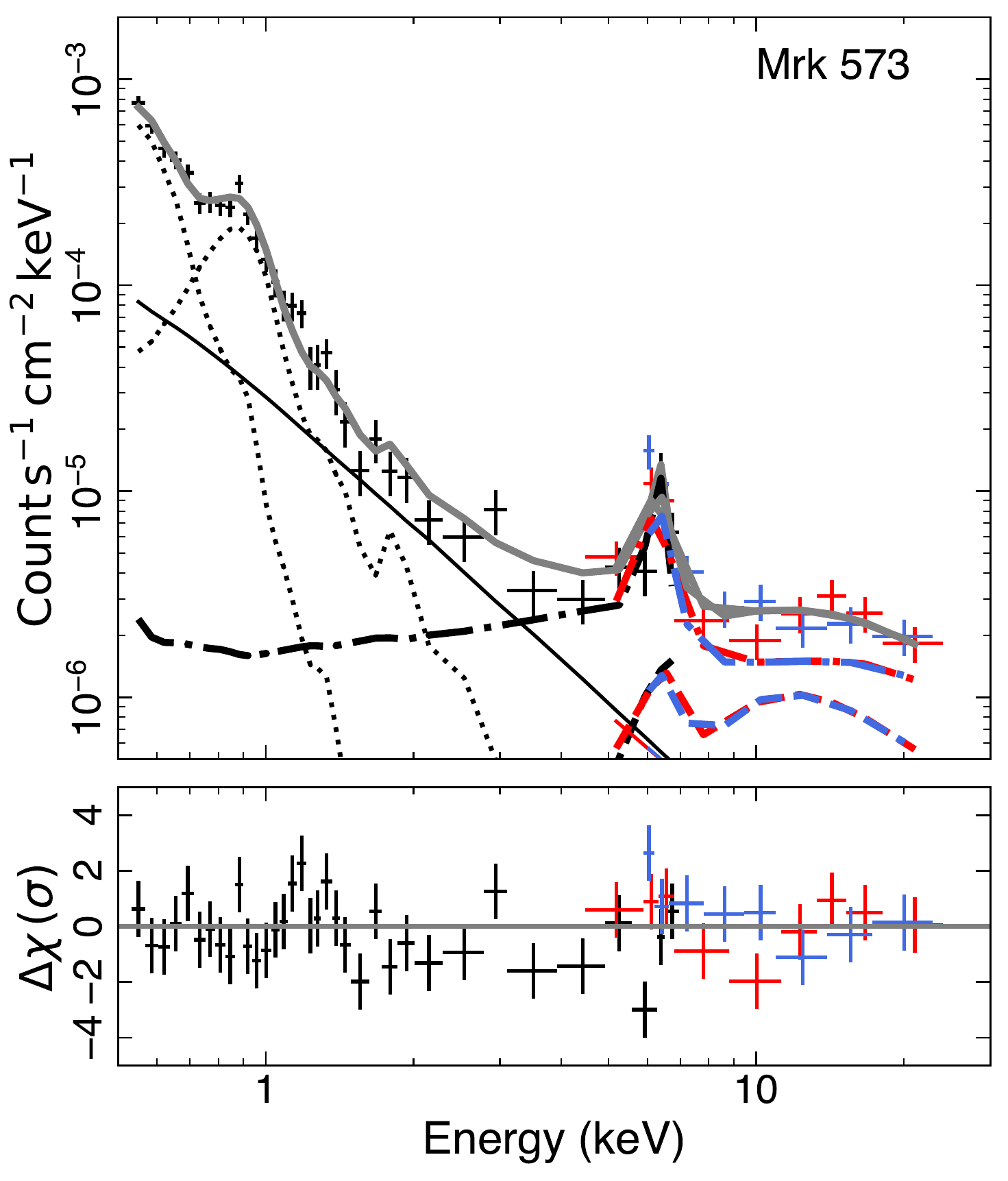}
    	\includegraphics[width = 0.3\textwidth]{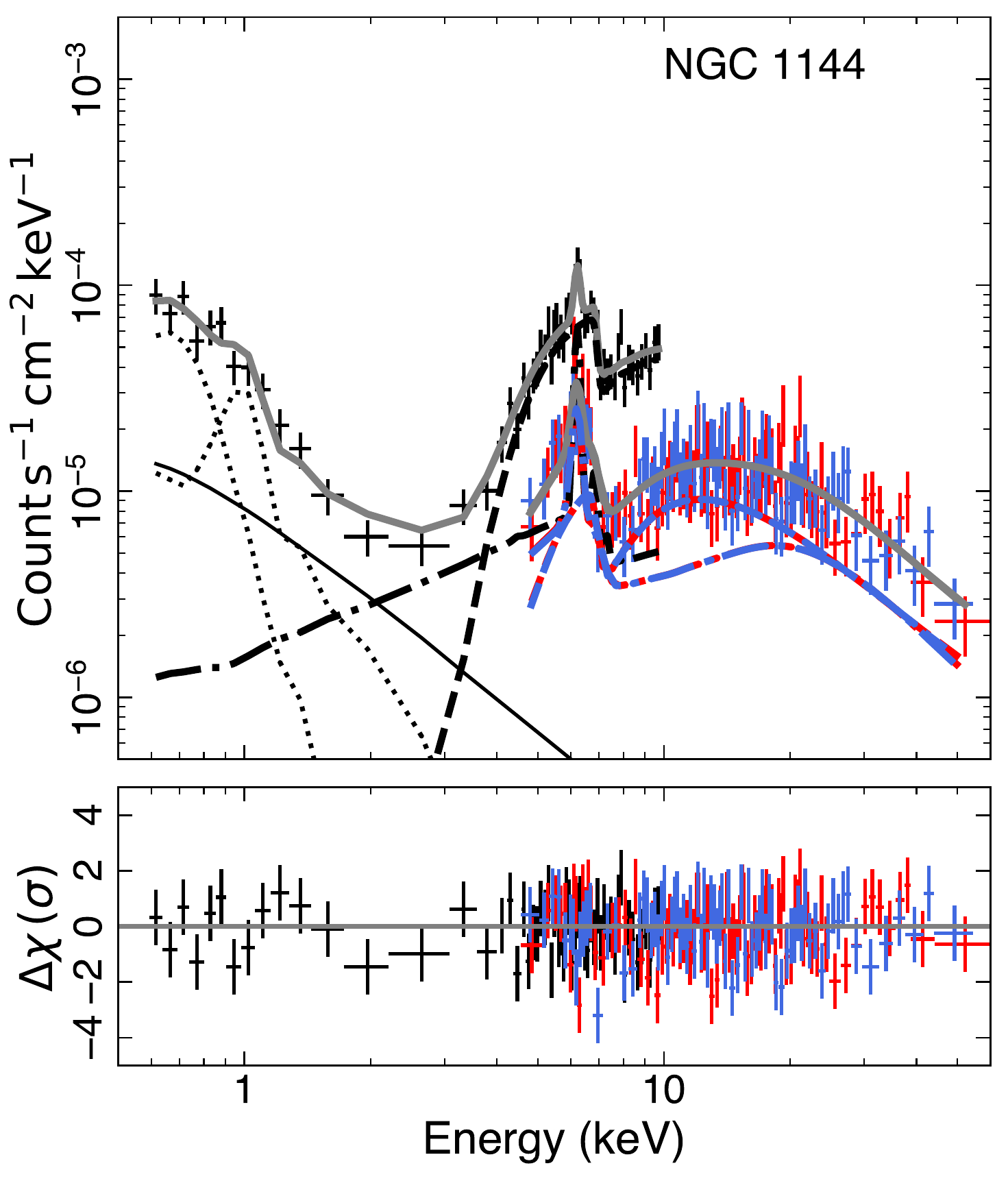}
	\includegraphics[width = 0.3\textwidth]{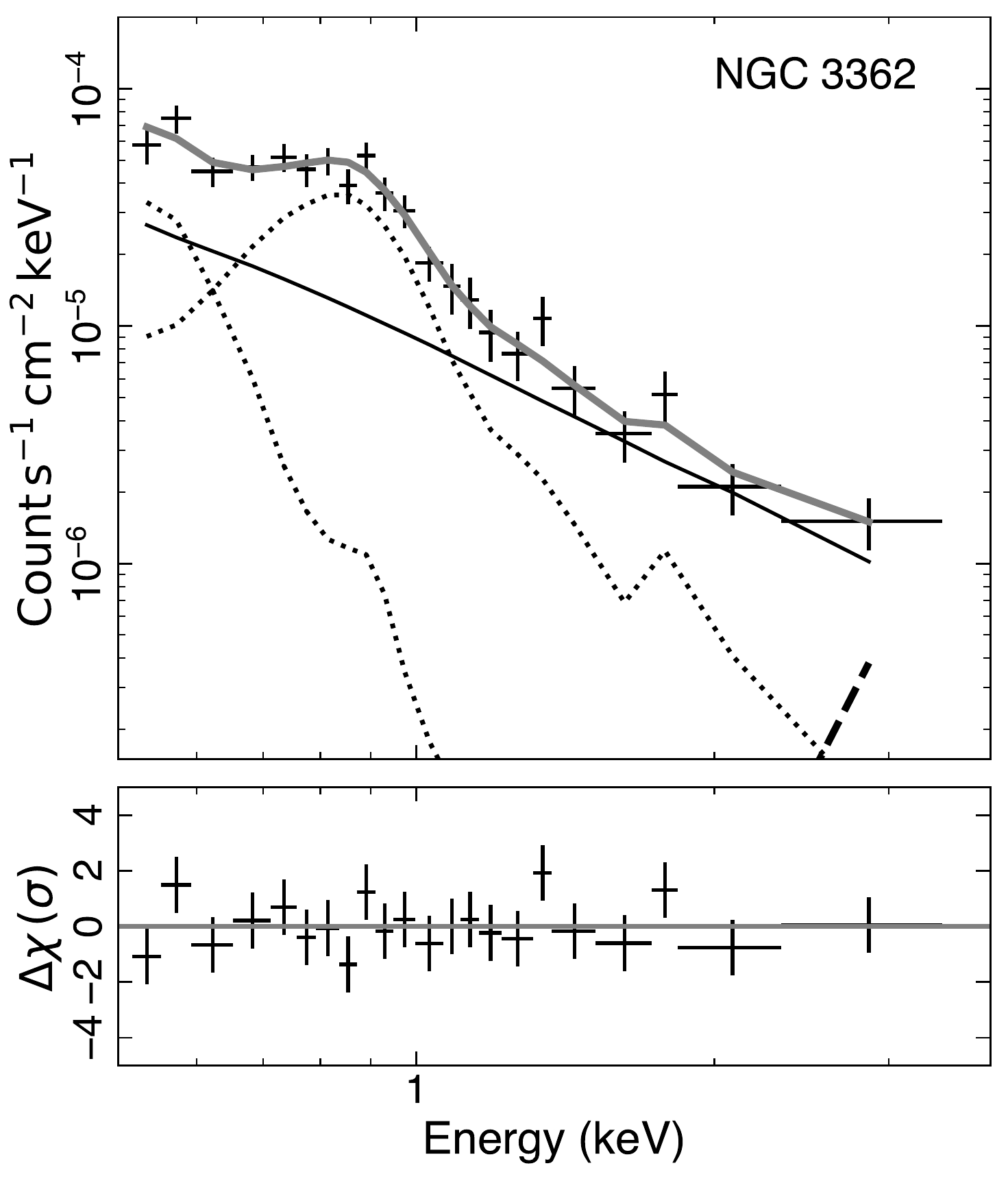}\\
	
	\includegraphics[width = 0.3\textwidth]{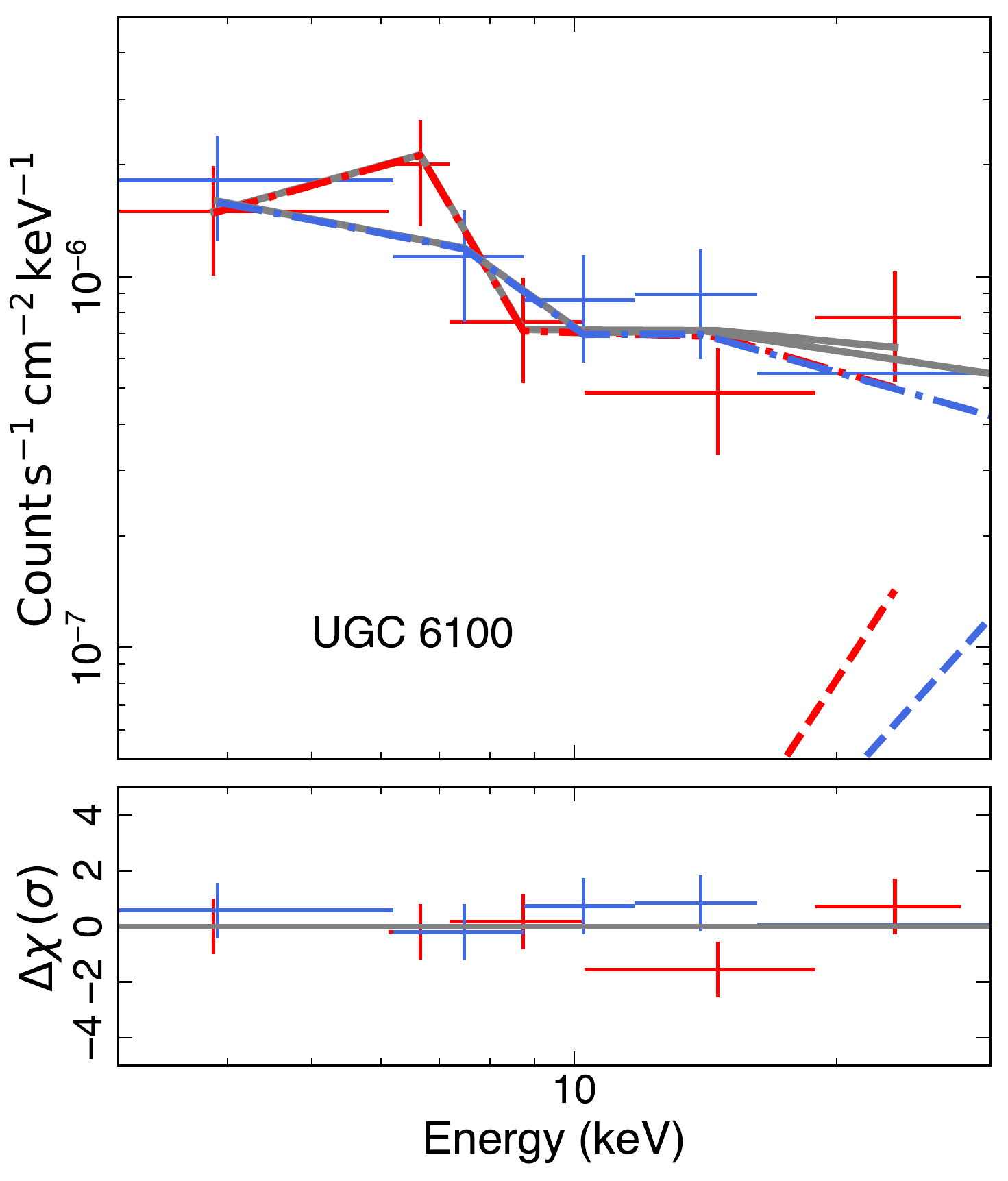}
    	\includegraphics[width = 0.3\textwidth]{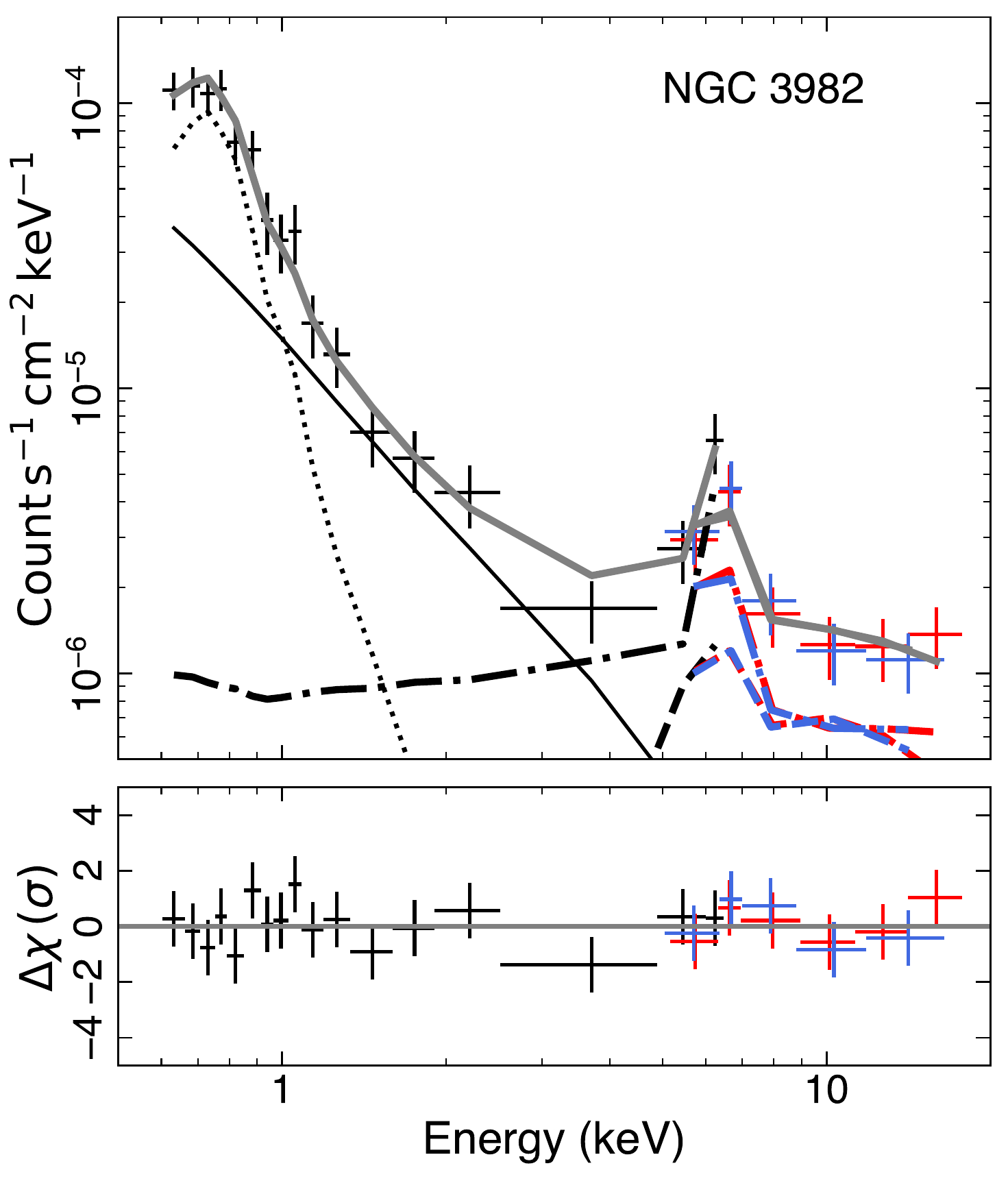}
	\includegraphics[width = 0.3\textwidth]{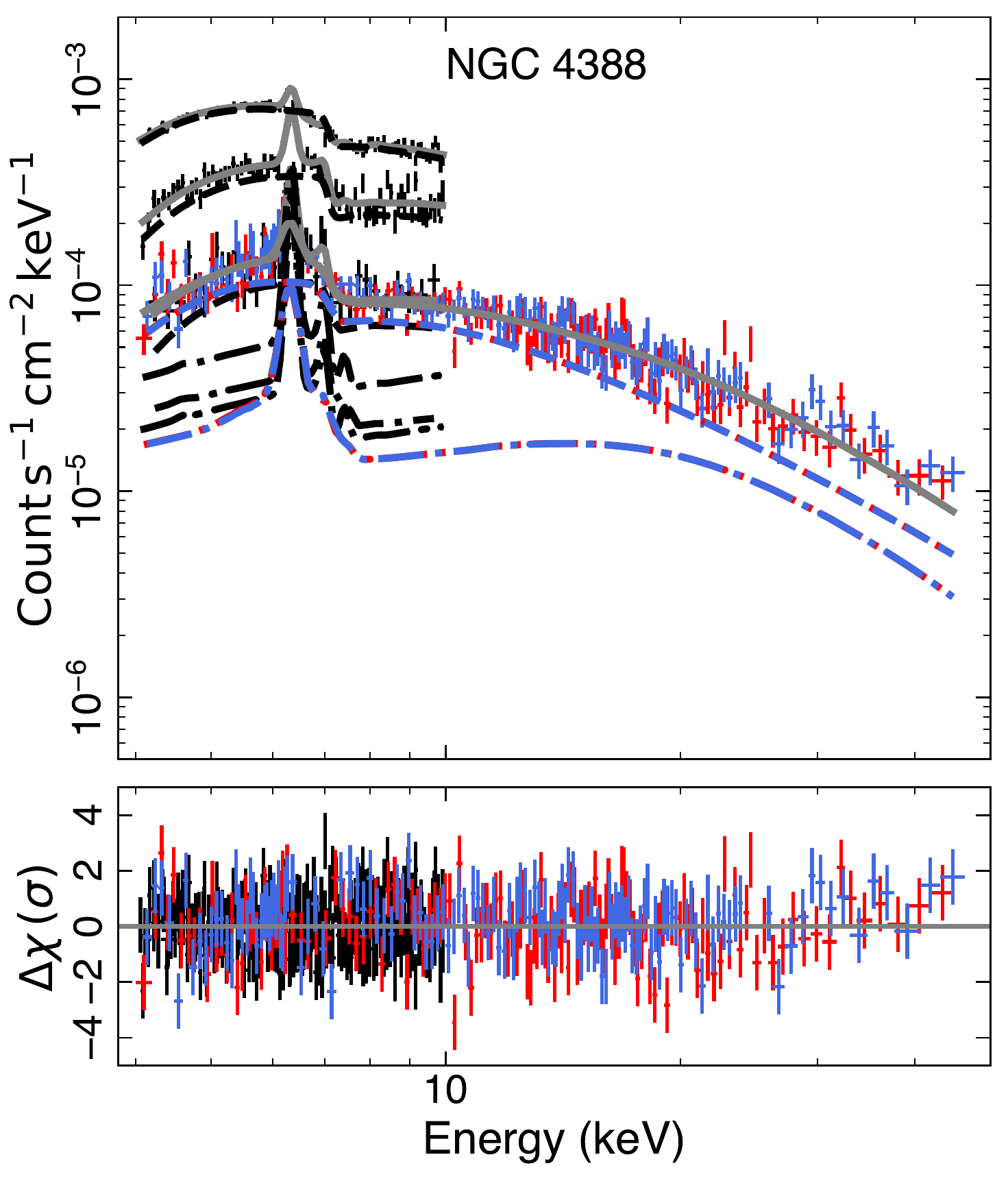}\\
	
	\includegraphics[width = 0.3\textwidth]{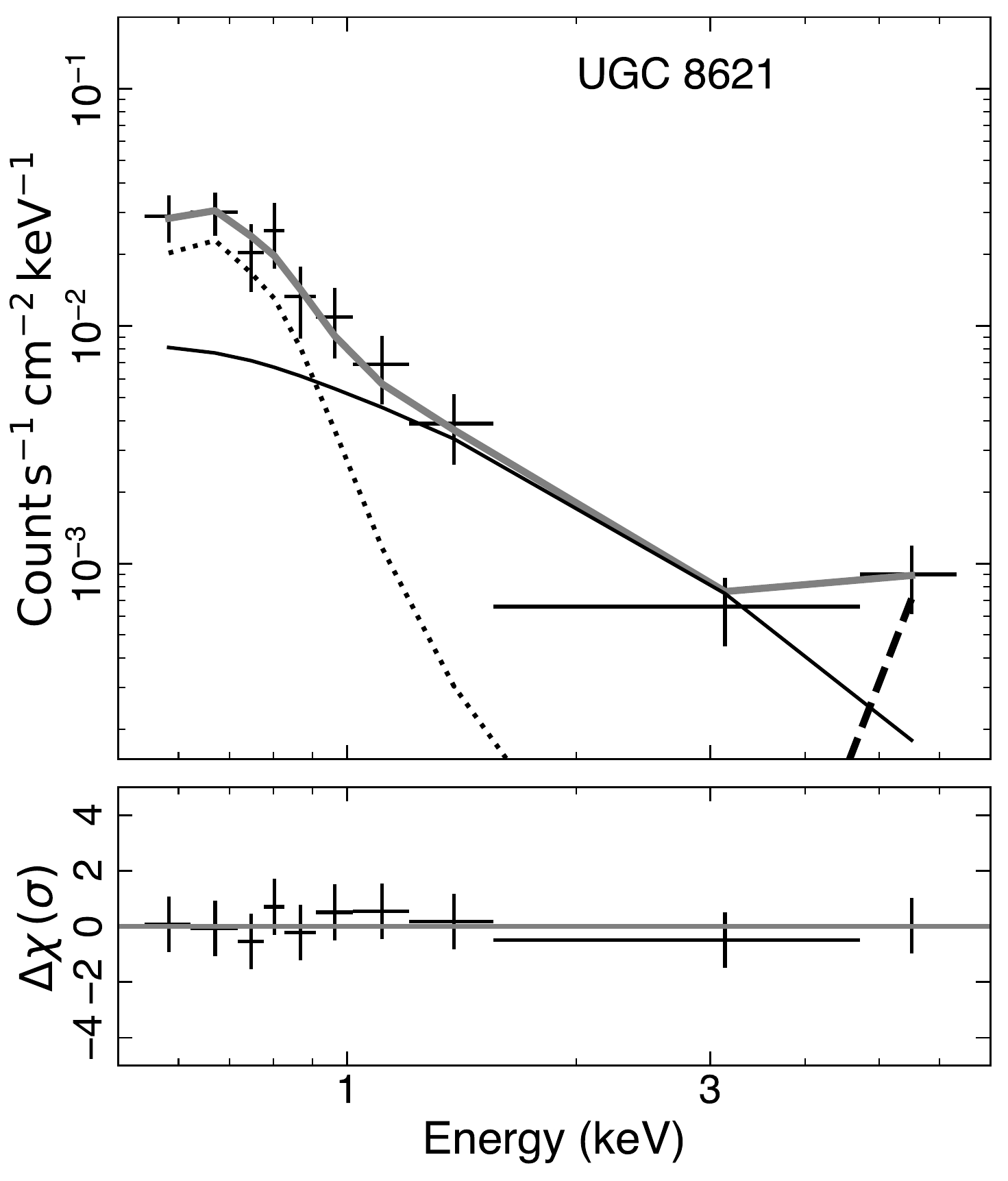}
    	\includegraphics[width = 0.3\textwidth]{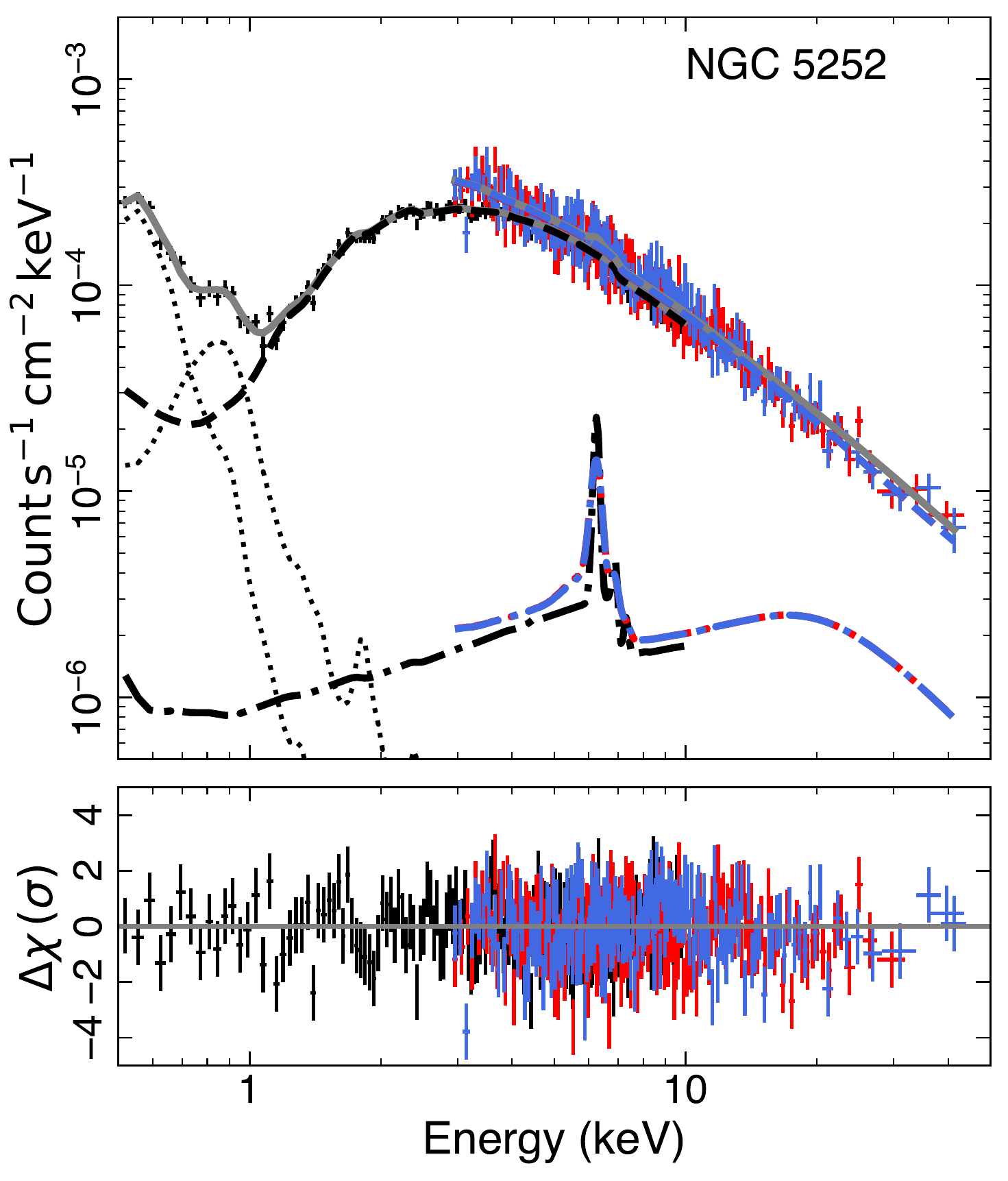}
	\includegraphics[width = 0.3\textwidth]{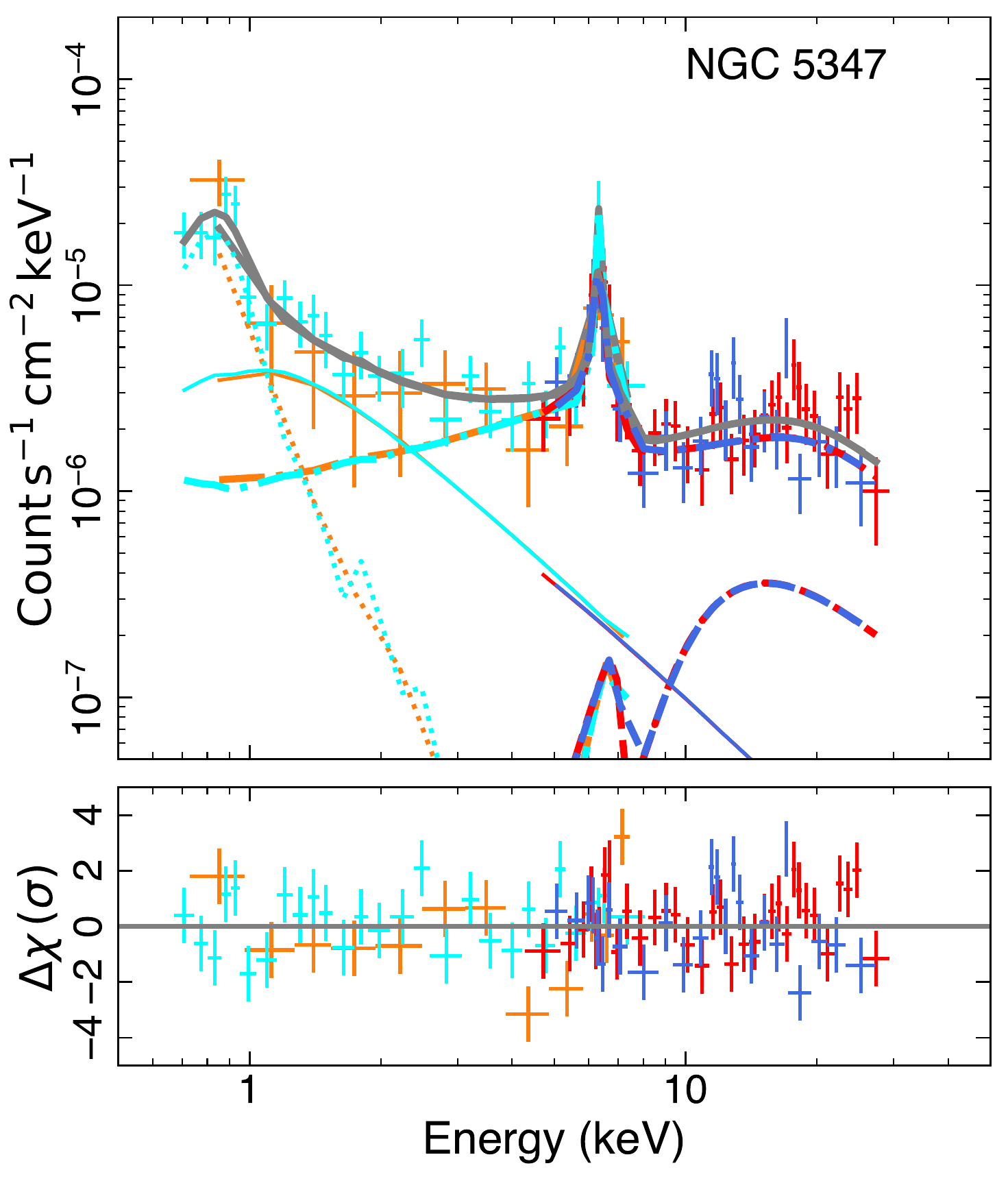}
\caption{The \nustar\ (red/blue for FPMA/FPMB), \xmm\ (black), \swift/XRT (green), \chandra\ (cyan), and \suzaku\ (orange) spectra of the 19 Seyfert galaxies studies in this work. The best-fit ``pexmon" model are shown in grey. We also show the different spectral components: primary power law (dashed lines), scattered power law (solid thin line), reflection (dash-dotted lines), {\tt `Apec'} and Gaussian lines (dotted lines). The lower panels show the corresponding residuals defined as $\Delta \chi = \rm (data - model)/\sigma$.}
\label{fig:spectra1}
\end{figure*}

\begin{figure*}
\centering

     \includegraphics[width = 0.28\textwidth]{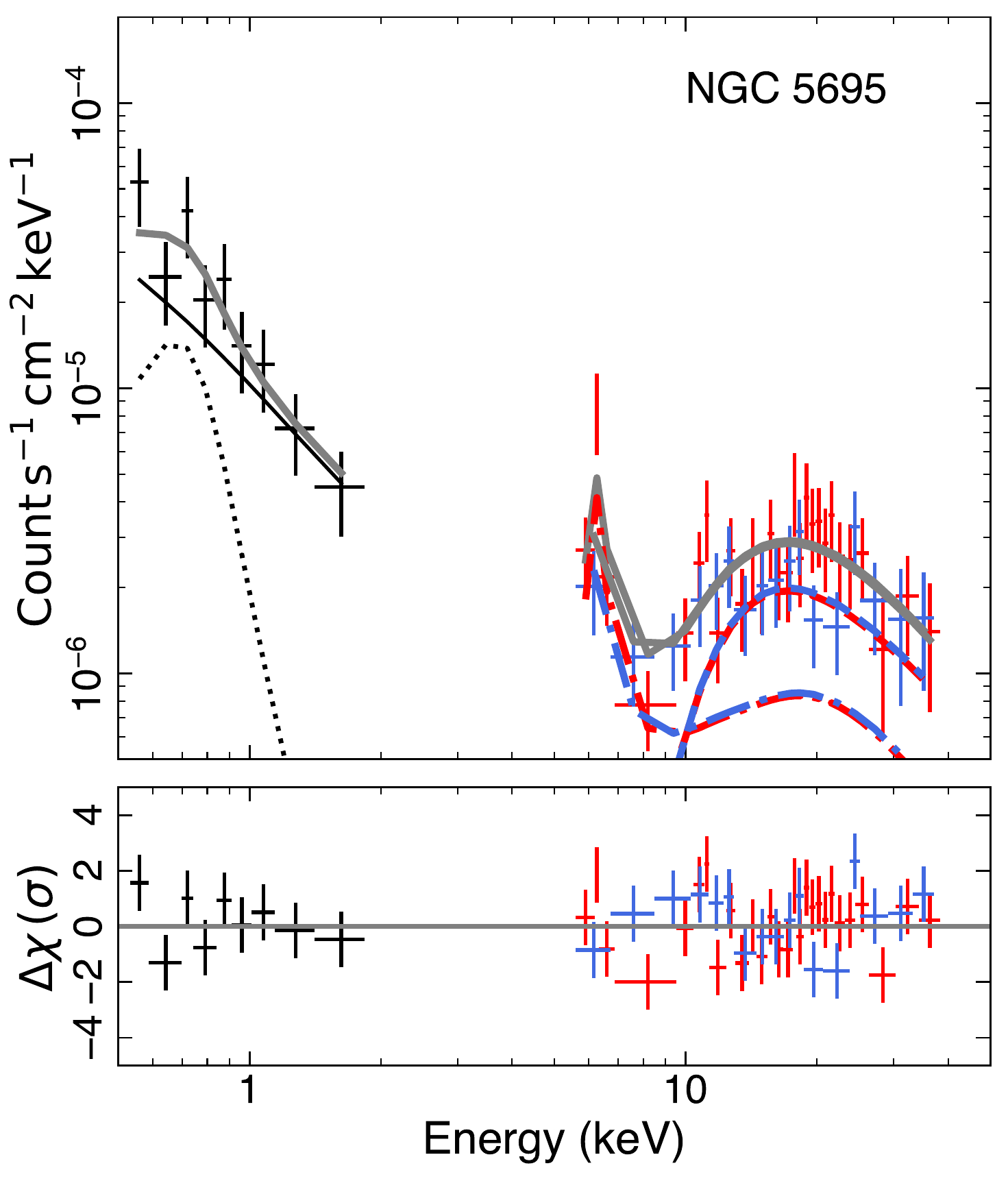}
    	\includegraphics[width = 0.28\textwidth]{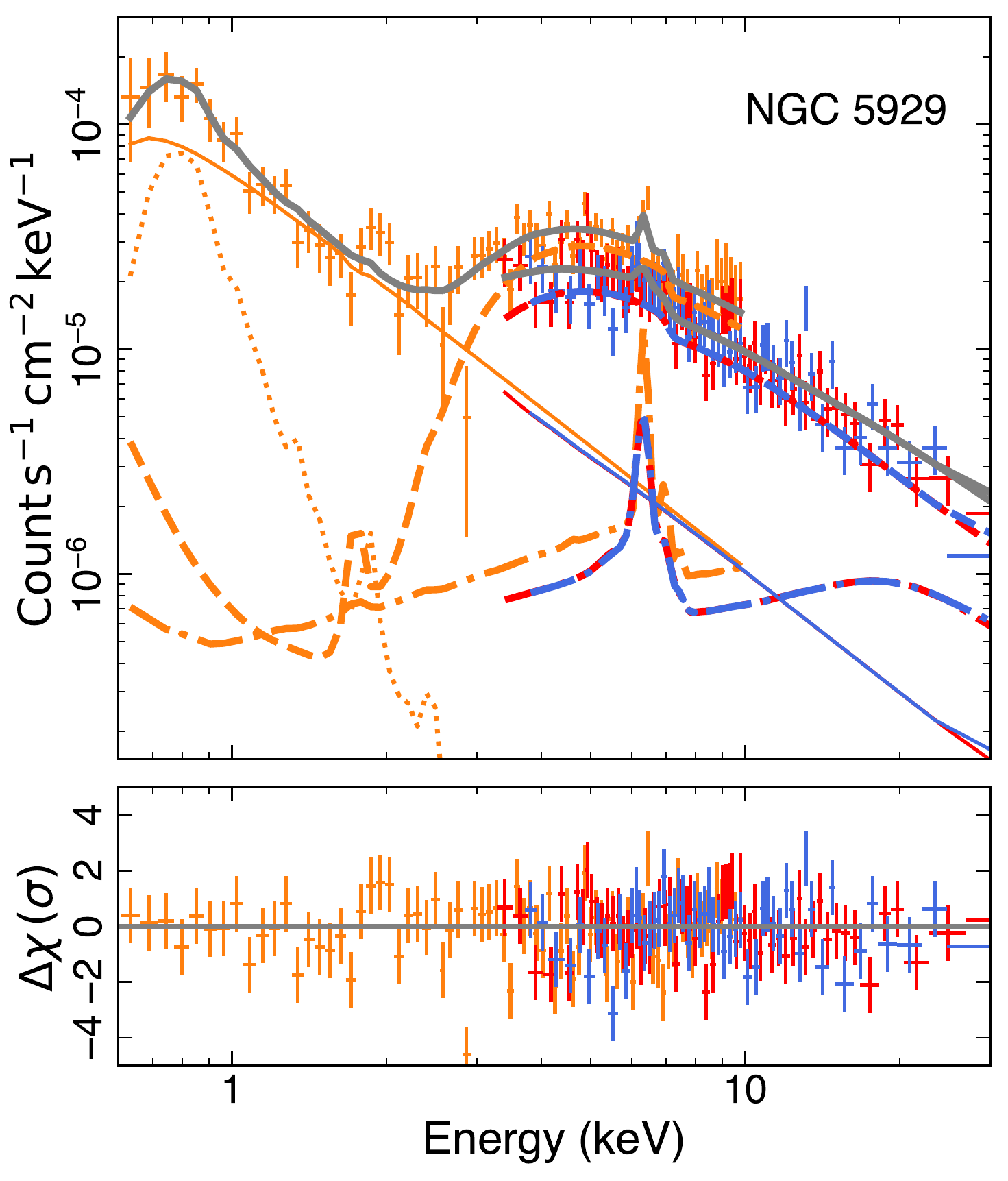}
	\includegraphics[width = 0.28\textwidth]{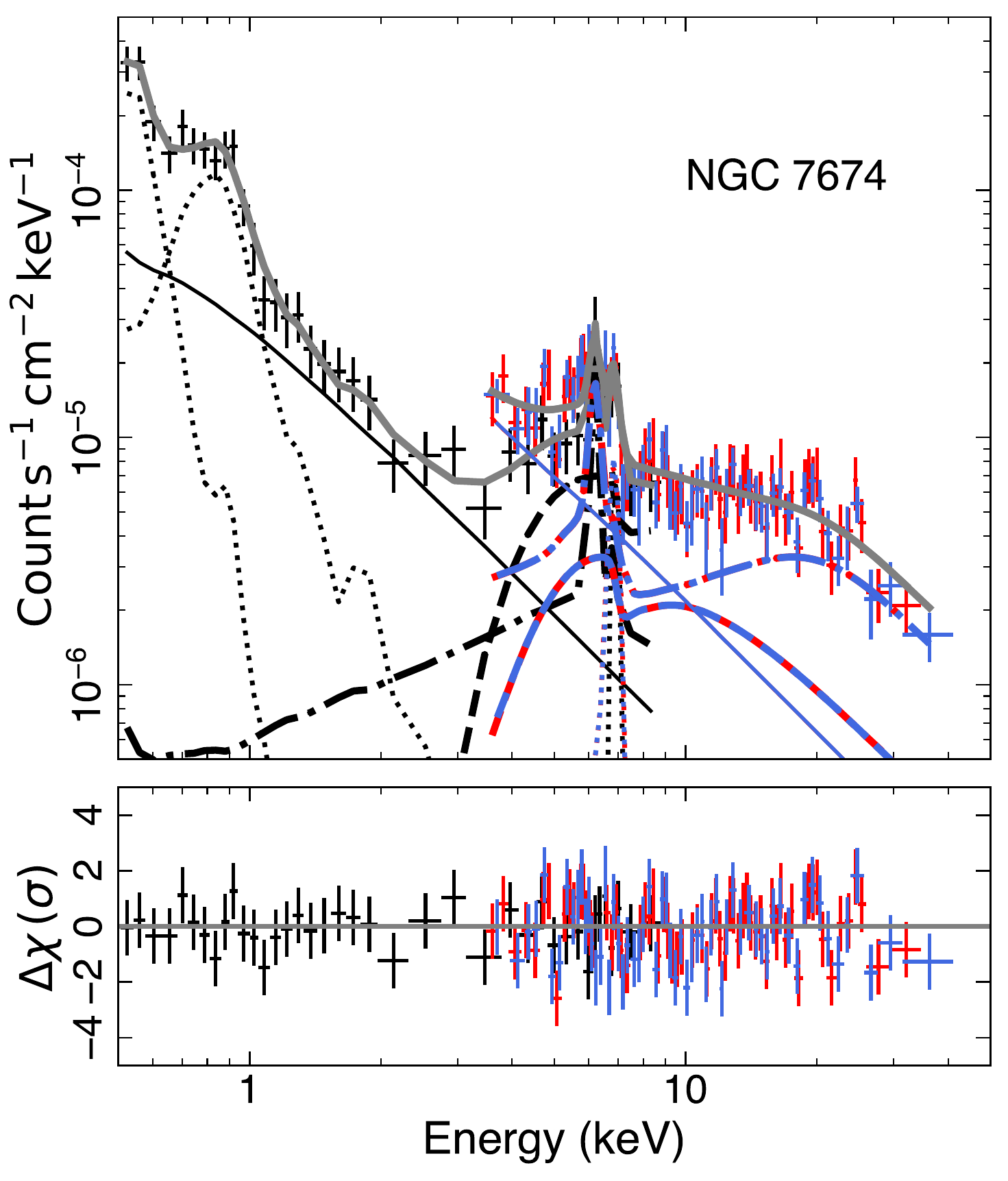}\\
	
	\includegraphics[width = 0.28\textwidth]{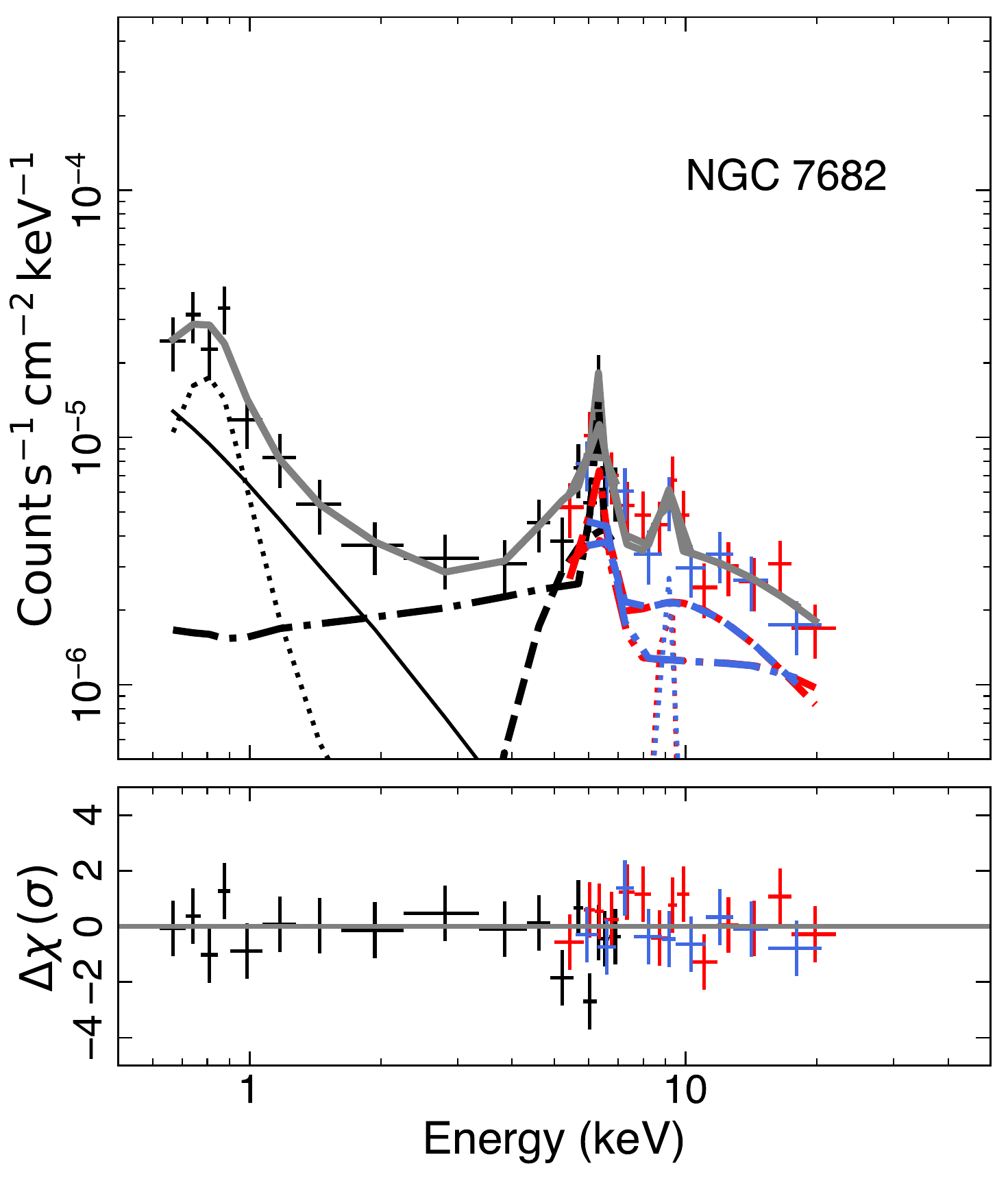}
    	\includegraphics[width = 0.28\textwidth]{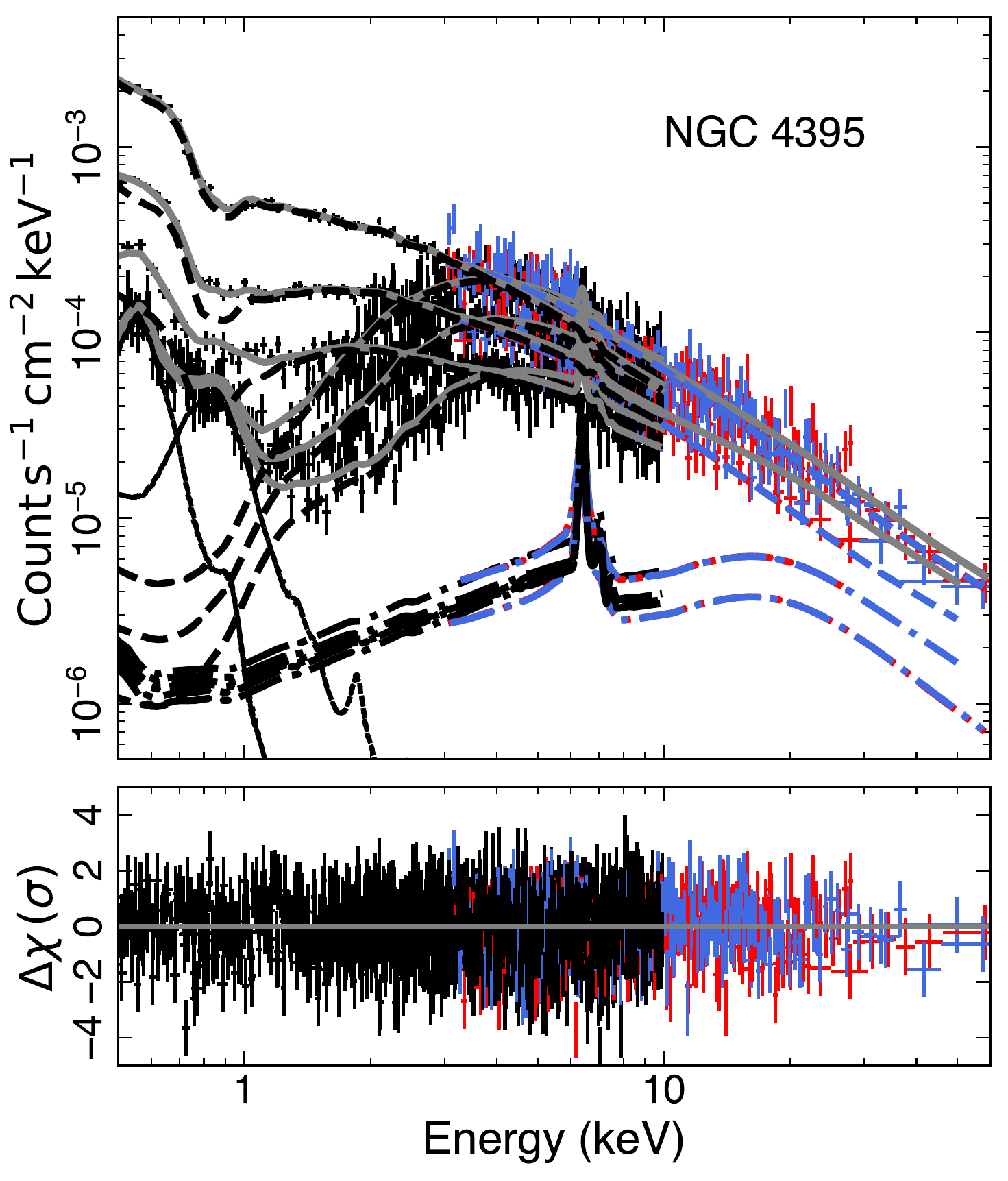}
	\includegraphics[width = 0.28\textwidth]{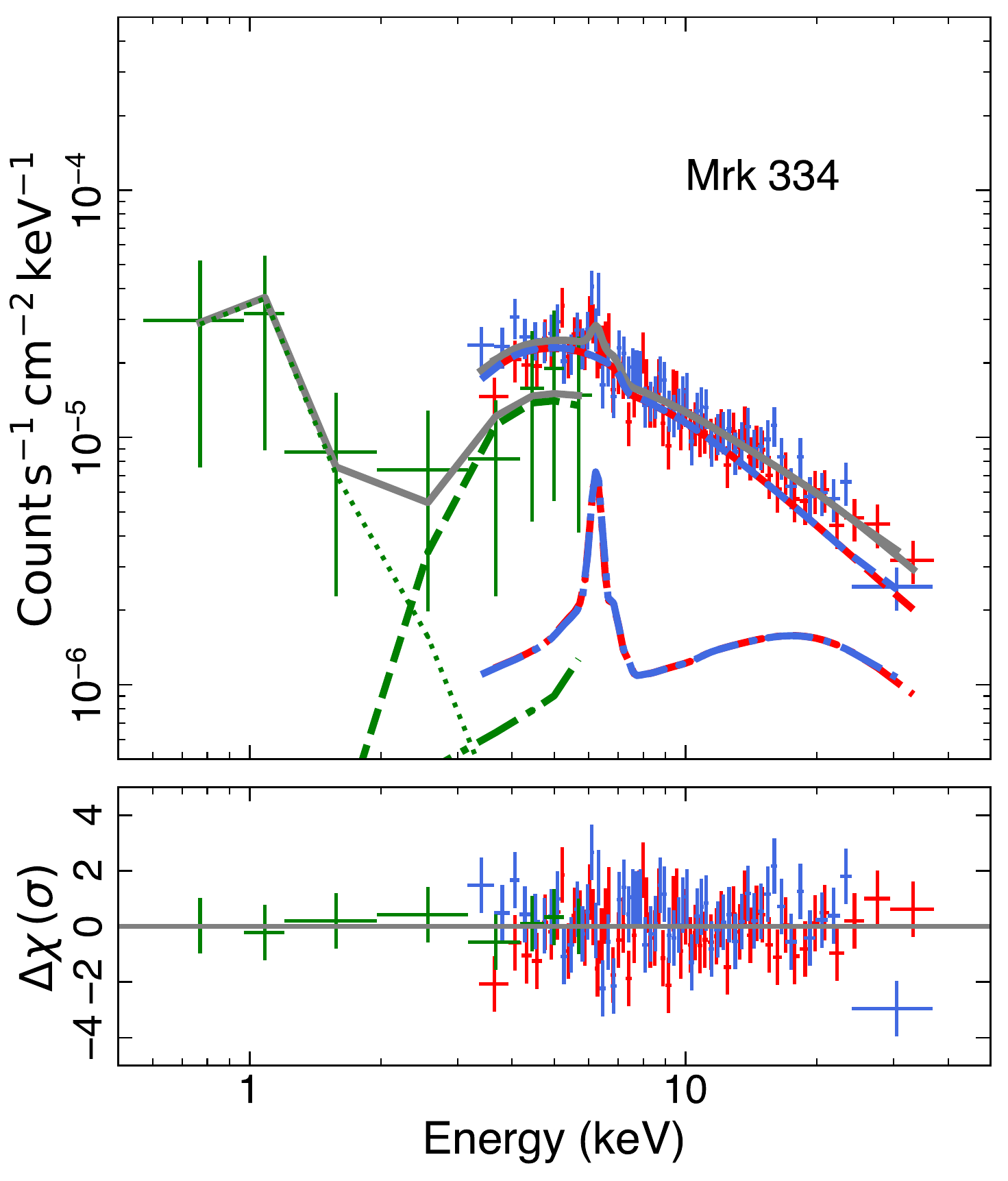}\\

	\includegraphics[width = 0.28\textwidth]{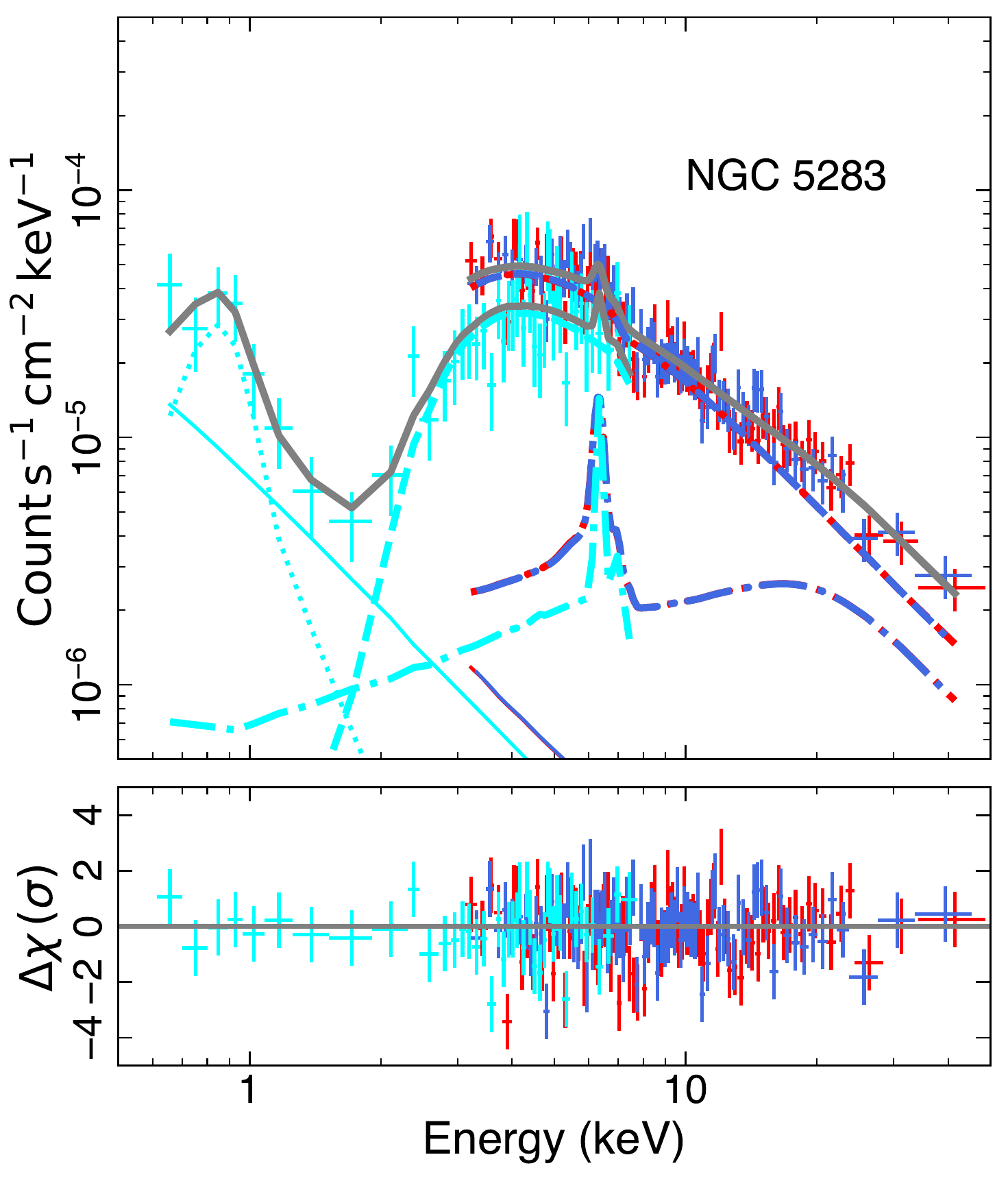}
    	\includegraphics[width = 0.28\textwidth]{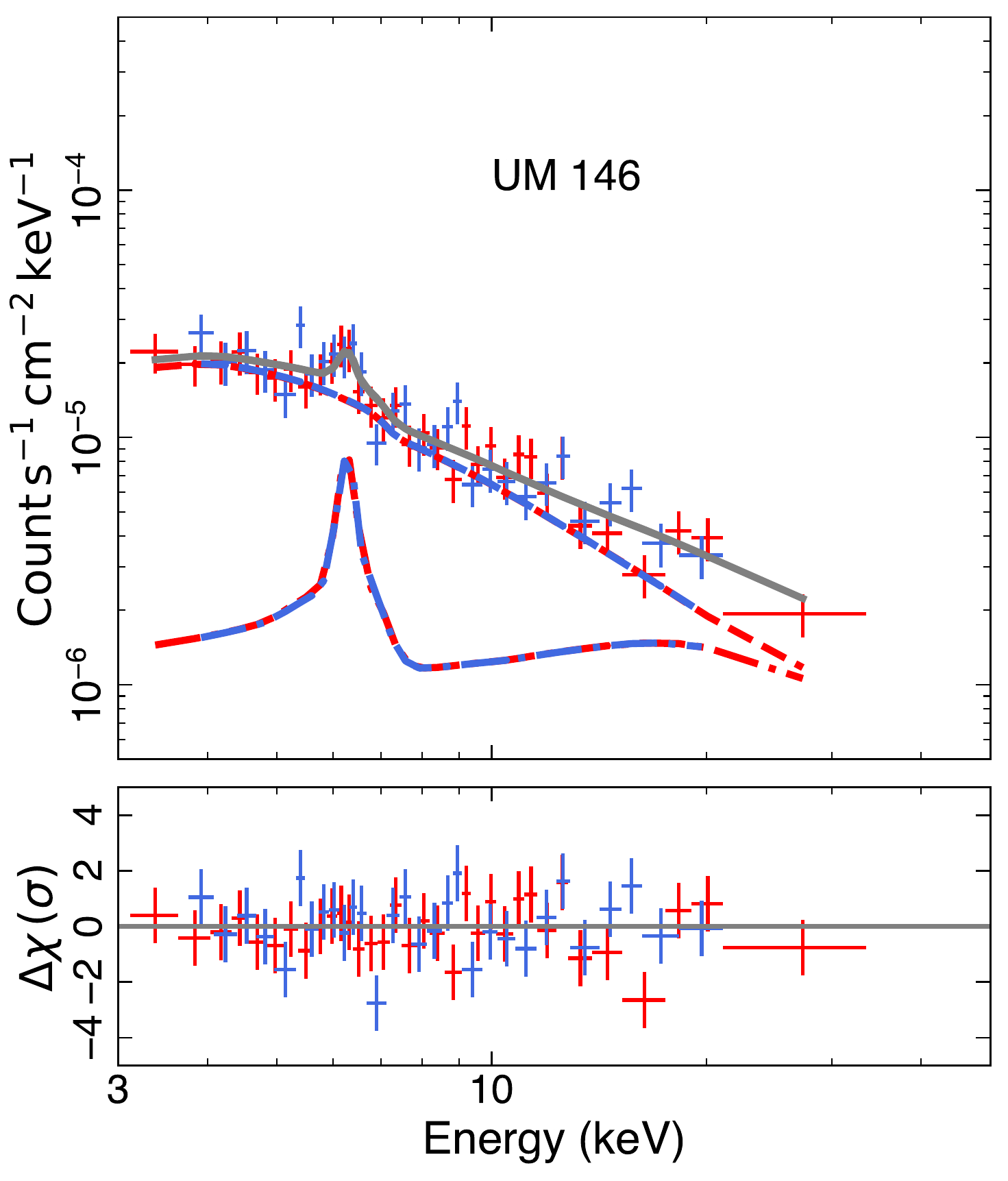}
	\includegraphics[width = 0.28\textwidth]{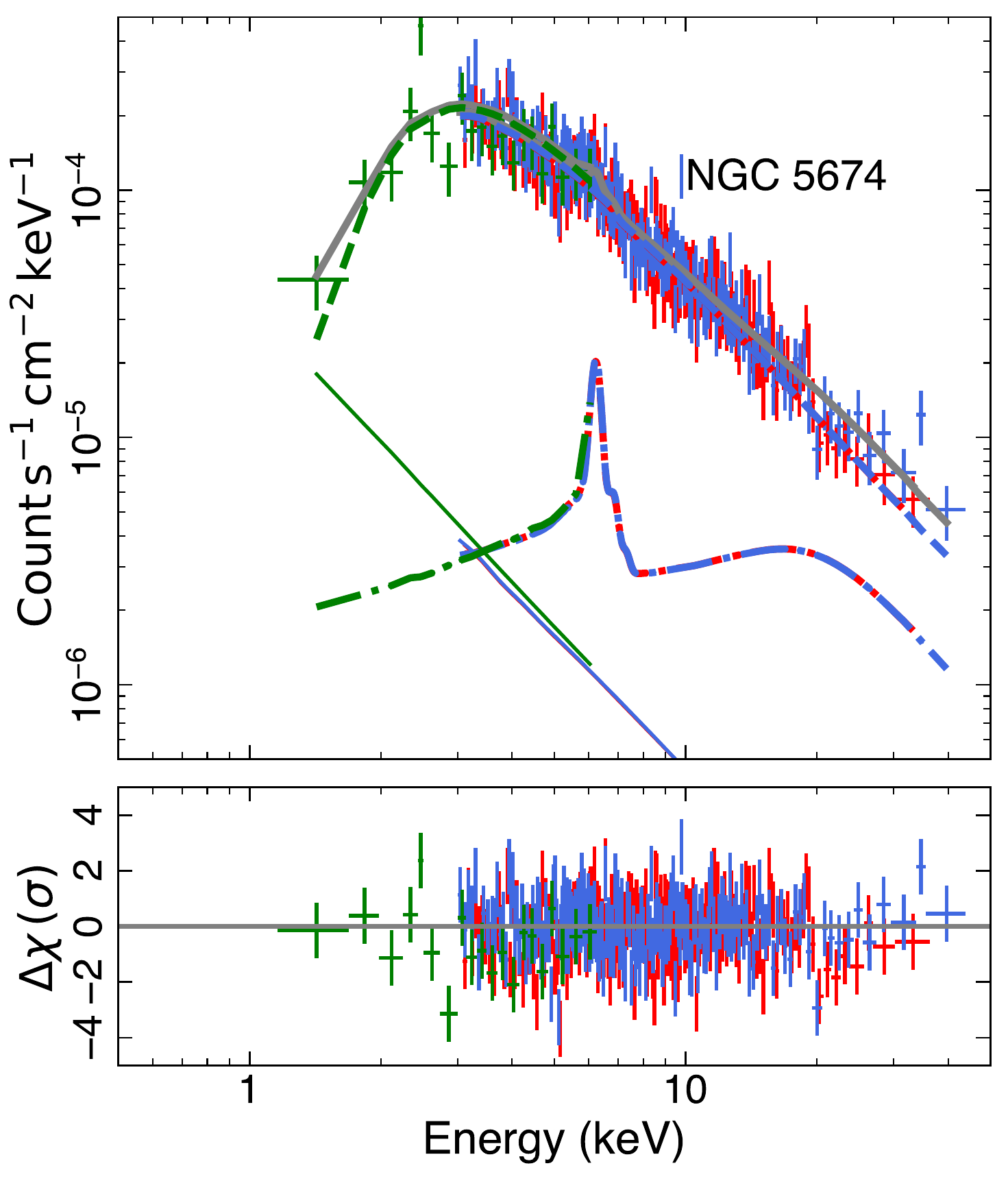}\\
	
    	\includegraphics[width = 0.28\textwidth]{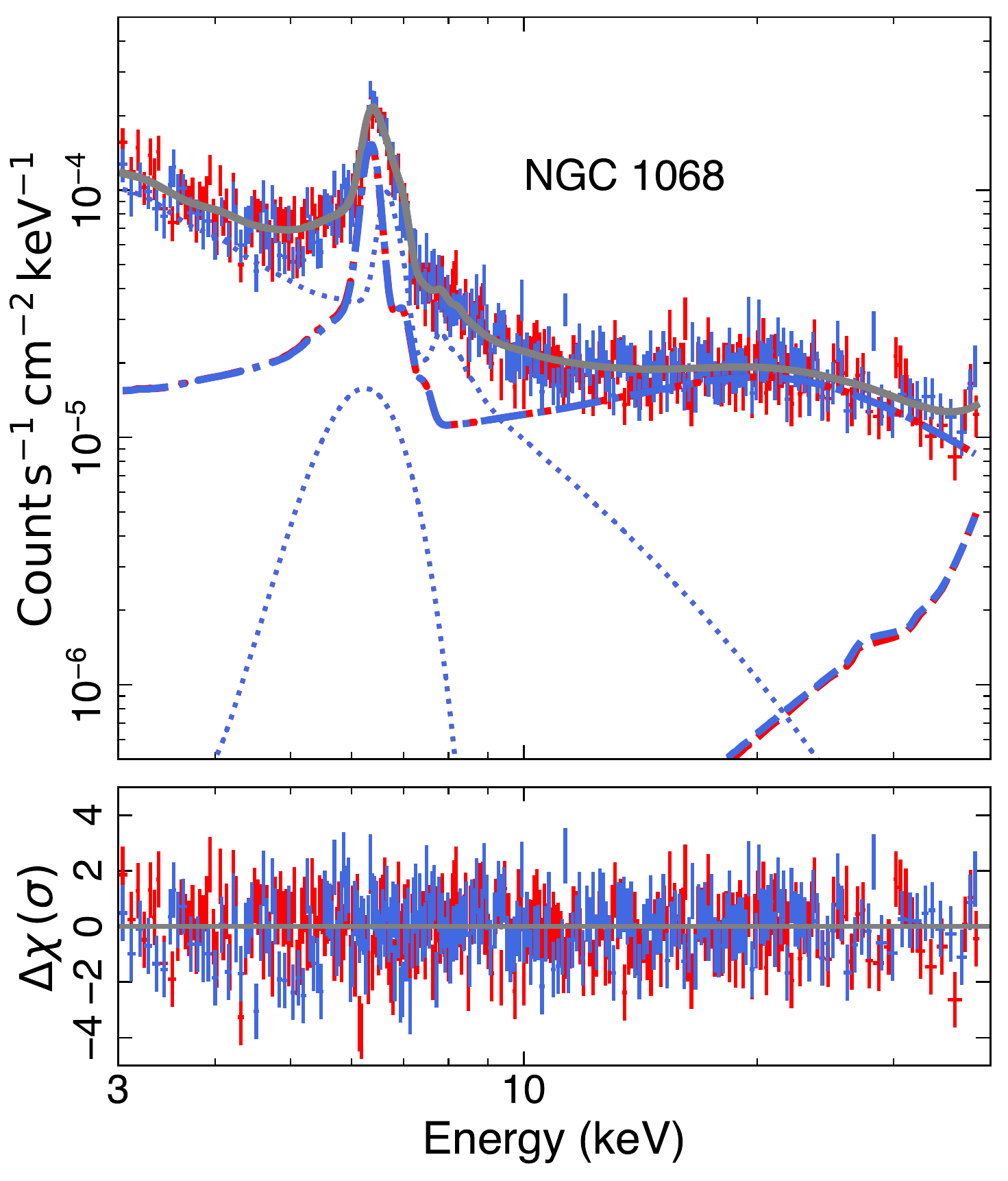}
\caption{Same as Figure~\ref{fig:spectra1}. We note that the gap seen in the  $\sim 2-4$~keV range of the NGC~5695 spectra is due to background domination.}
\label{fig:spectra2}
\end{figure*}

\begin{figure*}
\centering
	\includegraphics[width = 0.24\textwidth]{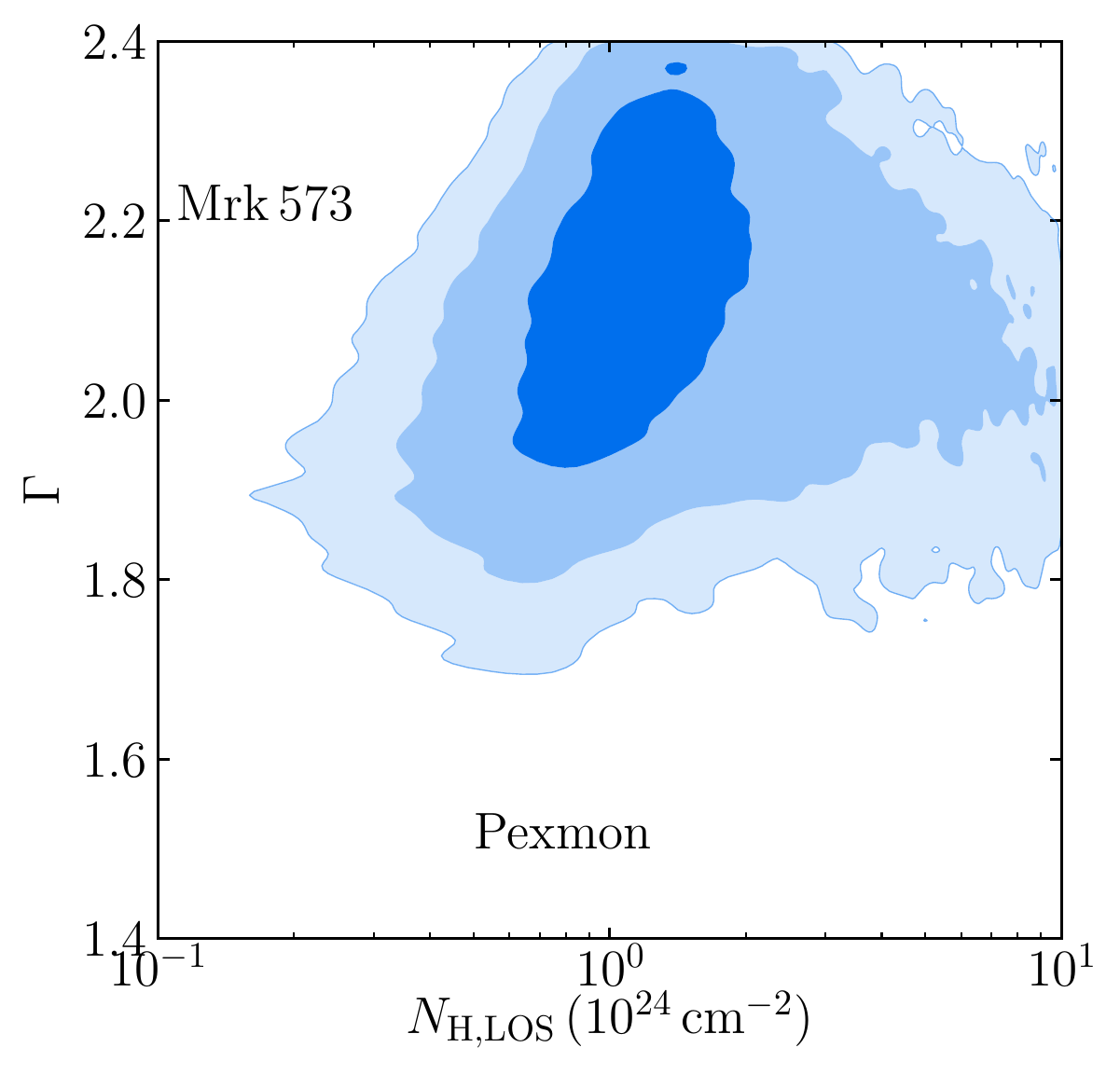}
	\includegraphics[width = 0.24\textwidth]{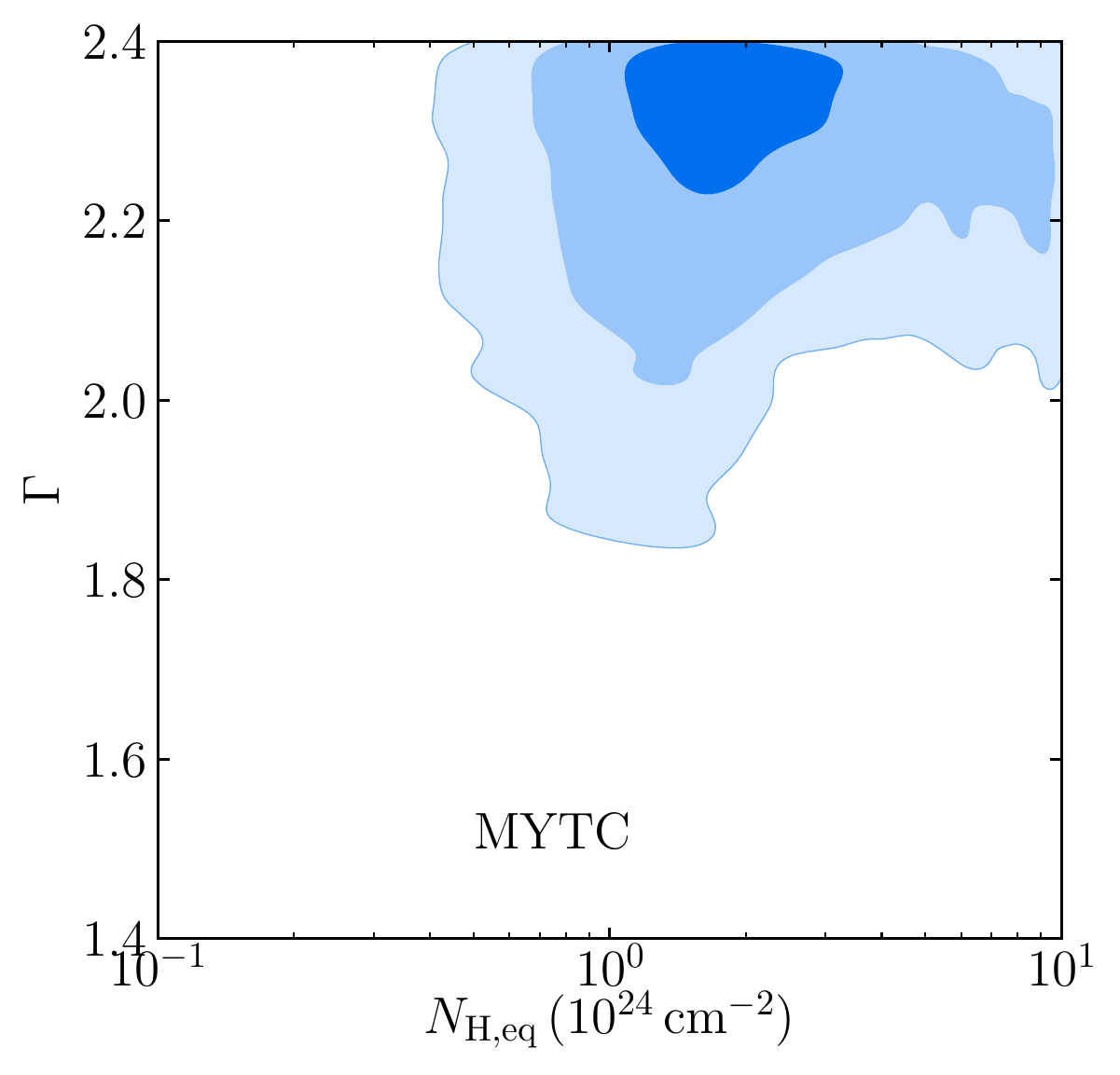}
	\includegraphics[width = 0.24\textwidth]{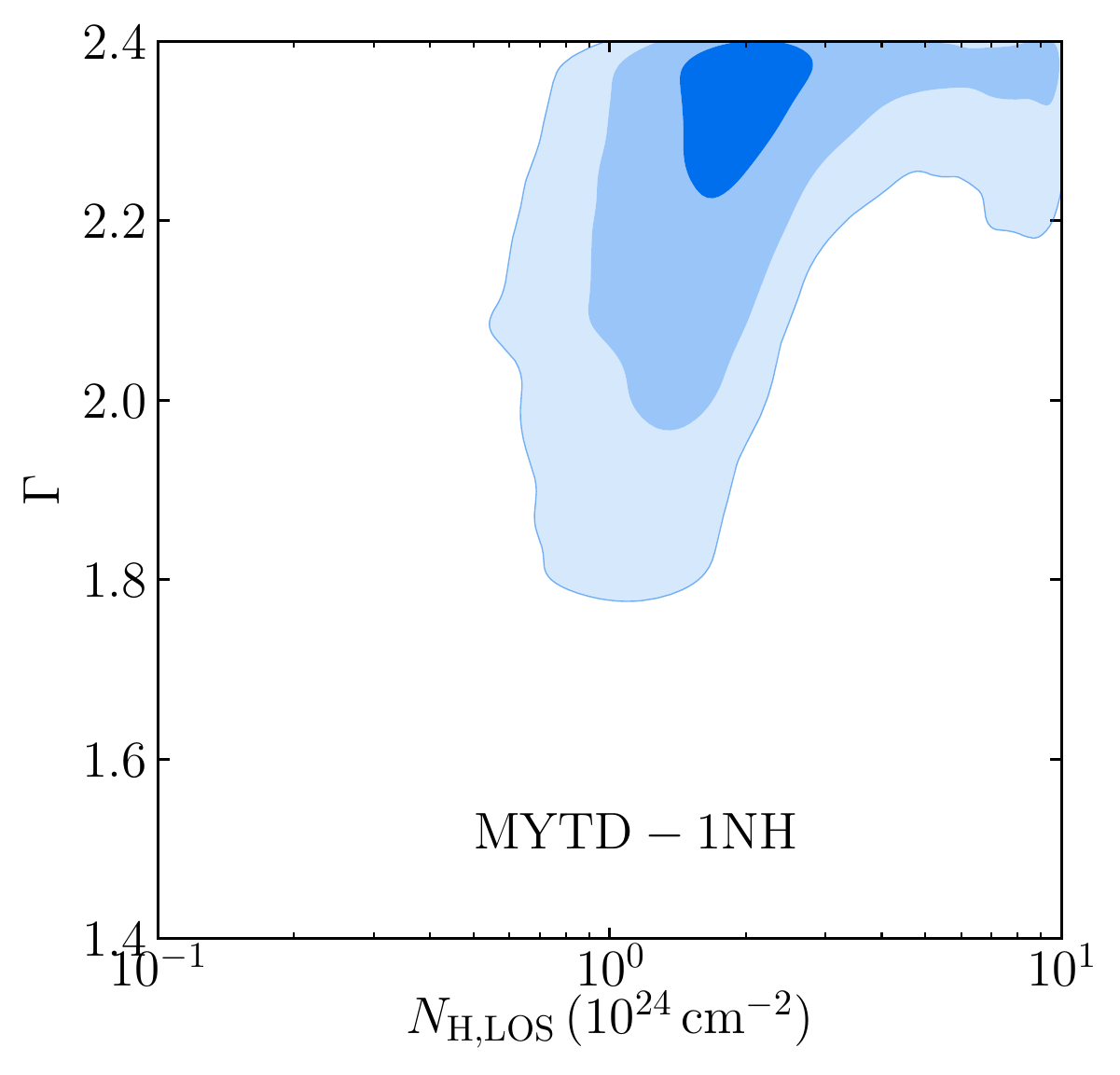}
	\includegraphics[width = 0.24\textwidth]{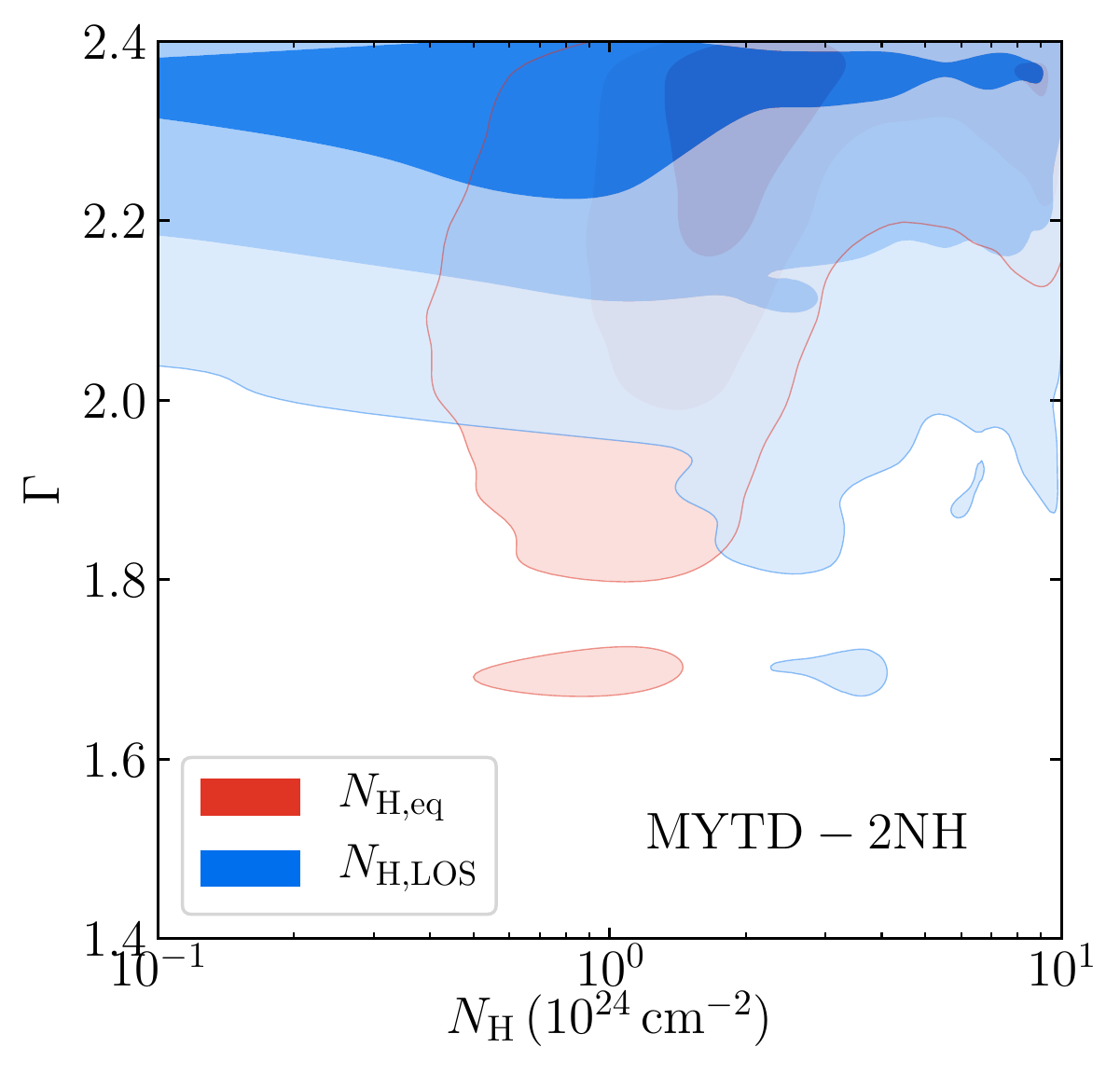}\\
	
	\includegraphics[width = 0.24\textwidth]{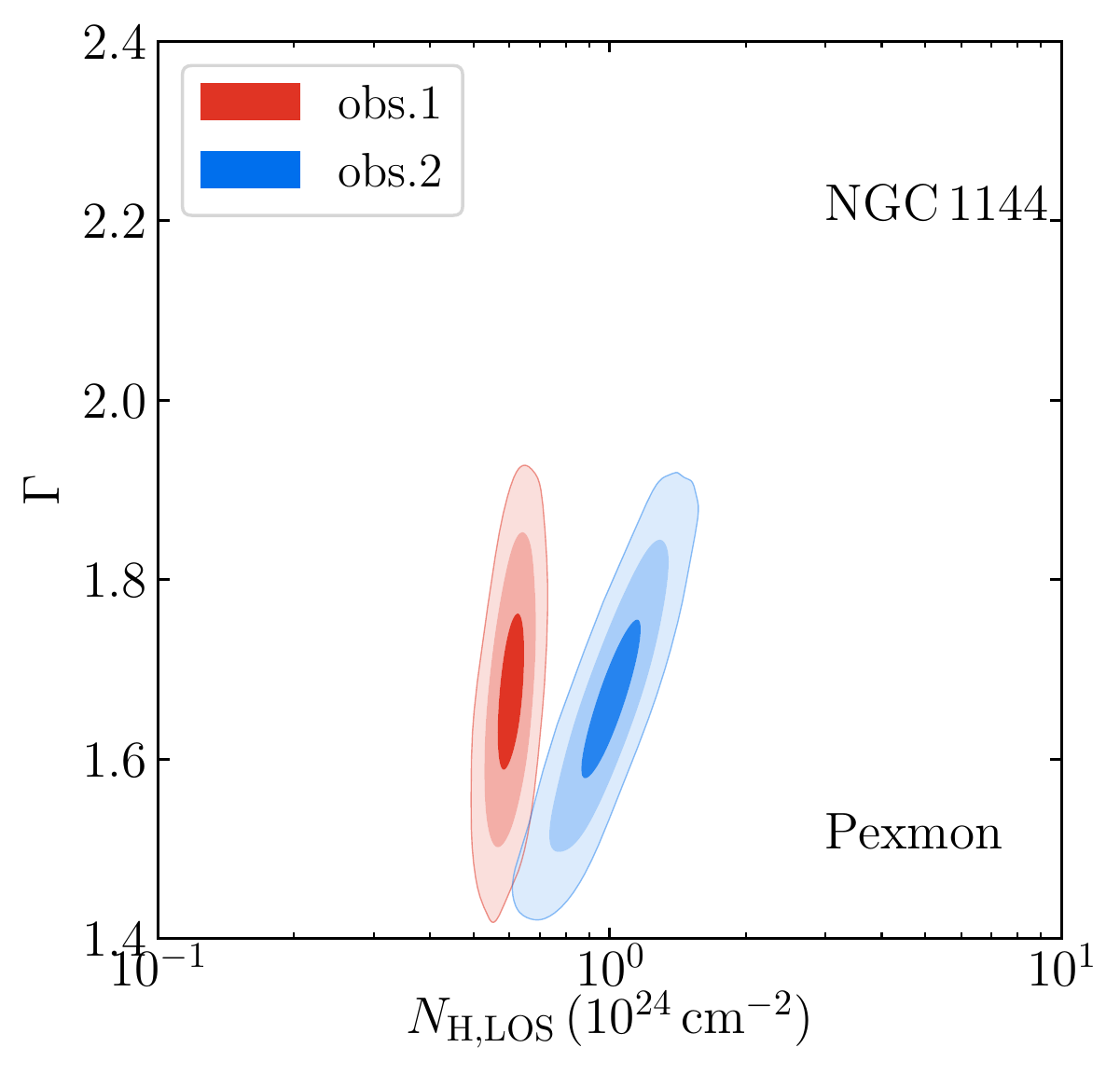}
	\includegraphics[width = 0.24\textwidth]{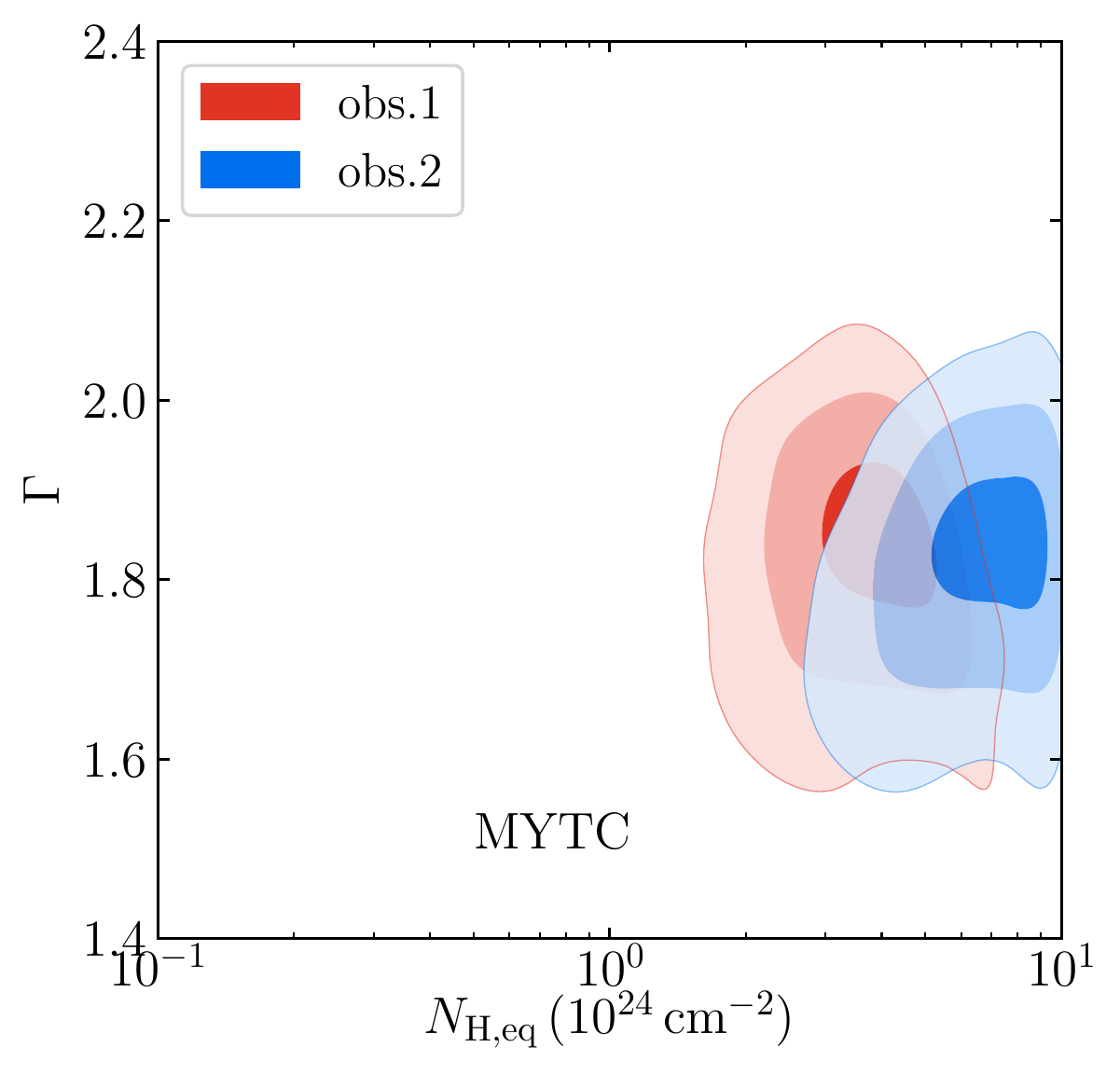}
	\includegraphics[width = 0.24\textwidth]{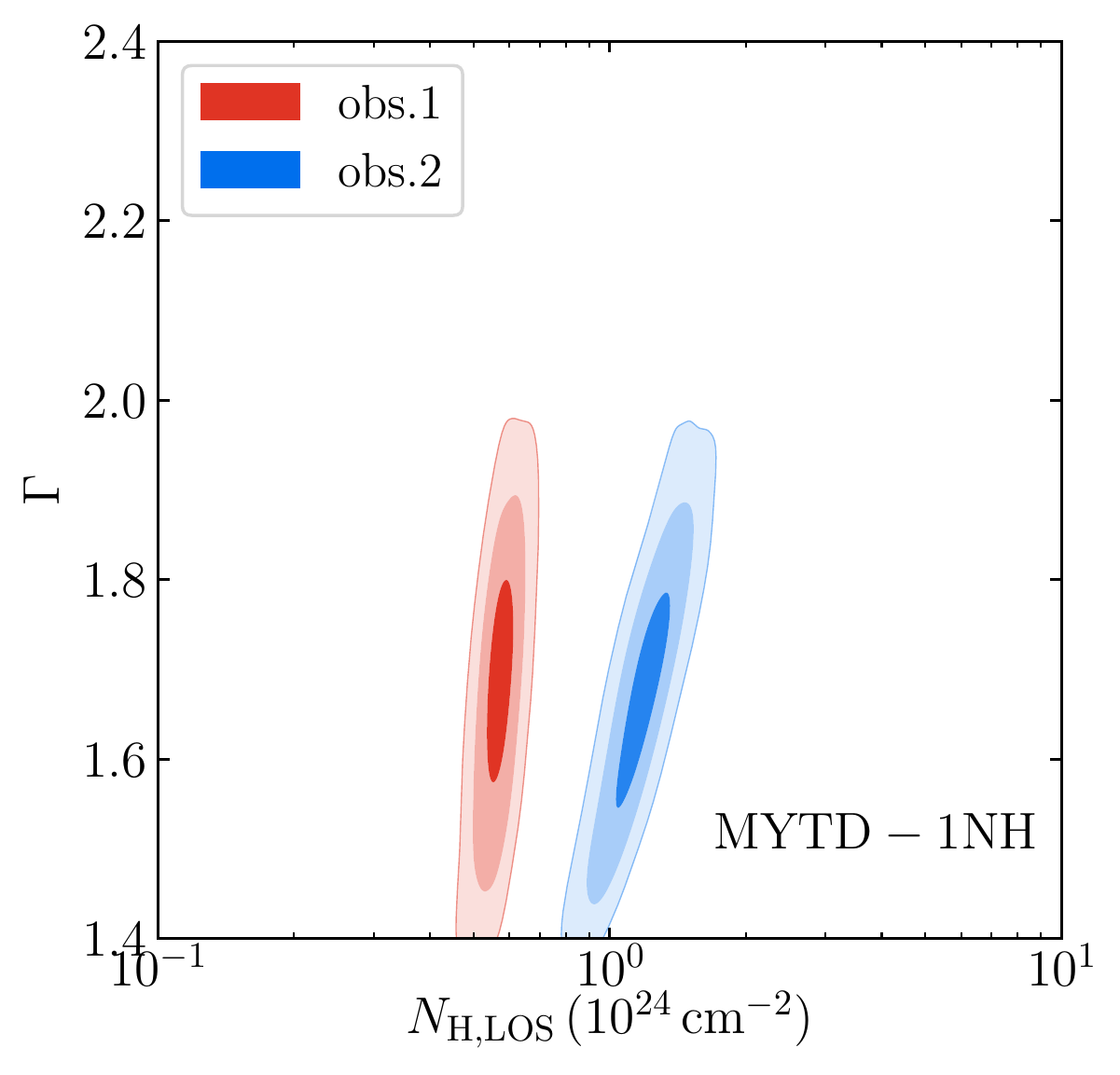}
	\includegraphics[width = 0.24\textwidth]{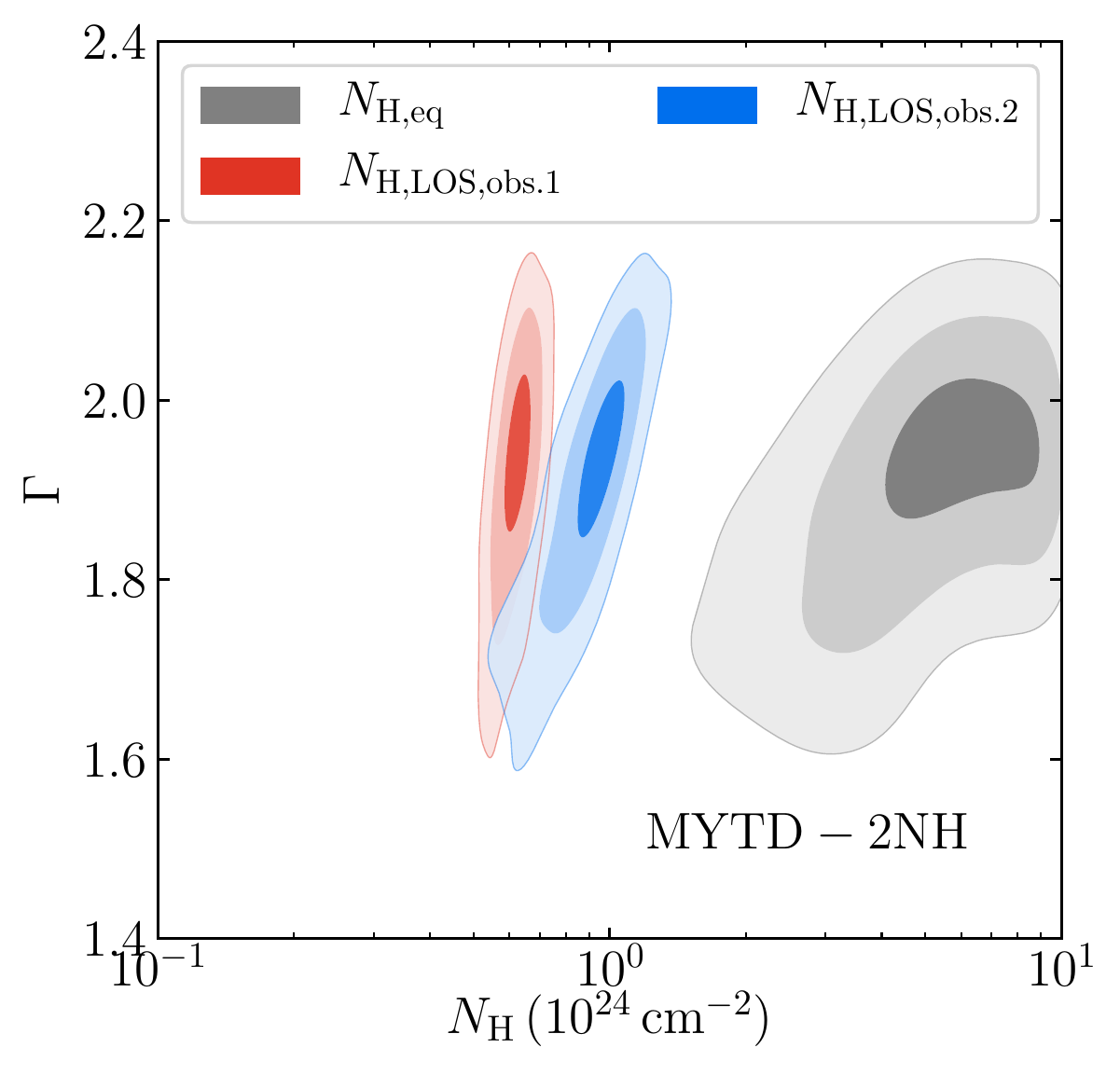}\\
	
	\includegraphics[width = 0.24\textwidth]{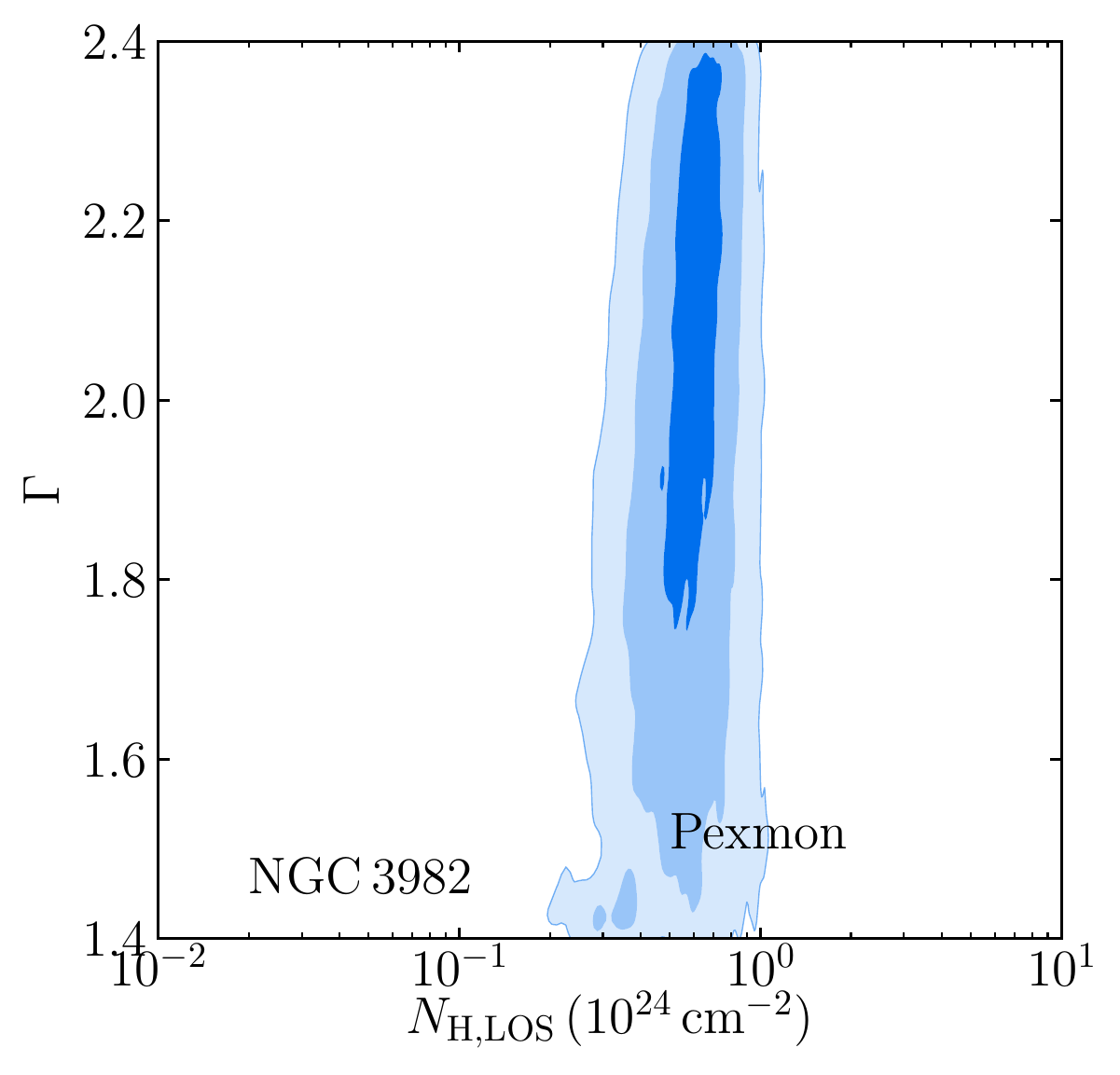}
	\includegraphics[width = 0.24\textwidth]{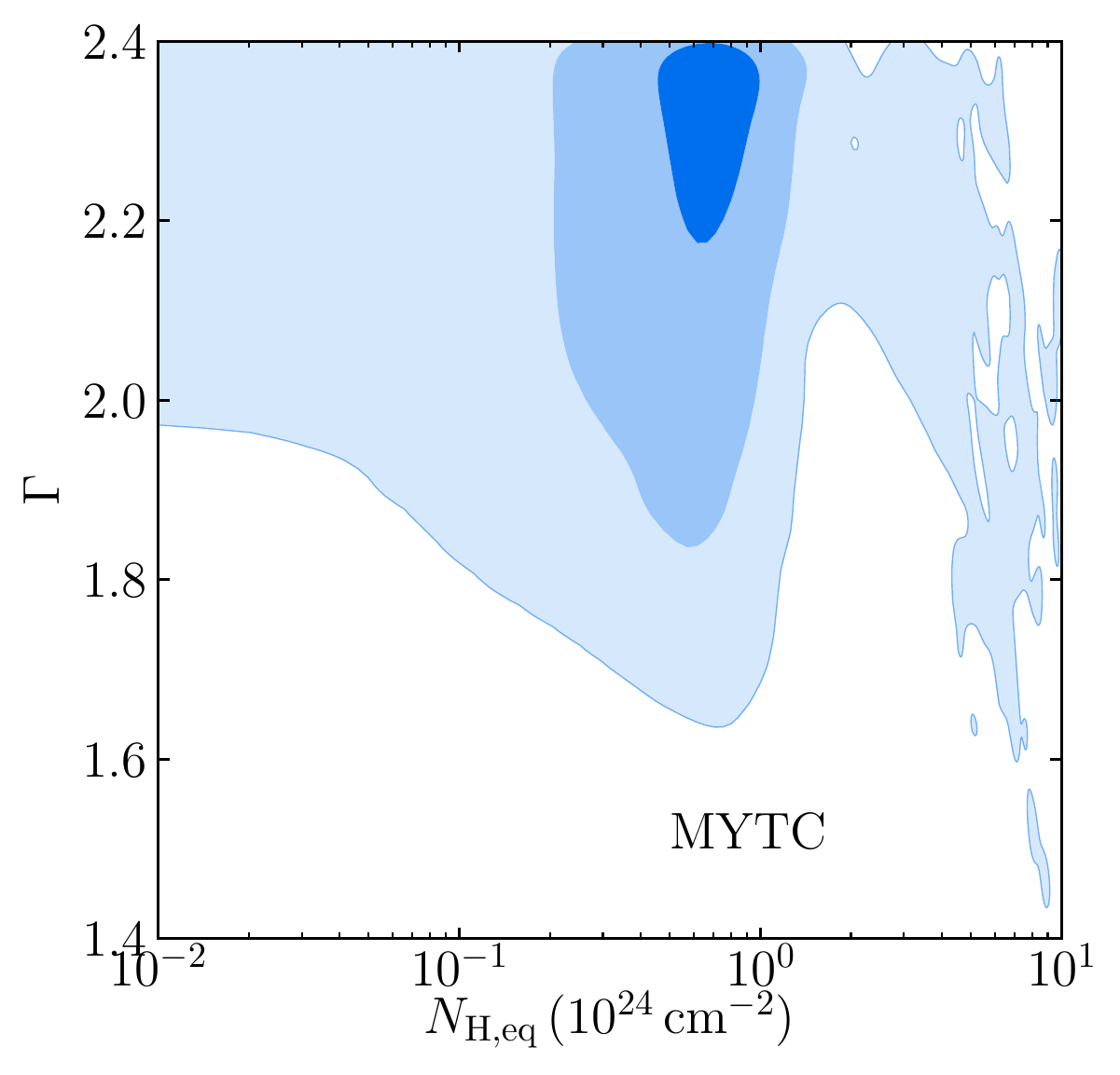}
	\includegraphics[width = 0.24\textwidth]{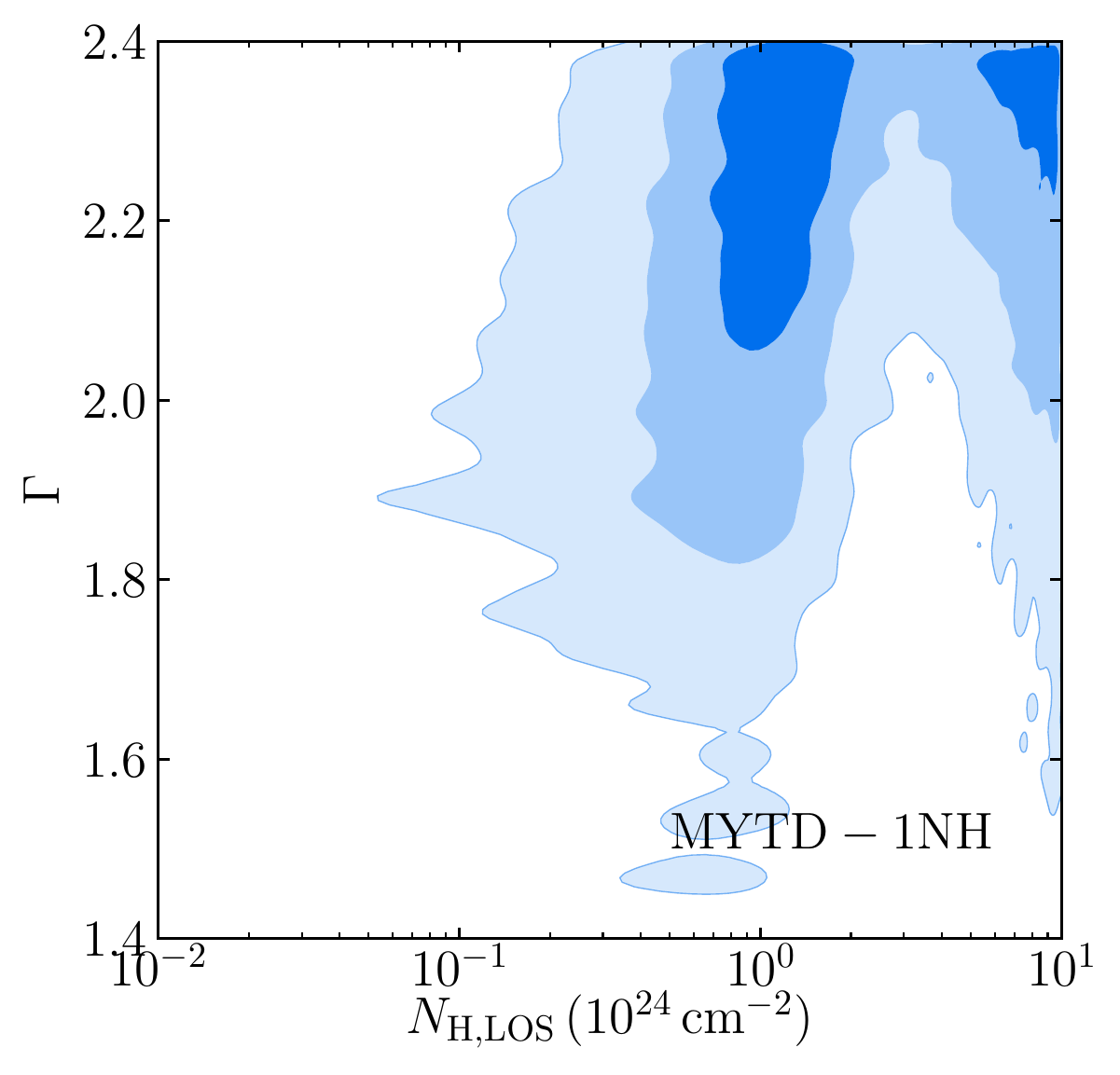}
	\includegraphics[width = 0.24\textwidth]{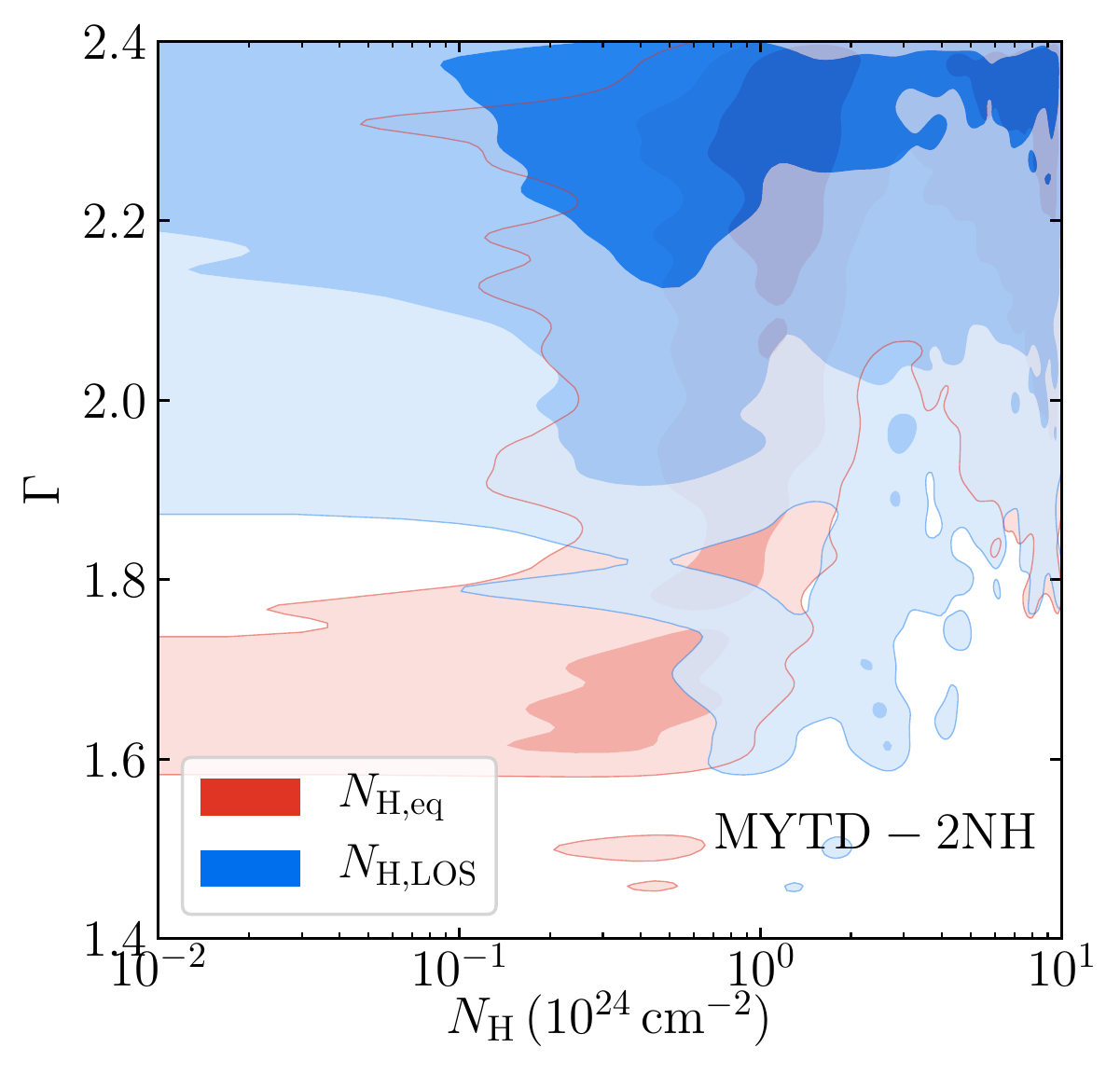}\\	

	\includegraphics[width = 0.24\textwidth]{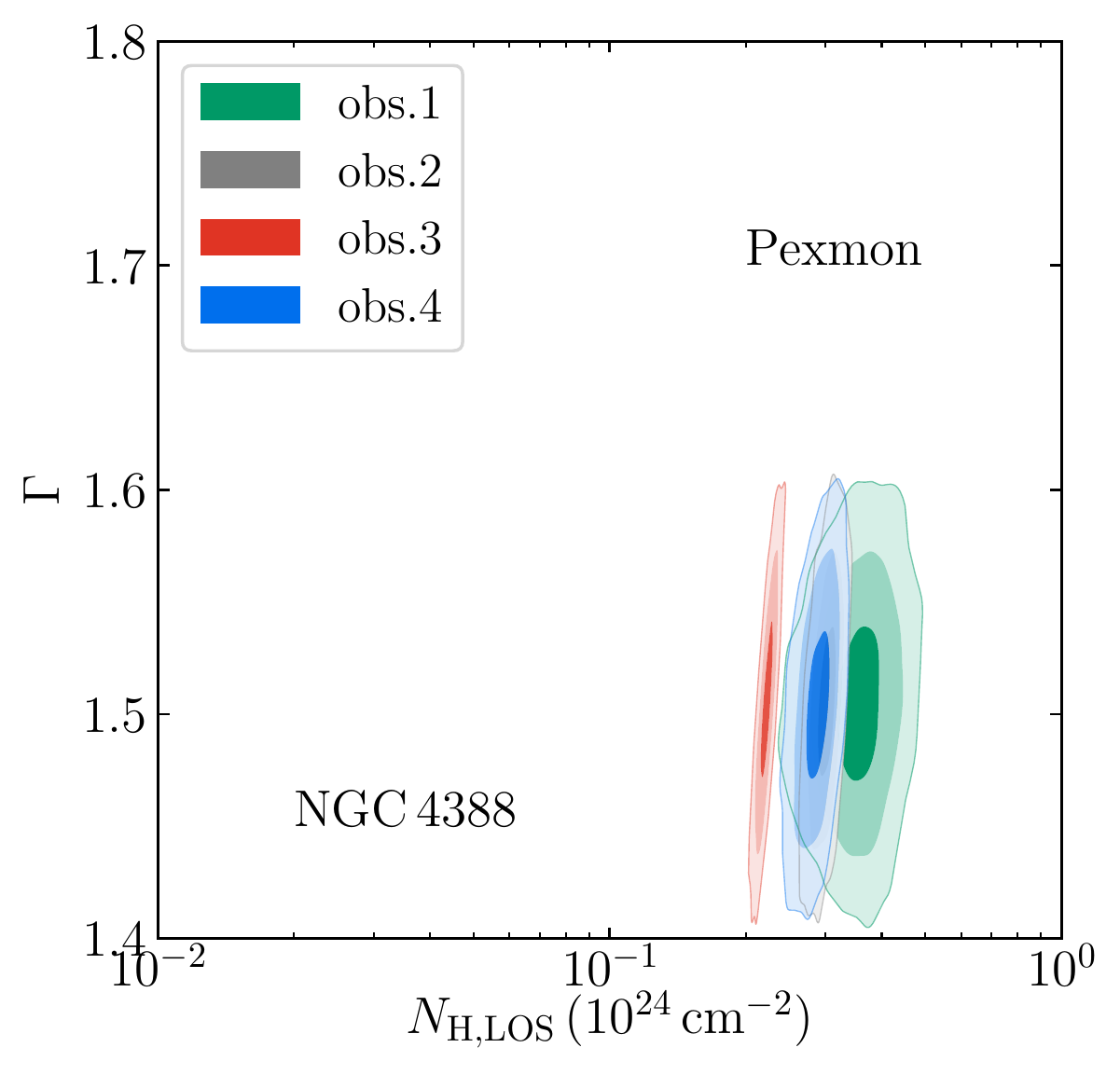}
	\includegraphics[width = 0.24\textwidth]{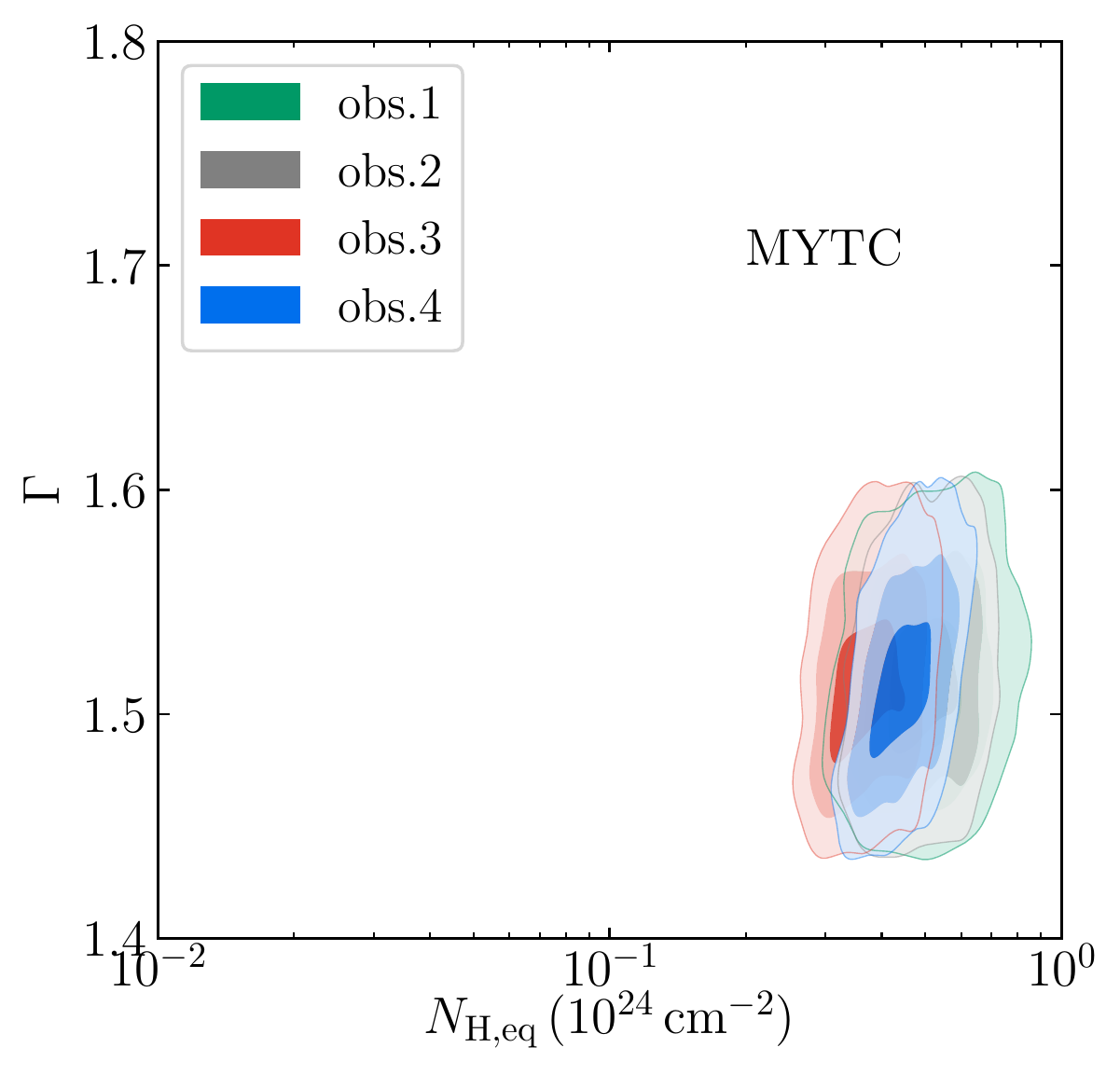}
	\includegraphics[width = 0.24\textwidth]{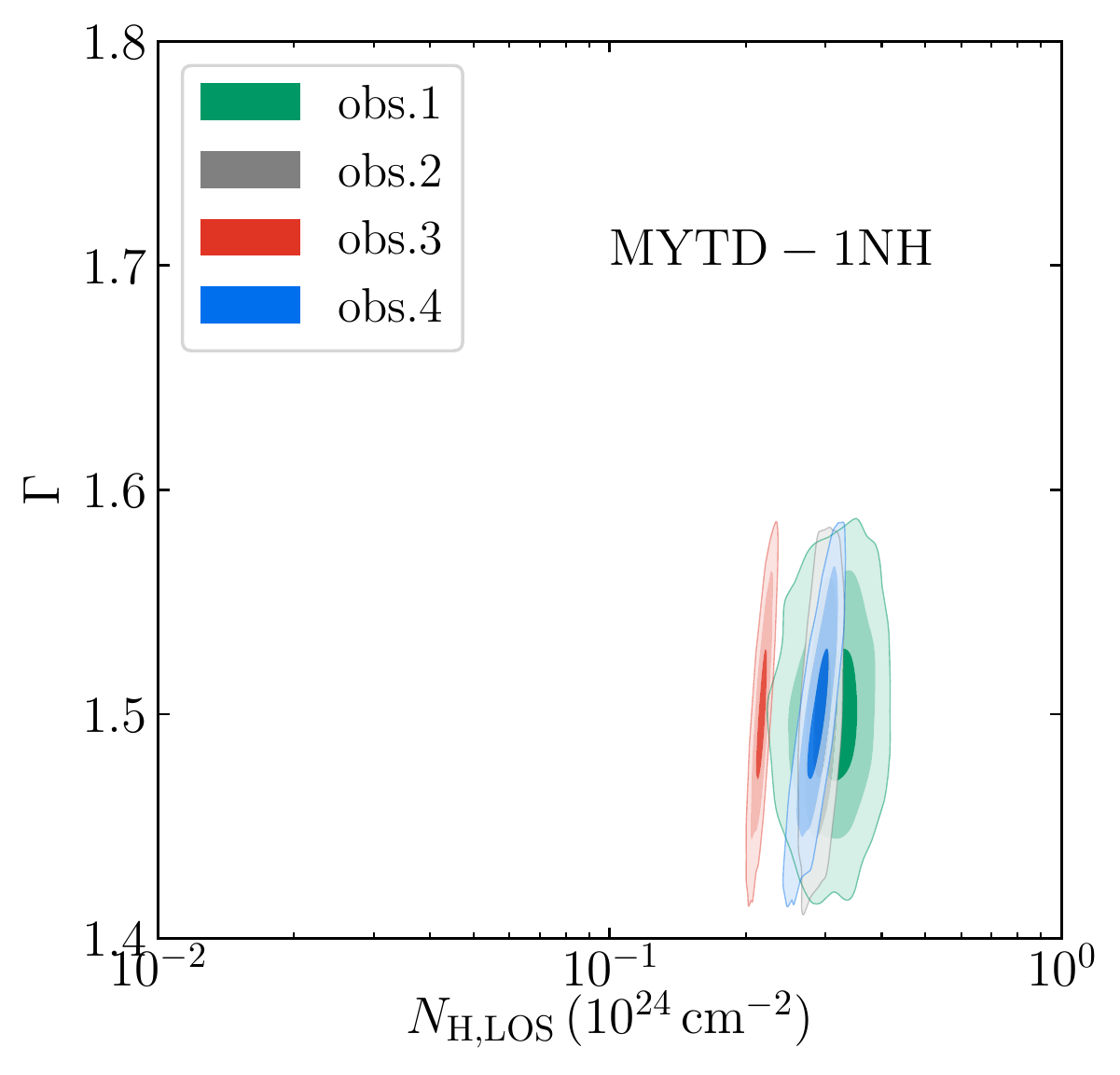}
	\includegraphics[width = 0.24\textwidth]{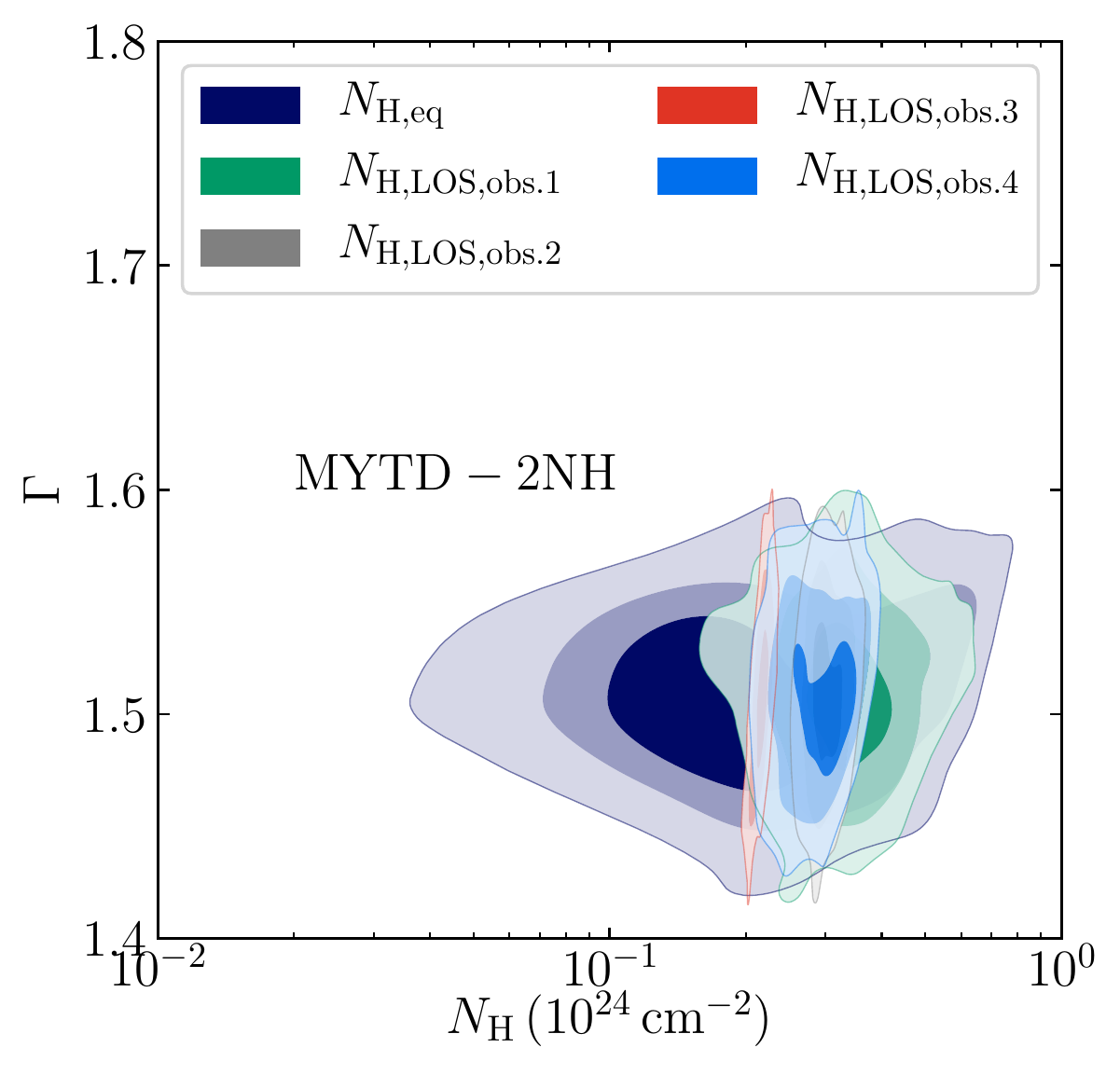}\\	
		
	\includegraphics[width = 0.24\textwidth]{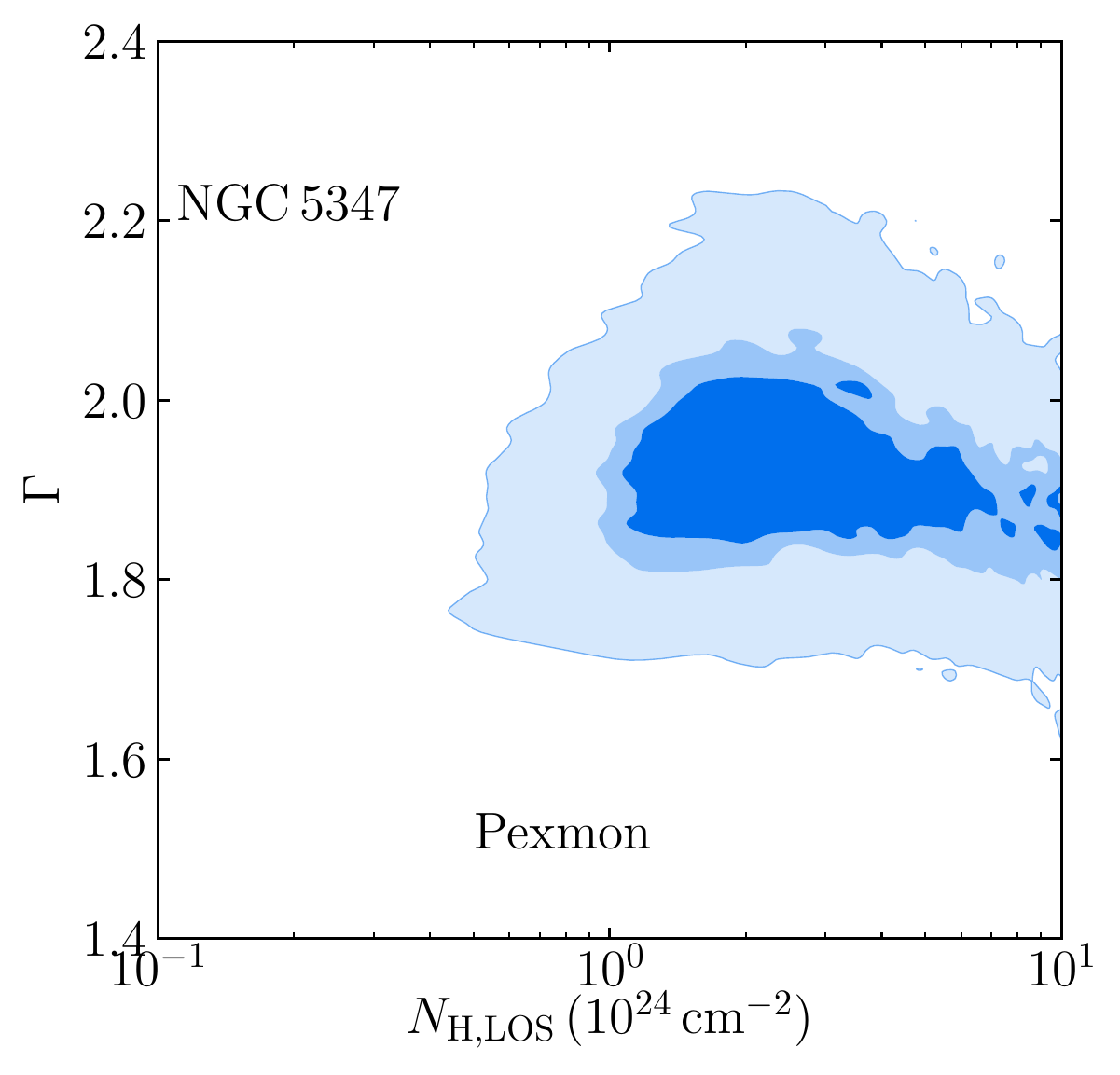}
	\includegraphics[width = 0.24\textwidth]{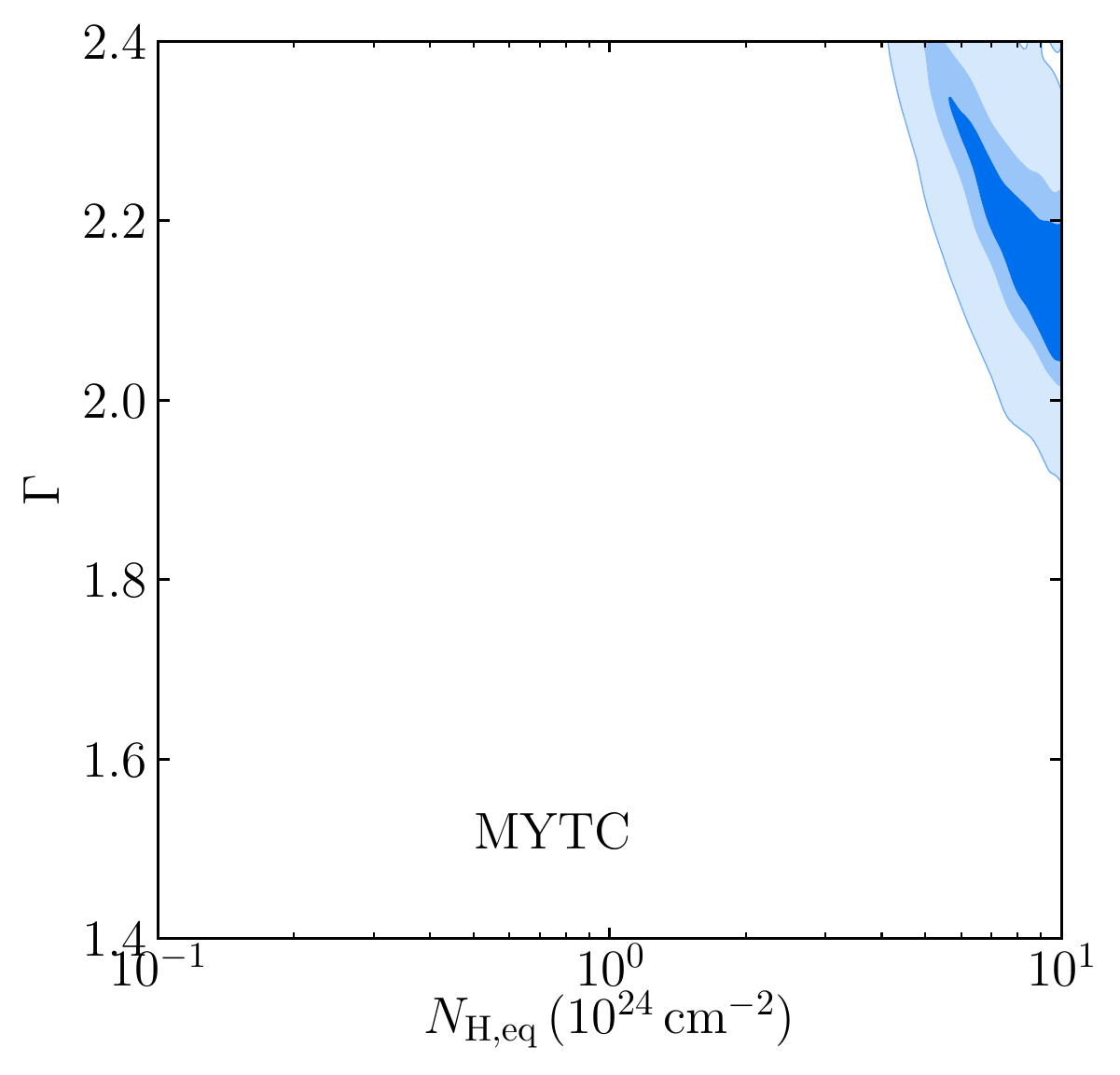}
	\includegraphics[width = 0.24\textwidth]{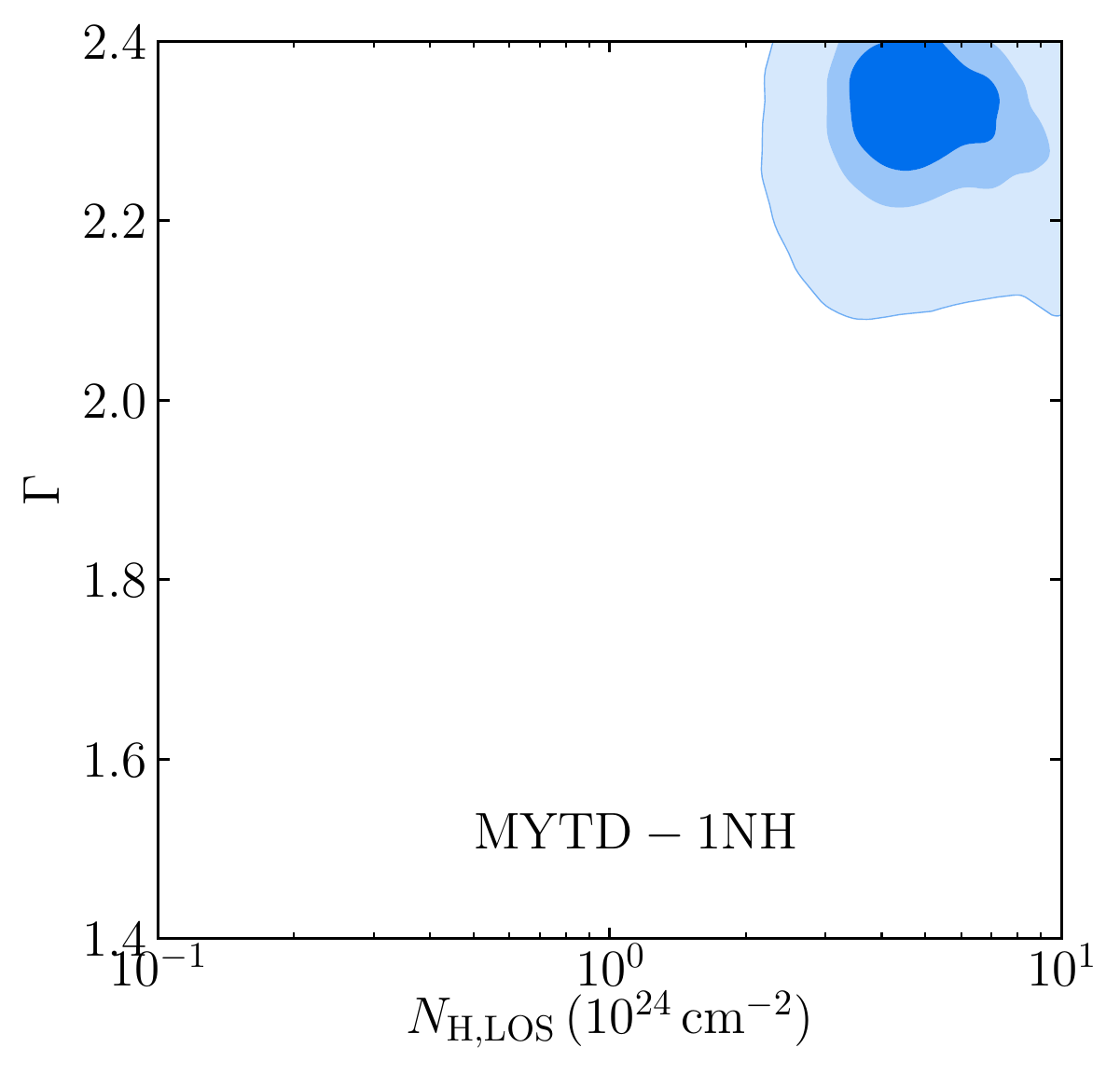}
\caption{$\Gamma-N_{\rm H}$ confidence contour for the different models used in this work obtained from the MCMC analysis. For the sources where $\Gamma$ was fixed (see text), we show only the 1D probability density of $N_{\rm H}$.}
\label{fig:mcmc1}
\end{figure*}

\begin{figure*}
\centering
	\includegraphics[width = 0.24\textwidth]{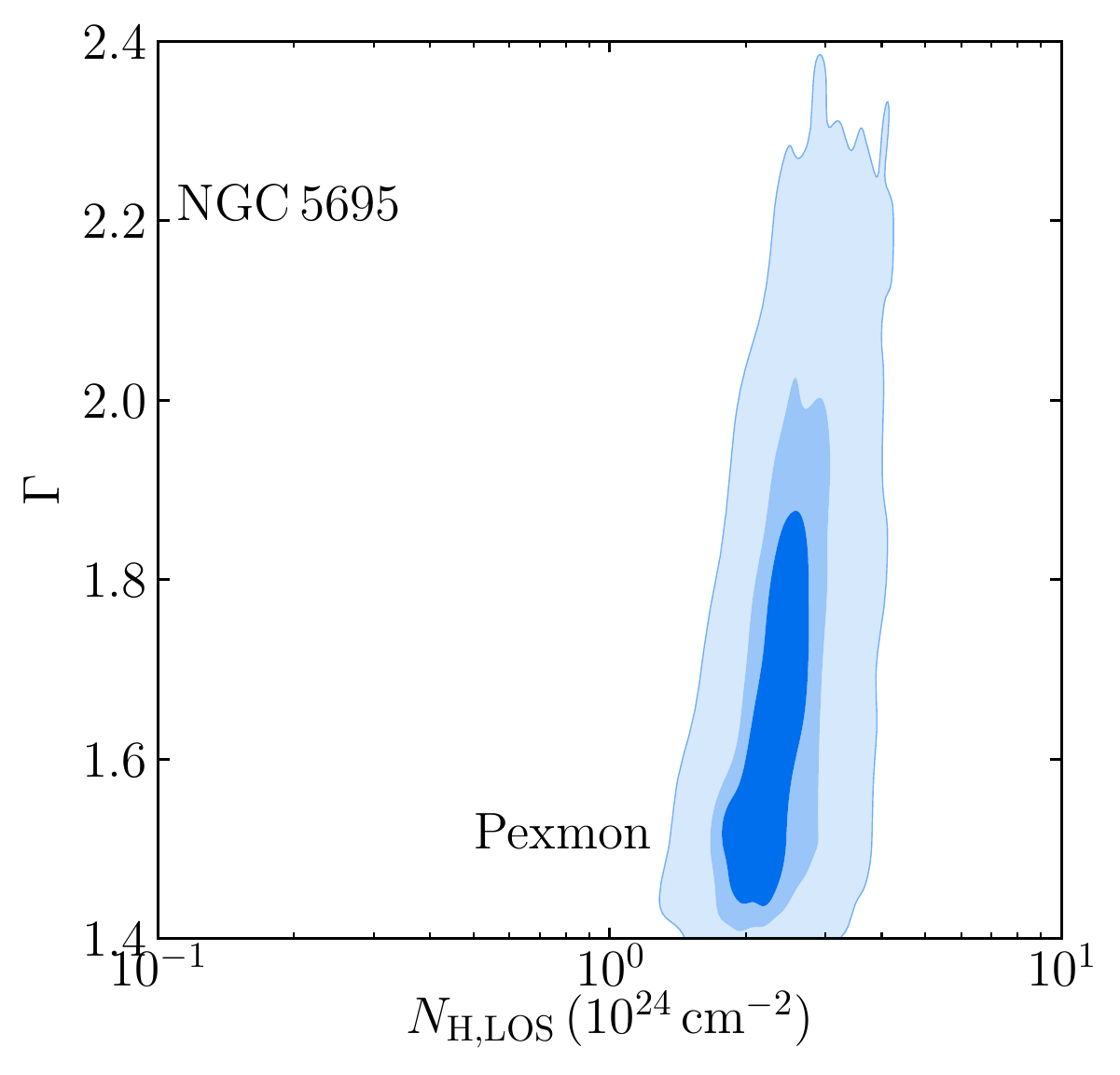}
	\includegraphics[width = 0.24\textwidth]{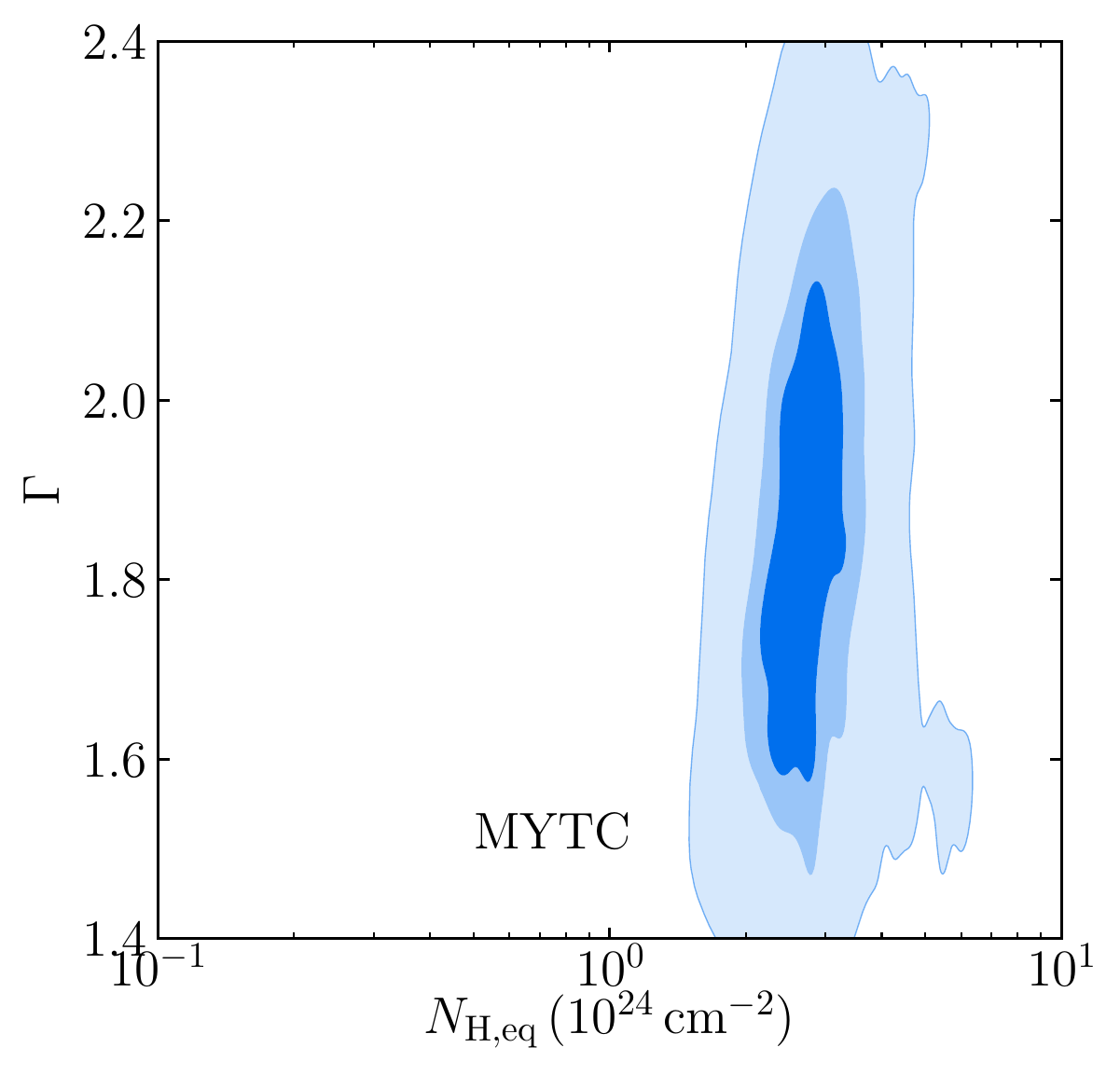}
	\includegraphics[width = 0.24\textwidth]{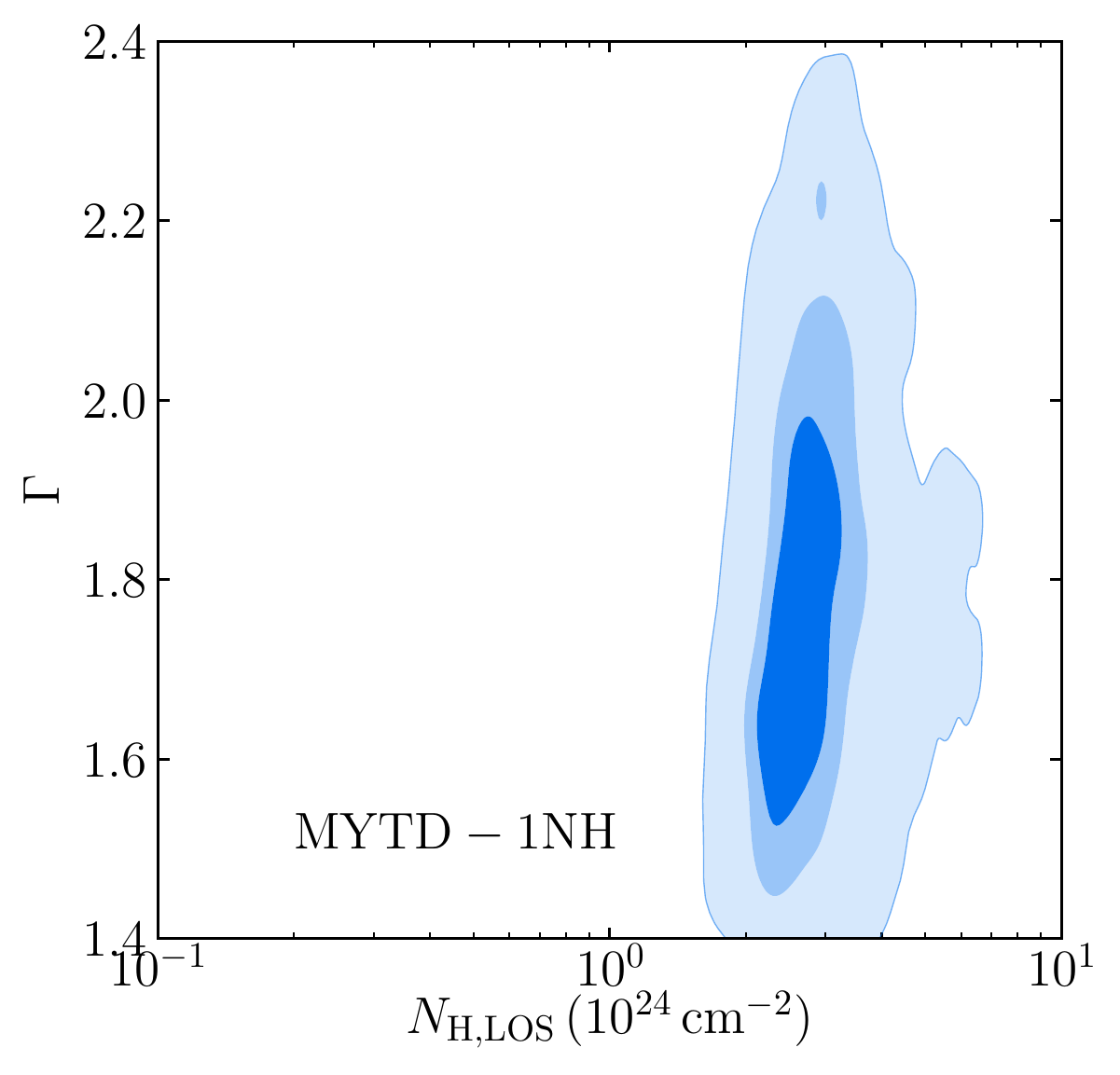}
	\includegraphics[width = 0.24\textwidth]{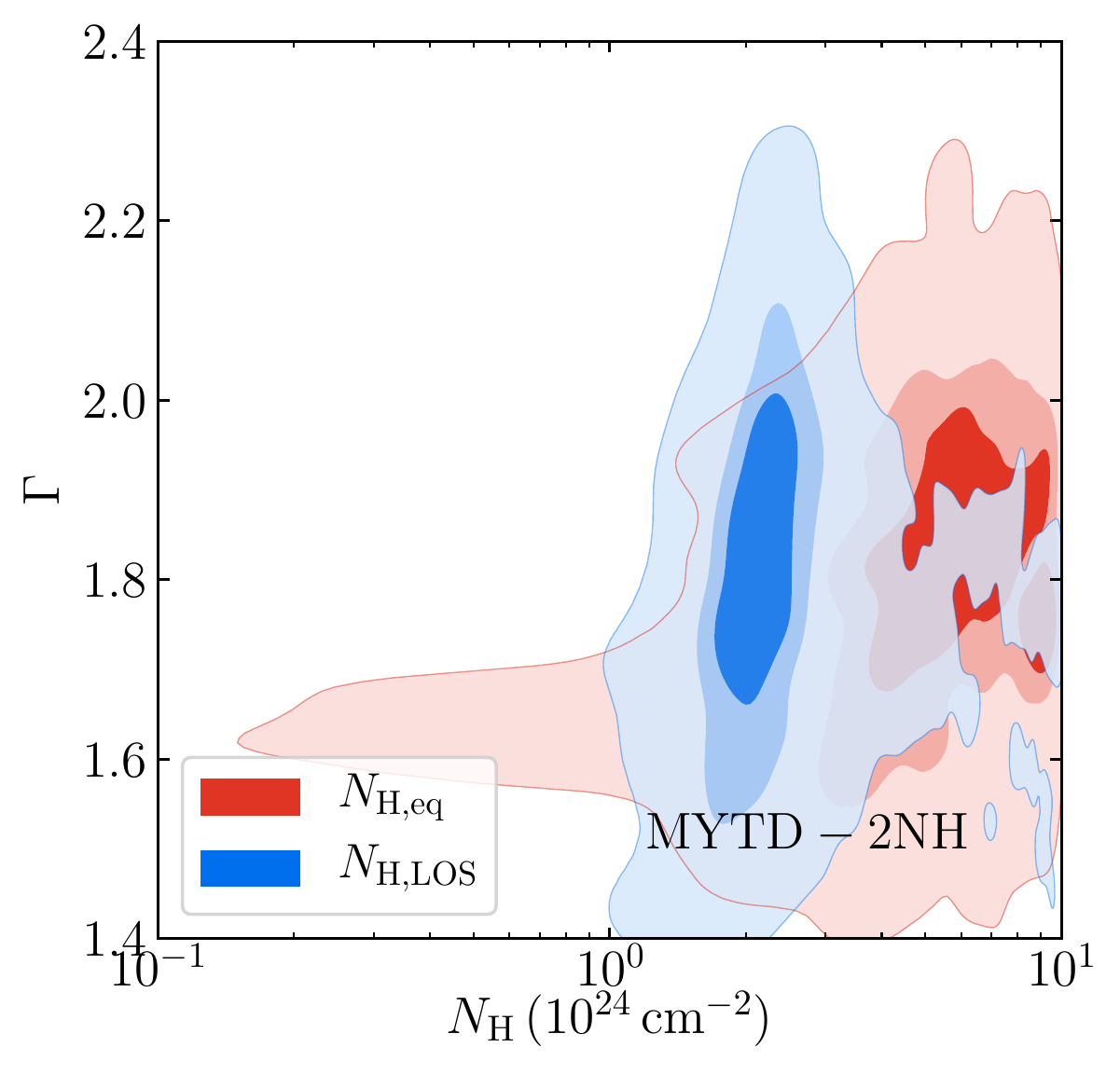}\\
	
	\includegraphics[width = 0.24\textwidth]{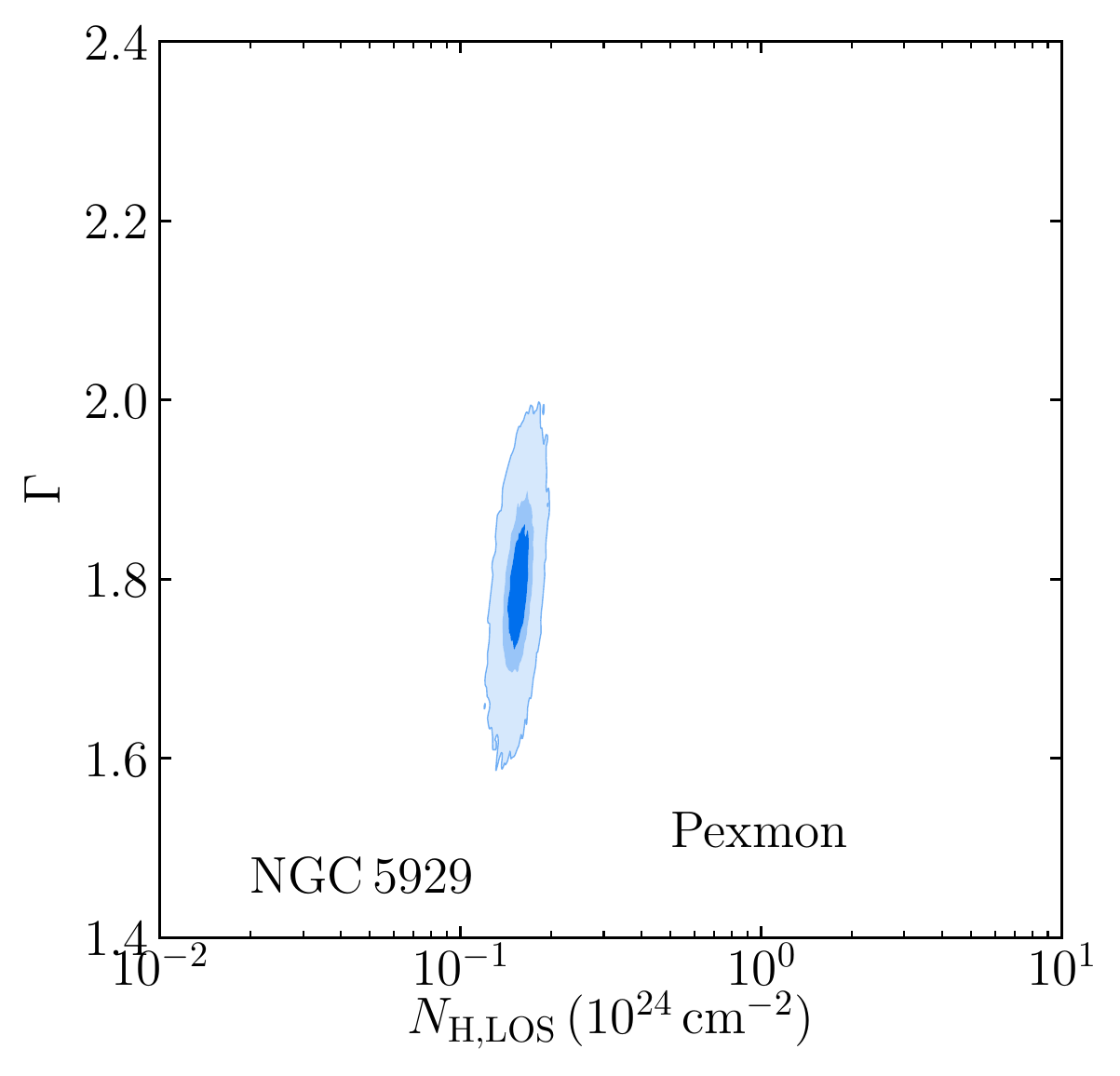}
	\includegraphics[width = 0.24\textwidth]{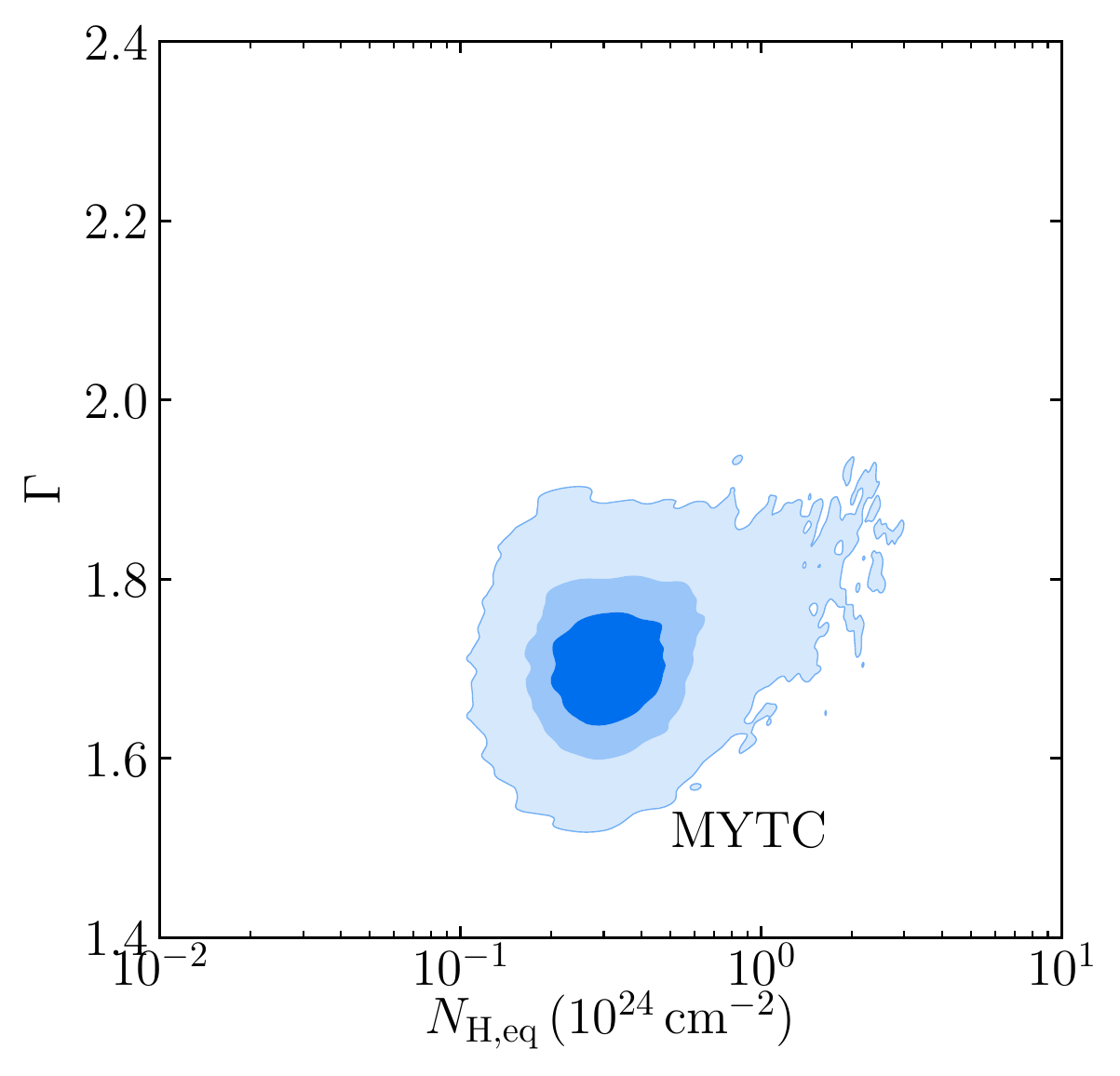}
	\includegraphics[width = 0.24\textwidth]{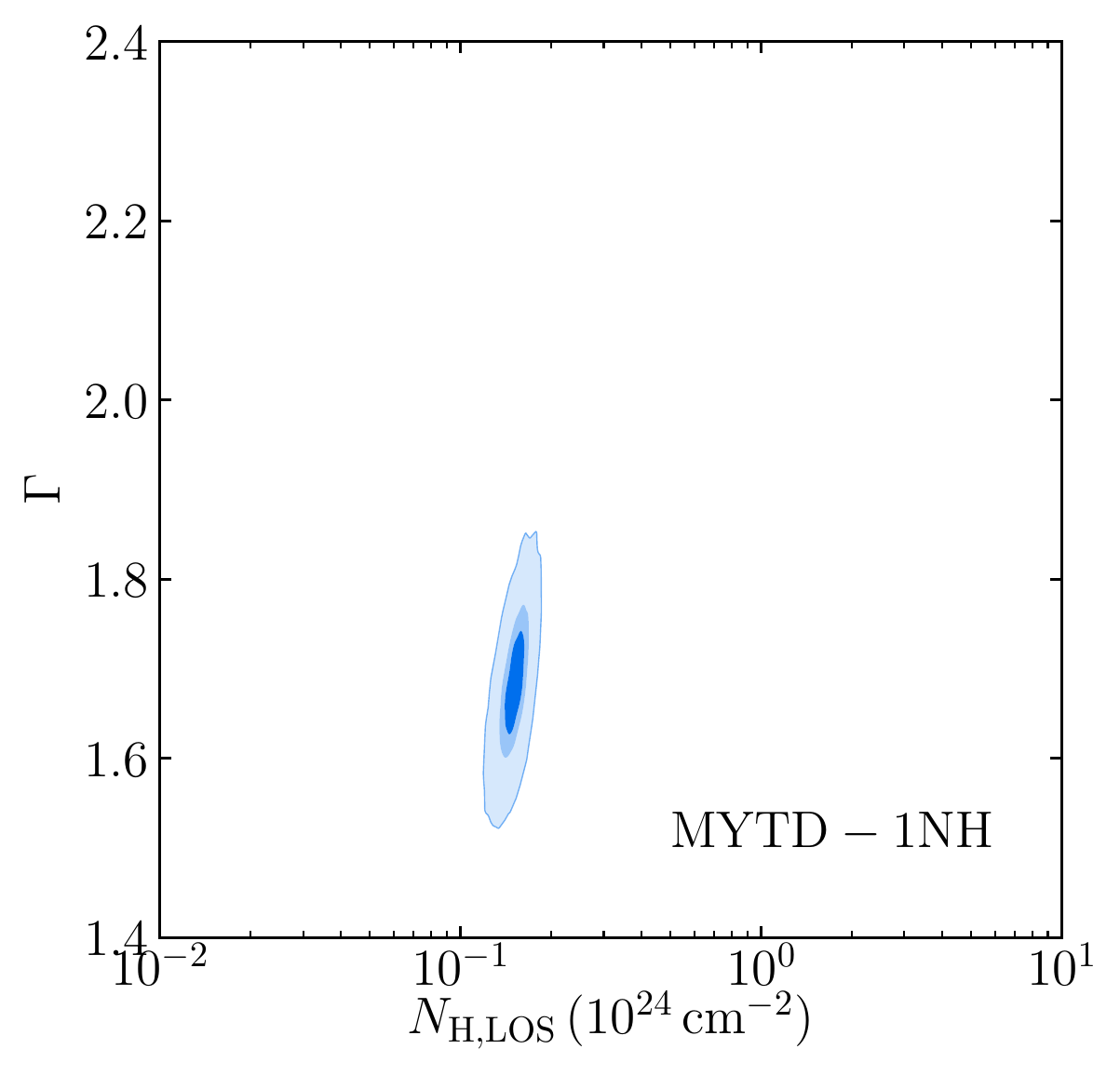}
	\includegraphics[width = 0.24\textwidth]{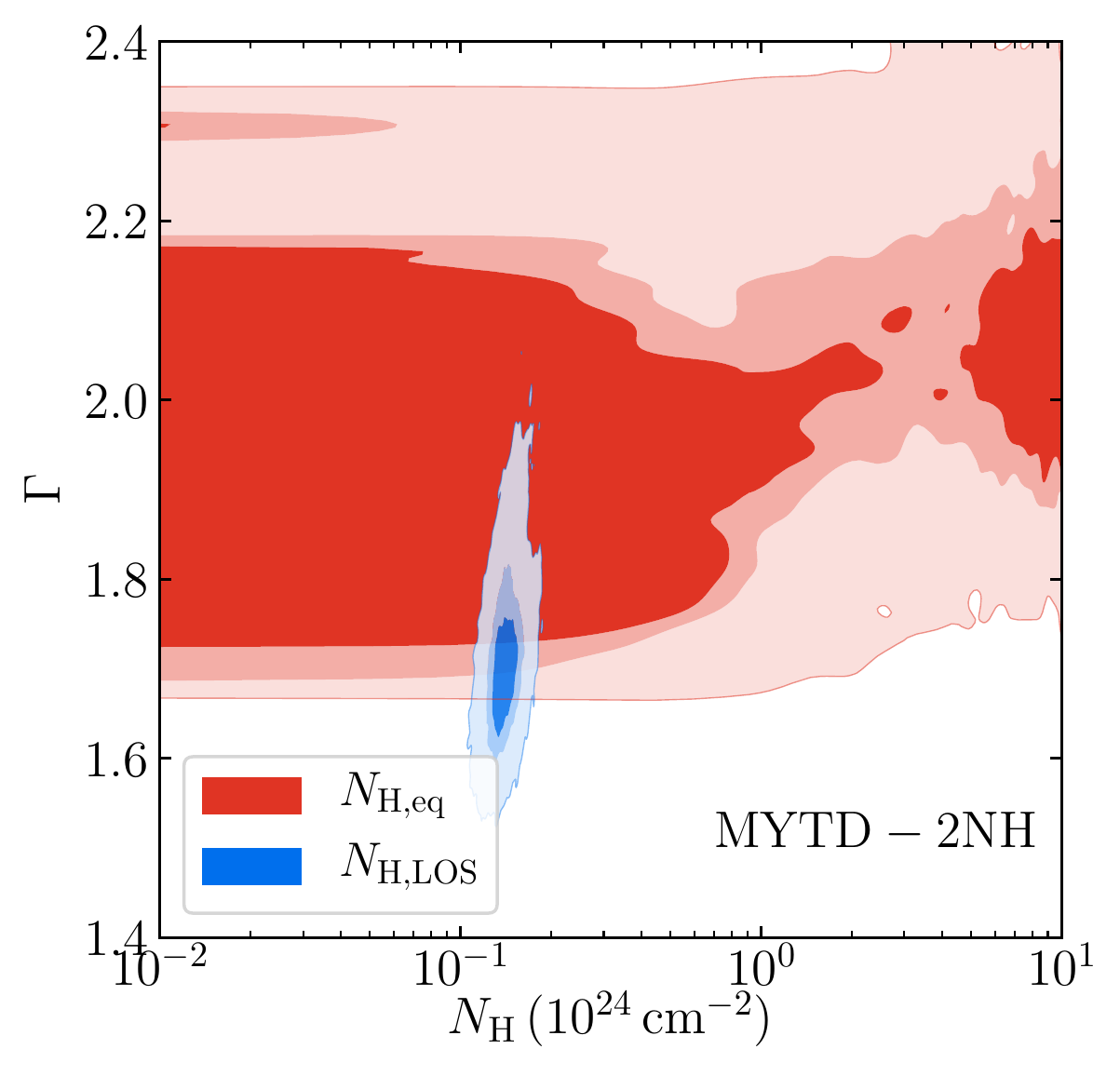}\\
	
	\includegraphics[width = 0.24\textwidth]{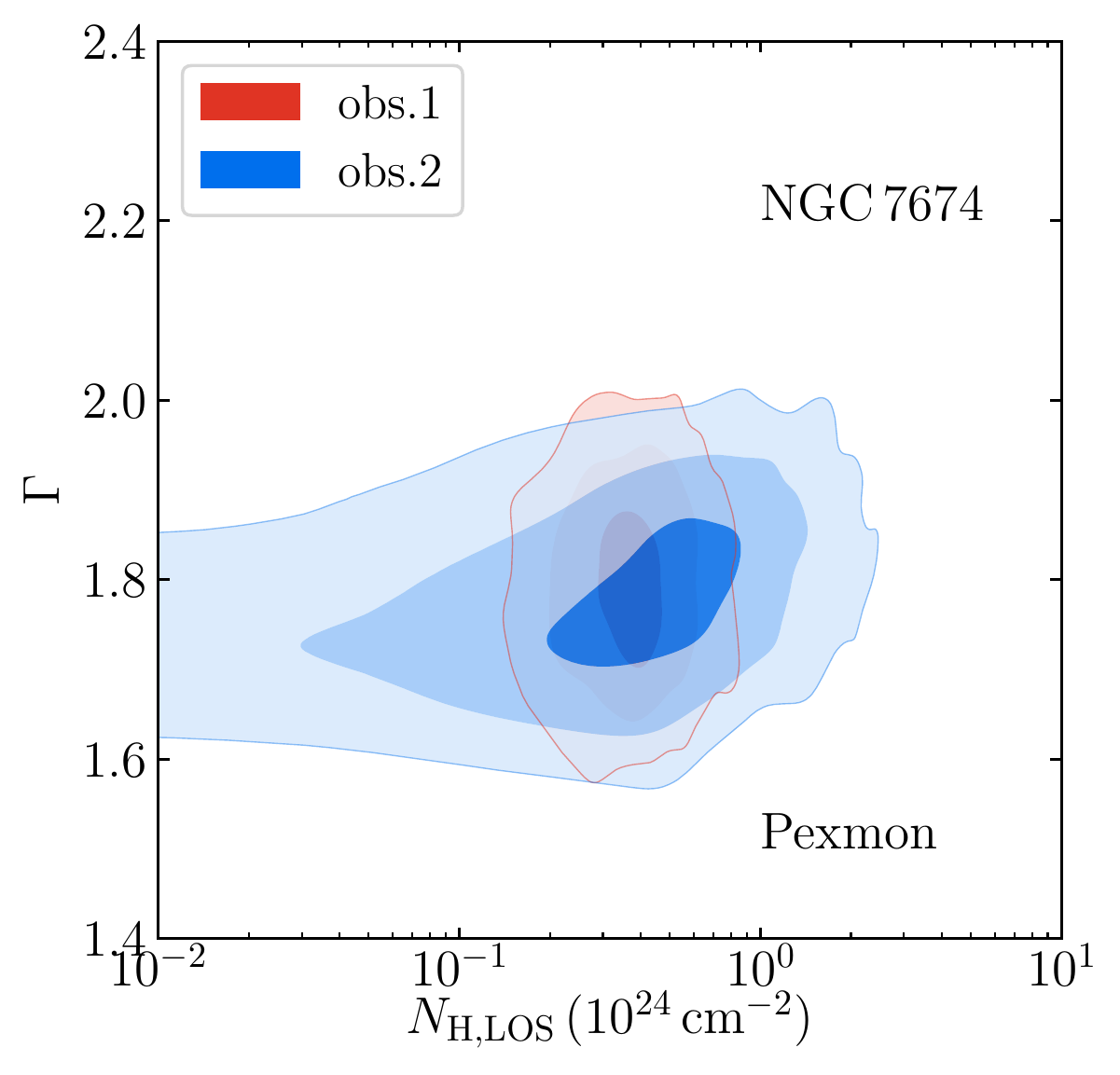}
	\includegraphics[width = 0.24\textwidth]{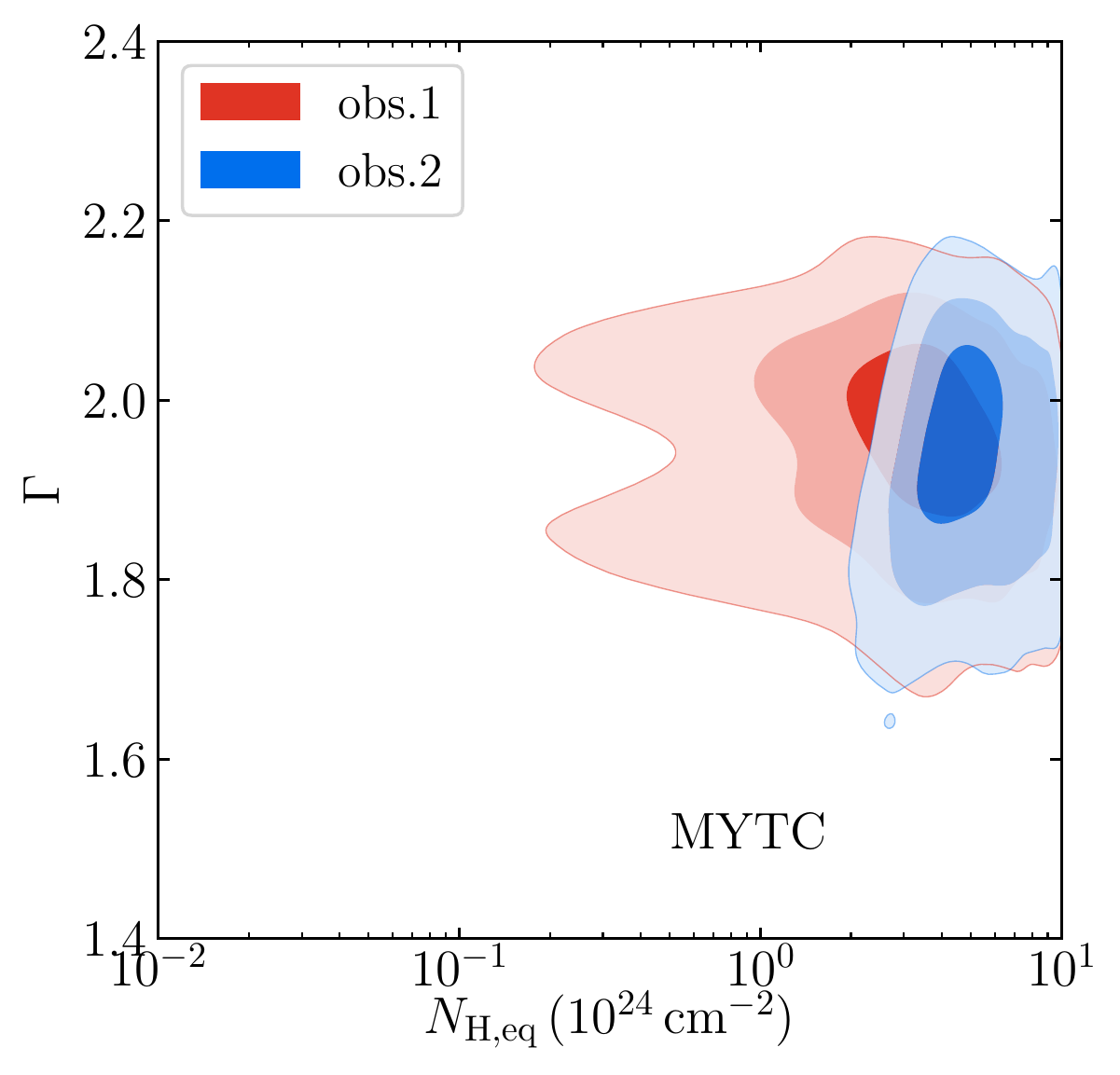}
	\includegraphics[width = 0.24\textwidth]{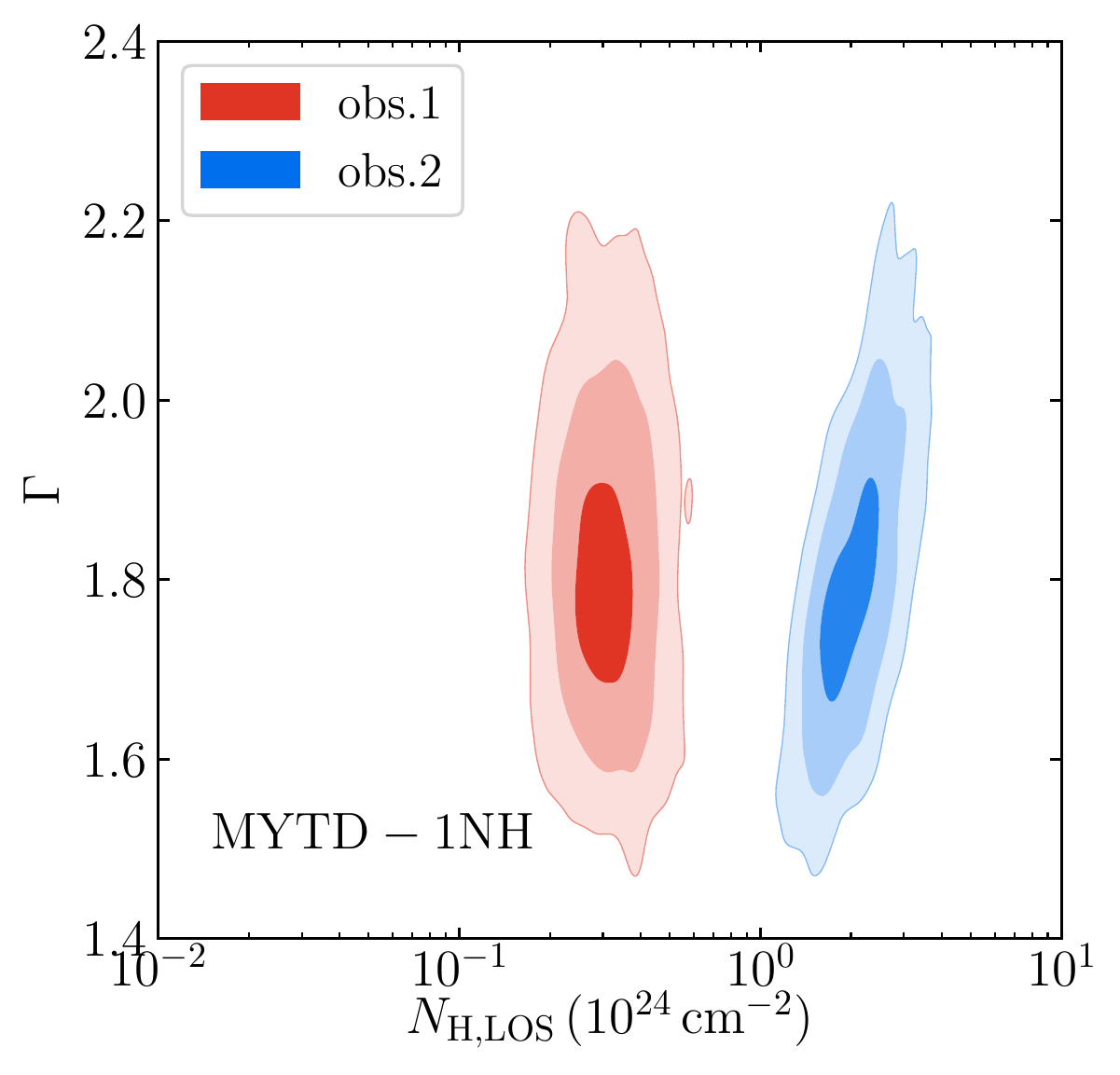}
	\includegraphics[width = 0.24\textwidth]{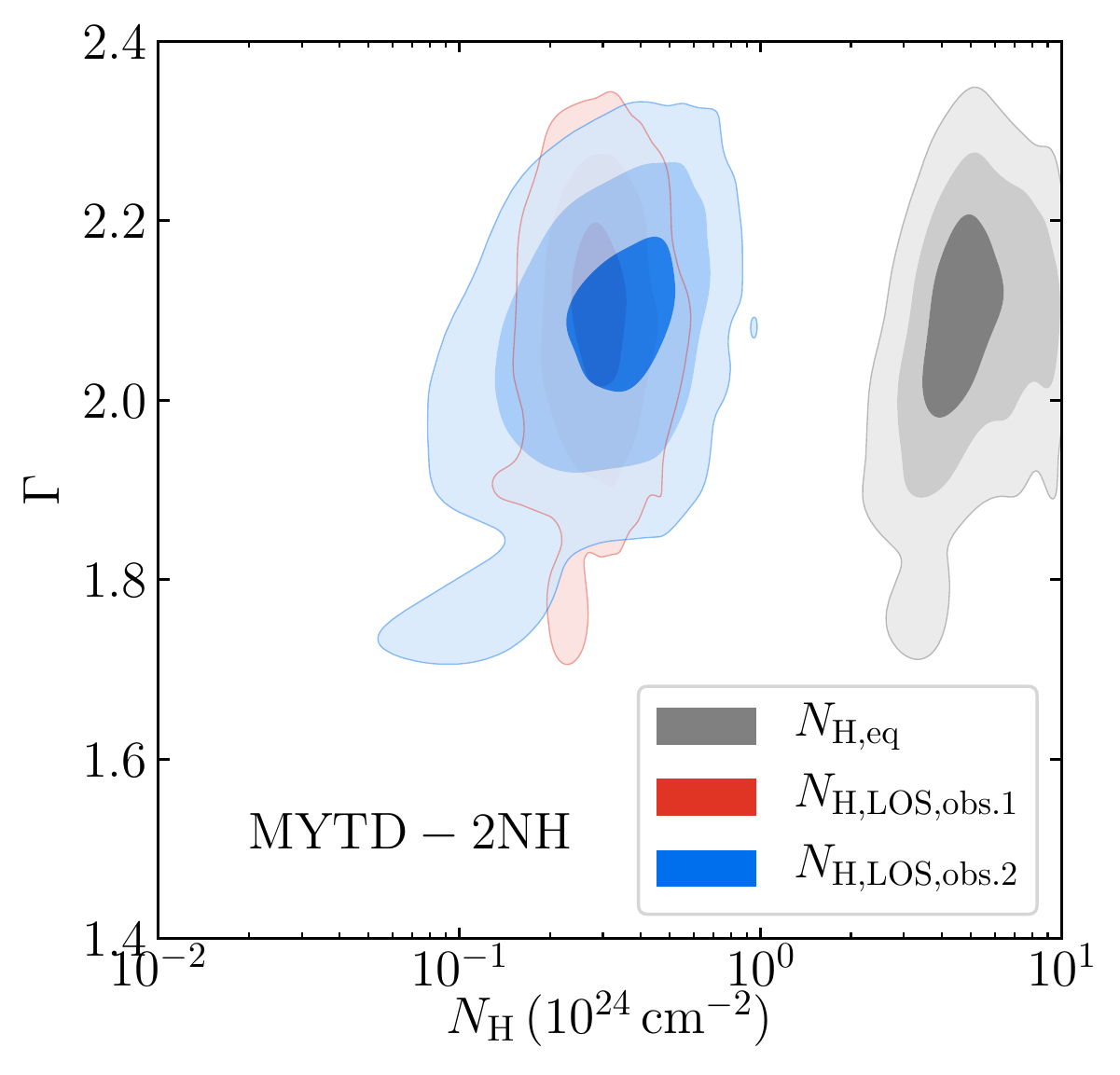}\\	

	\includegraphics[width = 0.24\textwidth]{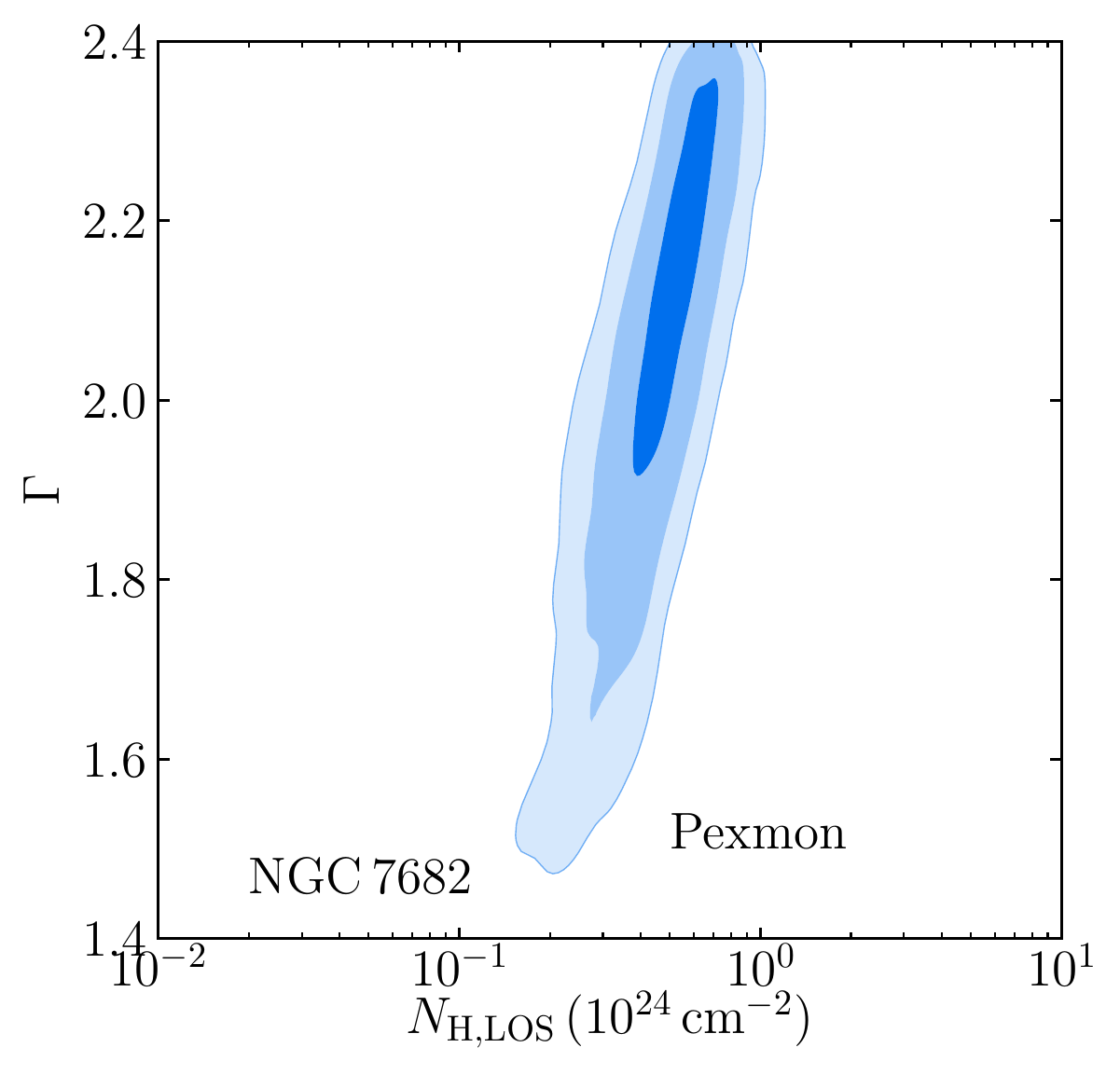}
	\includegraphics[width = 0.24\textwidth]{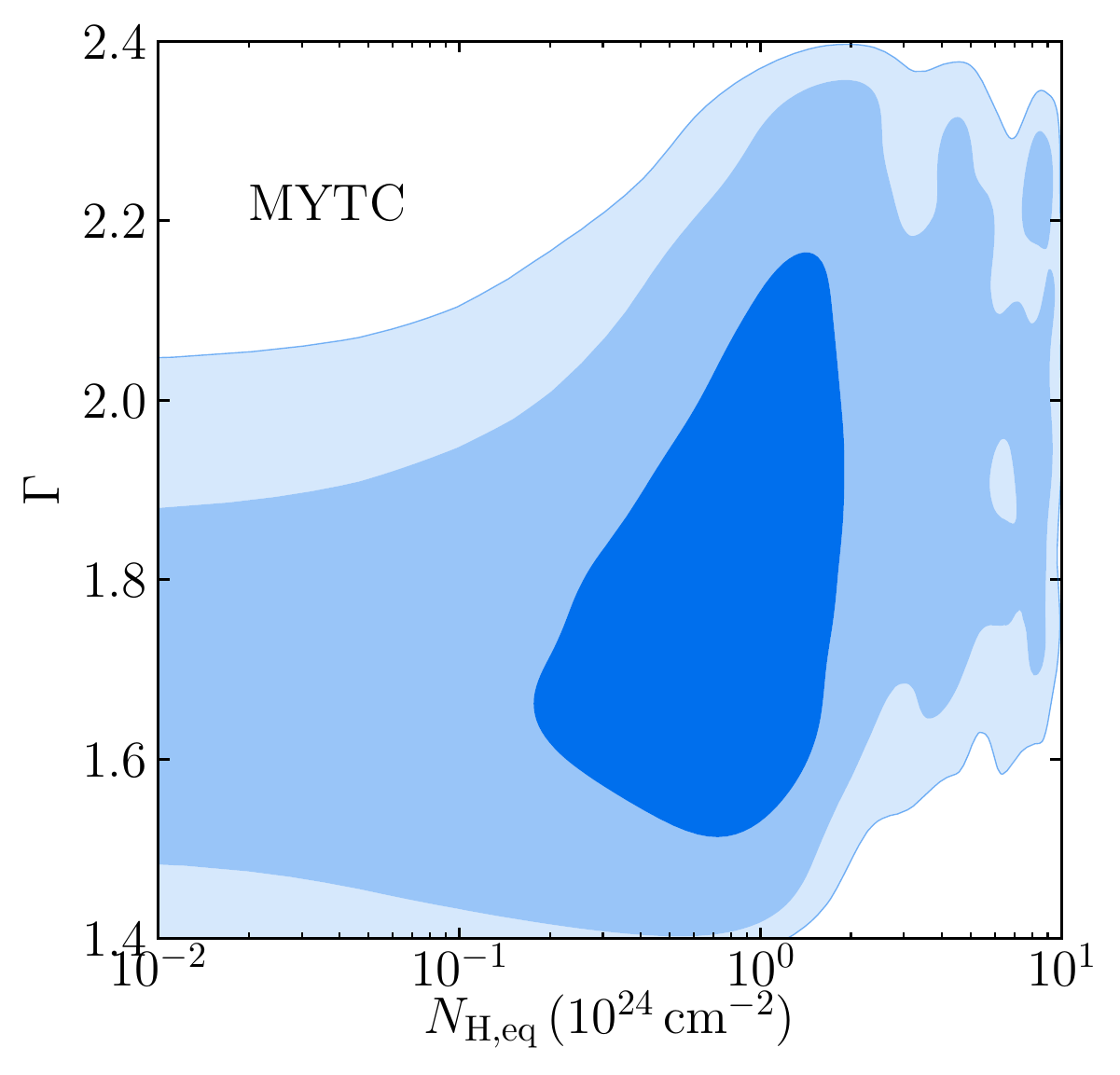}
	\includegraphics[width = 0.24\textwidth]{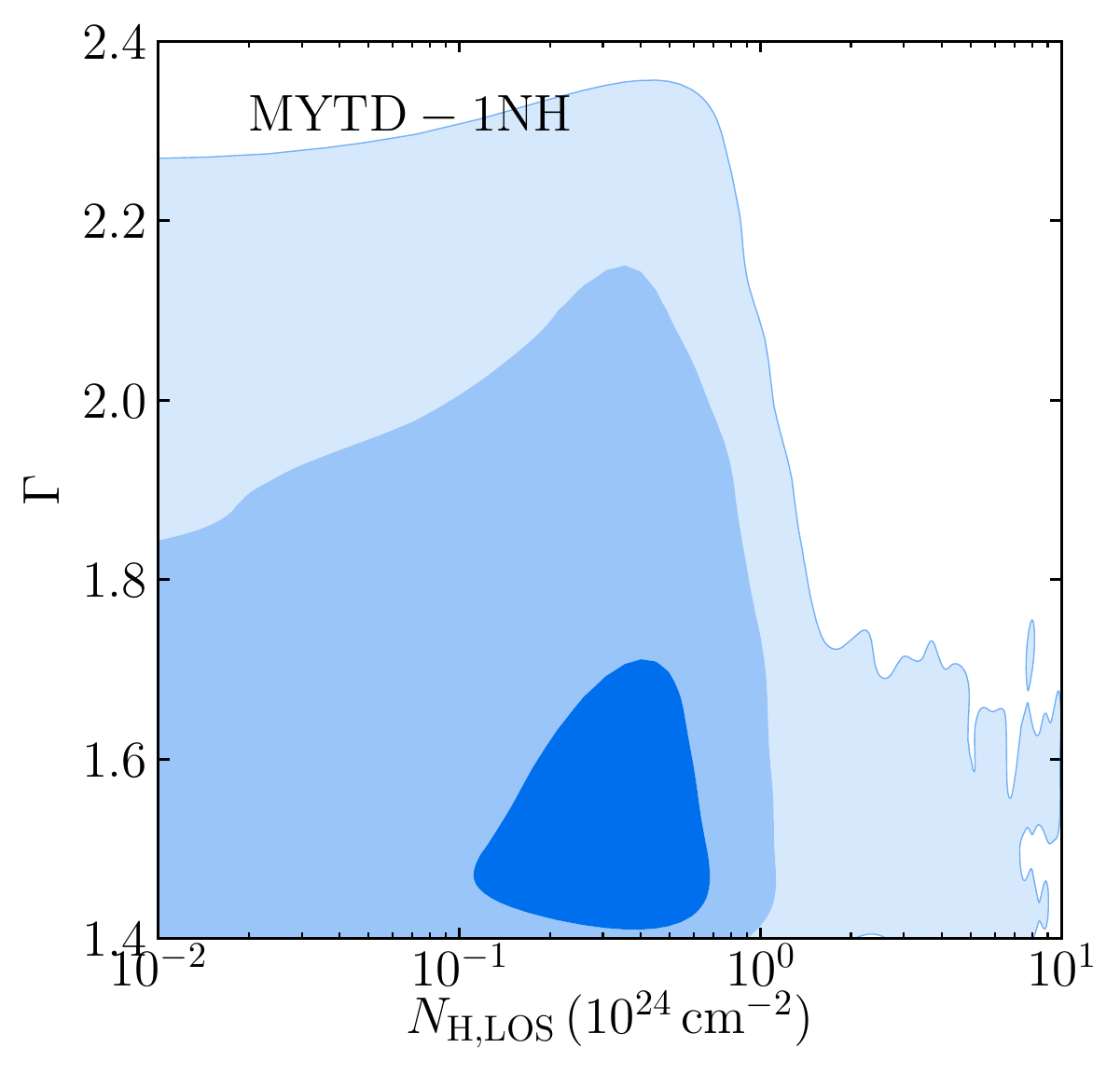}
	\includegraphics[width = 0.24\textwidth]{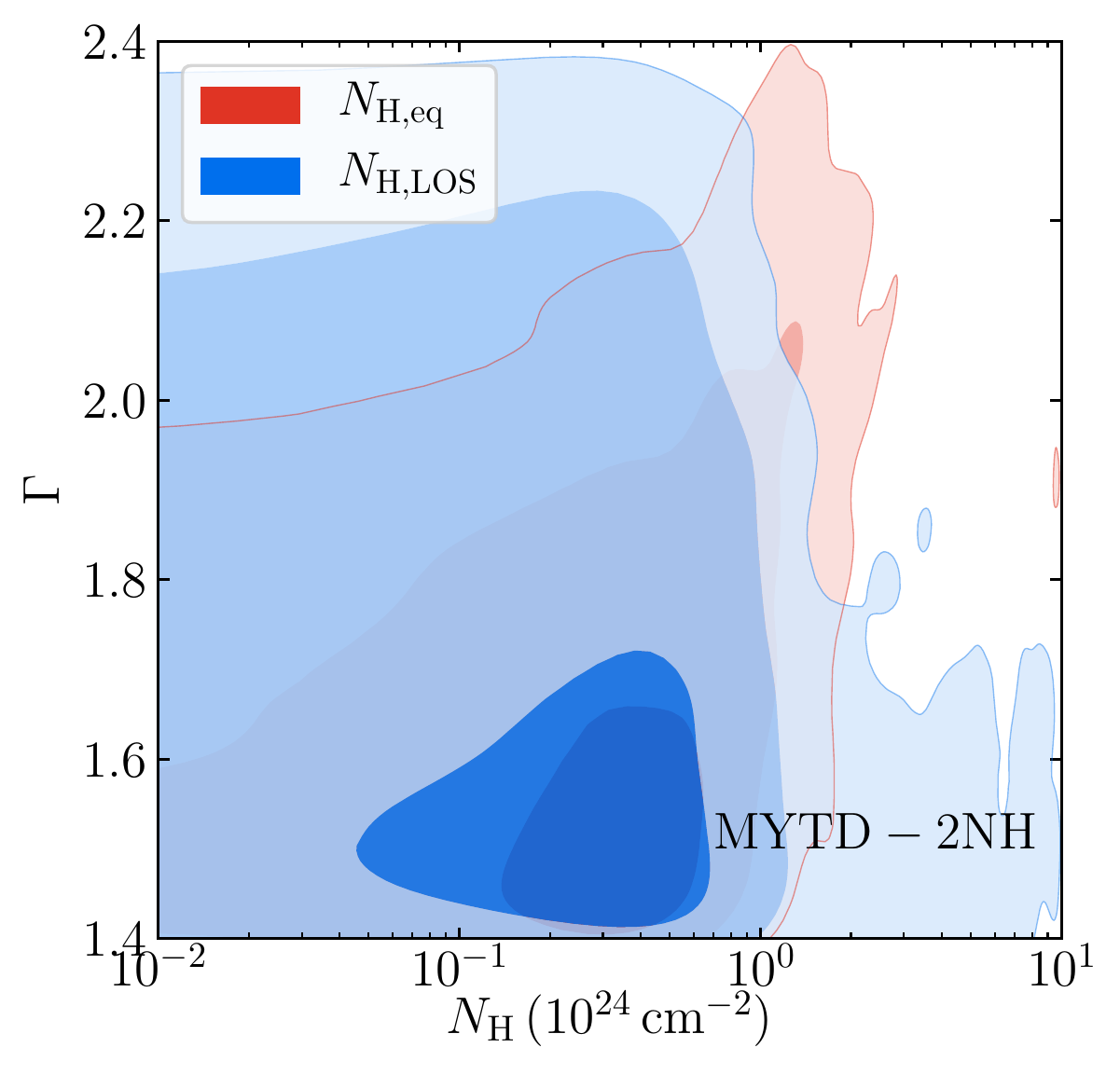}\\	
		
	\includegraphics[width = 0.24\textwidth]{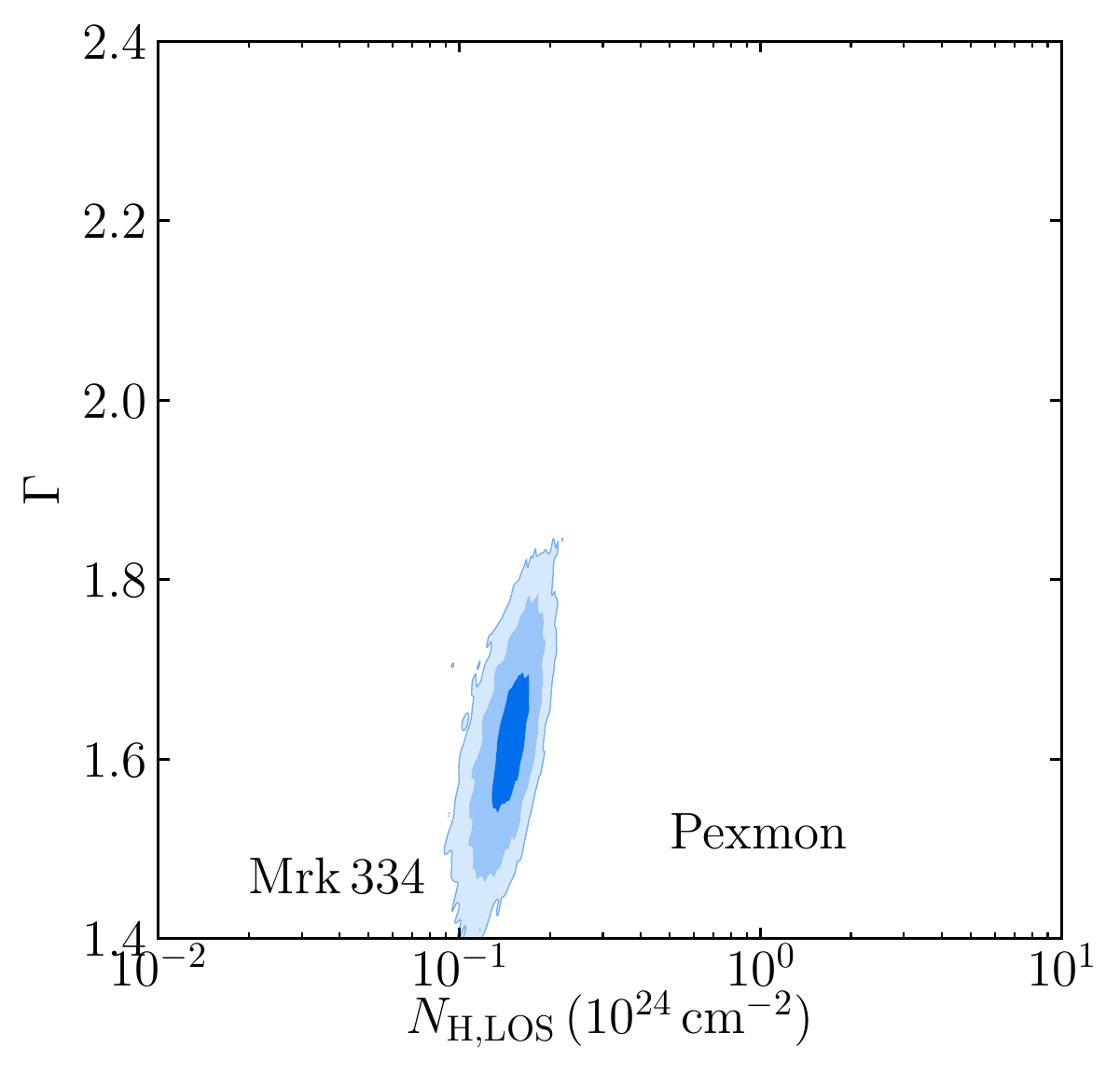}
	\includegraphics[width = 0.24\textwidth]{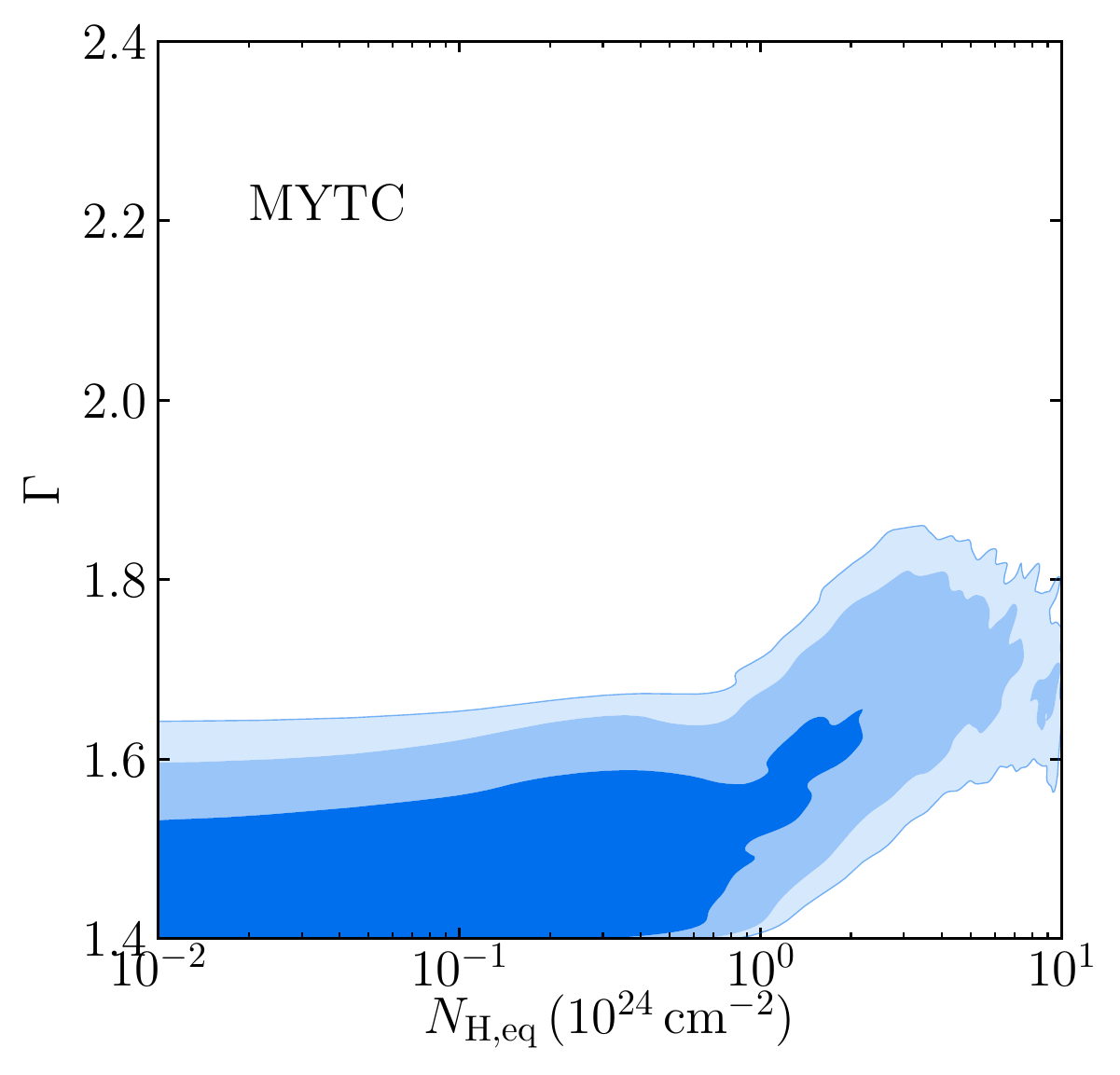}
	\includegraphics[width = 0.24\textwidth]{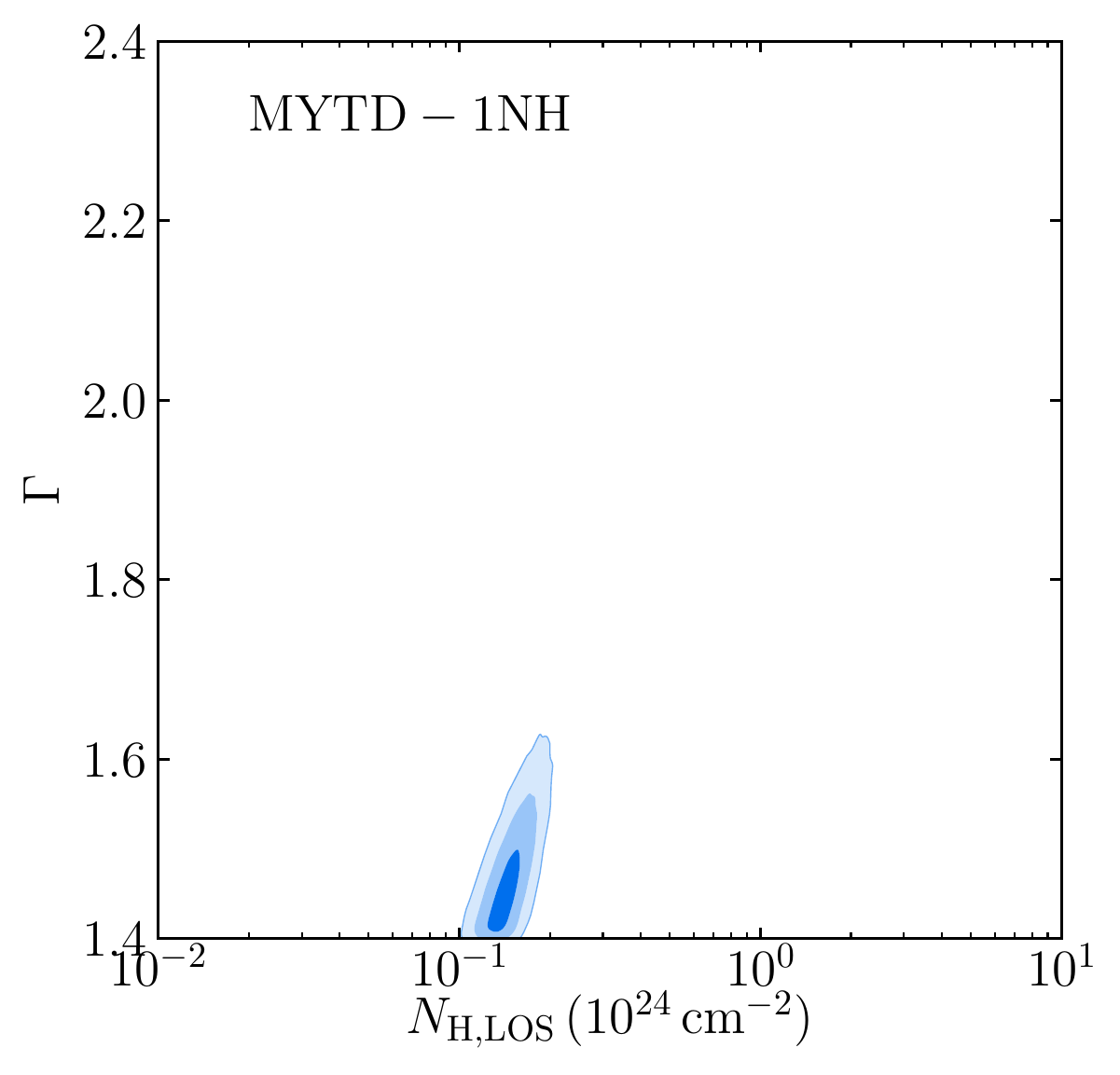}
	\includegraphics[width = 0.24\textwidth]{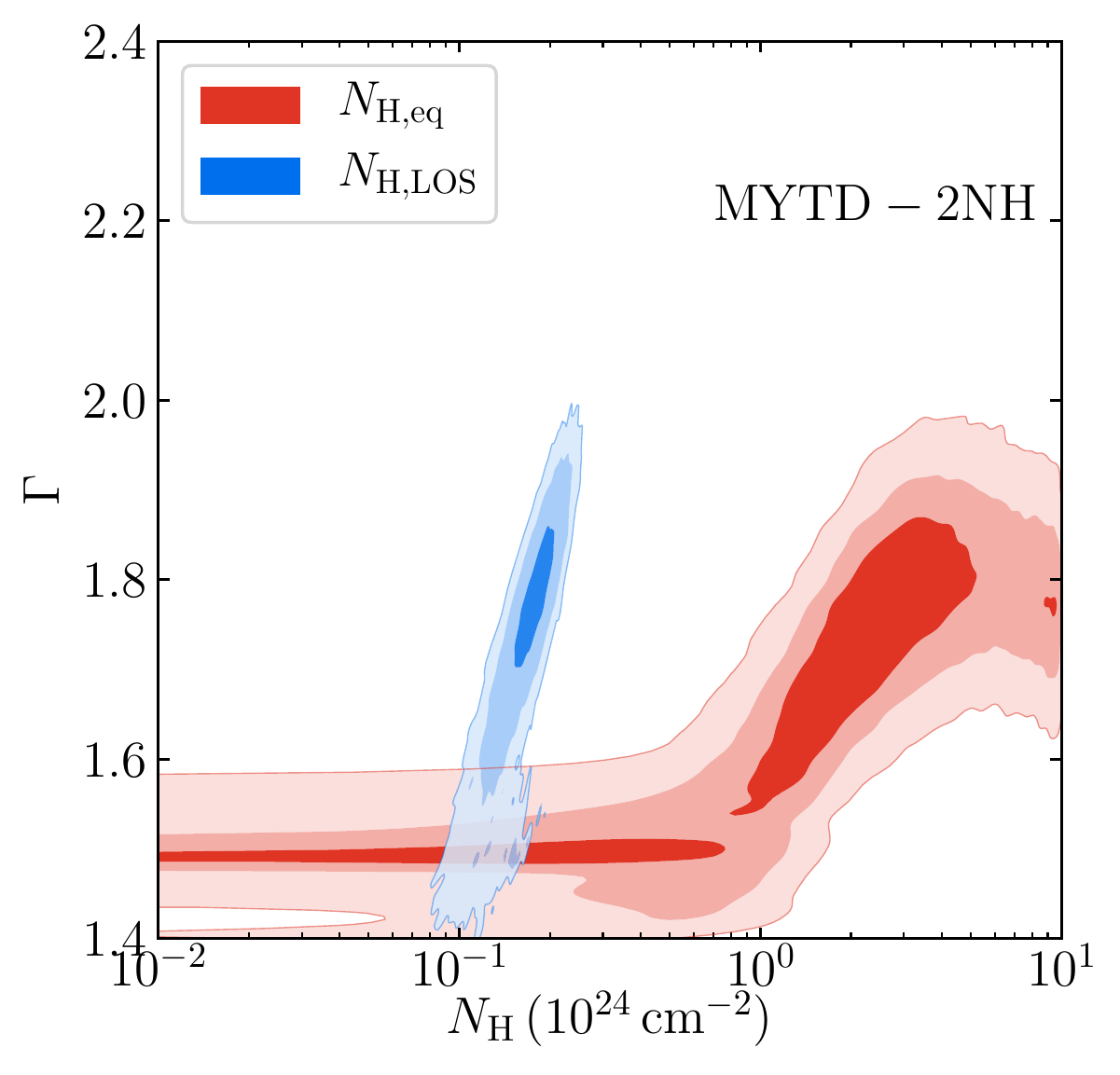}

\caption{Same as Figure~\ref{fig:mcmc1}.}
\label{fig:mcmc2}
\end{figure*}

\begin{figure*}
\centering
	\includegraphics[width = 0.24\textwidth]{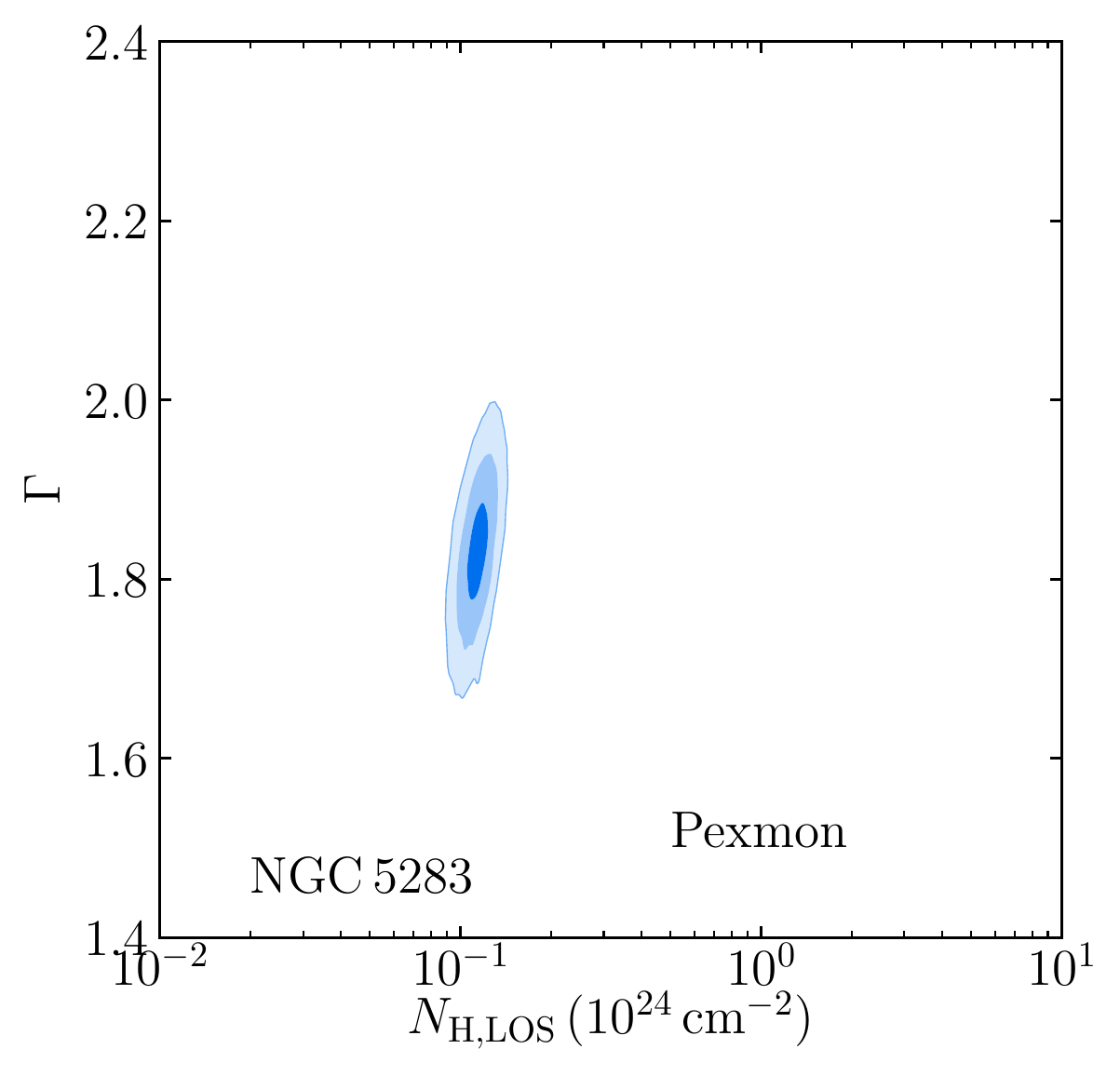}
	\includegraphics[width = 0.24\textwidth]{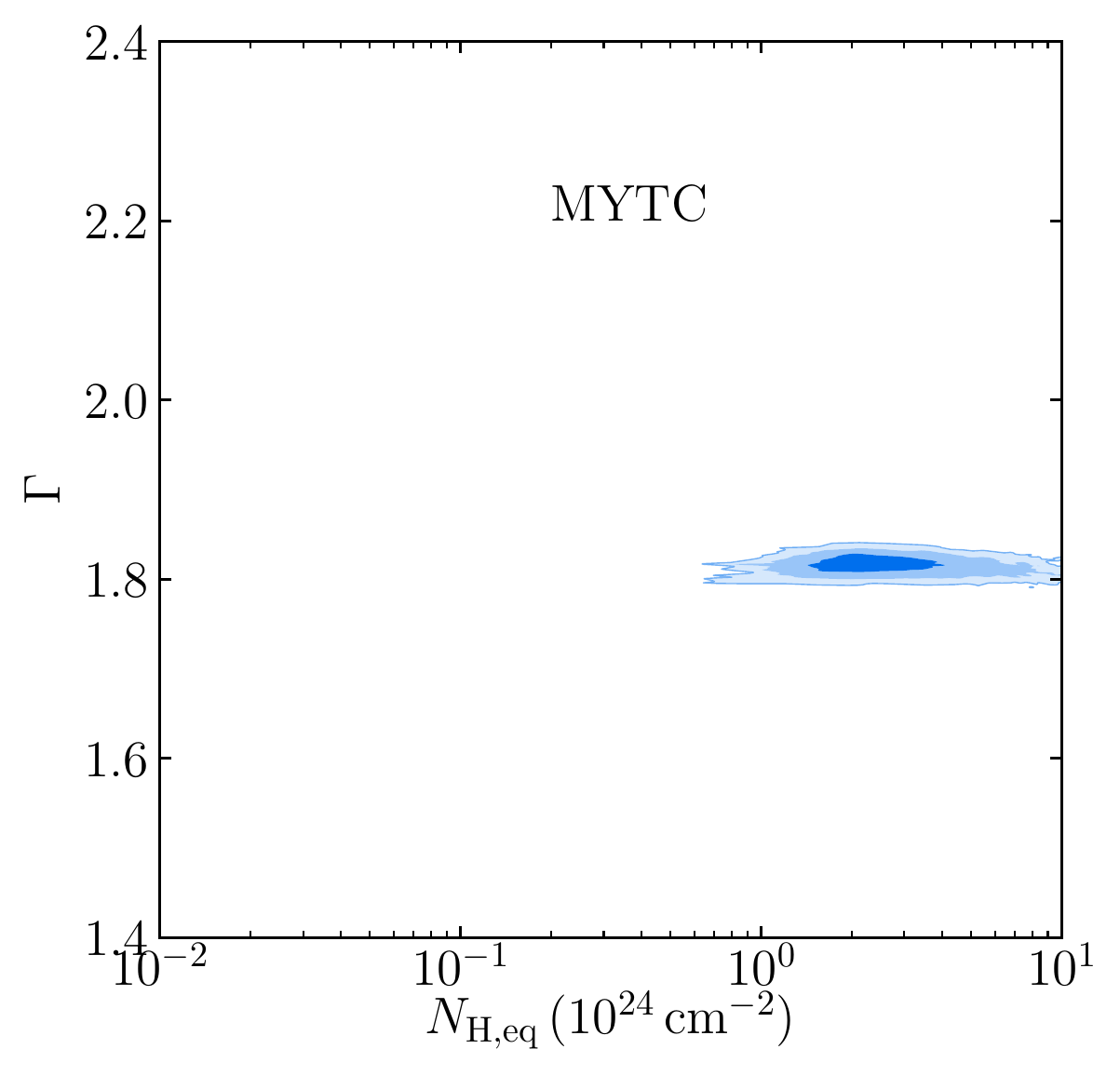}
	\includegraphics[width = 0.24\textwidth]{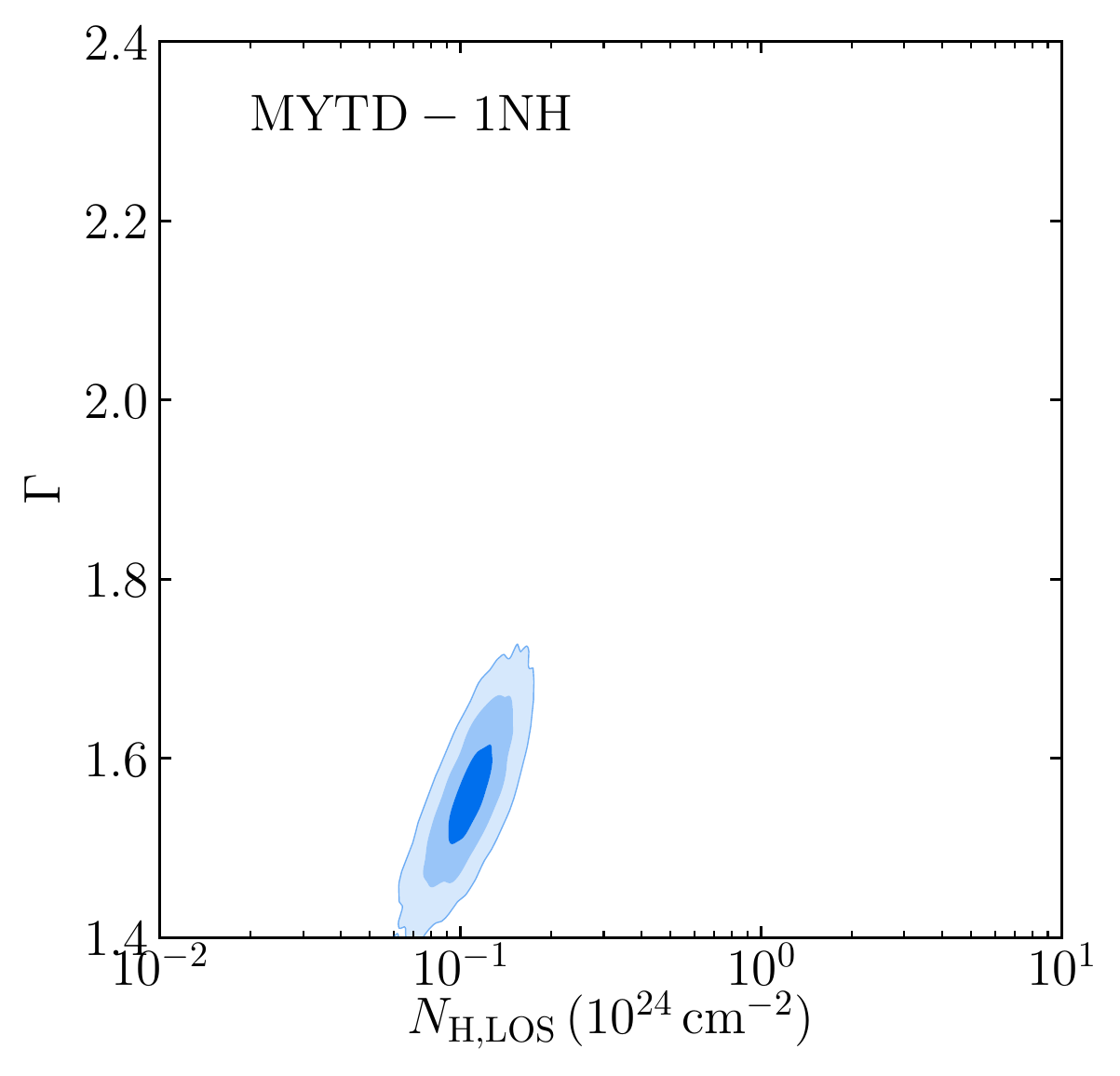}
	\includegraphics[width = 0.24\textwidth]{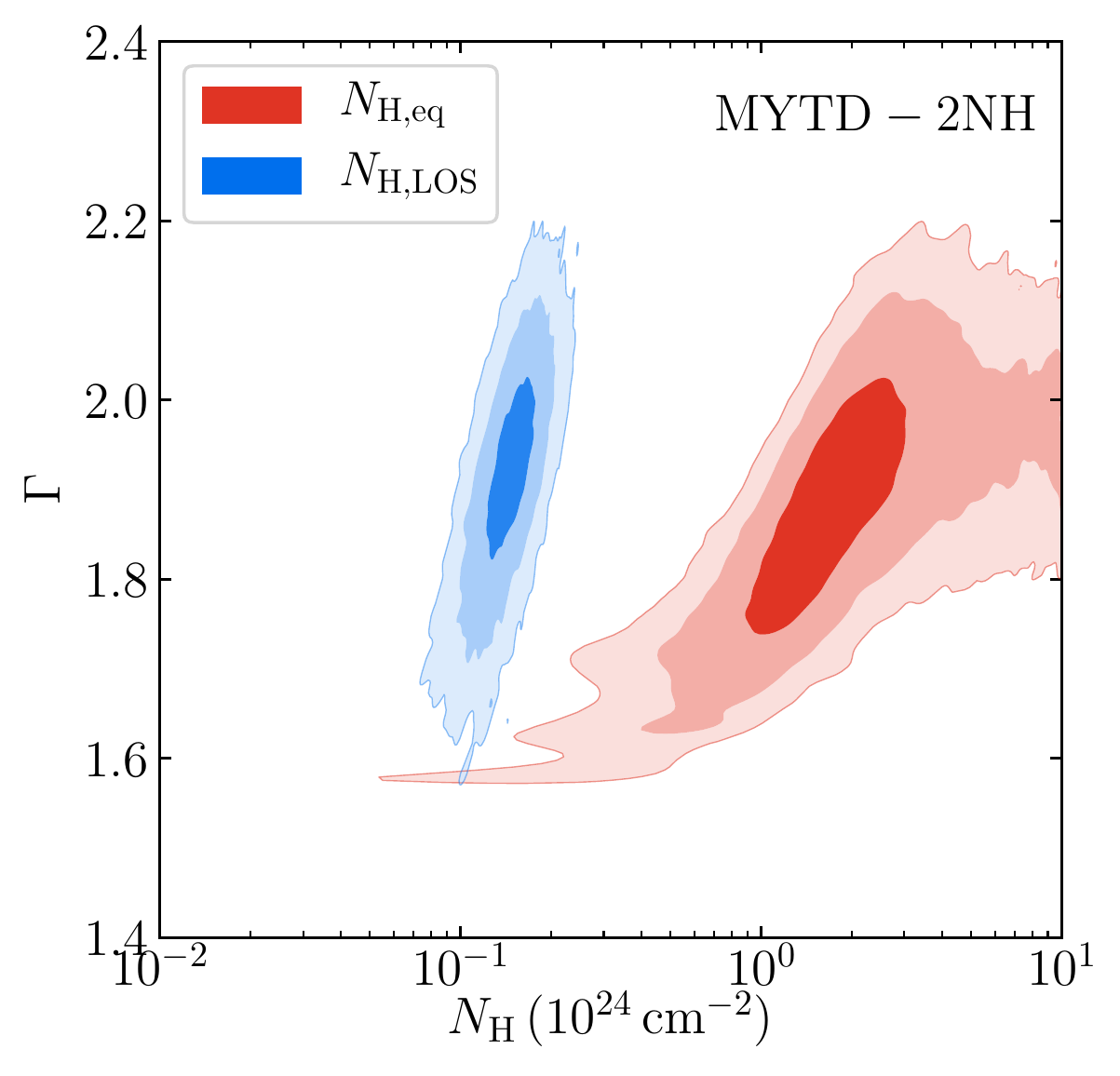}\\
	
	\includegraphics[width = 0.24\textwidth]{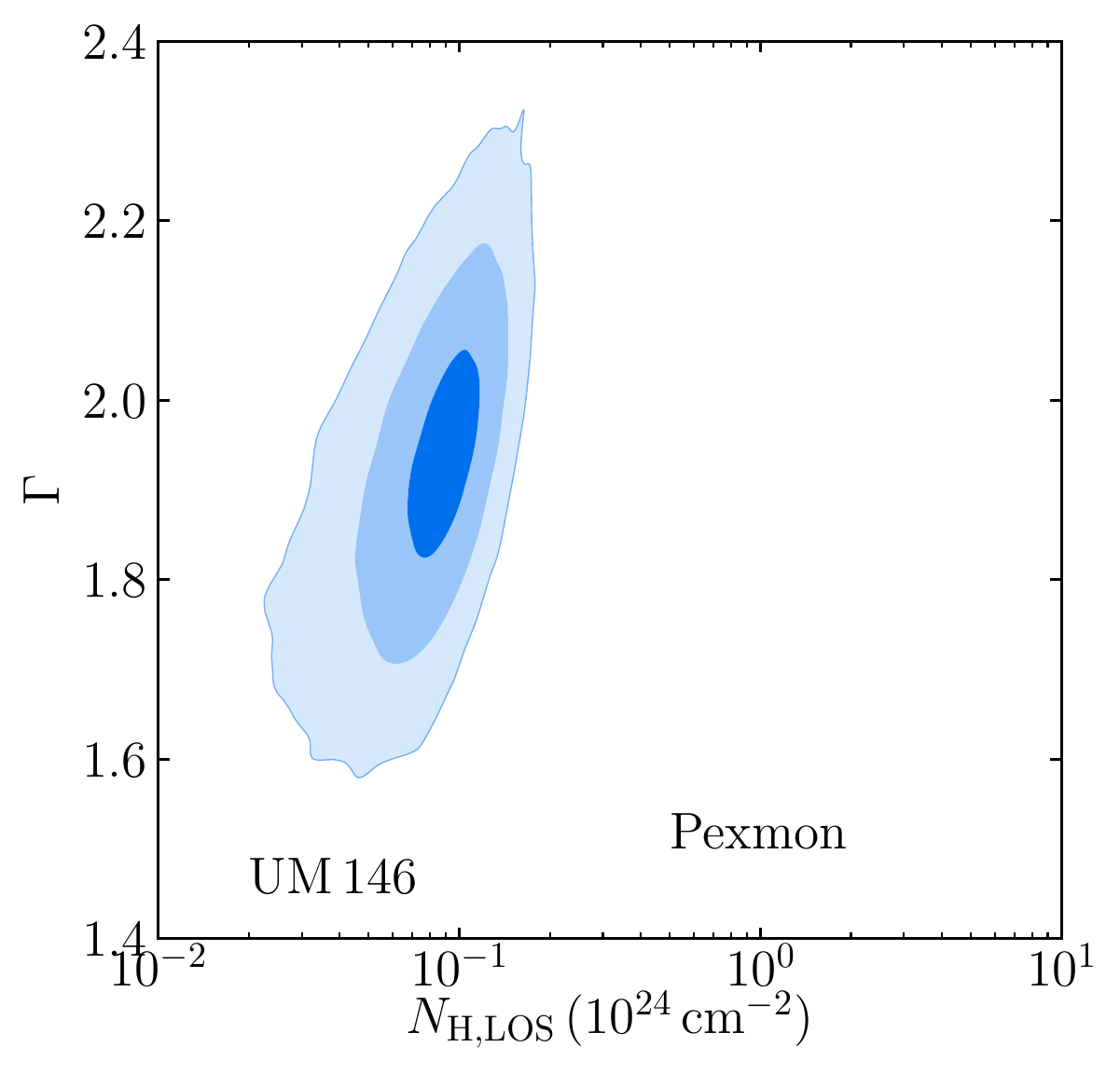}
	\includegraphics[width = 0.24\textwidth]{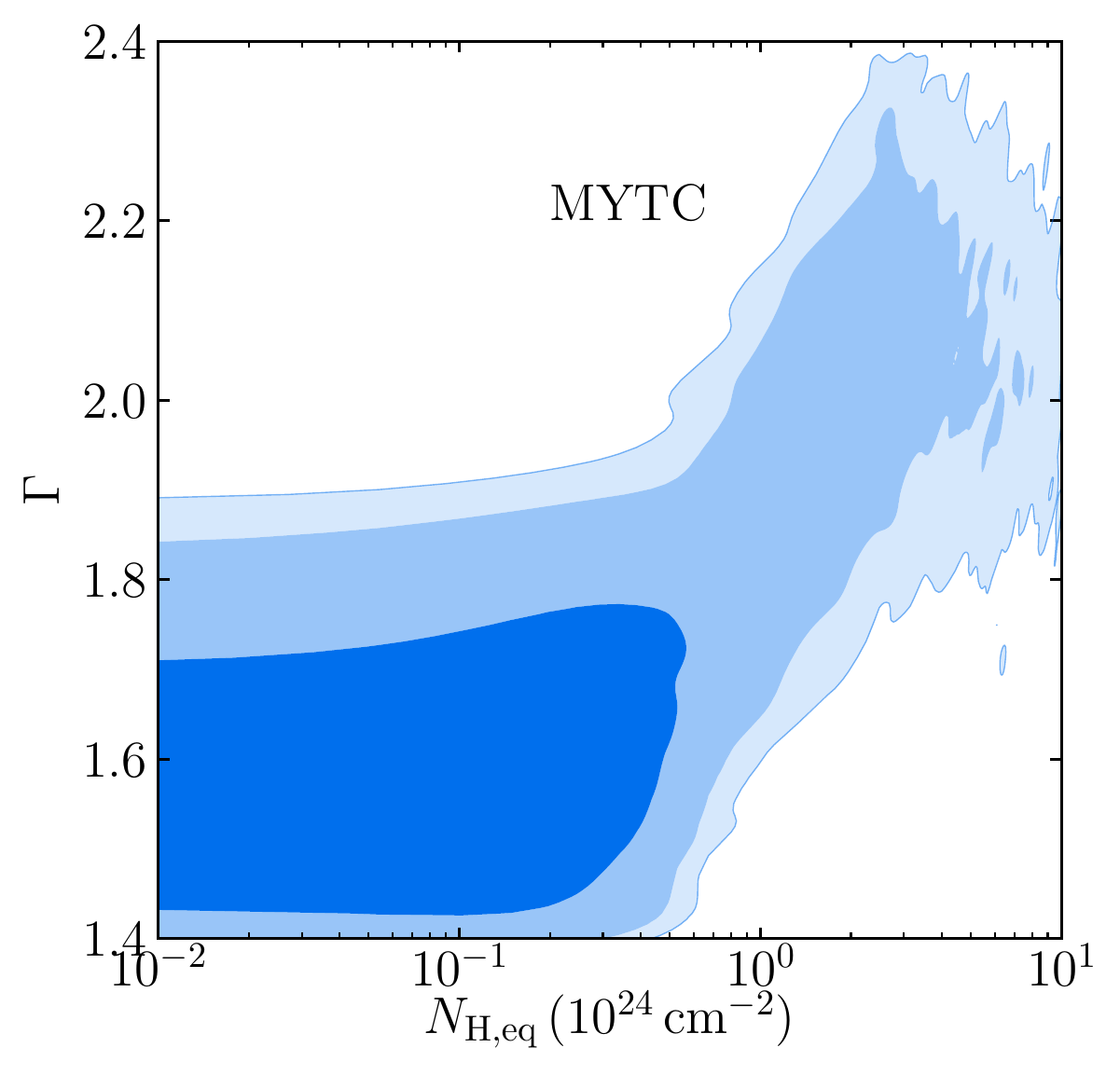}
	\includegraphics[width = 0.24\textwidth]{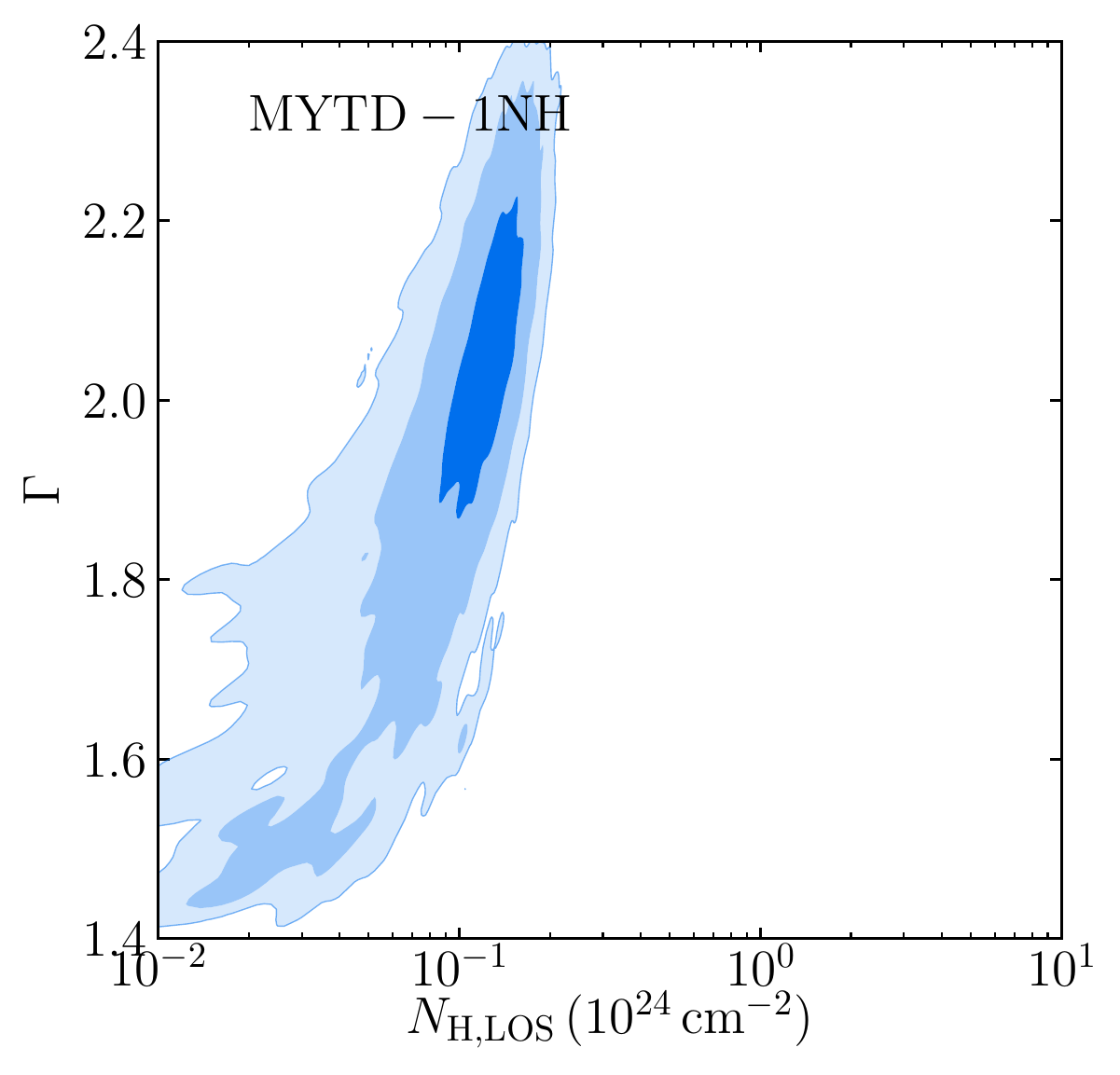}
	\includegraphics[width = 0.24\textwidth]{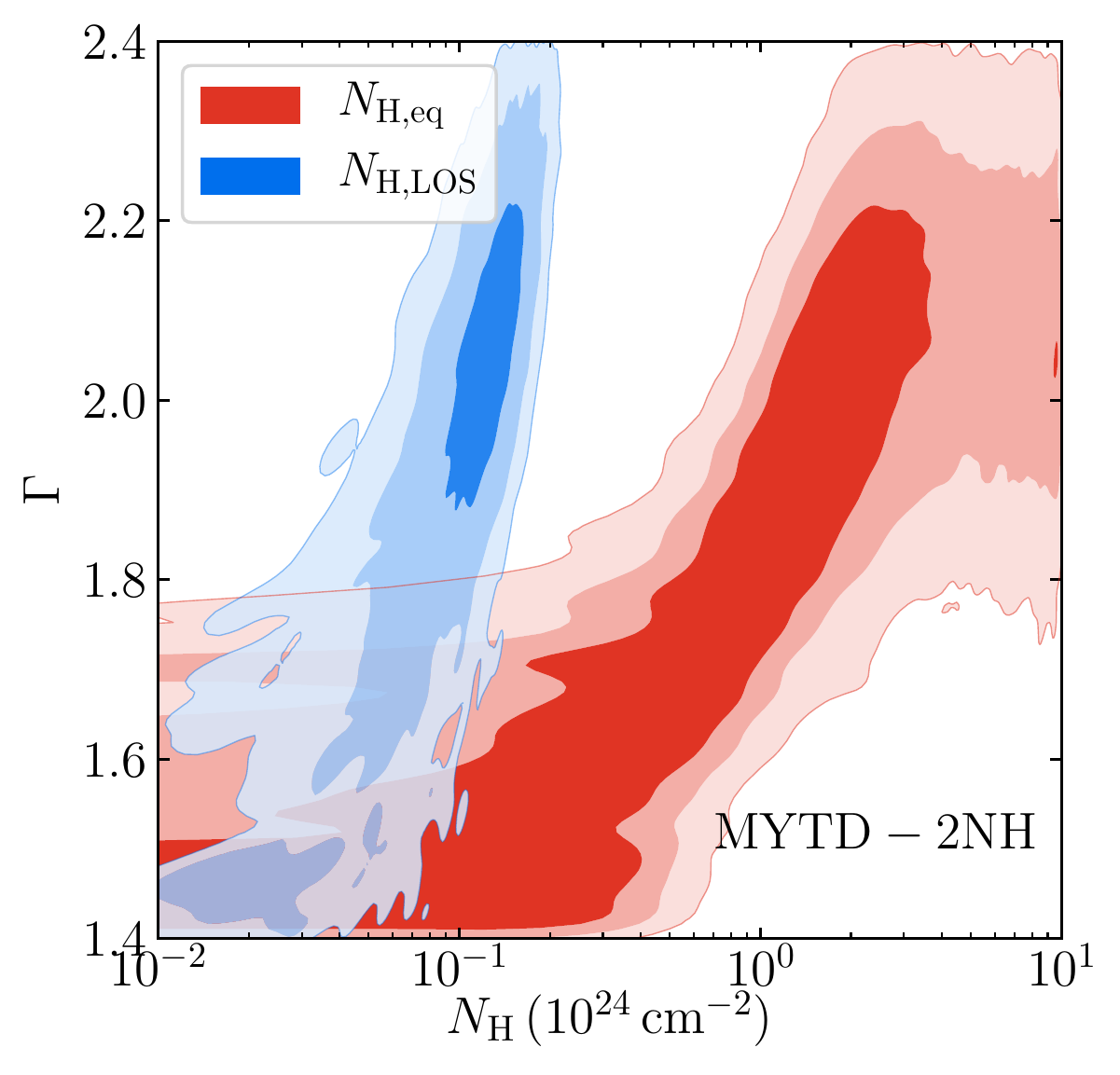}\\

	\includegraphics[width = 0.24\textwidth]{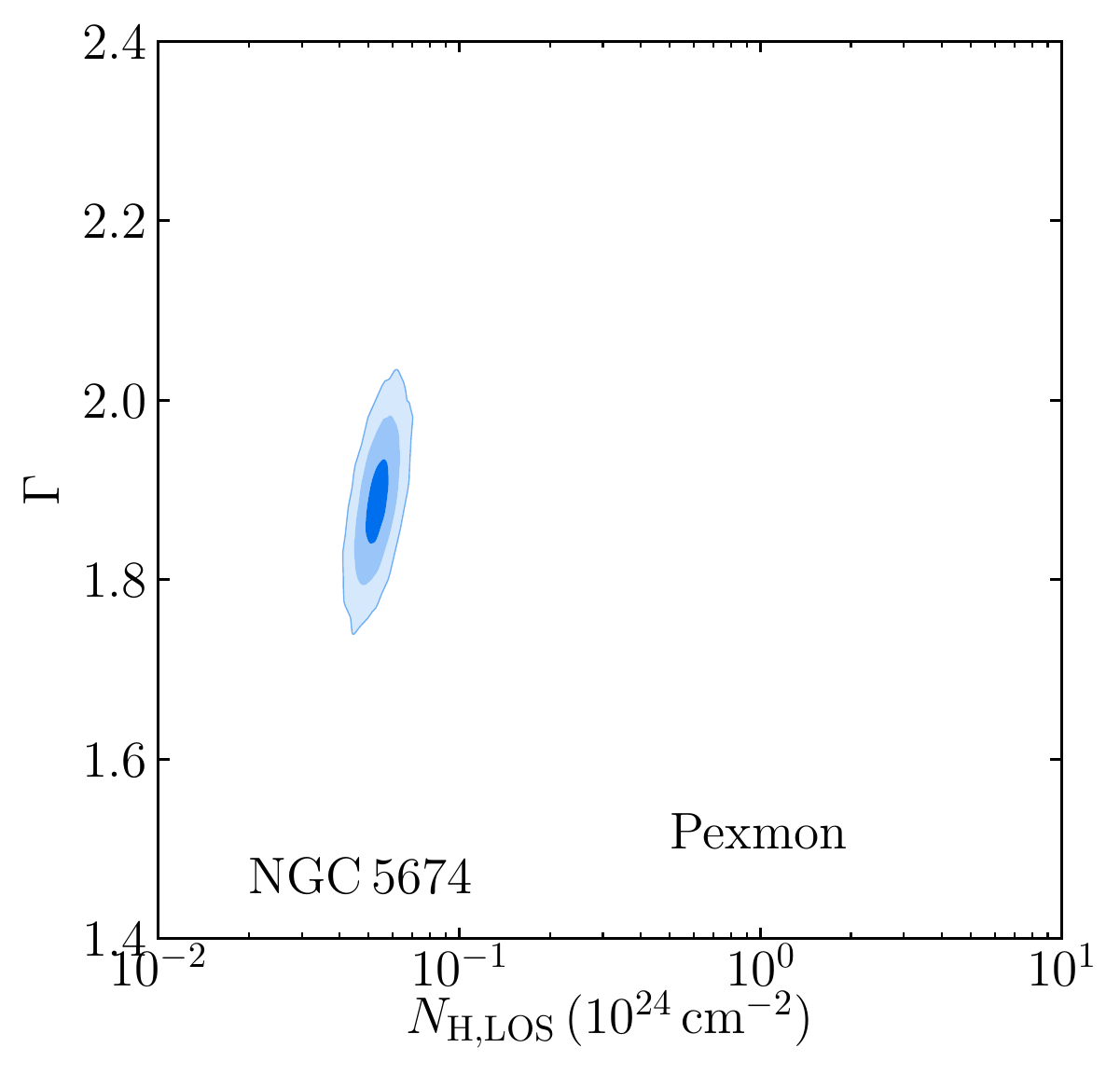}
	\includegraphics[width = 0.24\textwidth]{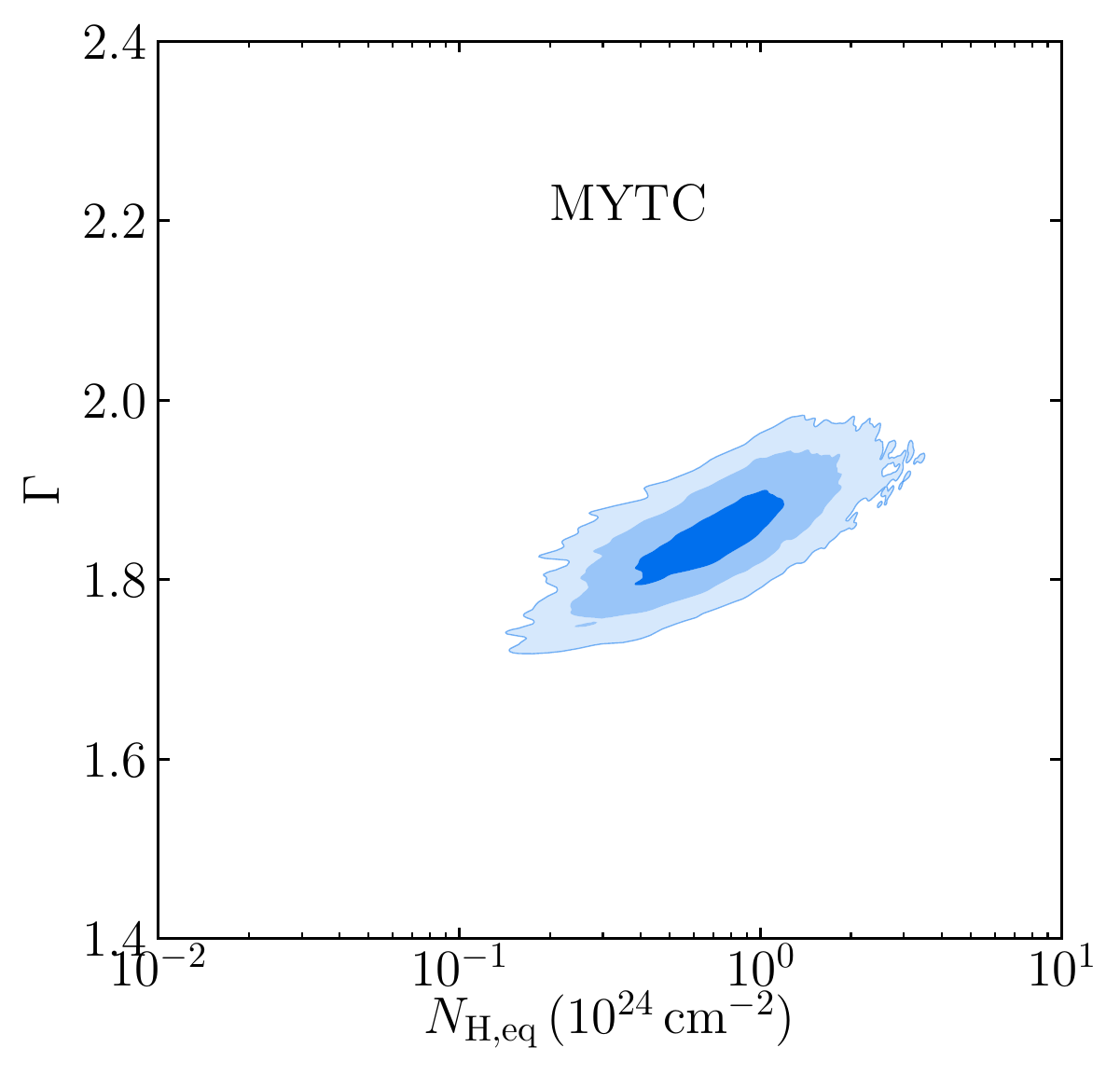}
	\includegraphics[width = 0.24\textwidth]{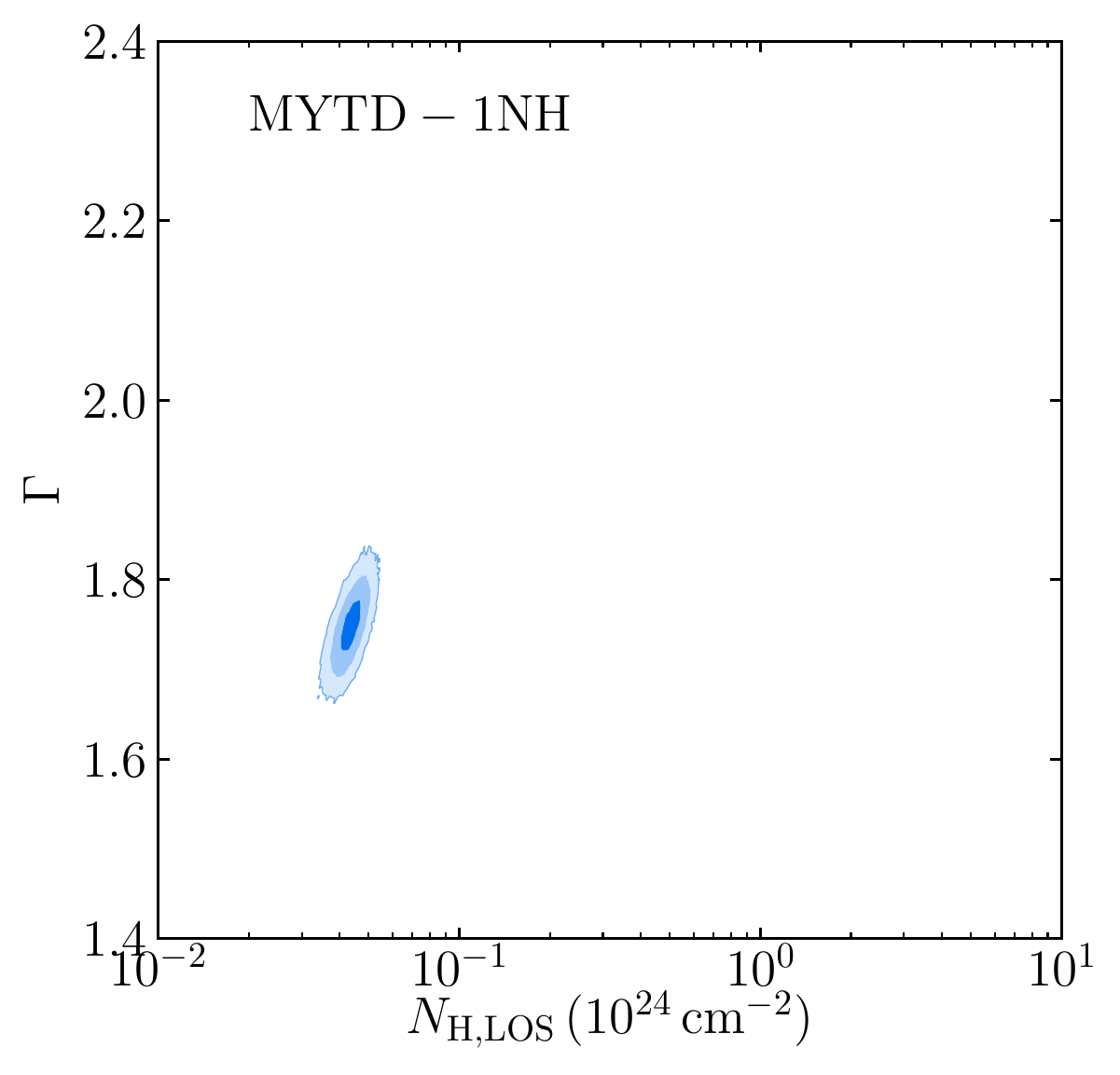}
	\includegraphics[width = 0.24\textwidth]{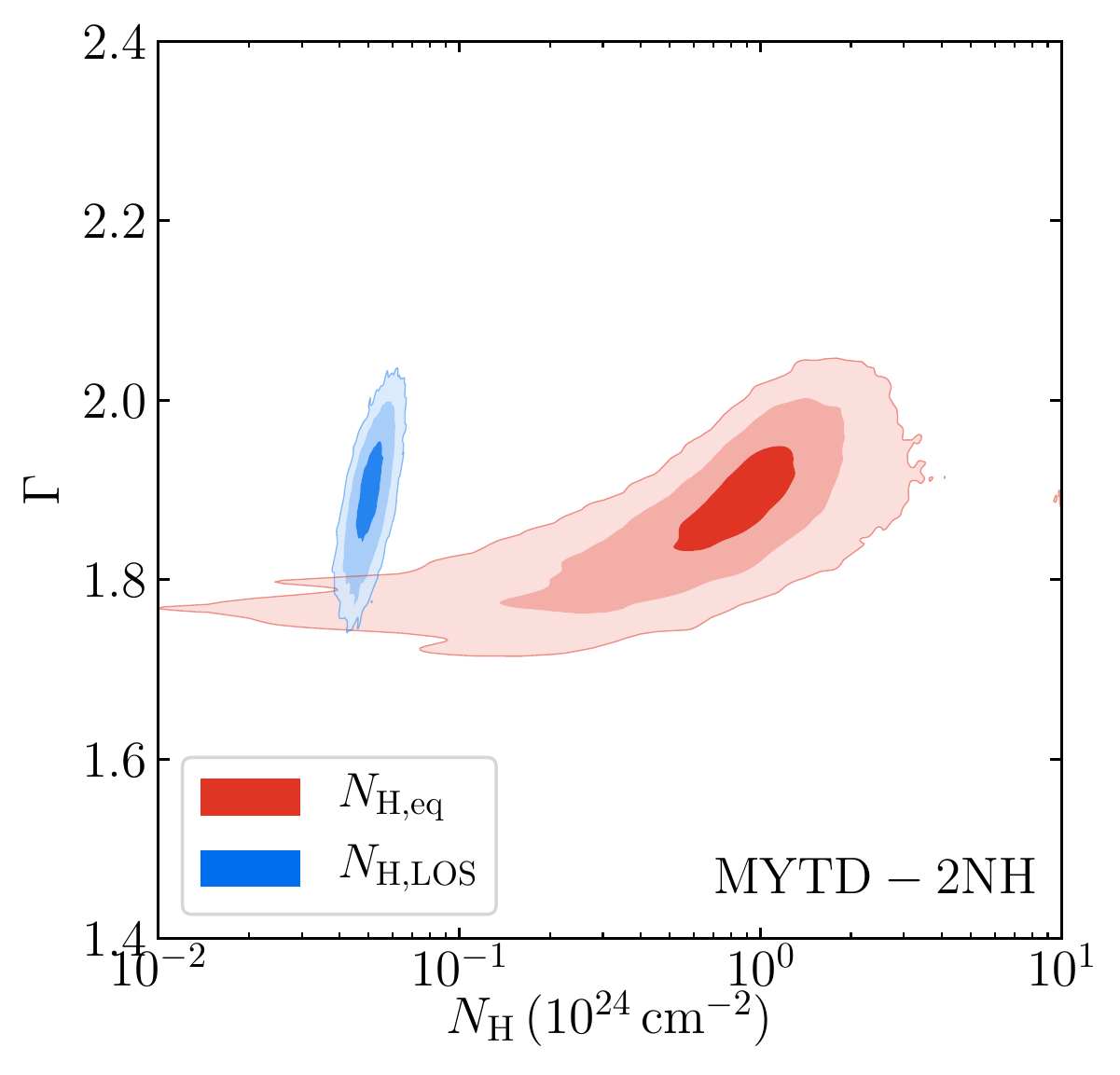}\\	
		
	\includegraphics[width = 0.24\textwidth]{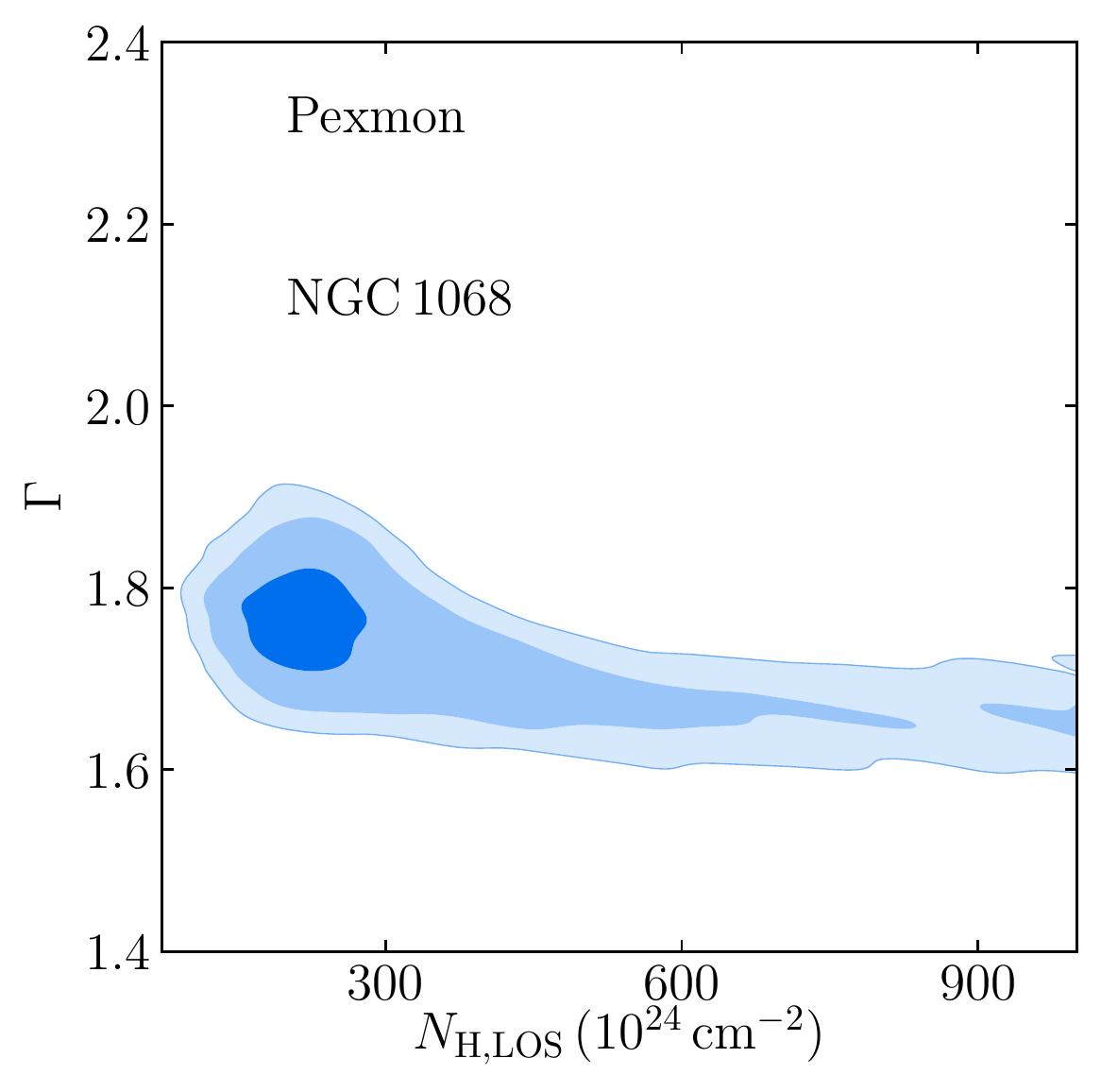}
	\includegraphics[width = 0.24\textwidth]{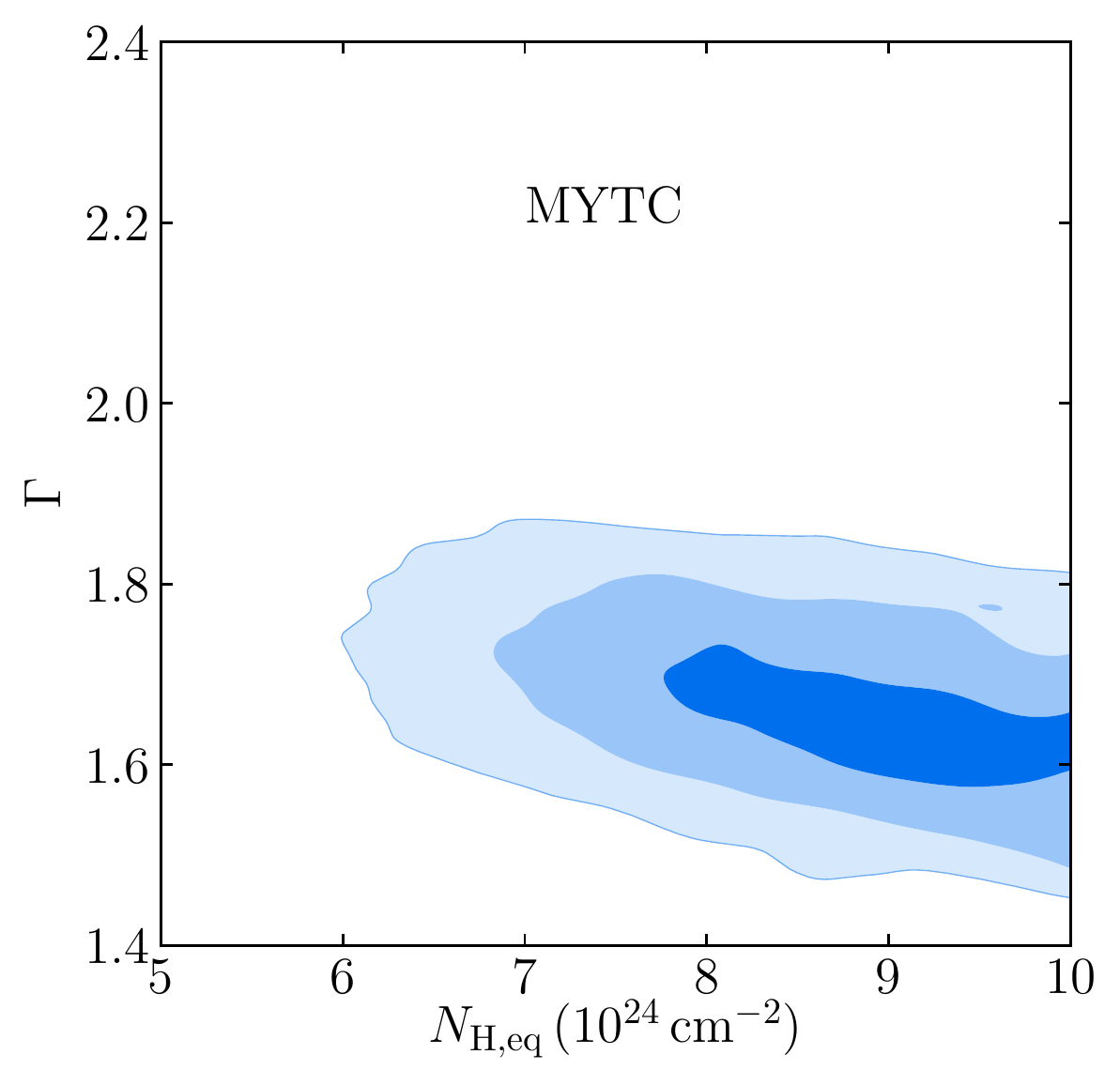}

\caption{Same as Figure~\ref{fig:mcmc1}.}
\label{fig:mcmc3}
\end{figure*}

\begin{figure*}
\centering
	\includegraphics[width = 0.24\textwidth]{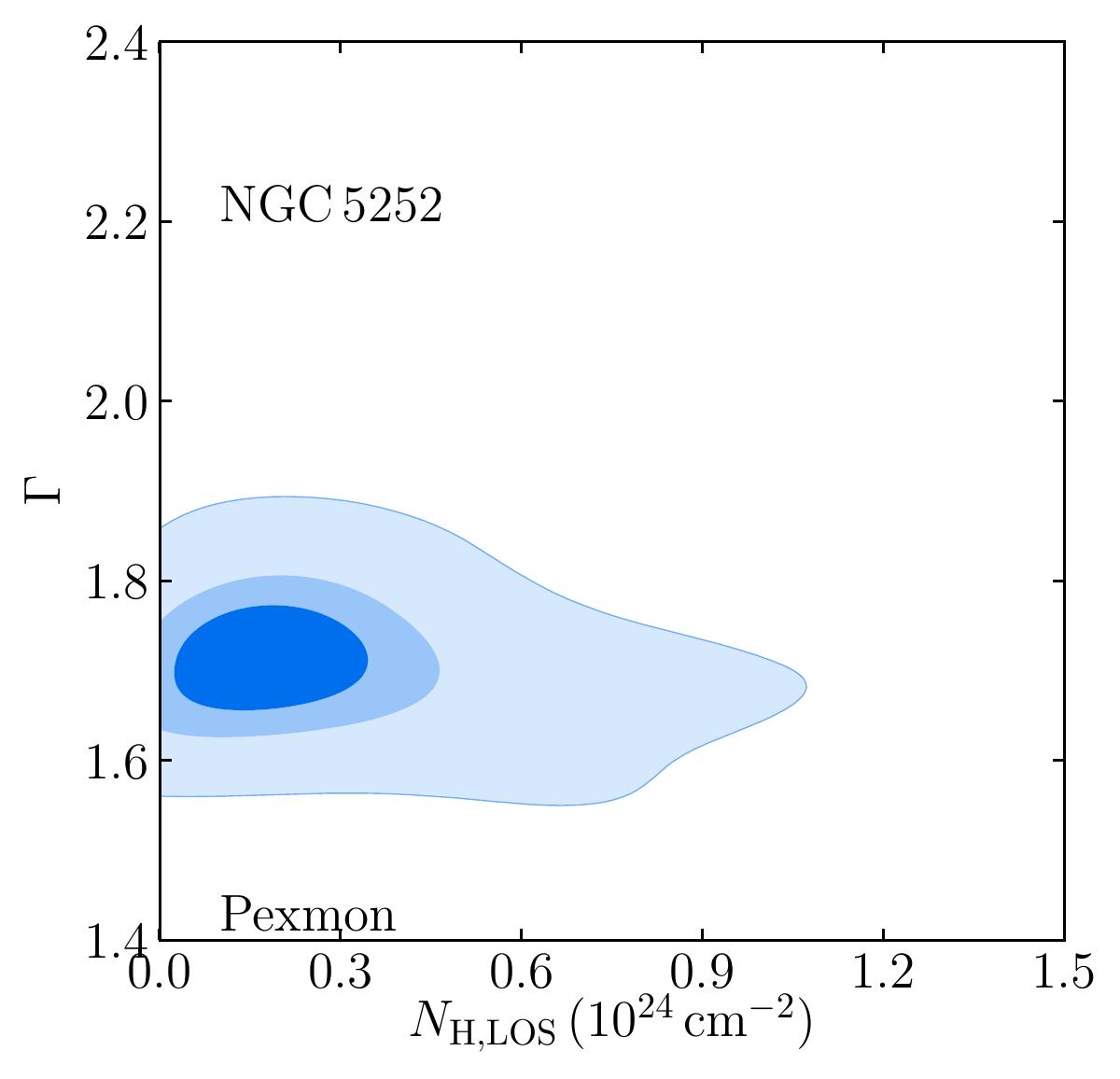}
	\includegraphics[width = 0.24\textwidth]{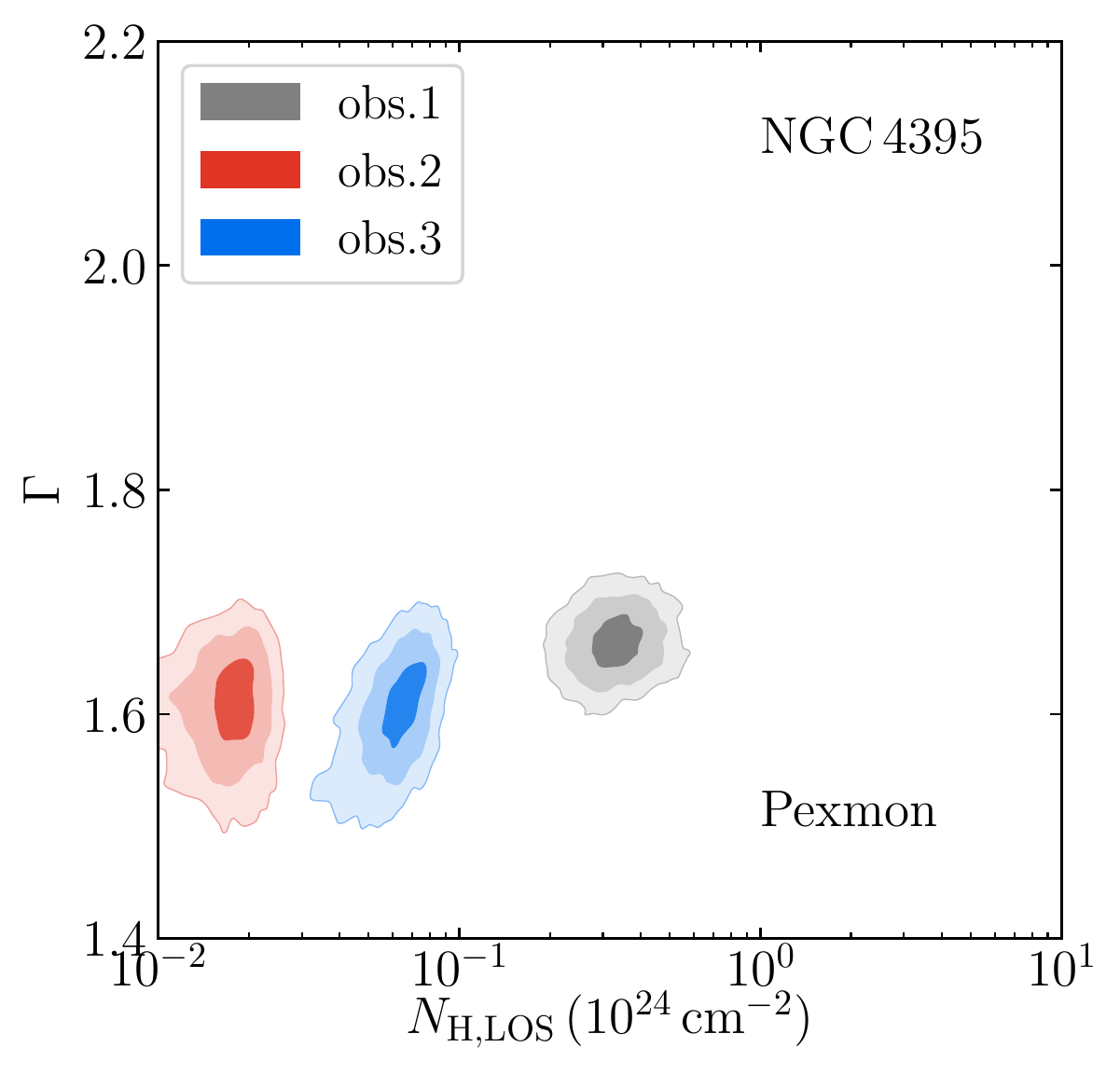}
	\includegraphics[width = 0.24\textwidth]{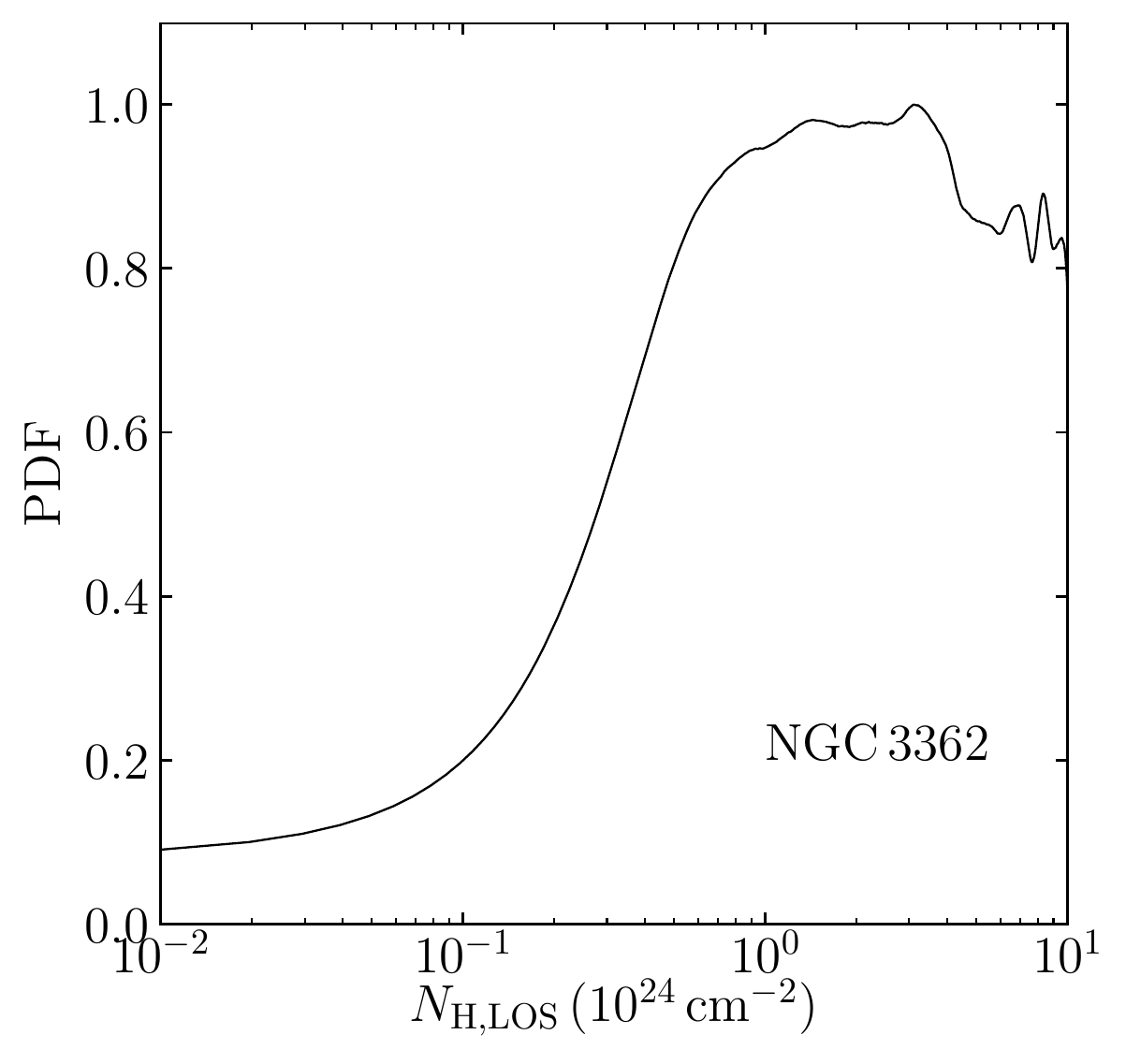}\\
	\includegraphics[width = 0.24\textwidth]{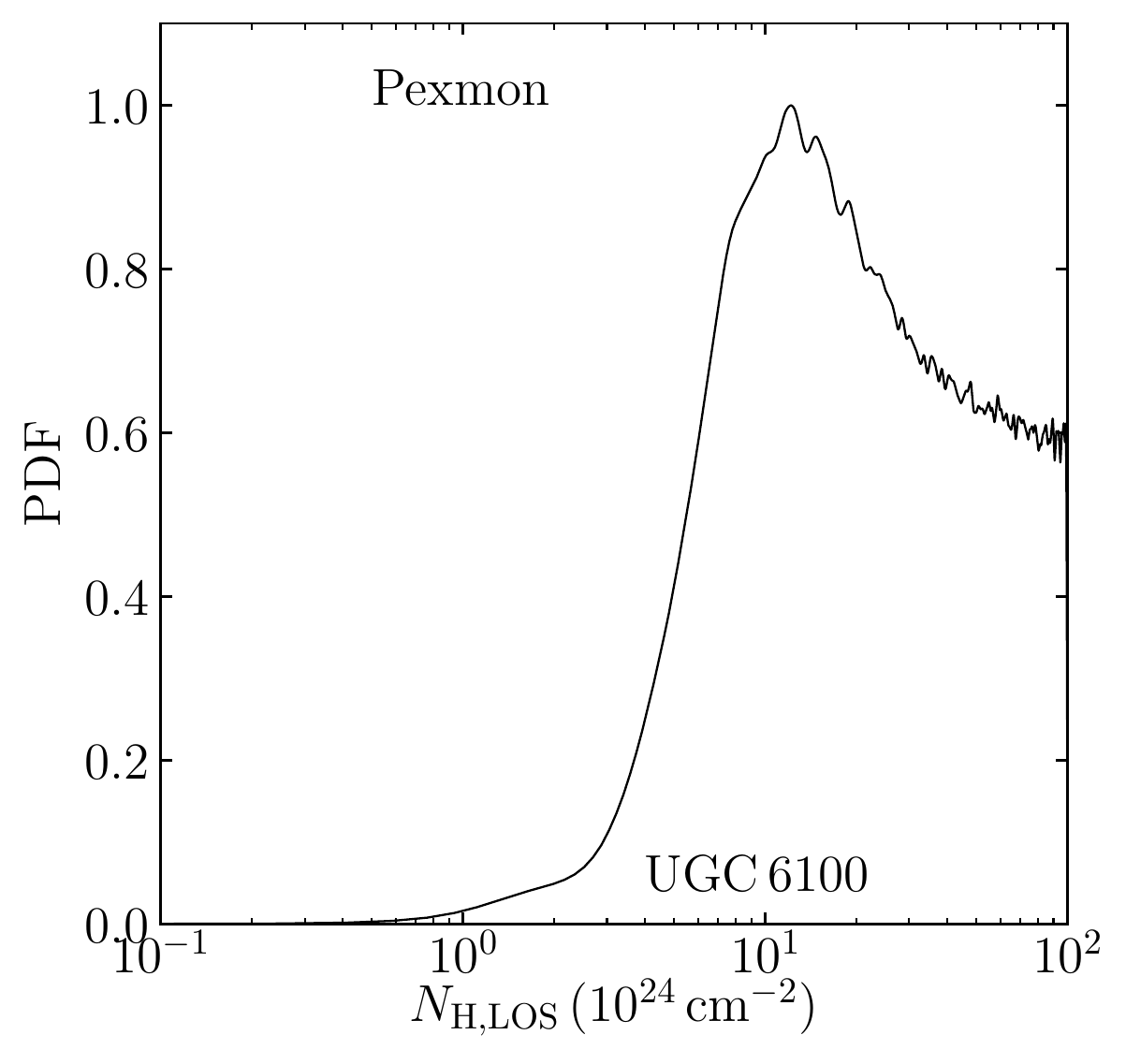}
	\includegraphics[width = 0.24\textwidth]{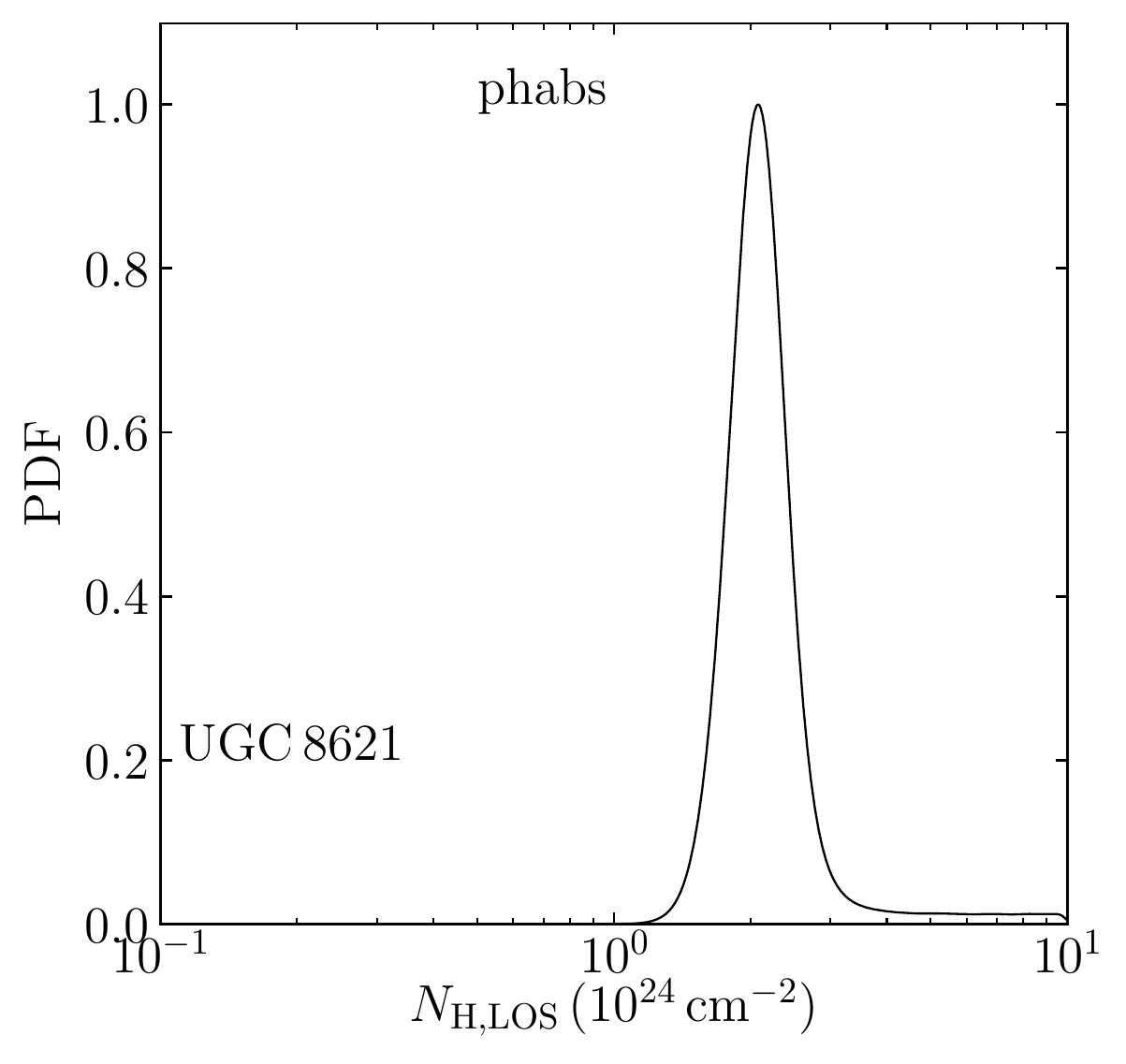}
\caption{Same as Figure~\ref{fig:mcmc1} but for the sources where we could not apply the MYTorus models.}
\label{fig:mcmc4}
\end{figure*}

In this section we present details on the spectral analysis of each of the sources studied in this paper. The spectra fitted with the `Pexmon' model are shown in Figures~\ref{fig:spectra1}-\ref{fig:spectra2}. The best-fit parameters obtained from the various models are shown in Tables~\ref{table:pexmon}-\ref{table:mytd2nh}. The $\Gamma-N_{\rm H}$ confidence contours for each model are shown in Figures~\ref{fig:mcmc1}-\ref{fig:mcmc4}.

\textbf{Mrk~573:} We fitted the \xmm\ EPIC-pn and the \nustar\ FPMA/FPMB spectra of this source using the four models that are described in Section~\ref{sec:spectralfit}. The source does not show any variability, thus we kept all the parameters tied for both observations. All models result in a CT column density in the LOS, in agreement with the results by \cite{Zhao20}. We note that MYTC does not provide a good fit using the standard configuration, i.e., by fixing the weight $A$ to unity. Instead, it requires a large value of $A=69_{-60}^{+20}$. The soft X-rays are fitted using two {\tt Apec} components with temperatures of 0.17 and 0.88~keV.

\textbf{NGC~1144:} We fitted the \xmm\ EPIC-pn and the \nustar\ FPMA/FPMB spectra of this source using the four models that are described in Section~\ref{sec:spectralfit}. The source shows variability. Hence, we fitted both observations assuming the same power law slope but we let the power law normalization, the column densities, and the reflection fraction (in the case of Pexmon) free to vary. The fits indicate changes in the intrinsic power law luminosity and the absorbing $N_{\rm H}$, confirmed by all models. The column density increases from $\sim 6\times 10^{23}~\rm cm^{-2}$ in the \xmm\ observation \citep[consistent with the results found by][]{Winter08, Winter09} to $\sim 10^{24}~\rm cm^{-2}$ in the \nustar\ observation, at the lower limit of CT. The soft X-rays are fitted using two {\tt Apec} components with temperatures of 0.33 and 1.46~keV.

\textbf{NGC~3362:} The source was not detected by \nustar. However, it was detected by \xmm\ below 4~keV. Thus we could not use neither the Pexmon nor the MYTorus configurations. We fitted the \xmm\ spectrum assuming an absorbed power law plus a scattered component and two {\tt Apec} components. We were not able to constrain the power law photon index, hence we fixed it to 1.8 and 2.2. Both values result in consistent results for the absorbing column density being CT. We adopt the value obtained for $\Gamma =1.8$, being $N_{\rm H} = (5 \pm 3) \times 10^{24}~\rm cm^{-2} $. We also fixed the fraction of the scattered component to its best-fit value of $C_{\rm sc} = 0.08$. The temperatures of the {\tt Apec} components are 0.05 and 0.79~keV. 

\textbf{UGC~6100:} The source was observed but not detected by \xmm. However, it was detected by \nustar. We fitted the FPMA/FPMB spectra using an absorbed power law plus Pexmon, with the reflection fraction fixed at unity. The power law photon index was not constrained, so we fixed it to 1.8 and 2.2. Both values result in good fits and consistent results for the absorbing column density being very high. We adopt the value obtained for $\Gamma =1.8$, being $N_{\rm H} = 2.8_{-1.1}^{+5.4}\times 10^{25}~\rm cm^{-2}$. We also applied different binning schemes to test the validity of our results, and all resulted in consistent values of $N_{\rm H}$. Due to the low quality of the spectra, MYTorus models result in no constraints.

\textbf{NGC~3982:} We fitted the \xmm\ EPIC-pn and the \nustar\ FPMA/FPMB spectra of this source using the four models that are described in Section~\ref{sec:spectralfit}. We tied all parameters between the spectra of the two observatories. The Pexmon and MYTC models suggest a Cth source with $N_{\rm H} = 6 \times 10^{23}~\rm cm^{-2}$. However, MYTD-1NH and MYTD-1NH suggest a CT column density with $N_{\rm H} = (5.3 \pm 3.3)\times 10^{24}~\rm cm^{-2}$ and $(4.5 \pm 3.1 )\times 10^{24}~\rm cm^{-2}$, respectively. We note that \cite{Panessa06} found that the source is a CT candidate based on the $F_{\rm X}/F_{\rm [OIII]}$ vs $F_{\rm [OIII]}/F_{\rm IR}$ diagnostic. The soft X-rays are modeled using a single {\tt Apec} component with $kT = 0.36~\rm keV$.

\textbf{NGC~4388:} The source has been observed three times with \xmm\ and one time with \nustar. We fit the spectra from the four observations above 3~keV in order to avoid the complexity of modeling the soft X-rays in this source \citep[see e.g.,][]{Miller19}. We considered total obscuration by fixing $C_{\rm sc}$ to zero in order to avoid overestimating the soft emission. We kept $\Gamma$ tied between the four observations and let the power law normalization and $N_{\rm H}$ vary. For MYTC the fit requires $A \neq 1$. We left $\mathcal{R}$ ($A$) free to vary between the different observations, for the Pexmon (MYTC) model. All four models indicate a Cth ($N_{\rm H} \sim 2.2-3.6 \times 10^{23}~\rm cm^{-2}$) source, in agreement with the results found by \cite{Masini16} and \cite{Miller19}. Our best fits show hints of a change in $N_{\rm H}$ by $\sim 1.5 \times 10^{23}~\rm cm^{-2}$. In addition, our modeling suggests variability in both intrinsic and reprocessed emission. The \xmm\ observation with the highest flux indicates the presence of absorption lines, which we modeled using three Gaussian lines in absorption. The best-fit suggests absorption lines at $6.61^{+0.11}_{-0.07}$~keV, $6.92 \pm 0.03$~keV, and $7.78^{+0.12}_{-0.08}$~keV. 

\textbf{UGC~8621:} The source was not detected by \nustar. However, it was observed and detected by \xmm\ below $\sim 6$~keV. Being unable to fit the spectrum in the Fe K range, we were neither able to use the Pexmon component, nor any of the MYTorus configurations. We fitted the spectrum assuming an absorbed power law plus a scattered component and a single {\tt Apec}. We were not able to get any constrains on the photon index so we fixed it to 1.8. The scattered fraction could not be constrained either, so we fixed it to $C_{\rm sc} = 0.002$. The best-fit model suggests a CT column density with $N_{\rm H} = 2.54_{-1.0}^{+0.3}\times 10^{24}~\rm cm^{-2}$.

\textbf{NGC~5252:} We analyzed the \xmm\ and \nustar\ spectra of the source. The source showed intrinsic and absorption variability, requiring both ionized and neutral partial covering absorption. The best-fit model, in XSPEC parlance, can be written as: $\rm zpcfabs \times zxipcf \times zcutoffpl + pexmon + Apec + Apec$. For the \nustar\ observation, due to the lack of good-quality simultaneous soft X-ray observations, we cannot constrain the parameters of both absorption components. Hence, we removed the ionized absorption and linked the neutral $N_{\rm H}$ to the one of \xmm, and let its covering fraction vary freely. The best-fit results for the neutral absorption are $N_{\rm H} = 1.48^{+0.15}_{-0.31} \times 10^{23}~\rm cm^{-2}$ and $(4.19\pm 0.93) \times 10^{22}~\rm cm^{-2}$, with a covering fraction of  $0.38^{+0.06}_{-0.04} ~(0.60\pm 0.05)$  for the \xmm\ (\nustar) observation. As for the ionized absorption, the best-fit results are $N_{\rm H} =3.26^{+0.92}_{-0.75}  \times 10^{22}~\rm cm^{-2}$, $\log \left( \xi/\rm erg~cm~s^{-1} \right) = -0.16^{+0.39}_{-0.22}$, $f_{\rm cov} =0.98\pm 0.01$. We tied the photon indices between  the two observations. We find a best-fit value of $\Gamma = 1.83^{+0.03}_{-0.04}$. Our results are in contradiction with the ones by \cite{Dadina10} who analyzed the same \xmm\ observation presented in the current work, in addition to a \chandra\ observation of the source. The authors modeled the spectra of this source with an extremely flat photon index of $\Gamma = 1.00^{+0.08}_{-0.06}$. Such a flat photon index can be ruled out by our data, thanks to the \nustar\ data.

\textbf{NGC~5347:} We fit the spectra obtained by \chandra, \suzaku, and \nustar\ using Pexmon, MYTC, and MYTD-1NH. We cannot get useful constraints on $N_{\rm H,eq}$  by letting it be free (i.e., MYTD-2NH model). A detailed analysis of this source is presented in \cite{Kammoun19n5347}, where we confirm the CT nature of the source. The only difference between the current work and  \cite{Kammoun19n5347} is that now we tie the Pexmon normalization to the one of the power law component and we let $\mathcal{R}$ free. This does not alter any of the previous results.

\textbf{NGC~5695:} We fit the spectra obtained by \xmm\ and \nustar. The \xmm\ spectra are background-dominated above 2~keV. We modeled the spectra assuming no variability. The soft X-rays are modeled invoking a single {\tt Apec} component. The results from the four models confirm a CT LOS column density in this source. This source was identified as a CT candidate based on its \xmm\ by \cite{Lamassa09}.

\textbf{NGC~5929:} We fit the \suzaku\ and \nustar\ FPMA/FPMB spectra for this source. All four models used in this analysis confirm a Cth column density during this observation. We note that variability in the intrinsic flux can be seen between the two observations. Letting the column density vary between the observations resulted in consistent results.

\textbf{NGC~7674:} We fit the non-simultaneous spectra obtained by \xmm\ and \nustar. We allow for both intrinsic and absorption-induced variability. We also allow the fraction of scattered (or unabsorbed) power law to vary between the two observations. The spectra showed an excess at $\sim 7$~keV that possibly corresponds to \ion{Fe}{26}. We modeled it by adding a Gaussian emission line. The soft X-rays are modeled using two {\tt Apec} components. The results from Pexmon and MYTD-2NH suggest that the LOS column density is consistent and Cth in both observations. However, the MYTD-1NH model suggests that the column density in the \xmm\ observation is Cth while the \nustar\ one is CT. We note that the best-fit MYTD-1NH returns $\chi^2/\rm dof = 175/156$ while the ones by Pexmon and MYTD-2NH return a better fit with $\chi^2/\rm dof = 166/155$ and $159/155$, respectively. All fits are statistically acceptable, however the MYD-1NH suggesting a CT solution is statistically worse than the other two models. Our results ($N_{\rm H, LOS} = 6.6_{-5}^{+2}, 3.9_{-1.6}^{+1.2}\times 10^{23}~\rm cm^{-2}$ using Pexmon and MYTD-2NH, respectively) are in agreement with the ones by \cite{Lamassa11} who also analyzed the \xmm\ observations of this source. \cite{Tanimoto20} fitted the \suzaku\ and \nustar\ spectra of NGC~7674 using XCLUMPU \citep{Tanimoto19} and found $N_{\rm H,LOS} = 2.4^{+2.2}_{-1.0}\times 10^{23}~\rm cm^{-2}$, which is consistent with our results. In contrast, \cite{Gandhi17} analyzed the \nustar, \suzaku, and \swift/XRT spectra of this source and claimed the presence of a CT column density, which we did not recover from our analysis. We also note that fitting the spectra using MYTC required the weights of the reprocessed components to be different than unity (standard configuration) with $A = 11.0_{-5.0}^{+2.4}$.

\textbf{NGC~7682:} We fit the \xmm\ and \nustar\ spectra of this source, and found no indication of variability. The source is detected out to $\sim 7$~keV in the \xmm\ observation. We fit the spectra using all four models. All models result in consistent and Cth column densities in our LOS. Both FPMA and FPMB spectra show an excess at $\sim 9~ \rm keV$ that we modeled using a Gaussian emission line (with a best-fit energy of $9.5 \pm 0.3~\rm keV$). The soft X-rays are modeled using a single {\tt Apec} component with $kT = 0.64$~keV.

\textbf{NGC~4395:} We fit the flux-resolved spectra obtained during two \xmm\ observations and a \nustar\ observation. Our model consists of a partial covering neutral absorber and two ionized partial covering absorbers. A detailed analysis is presented in \cite{Kammoun19n4395}.  The only difference between the current work and  \cite{Kammoun19n4395} is that now we tie the Pexmon normalization to the one of the power law component and we let $\mathcal{R}$ be free. This does not alter any of the previous results. Due to the complexity of the model invoking partial covering and ionized absorption, we were not able to fit the spectra with MYTorus.

\textbf{Mrk~334:} We fit the \swift/XRT and \nustar\ FPMA/FPMB spectra of this source, assuming an absorbed power law plus reprocessed emission and an {\tt Apec} component. The source shows flux variability. Due to the poor quality of the XRT spectrum we are not able to assess whether the variability is intrinsic or due to a change in absorption. Leaving $N_{\rm H}$ free to vary results in unconstrained results. Thus, we assumed a change in the intrinsic power law flux. The best-fit results indicate a change by a factor of $\sim 1.6$ in intrinsic flux.

\textbf{NGC~5283:} We fit the \chandra\ and \nustar\ spectra of this source. The spectra are fitted assuming an absorbed power law plus scattered component and reprocessed emission. All four models were applied and suggest consistent and Cth absorption in our LOS.  We note that variability in the intrinsic flux can be seen between the two observations. Letting the column density to vary between the observations resulted in consistent results.

\textbf{UM~146:} The source was detected with XRT only above $\sim 3$~keV (a total of 10 counts in 2~ks). Thus, we fit the \nustar\ FPMA/FPMB spectra only, assuming an absorbed power law plus reprocessed emission. All four models suggest a Cth column density in our LOS.

\textbf{NGC~5674:} We fit the \swift/XRT and \nustar\ spectra of this source. The XRT spectrum was fitted above 1~keV. The spectra are fitted assuming an absorbed power law plus scattered component and  a reprocessed emission. All four models were applied and suggested consistent and Cth absorption in our LOS. 

\textbf{NGC~1068:} Due to the complexity of the spectra of this source in the soft X-rays \cite[][]{Kallman14, Bauer15} we fit the FPMA/FPMB spectra extracted from only one \nustar\ observation. The source is known to be consistently highly CT. Detailed analysis of the multi-epoch, X-ray spectra of this source are presented in \cite{Bauer15} and \cite{Zaino20}. We modeled the spectra using an absorbed power law plus a high-temperature {\tt Apec} component and reprocessed emission. We applied the Pexmon and MYTC models only. The two models consistently confirm the CT nature of the source. We found in both models an excess in the $6-7$~keV that we model with a Gaussian emission line at the rest-frame of the source. We found a best-fit energy of 6.3~keV for both models. We note that \cite{Zaino20} report the presence of a similar excess in the \nustar\ spectra of this source. The authors model this feature with a Gaussian line fixed at 6~keV, attributing it to instrumental calibrations. Investigating the origin of this feature is beyond the scope of the current paper. It is also worth noting that the Pexmon model implies large LOS column densities that cannot be achieved by MYTorus (having an upper limit at $10^{25}~\rm cm^{-2}$. This resulted in discrepancy in the intrinsic luminosity that are implied by our best-fits being $4.7\times 10^{43}~\rm erg~s^{-1}$ and $2.3 \times 10^{42}~\rm erg~s^{-1}$ for Pexmon and MYTC, respectively. The $L_{2-10}$ value inferred by Pexmon is in agreement with the results by \cite{Marinucci16} and \cite{Zaino20}, favoring a high luminosity (a few times $10^{43}~\rm erg~s^{-1}$) by adding the constraints from the mid- IR and [\ion{O}{3}] observations of this source.


\clearpage

\startlongtable
\begin{deluxetable*}{lllllll}
\tablecaption{Log of the observations that we analyzed. The last column indicates the statistics used for the spectral fits. \label{table:log}}
\tablehead{\colhead{Source} &  \colhead{Instrument} & \colhead{ObsID} & \colhead{Obs date} & \colhead{Count rate$^\ast$} & \colhead{exp. time} & \colhead{Statistics} \\
\colhead{} &  \colhead{} & \colhead{} & \colhead{} & \colhead{($\rm count~ ks^{-1}$)} & \colhead{(ks)} & \colhead{}}
\startdata
Mrk~573	&	\xmm	&	200430701	&	2004-01-15	&$	190 \pm 5	$&	9.04	&	$\chi^2$	\\
	&	\nustar/FPMA	&	60360004002	&	2018-01-06	&$	9.15 \pm 0.6	$&	30.4	&		\\
	&	\nustar/FPMB	&	60360004002	&		&$	8.13 \pm 0.6	$&	31.6	&		\\
NGC~1144	&	\xmm	&	312190401	&	2006-01-28	&$	210 \pm 5	$&	8.8	&	$\chi^2$	\\
	&	\nustar/FPMA	&	60368001002	&	2017-10-14	&$	48 \pm 2	$&	19.7	&		\\
	&	\nustar/FPMB	&	60368001002	&		&$	44 \pm 2	$&	20.1	&		\\
NGC~3362	&	\xmm	&	405240901	&	2007-06-05	&$	32 \pm 1	$&	26.1	&	$C$-stat	\\
	&	\nustar	&	60465003002	&	2018-11-26	&$	N/D	$&	32	.0&		\\
UGC~6100	&	\xmm	&	301151101	&	2005-10-27	&$	N/D	$&	14.4	&	$C$-stat	\\
	&	\nustar/FPMA	&	60465004002	&	2019-05-20	&$	2.32 \pm 0.4	$&	30.7	&		\\
	&	\nustar/FPMB	&	60465004002	&		&$	2.76 \pm 0.4	$&	30.5	&		\\
NGC~3982	&	\xmm	&	204651201	&	2004-06-15	&$	50 \pm 2	$&	9.2	&	$C$-stat	\\
	&	\nustar/FPMA	&	60375001002	&	2017-12-06	&$	4.26 \pm 0.4	$&	29.9	&		\\
	&	\nustar/FPMB	&	60375001002	&		&$	3.81 \pm 0.4	$&	30.6	&		\\
NGC~4388	&	\xmm	&	110930301	&	2002-07-07	&$	400 \pm 10	$&	3.9	&	$\chi^2$	\\
	&	\xmm	&	110930701	&	2002-12-12	&$	1140 \pm 10	$&	7.3	&		\\
	&	\xmm	&	675140101	&	2011-06-17	&$	1950 \pm 8	$&	31.9	&		\\
	&	\nustar/FPMA	&	60061228002	&	2013-12-27	&$	320 \pm 4	$&	21.4	&		\\
	&	\nustar/FPMB	&	60061228002	&		&$	310 \pm 4	$&	21.4	&		\\
UGC~8621	&	\xmm	&	204651101	&	2005-01-07	&$	16 \pm 1	$&	9	&	$C$-stat	\\
	&	\nustar	&	60368003002	&	2017-12-06	&$	N/D	$&	12.6	&		\\
NGC~5252	&	\xmm	&	152940101	&	2003-07-18	&$	1040 \pm 5	$&	36.9	&	$\chi^2$	\\
	&	\nustar/FPMA	&	60061245002	&	2013-05-11	&$	412 \pm 5	$&	19	&		\\
	&	\nustar/FPMB	&	60061245002	&		&$	384 \pm 5	$&	19	&		\\
NGC~5347	&	\suzaku	&	703011010	&	2008-06-10	&$	5.11 \pm 0.5	$&	42	&	$C$-stat	\\
	&	\chandra	&	4867	&	2004-06-05	&$	11.9 \pm 0.6	$&	36.9	&		\\
	&	\nustar/FPMA	&	60001163002	&	2015-01-16	&$	7.2 \pm 0.47	$&	46.6	&		\\
	&	\nustar/FPMB	&	60001163002	&		&$	8.9 \pm 0.49	$&	46.6	&		\\
NGC~5695	&	\xmm	&	504100401	&	2007-12-16	&$	12.2 \pm 1.5	$&	7.5	&	$C$-stat	\\
	&	\nustar/FPMA	&	60368004002	&	2018-01-16	&$	9.9 \pm 0.6	$&	38.5	&		\\
	&	\nustar/FPMB	&	60368004002	&		&$	7.9 \pm 0.6	$&	37.9	&		\\
NGC~5929	&	\suzaku	&	708022010	&	2013-12-25	&$	49 \pm 1	$&	24.65	&	$\chi^2$	\\
	&	\nustar/FPMA	&	60465009002	&	2018-09-17	&$	45.8 \pm 1.4	$&	28.5	&		\\
	&	\nustar/FPMB	&	60465009002	&		&$	39.9 \pm 1.3	$&	28.4	&		\\
NGC~7674	&	\xmm	&	200660101	&	2004-06-02	&$	118 \pm 4	$&	7.6	&	$\chi^2$	\\
	&	\nustar/FPMA	&	60001151002	&	2014-09-30	&$	28.6 \pm 0.8	$&	52	&		\\
	&	\nustar/FPMB	&	60001151002	&		&$	28.9 \pm 0.8	$&	52	&		\\
NGC~7682	&	\xmm	&	301150501	&	2005-05-27	&$	31.4 \pm 2	$&	13.1	&	$\chi^2$	\\
	&	\nustar/FPMA	&	60368002002	&	2017-10-06	&$	12.1 \pm 0.8	$&	22.4	&		\\
	&	\nustar/FPMB	&	60368002002	&		&$	9.2 \pm 0.7	$&	21.7	&		\\
NGC~4395$^\dagger$	&	\xmm	&	142830101	&	2003-11-30	&$		$&	88.7	&	$\chi^2$	\\
	&	\xmm	&	744010101	&	2014-12-28	&$		$&	36.2	&		\\
	&	\xmm	&	744010201	&	2014-12-30	&$		$&	22.5	&		\\
	&	\nustar	&	60061322002	&	2013-05-10	&$		$&	18.9	&		\\
Mrk~334	&	\swift/XRT	&	882529	&	2019-05-21	&$	5.3 \pm 1.4	$&	2.3	&	$\chi^2$	\\
	&	\nustar/FPMA	&	60465001002	&	2019-05-21	&$	59.2 \pm 1.5	$&	32.2	&		\\
	&	\nustar/FPMB	&	60465001002	&		&$	58.9 \pm 1.5	$&	32	&		\\
NGC~5283	&	\chandra	&	4846	&	2003-11-24	&$	49.9\pm2.4	$&	8.9	&	$\chi^2$	\\
	&	\nustar/FPMA	&	60465006002	&	2018-11-17	&$	86.8 \pm 1.7	$&	32.7	&		\\
	&	\nustar/FPMB	&	60465006002	&		&$	83.1 \pm 1.7	$&	32.9	&		\\
UM~146	&	\nustar/FPMA	&	60465002002	&	2019-02-05	&$	36.6 \pm 1.2	$&	28.5	&	$\chi^2$	\\
	&	\nustar/FPMB	&	60465002002	&		&$	35.1 \pm 1.3	$&	28.7	&		\\
NGC~5674	&	\swift/XRT	&	80672	&	2014-07-10	&$	56.1 \pm 3	$&	6.3	&	$\chi^2$	\\
	&	\nustar/FPMA	&	60061337002	&	2014-07-10	&$	211.9 \pm 3.3	$&	20.4	&		\\
	&	\nustar/FPMB	&	60061337002	&		&$	201.5 \pm 3.3	$&	20.5	&		\\
NGC~1068	&	\nustar/FPMA	&	60002030002	&	2012-12-18	&$	214 \pm 2	$&	57.9	&	$\chi^2$	\\
	&	\nustar/FPMB	&	60002030002	&		&$	197 \pm 2	$&	57.8	&		\\
\enddata
\tablecomments{$\ast$: count rates are reported in the full bands analyzed for each instrument. $\dagger$: we analyze the flux-resolved spectra of this source. The corresponding count rates are reported in \cite{Kammoun19n4395}. }
\end{deluxetable*}

\startlongtable
\begin{deluxetable*}{lllllllllll}
\tablecaption{The best-fit parameters obtained by fitting the `Pexmon' model. \label{table:pexmon}}

\tablehead{\colhead{Source} &  \colhead{$N_{\rm H,LOS}$} & \colhead{$\Gamma$} & \colhead{$ \rm Norm_{\rm PL}$} & \colhead{$C_{sc}$} & \colhead{$\mathcal{R}$} & \colhead{$kT_1$} & \colhead{$\rm Norm_{\rm Apec,1}$} &  \colhead{$kT_2$} & \colhead{$\rm Norm_{\rm Apec,2}$} & \colhead{$L_{2-10}$}}
\startdata
Mrk 573	&	$2.76^{+0.25}_{-2.20} $	&	$2.11\pm 0.13 $	&	$3.7^{+0.6}_{-2.9}$	&	$0.12^{+0.04}_{-0.09} $	&	$5.0_{-3.2}^{+1.7} $	&	$0.17^{+0.01}_{-0.01} $	&	$1.48^{+0.09}_{-0.21}$	&	$0.88^{+0.04}_{-0.05} $	&	$4.8\pm 0.4$	&	0.52	\\
NGC 1144	&	$0.61\pm 0.04$	&	$1.68\pm 0.09$	&	$41.7^{+7.2}_{-11}$	&	$0.002^{+0.001}_{-0.002}$	&	$0.33_{-0.08}^{+0.06}$	&	$0.32^{+0.06}_{-0.09} $	&	$0.21^{+0.03}_{-0.07}$	&	$1.46^{+0.34}_{-0.27} $	&	$2.43^{+0.76}_{-1.3}$	&	49.6	\\
	&	$1.06^{+0.1}_{-0.2} $	&	$t$	&	$12.4^{+2.3}_{-4.8}$	&	$t$	&	$0.92_{-0.26}^{+0.16}$	&	$t$	&	$t$	&	$t$	&	$t$	&	14.7	\\
NGC 3362	&	$5 \pm 3$	&	$1.8^f$	&	$1.33\pm 0.16$	&	$0.08^f$	&	$-$	&	$0.05^{-0.01}_{-0.02}$	&	 $7.9 \pm 3$	&	$0.79\pm 0.04$	&	$0.8\pm 0.1$	&	0.49	\\
UGC 6100	&	$28_{-11}^{+54}$	&	$1.8^f$	&	$6.32\pm 0.68 $	&	$-$	&	$1^f$	&	$-$	&	$-$	&	$-$	&	$-$	&	4.16	\\
NGC 3982	&	$0.6^{+0.1}_{-0.2}$	&	$1.97^{+0.35}_{-0.20}$	&	$3.3^{+1.1}_{-2.8}$	&	$0.08^{+0.01}_{-0.06}$	&	$3.52_{-0.89}^{+0.45}$	&	$0.36^{+0.03}_{-0.06}$	&	$0.2\pm 0..4$	&	$-$	&	$-$	&	0.03	\\
NGC 4388	&	$0.36^{+0.38}_{-0.43}$	&	$1.51\pm 0.03$	&	$36.7^{+4.7}_{-5.8}$	&	$-$	&	$2.15_{-0.65}^{+0.40}     $	&	$-$	&	$-$	&	$-$	&	$-$	&	3.13	\\
	&	 $0.30\pm 0.13$	&	$t$	&	$97.1^{+7.7}_{-9.2}$	&	$-$	&	$1.09_{-0.14}^{+0.12}$	&	$-$	&	$-$	&	$-$	&	$-$	&	8.28	\\
	&	$0.22\pm 0.08$ 	&	$t$	&	$172^{+12}_{-14}$	&	$-$	&	$0.369_{-0.08}^{+0.03}$	&	$-$	&	$-$	&	$-$	&	$-$	&	14.7	\\
	&	$0.29\pm 0.17$	&	$t$	&	$34.1^{+2.6}_{-3.1}$	&	$-$	&	$1.61_{-0.22}^{+0.18}$	&	$-$	&	$-$	&	$-$	&	$-$	&	2.91	\\
UGC 8621	&	$2.5^{+0.3}_{-1.0}$	&	$1.8^f$	&	 $30.8^{+6.0}_{-6.8}$	&	$0.002^f$	&	$-$	&	$0.28^{+0.03}_{-0.05}$	&	$0.07^{+0.02}_{-0.02}$	&	$-$	&	$-$	&	9.35	\\
NGC 5252	&	$0.15^{+0.02}_{-0.03}$	&	$1.83^{+0.03}_{-0.04}$	&	$49.3^{+3.2}_{-4.5}$	&	$-$	&	$0.14\pm 0.04$ 	&	$0.16^{+0.02}_{-0.01}$	&	$0.46\pm 0.09$	&	$0.91^{+0.07}_{-0.16}$	&	$0.76^{+0.51}_{-0.45}$	&	18.8	\\
	&	$t$	&	$t$	&	$57.2^{+4.2}_{-5.9}$	&	$-$	&	$t$	&	$t$	&	$t$	&	$t$	&	$t$	&	21.9	\\
NGC 5347	&	$4.8^{+2.0}_{-2.6}$	&	$1.91^{+0.08}_{-0.10}$	&	 $1.63^{+0.03}_{-1.01}$	&	$0.089^{+0.031}_{-0.062} $	&	$6.3_{-1.7}^{+3.2}$	&	$0.79\pm 0.07$	&	$0.04\pm0.01$	&	$-$	&	$-$	&	0.06	\\
NGC 5695	&	$2.55^{+0.4}_{-0.5}$ 	&	$1.72^{+0.14}_{-0.26}$	&	$6.62^{+0.74}_{-5.5}$	&	$0.03^{+0.01}_{-0.02} $	&	$0.40_{-0.21}^{+0.07}$	&	$0.08^{-0.02}_{-0.06}$	&	$27.7_{-14}^{+15}$	&	$-$	&	$-$	&	1.12	\\
NGC 5929	&	$0.16\pm 0.01$	&	$1.8\pm 0.06$	&	$9.3^{+1.1}_{-1.5}$	&	$0.07^{+0.01}_{-0.01} $	&	 $0.44_{-0.17}^{+0.13}$	&	$0.52^{+0.16}_{-0.21}$ 	&	$0.18^{+0.02}_{-0.09}$	&	$-$	&	$-$	&	0.49	\\
	&	$t$	&	$t$	&	$6.5^{+0.8}_{-1.1}$	&	$t$	&	$t$	&	$t$	&	$t$	&	$-$	&	$-$	&		0.34\\
NGC 7674	&	$0.41^{+0.08}_{-0.41}$ 	&	$1.79\pm 0.07$	&	$3.72^{+0.78}_{-1.8}$	&	$0.1^{+0.02}_{-0.04}$	&	$2.07_{-1.4}^{+0.51}$	&	$0.11^{+0.02}_{-0.03} $	&	$5.07^{+0.92}_{-4.7}$	&	$0.78^{+0.04}_{-0.05}$	&	$2.86^{+0.33}_{-0.37}$	&	2.37	\\
	&	$0.66^{+0.2}_{-0.5}$ 	&	$t$	&	$2.51^{+0.43}_{-1.1}$	&	$0.6^{+0.17}_{-0.21}$	&	$4.09_{-1.8}^{+0.82}$ 	&	$t$	&	$t$	&	$t$	&	$t$	&	1.6	\\
NGC 7682	&	$0.54^{+0.1}_{-0.2}$	&	$2.11^{+0.24}_{-0.12}$	&	$5.4^{+1.6}_{-4.1}$	&	$0.02^{+0.004}_{-0.02} $	&	$2.67_{-1.6}^{+0.51}$ 	&	$0.64^{+0.17}_{-0.11}$	&	$0.04^{+0.01}_{-0.02}$	&	$-$	&	$-$	&	0.76	\\
NGC 4395	&	$0.018 \pm 0.003$	&	$1.61^{+0.04}_{-0.03}$	&	$11.92^{+1.0}_{-1.1}$	&	$-$	&	$0.61_{-0.15}^{+0.11} $	&	$0.15^{+0.007}_{-0.01}$	&	$0.57^{+0.06}_{-0.07}$	&	$0.76\pm 0.03$	&	$1.6\pm 0.1$	&	0.01	\\
	&	$t$	&	$t$	&	 $20.5^{+1.4}_{-1.5}$	&	$-$	&	$0.29_{-0.08}^{+0.07}$	&	$t$	&	$t$	&	$t$	&	$t$	&	0.02	\\
	&	$t$	&	$t$	&	$29.9\pm 2.1$	&	$-$	&	 $0.32_{-0.13}^{+0.09}$	&	$t$	&	$t$	&	$t$	&	$t$	&	0.03	\\
	&	$0.06\pm 0.01$ 	&	$t$	&	$14.1^{+1.0}_{-1.2}$	&	$-$	&	$0.71_{-0.19}^{+0.15}$	&	$t$	&	$t$	&	$t$	&	$t$	&	0.02	\\
	&	$t$	&	$t$	&	$27.8\pm 1.9 $	&	$-$	&	$0.25_{-0.10}^{+0.09}$ 	&	$t$	&	$t$	&	$t$	&	$t$	&	0.03	\\
	&	$0.35^{+0.06}_{-0.08}$ 	&	$1.66\pm 0.02 $	&	$11.01^{+0.9}_{-1.1}$	&	$-$	&	$0.68_{-0.17}^{+0.13}$	&	$t$	&	$t$	&	$t$	&	$t$	&	0.01	\\
	&	$t$	&	$t$	&	$17.3\pm 1.2$	&	$-$	&	$0.46_{-0.10}^{+0.08}$	&	$t$	&	$t$	&	$t$	&	$t$	&	0.02	\\
	&	$t$	&	$t$	&	$24.8^{+1.5}_{-1.7}$	&	$-$	&	$0.28_{-0.08}^{+0.07}$	&	$t$	&	$t$	&	$t$	&	$t$	&	0.03	\\
Mrk 334	&	$0.15\pm 0.02$	&	$1.62\pm 0.08$	&	$6.11^{+0.85}_{-1.2}$	&	$-$	&	 $0.91_{-0.72}^{+0.07}$	&	$1.38_{-0.38}^{+1.5 } $	&	$0.38_{-0.2}^{+0.4}$	&	$-$	&	$-$	&	2.95	\\
	&	$t$	&	$t$	&	$3.1_{-1}^{+1.5} $	&	$-$	&	$t$	&	$t$	&	$t$	&	$-$	&	$-$	&	2.95	\\
NGC 5283	&	$0.11\pm 0.09$	&	$1.83 \pm 0.05 $	&	$13.8^{+1.3}_{-1.6}$	&	$0.01^{+0.004}_{-0.005}$	&	$0.71_{-0.21}^{+0.16}$	&	$0.80^{+0.22}_{-0.14}$	&	$0.08^{+0.01}_{-0.04}$	&	$-$	&	$-$	&	1.09	\\
	&	$t$	&	$t$	&	$8.74^{+0.93}_{-1.2}$	&	$t$	&	$t$	&	$t$	&	$t$	&	$t$	&	$t$	&	0.69	\\
UM 146	&	$0.09^{+0.02}_{-0.04}$ 	&	$1.95\pm 0.12 $	&	$6.5^{+1.1}_{-1.9}$	&	$-$	&	$1.16_{-0.52}^{+0.31}$ 	&	$-$	&	$-$	&	$-$	&	$-$	&	1.19	\\
NGC 5674	&	$0.054^{+0.04}_{-0.05}$	&	$1.89\pm 0.05$	&	$39.0^{+3.0}_{-3.6}$	&	$0.01^{+0.005}_{-0.01} $	&	$0.35_{-0.13}^{+0.10}$	&	$-$	&	$-$	&	$-$	&	$-$	&	23.2	\\
NGC 1068	&	$316^{+146}_{-129}$	&	$1.74^{+0.06}_{-0.05}$	&	$3960^{+930}_{-3200}$	&	$-$	&	$0.02 \pm 0.01$	&	$7.49\pm 0.41$ 	&	$20.9\pm 0.6$	&	$-$	&	$-$	&	47.7	\\
\enddata
\tablecomments{$t$: tied. $f$: fixed. The column density is in units of $10^{24}~\rm cm^{-2}$. $\rm Norm_{PL}/\rm Norm_{Apec,1}$ and $\rm Norm_{Apec,2}$ are in units of $10^{-4}$ and $10^{-5}\rm ~ photon~s^{-1}~keV^{-1}~cm^{-2}$. The luminosity is in unit of $10^{42}~\rm erg~s^{-1}$.}
\end{deluxetable*}

\startlongtable
\begin{deluxetable*}{lllllllll}
\tablecaption{The best-fit parameters obtained by fitting the `MYTC' model. We used eq.~\ref{eq:NHangle} to estimate the $N_{\rm H,LOS}$ values. \label{table:mytc}}
\tablehead{\colhead{Source} &  \colhead{$N_{\rm H,eq}$} &\colhead{$\theta$} & \colhead{$N_{\rm H,LOS}$}& \colhead{$\Gamma$} & \colhead{$ \rm Norm_{\rm PL}$} & \colhead{$C_{sc}$} & \colhead{$A$} & \colhead{$L_{2-10}$}}
\startdata
Mrk 573	&	$3.29^{+0.25}_{-2.3}$	&	$68.4^{+3.5}_{-6.9}$	&	2.23	&	$2.28^{+0.11}_{-0.04}$	&	$2.65^{+2.4}_{-2.3}$	&	$0.27^{+0.08}_{-0.24}$	&	$69^{+20}_{-60}$	&	0.29	\\
NGC 1144	&	$4.13^{+0.97}_{-1.2}$	&	$60.67^{+0.10}_{-0.23}$	&	0.83	&	$1.83^{+0.1}_{-0.08}$	&	$87^{+16}_{-22}$	&	$9.1^{+3.3}_{-7.8}\times 10^{-4}$	&	1$^f$	&	52.30	\\
	&	$7.2^{+2.3}_{-1.4}$	&	$t$	&	1.44	&	$t$	&	$49^{+10}_{-22}$	&	$t$	&	1$^f$	&	29.50	\\
NGC 3982	&	$0.98^{+0.17}_{-0.50}$	&	$69.9^{+3.4}_{-9.7} $	&	0.71	&	$2.22^{+0.17}_{-0.061}$	&	$10.9^{+1.3}_{-6.8}$	&	$0.025^{+0.008}_{-0.01} $	&	1$^f$	&	0.06	\\
NGC 4388	&	$0.52^{+0.08}_{-0.11}$	&	$66.0^{+1.3}_{-2.6}$	&	0.30	&	$1.516\pm 0.03$	&	$35.7^{+5.7}_{-8.4}$	&	$-$	&	$3.63^{+0.67}_{-1.5}$	&	3.01	\\
	&	$0.51^{+0.08}_{-0.09}$	&	$t$	&	0.30	&	$t$	&	$118.9^{+9.5}_{-12}$	&	$-$	&	 $1.42^{+0.17}_{-0.24}$	&	10.00	\\
	&	$0.39^{+0.05}_{-0.07}$	&	$t$	&	0.22	&	$t$	&	$204^{+14}_{-17}$	&	$-$	&	$0.47^{+0.06}_{-0.08}$	&	17.20	\\
	&	$0.46^{+0.06}_{-0.07}$	&	$t$	&	0.27	&	$t$	&	$33.7^{+3.4}_{-4.2}$	&	$-$	&	$3.02^{+0.38}_{-0.56}$	&	2.85	\\
NGC 5347	&	$7.8^{+1.9}_{-1.0}$	&	$68.8^{+3.5}_{-5.3}$	&	5.39	&	$2.20\pm 0.1$	&	$98^{+13}_{-52}$	&	$0.001\pm 0.001$	&	1$^f$	&	2.49	\\
NGC 5695	&	$2.94^{+0.45}_{-0.70}$	&	$77.8^{+4.5}_{-3.2}$	&	2.66	&	$1.86^{+0.21}_{-0.26}$	&	$36.5^{+4.8}_{-33}$	&	$0.008^{+0.001}_{-0.007}$	&	1$^f$	&	4.97	\\
NGC 5929	&	$0.52^{+0.01}_{-0.32}$	&	 $63.74^{+0.50}_{-3.5}$	&	0.24	&	$1.72^{+0.06}_{-0.08}$ 	&	$8.9^{+1}_{-1.7}$	&	$0.076^{+0.009}_{-0.01}$	&	1$^f$	&	0.53	\\
	&	$t$	&	$t$	&	$t$	&	$t$	&	$6.36^{+0.8}_{-1.3}$	&	$t$	&	$t$	&	0.38	\\
NGC 7674	&	$4.8^{+1.6}_{-2.8}$	&	$60.34^{+0.07}_{-0.19}$	&	0.69	&	$1.94^{+0.10}_{-0.079}$	&	$4.2^{+0.1}_{-1.8}$	&	$0.09^{+0.02}_{-0.04}$	&	$11.1^{+2.4}_{-5.0}$	&	2.15	\\
	&	$5.7^{+1.1}_{-2.5}$	&	$t$	&	0.82	&	$t$	&	$4.98^{+0.84}_{-2.4}$	&	$0.34^{+0.1}_{-0.17}$	&	$t$	&	2.40	\\
NGC 7682	&	$2.9^{+5.4}_{-2.7}$	&	$62.88^{+0.02}_{-2.8}$	&	1.19	&	$1.89^{+0.19}_{-0.26}$	&	$7.1^{+1.2}_{-5.3}$	&	$0.03^{+0.01}_{-0.02}$	&	1$^f$	&	1.38	\\
Mrk 334	&	$2.32^{+0.66}_{-2.4}$	&	$63.7^{-1.5}_{-3.7}$	&	1.08	&	$1.59^{+0.11}_{-0.14}$	&	$6.8^{+1.1}_{-2.5}$	&	$-$	&	1$^f$	&	3.45	\\
	&	$t$	&	$t$	&	$t$	&	$t$	&	$3.37_{-1.1}^{+2.4}$	&	$-$	&	1$^f$	&	2.07	\\
NGC 5283	&	$3.71^{+0.68}_{-2.3}$	&	$60.174^{+0.045}_{-0.11}$	&	0.38	&	$1.82\pm 0.01$	&	$15.2^{+3.2}_{-3.2}$	&	$0.01^{+0.003}_{-0.005}$	&	1$^f$	&	1.25	\\
	&	$t$	&	$t$	&	$t$	&	$t$	&	 $9.86\pm 0.59$	&	$t$	&	$t$	&	0.81	\\
UM 146	&	$1.52^{+0.008}_{-1.5}$	&	$65.18^{+0.69}_{-5.2} $	&	0.83	&	$1.78\pm 0.23$	&	$5.6^{+1.1}_{-3.7}$	&	$-$	&	1$^f$	&	1.21	\\
NGC 5674	&	$0.96^{+0.23}_{-0.55}$	&	$60.28^{+0.05}_{-0.15}$	&	0.12	&	$1.86\pm 0.04 $	&	$37.6^{+3.2}_{-4.3}$	&	$0.012^{+0.005}_{-0.01}$	&	1$^f$	&	16.10	\\
NGC 1068	&	$8.60^{+1.1}_{-0.66}$	&	$69.2^{+2.9}_{-3.7}$	&	6.05	&	$1.67\pm 0.07$	&	$172^{+28}_{-86}$	&	$-$	&	$2.52^{+0.16}_{-0.20}$	&	2.31	\\
\enddata
\tablecomments{$t$: tied. $f$: fixed. The column density is in unit of $10^{24}~\rm cm^{-2}$. $\rm Norm_{PL}$ is in units of $10^{-4}\rm ~ photon~s^{-1}~keV^{-1}~cm^{-2}$. The luminosity is in units of $10^{42}~\rm erg~s^{-1}$.}
\end{deluxetable*}

\startlongtable
\begin{deluxetable*}{llllllll}
\tablecaption{The best-fit parameters obtained by fitting the `MYTD-1NH' model. \label{table:mytd1nh}}
\tablehead{\colhead{Source} &  \colhead{$N_{\rm H, LOS}$} & \colhead{$\Gamma$} & \colhead{$ \rm Norm_{\rm PL}$} & \colhead{$C_{sc}$} & \colhead{$A_0$} & \colhead{$A_{90}$} & \colhead{$L_{2-10}$}}
\startdata
Mrk 573	&	$2.83^{+0.30}_{-1.6}$	&	$2.29^{+0.11}_{-0.03}$	&	$3.26^{+0.32}_{-2.8}$	&	$0.23^{+0.074}_{-0.20}$	&	$29^{+8}_{-30}$	&	$A_0$	&	0.35	\\
NGC 1144	&	$0.57\pm 0.04$	&	$1.68^{+0.12}_{-0.11}$	&	$62^{+13}_{-22}$	&	$0.003^{+0.001}_{-0.001}$	&	$0.41^{+0.08}_{-0.12}$	&	$A_0$	&	47.1	\\
	&	$1.21^{+0.15}_{-0.17}$	&	$t$	&	$34.4^{+7.8}_{-18}$	&	$0.68^{+0.15}_{-0.18}$	&	$t$	&	$t$	&	26.1	\\
NGC 3982	&	$5.3\pm 3.3$	&	$2.22^{+0.18}_{-0.057}$	&	$2.2^{-0.66}_{-2.1}$	&	$0.26^{+0.08}_{-0.26}$	&	$39^{+20}_{-39}$	&	$A_0$	&	0.01	\\
NGC 4388	&	$0.32\pm 0.03$	&	$1.50^{+0.03}_{-0.03}$	&	$37.5^{+5.5}_{-7.5}$	&	$-$	&	$1.97^{+0.38}_{-0.75}$	&	$A_0$	&	3.25	\\
	&	$0.29\pm 0.01$	&	$t$	&	$119.9^{+9.3}_{-11}$	&	$-$	&	$0.82^{+0.10}_{-0.12}$	&	$A_0$	&	10.4	\\
	&	$0.22\pm 0.006$	&	$t$	&	$199^{+12}_{-16}$	&	$-$	&	$0.309\pm 0.037$	&	$A_0$	&	17.3	\\
	&	$0.29\pm 0.02$	&	$t$	&	 $36.4^{+3.3}_{-4.4}$	&	$-$	&	$1.60^{+0.19}_{-0.25}$	&	$A_0$	&	3.16	\\
NGC 5347	&	$5.9^{+1.5}_{-2.6}$	&	$2.29^{+0.09}_{-0.04} $	&	$64\pm 13$	&	$0.003^{+0.0006}_{-0.001}$	&	$0.66^{+0.07}_{-0.04} $	&	$0^f$	&	1.40	\\
NGC 5695	&	$2.99^{+0.32}_{-0.78}$	&	$1.79^{+0.16}_{-0.24}$	&	$55.2^{+4.1}_{-52}$	&	$0.007^{+0.001}_{-0.006}$	&	$0.25^{-0.02}_{-0.20}$	&	$A_0$	&	8.36	\\
NGC 5929	&	$0.15\pm 0.01$	&	$1.69\pm 0.06$	&	$8.11^{+0.93}_{-1.3}$	&	$0.08^{+0.01}_{-0.01} $	&	$1^f$	&	$A_0$	&	0.51	\\
	&	$t$	&	$t$	&	$5.8^{+0.70}_{-0.96}$	&	$t$	&	$t$	&	$t$	&	0.36	\\
NGC 7674	&	$0.33^{+0.05}_{-0.08}$	&	$1.81^{+0.11}_{-0.13}$	&	$3.09^{+0.52}_{-1.3}$	&	$0.13^{+0.03}_{-0.05}$	&	$4.9^{+1.2}_{-2.5} $	&	$A_0$	&	1.93	\\
	&	$2.15^{+0.32}_{-0.52}$	&	$t$	&	$4.95^{+0.88}_{-2.9}$	&	$0.56^{+0.15}_{-0.28}$	&	$t$	&	$A_0$	&	3.09	\\
NGC 7682	&	$0.71^{-0.15}_{-0.54}$	&	$1.62^{+0.08}_{-0.21}$	&	$1.79^{+0.25}_{-1.4}$	&	$0.09^{+0.02}_{-0.07}$	&	$1^f$	&	$A_0$	&	0.53	\\
Mrk 334	&	$0.15^{+0.01}_{-0.02}$	&	$1.47^{+0.03}_{-0.06}$	&	$4.85^{+0.36}_{-0.81}$	&	$-$	&	$1^f$	&	$A_0$	&	2.98	\\
	&	$t$	&	$t$	&	$2.23_{-0.9}^{+1.2}$	&	$-$	&	$t$	&	$A_0$	&	1.60	\\
NGC 5283	&	$0.1\pm 0.01$	&	$1.62\pm 0.04$	&	$9.4^{+1.0}_{-1.4}$	&	$0.013^{+0.005}_{-0.006}$	&	$2.13^{+0.59}_{-0.88}$	&	$A_0$	&	1.03	\\
	&	$t$	&	$t$	&	$6.03^{+0.74}_{-1.0}$	&	$t$	&	$t$	&	1$^f$	&	0.66	\\
UM 146	&	$0.12^{+0.04}_{-0.03}$	&	$2.00^{+0.22}_{-0.13}$	&	$9.4^{+3.4}_{-4.7}$	&	$-$	&	$2.02^{+0.56}_{-0.99}$	&	$A_0$	&	1.59	\\
NGC 5674	&	$0.044\pm 0.003$	&	$1.75\pm 0.03$	&	$28.1^{+2.1}_{-2.5}$\	&	$-$	&	$4.4^{+1.2}_{-1.6}$	&	$A_0$	&	14.3	\\
\enddata
\tablecomments{$t$: tied. $f$: fixed. The column density is in unit of $10^{24}~\rm cm^{-2}$. $\rm Norm_{PL}$ is in units of $10^{-4}\rm ~ photon~s^{-1}~keV^{-1}~cm^{-2}$. The luminosity is in units of $10^{42}~\rm erg~s^{-1}$.}
\end{deluxetable*}

\startlongtable
\begin{deluxetable*}{lllllllll}
\tablecaption{The best-fit parameters obtained by fitting the `MYTD-2NH' model. \label{table:mytd2nh}}
\tablehead{\colhead{Source} &  \colhead{$N_{\rm H,LOS}$} &  \colhead{$N_{\rm H,eq}$} & \colhead{$\Gamma$} & \colhead{$ \rm Norm_{\rm PL}$} & \colhead{$C_{sc}$} & \colhead{$A_0$} & \colhead{$A_{90}$} & \colhead{$L_{2-10}$}}
\startdata
Mrk 573	&	$3.9\pm 3.0$	&	$4.0^{+5.8}_{-2.9}$	&	$2.29^{+0.11}_{-0.03}$	&	$3.6^{+1}_{-2.9}$	&	$0.18^{+0.04}_{-0.14}$	&	$23^{+6}_{-20}$	&	$A_0$	&	0.39	\\
NGC 1144	&	$0.63\pm 0.04$	&	$6.3\pm 2.0$ 	&	$1.93^{+0.10}_{-0.08}$	&	$122^{+27}_{-36}$	&	$12.1^{+5.1}_{-6.7}\times 10^{-4}$	&	$0.68^{+0.13}_{-0.19} $	&	$A_0$	&	63.0	\\
	&	$0.95\pm 0.12$	&	$6.3\pm 2.0$ 	&	$1.93^{+0.10}_{-0.08}$	&	$47^{+14}_{-21}$	&	$t$	&	$1.30^{+0.23}_{-0.45}$	&	$A_0$	&	24.3	\\
NGC 3982	&	$4.5\pm 3.1$	&	$5.3\pm 3.3 $	&	$2.22^{+0.18}_{-0.05}$	&	$2.15^{+0.34}_{-2}$	&	$0.22^{+0.08}_{-0.21}$	&	$35^{+10}_{-30}$	&	$A_0$	&	0.01	\\
NGC 4388	&	$0.35^{+0.06}_{-0.08}$	&	 $0.3^{+0.09}_{-0.19}$	&	$1.51\pm 0.03$	&	$42.1^{+7}_{-11}$	&	$-$	&	$2.09^{+0.44}_{-0.80}$	&	$A_0$	&	3.59	\\
	&	$0.30^{+0.02}_{-0.02}$	&	$t$	&	$t$	&	$125^{+10}_{-12}$	&	$-$	&	$0.92^{+0.14}_{-0.30}$	&	$A_0$	&	10.6	\\
	&	$0.22\pm 0.006$	&	$t$	&	$t$	&	$202\pm 14$	&	$-$	&	$0.30^{+0.05}_{-0.11}$	&	$A_0$	&	17.3	\\
	&	$0.29^{+0.04}_{-0.05}$	&	$t$	&	$t$	&	$37.5^{+4.4}_{-5}$	&	$-$	&	 $1.81^{+0.26}_{-0.54}$	&	$A_0$	&	3.19	\\
NGC 5695	&	$2.42^{+0.20}_{-0.68}$	&	$6.2^{+2.7}_{-2.1}$	&	$1.81^{+0.18}_{-0.15}$	&	$30.1^{+6.1}_{-27}$	&	$0.01^{+0.002}_{-0.008}$	&	$0.515^{+0.029}_{-0.45}$	&	$A_0$	&	4.42	\\
NGC 5929	&	$0.14^{+0.01}_{-0.02}$	&	$0.42^{+0.04}_{-0.31}$	&	$1.71^{+0.05}_{-0.08}$	&	$8.08^{+0.89}_{-1.6}$	&	$0.08\pm 0.01$	&	$1^f$	&	$A_0$	&	0.49	\\
	&	$t$	&	$t$	&	$t$	&	$5.81^{+0.65}_{-1.2}$	&	$t$	&	$t$	&	$A_0$	&	0.35	\\
NGC 7674	&	$0.31^{+0.05}_{-0.07}$	&	$5.7^{+1.0}_{-2.3} $	&	$2.09^{+0.09}_{-0.08}$	&	$6.8^{+1.4}_{-2}$	&	$0.07^{+0.01}_{-0.02}$	&	$6.9^{+1.2}_{-2.4}$	&	$A_0$	&	2.77	\\
	&	$0.39^{+0.12}_{-0.16}$	&	$t$	&	$t$	&	 $8.1^{+1.8}_{-3.1}$	&	$0.25^{+0.077}_{-0.10}$	&	$6.9^{+1.2}_{-2.4}$	&	$A_0$	&	3.31	\\
NGC 7682	&	$0.76^{+0.24}_{-0.63}$	&	$0.6^{+0.08}_{-0.46}$ 	&	$1.65^{+0.08}_{-0.24}$	&	$1.59^{+0.29}_{-1.4}$	&	 $0.122^{+0.022}_{-0.11}$	&	 $9.69^{-0.19}_{-8.5}$	&	$A_0$	&	0.45	\\
Mrk 334	&	$0.15^{+0.01}_{-0.02}$	&	$1.09\pm 0.04$	&	$1.47^{+0.03}_{-0.06}$	&	$4.85^{+0.36}_{-0.81}$	&	$-$	&	$1^f$	&	$A_0$	&	2.98	\\
	&	$t$	&	$t$	&	$t$	&	$2.17^{+0.9}_{-0.7}$	&	$-$	&	$1^f$	&	$A_0$	&	1.75	\\
NGC 5283	&	$0.12\pm 0.01$	&	$3.83^{+0.52}_{-2.5}$	&	$1.94\pm 0.07$	&	$19.9^{+2.9}_{-3.5}$	&	$0.01^{+0.003}_{-0.004}$	&	$1.18^{+0.27}_{-0.37}$	&	$A_0$	&	1.32	\\
	&	$t$	&	$t$	&	$t$	&	$12.6^{+1.9}_{-2.5}$	&	$t$	&	$t$	&	$A_0$	&	0.84	\\
UM 146	&	$0.12^{+0.05}_{-0.03}$	&	$4.0^{+5.3}_{-3.9} $	&	$1.99^{+0.23}_{-0.13}$	&	$9.4^{+3.8}_{-4.5}$	&	$-$	&	$2.14^{+0.47}_{-1.1}$	&	$A_0$	&	0.57	\\
NGC 5674	&	$0.051^{+0.004}_{-0.006}$	&	$1.14^{+0.24}_{-0.60}$	&	$1.89^{+0.06}_{-0.05}$	&	$39.8^{4.1}_{-4.9}$	&	$0.012^{+0.004}_{-0.01}$	&	$0.73^{+0.22}_{-0.12}$	&	$A_0$	&	16.2	\\
\enddata
\tablecomments{$t$: tied. $f$: fixed. The column densities are in unit of $10^{24}~\rm cm^{-2}$. $\rm Norm_{PL}$ is in units of $10^{-4}\rm ~ photon~s^{-1}~keV^{-1}~cm^{-2}$. The luminosity is in units of $10^{42}~\rm erg~s^{-1}$.}
\end{deluxetable*}


	\bibliographystyle{aasjournal}
	\bibliography{references}

\end{document}